\documentclass[12pt]{book} 

\usepackage[
a4paper,
vmargin=2.5cm,
hmargin=3cm
]{geometry}

\usepackage{amsmath,amssymb,amsfonts,amsthm}

\usepackage{lmodern}
\usepackage[T1]{fontenc}
\usepackage[utf8]{inputenc}

\usepackage[english]{babel} 
\usepackage[babel, english=american]{csquotes}
\AtBeginEnvironment{quote}{\small}

\usepackage{titlesec}
\usepackage{titletoc}
\titleformat{\chapter}[block]
{\large\bfseries\filcenter}
{\thechapter.}%
{5pt}
{\MakeUppercase}
{}
\titlespacing{\chapter}{0pt}{0pt}{*3}
\titlecontents{chapter}
[0pt]                                               
{}
{\contentsmargin{0pt}\thecontentslabel.\enspace\MakeUppercase}
{\contentsmargin{0pt}\MakeUppercase}                        
{\titlerule*[.7pc]{.}\contentspage}                 

\titleformat{\section}
{\bfseries}
{\thesection.}
{5pt}
{}
\titlecontents{section}
[5pt]                                               
{}
{\contentsmargin{0pt}\thecontentslabel.\enspace}
{\contentsmargin{0pt}}
{\titlerule*[.7pc]{.}\contentspage}

\titleformat{\subsection}
{\normalsize\bfseries}
{\thesubsection.}
{5pt}
{}
\titlecontents{subsection}
[10pt]                                               
{}
{\contentsmargin{0pt}                          
	\thecontentslabel.\enspace}
{\contentsmargin{0pt}}                        
{\titlerule*[.7pc]{.}\contentspage}  

\usepackage{chngcntr}
\counterwithout{footnote}{chapter}

\usepackage{graphicx}
\graphicspath{{imagenes/}} 

\usepackage{geometry}
\usepackage[utf8]{inputenc}
\usepackage{physics}
\usepackage{titlesec}

\newcommand{\MYTITLE}{Fast and accurate calculation of the bootstrap current and radial neoclassical transport in low collisionality stellarator plasmas}
\usepackage{hyperref}
\hypersetup{
	pdfa,
	colorlinks,
	citecolor=blue,
	filecolor=black,%
	linkcolor=blue,
	pdfauthor={Francisco Javier Escoto López},
	pdftitle={\MYTITLE},
	pdfsubject={Doctoral thesis Universidad Carlos III de Madrid. Plasmas and Nuclear Fusion}, 
	pdfkeywords={stellarator; neoclassical transport; bootstrap current}
}
\usepackage{amsmath}
\usepackage{amsfonts}
\usepackage{amssymb}
\usepackage{mathtools}
\usepackage{mleftright}
\usepackage{algorithm}
\usepackage{algpseudocode}

\usepackage{enumitem}

\usepackage{etoolbox}
\preto{\chapter}{}
\preto{\section}{}
\preto{\subsection}{}
\preto{\subsubsection}{}

\usepackage{fancyhdr}

\usepackage{siunitx}
\usepackage{url}
\usepackage[dvipsnames]{xcolor}

\usepackage{graphicx}
\usepackage{caption}
\usepackage{subcaption}

\usepackage{makecell}
\usepackage{float}
\usepackage[section]{placeins} %
\usepackage{enumitem}
\usepackage{xifthen}

\usepackage{tikz}
\usetikzlibrary{plotmarks}
\usepackage{pgfplots}
\pgfplotsset{compat=newest}
\usetikzlibrary{calc,intersections}
\usetikzlibrary{arrows,arrows.meta,positioning,shapes.geometric,shapes}
\usetikzlibrary{decorations.markings, decorations.text}
\usetikzlibrary{matrix}
\usepgfplotslibrary{colorbrewer}
\usepgfplotslibrary{colormaps} 
\usepgfplotslibrary{external}
\usepgfplotslibrary{fillbetween}
[
]
\pgfplotsset{
	every tick label/.append style={font=\small},
	every axis/.append style={
		width = 0.45\textwidth,
		mark size = 2.85pt,
		axis line style = ultra thick,
		legend style={cells={align=left}},
		legend style={draw=none},
		legend style={font=\small},
		legend style={fill=none, text opacity=1}, 
		colormap/jet
	},
	every axis plot/.append style={very thick} 
}

\usepackage{pdfpages}
\usepackage{booktabs}
\usepackage{longtable}
\usepackage{appendix}
\usepackage{float}
\usepackage{slashbox}
\usepackage{subcaption}

\DeclarePairedDelimiterX{\mean}[1]{\langle}{\rangle}{#1}
  
\newcommand{\mmean}[1]{\mean*{\!\mean*{#1}\!}} 
\newcommand{\Mmean}[1]{\mean*{\!\!\mean*{#1}\!\!}}

\newcommand{\VV}{\mathcal{V}}
\newcommand{\Lorentz}{\mathcal{L}}
\newcommand{\MONKES}{{\texttt{MONKES}}}
\newcommand{\DKES}{{\texttt{DKES}}}
\newcommand{\SFINCS}{{\texttt{SFINCS}}}
\newcommand{\ii}{\text{i}}

\newcommand{\vect}[1]{\vb*{#1}}

\newcommand{\Frac}[2]
{
	\left.{#1} \right/ {#2}
}

\newcommand{\MM}{\mathcal{M}}
\newcommand{\CC}{\mathcal{C}}
\newcommand{\DD}{\mathcal{D}}
\newcommand{\Fsmooth}{\mathcal{F}_{\MM}}

\newcommand{\rhostar}[1]{\rho_{#1*}}
\newcommand{\vth}[1]{v_{\text{t}#1}}
\newcommand{\vp}{v_{\parallel}}

\newcommand{\lambdac}{ \lambda_{\text{c}} }
\newcommand{\Bmax}{B_{\text{max}}}
\newcommand{\Bmin}{B_{\text{min}}}
\newcommand{\Er}{\widehat{E}_{r}}
\newcommand{\Epsi}{\widehat{E}_{\psi}}
\newcommand{\Dij}[1]{\widehat{D}_{#1}}
\newcommand{\dij}[1]{\widehat{d}_{#1}}
\newcommand{\epseff}{\epsilon_{\text{eff}}}
\newcommand{\qmarks}[1]{``#1''}
\newcommand{\GammaC}{\Gamma_{\text{c}}}
\newcommand{\GammaAlpha}{\Gamma_{\alpha}}
\newcommand{\psilcfs}{\psi_{\text{lcfs}}}
\newcommand{\Nfp}{N_{\text{fp}}}
\newcommand{\Nfs}{N_{\text{fs}}}
\newcommand{\BpwO}{B_{\text{pwO}}}

\newcommand{\Matrix}[2]
{
	\mleft[
	\begin{array}{#1}
		#2
	\end{array}
	\mright]
}
\newcommand{\Eval}[1]
{
	#1|
}

\newcommand{\vaverage}[2]
{ 
  \mean*{#2}_{\vb*{v},#1}
}

\newcommand{\gyroav}[1]
{
	\mean*{#1}_{\gamma}
}

\newcommand{\orbav}[1]
{
	\mean*{#1}_{\text{o}}
}

\newcommand{\Orbav}[1]
{
	\mean{#1}_{\text{o}}
}

\newcommand{\vmoment}[1]
{ 
\int{#1}\dd[3]{\vb*{v}}
}

\newcommand{\includepwOMagneticField}[3]
{%
	\begin{subfigure}[t]{0.35\textwidth} 
		
		\includegraphics{pwQI_B_#2_#1} 
		\caption{}
		\label{subfig:pwOMagneticField_pow_#1_wa#2pi}
	\end{subfigure}%
}

\newcommand{\includepwOMagneticFieldIsolines}[3]
{%
	\begin{subfigure}[t]{0.39\textwidth} 
		
		\includegraphics{pwQI_B_#2_#1_small_D31}    
		
		\caption{}
		\label{subfig:pwOMagneticField_Isolines_pow_#1_wa#2pi}
	\end{subfigure}%
}

\newcommand{\includepwOMagneticFieldAppendix}[2]
{%
	\begin{subfigure}[t]{0.24\textwidth}  
		\includegraphics{Appendix_pwQI_B_#2_#1.pdf} 
		\caption{}
		\label{subfig:Appendix_pwOMagneticField_pow_#1_wa#2pi}
	\end{subfigure}%
}

\begin{document}
	
	\pagestyle{empty}
	\pagenumbering{roman} 
	
	\begin{titlepage}
		\begin{center}
			\begin{Huge}
				\MYTITLE\\			
				\bigskip
				by\\
				\bigskip
				Francisco Javier Escoto López\\
			\end{Huge}
			\vspace{2cm}	
			\begin{Large}
				A dissertation submitted in partial fulfillment of the requirements for the degree of Doctor of Philosophy in\\
				\bigskip
				Plasmas and Nuclear Fusion\\
				\vspace{2cm}
				Universidad Carlos III de Madrid\\
				\vspace{2cm}
				Advisors:\\
				\bigskip
				Dr. José Luis Velasco Garasa\\
				Prof. Dr. Iván Calvo Rubio\\ 
				\bigskip
				Tutor:\\
				\bigskip
				Prof. Dr. Victor Tribaldos Macía\\
				\bigskip
				July 14, 2025\\
			\end{Large}
			
		\end{center}
	\end{titlepage}
	
%
%
%
%
	
	\vspace*{\fill}
	\begin{center}
		This thesis is distributed under license ``Creative Commons \textbf{Atributtion - Non Commercial - Non Derivatives}''.\\
		\medskip
		\includegraphics[width=4.2cm]{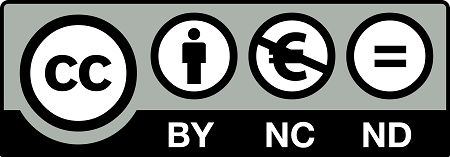} 
	\end{center}

	
	\setcounter{page}{0}
	\newpage
	\begingroup
	\pagestyle{plain}
	
	\begin{flushright} 
		\qmarks{It is open to every man to choose the direction of his striving; and also every man may draw comfort from Lessing's saying, that the search for truth is more precious than its possession.} -- Albert Einstein.
		\vspace{2cm}
		
		\qmarks{Muchas veces me ha pasado eso: luchar incensantemente contra un obstáculo que me impide hacer algo que juzgo necesario o conveniente, aceptar con rabia la derrota y finalmente, un tiempo después, comprobar que el destino tenía razón.} -- Ernesto Sábato
		\vspace{2cm}
		
		\qmarks{La ciencia es una escuela de modestia, de valor intelectual y de tolerancia: muestra que el pensamiento es un proceso, que no hay gran hombre que no se haya equivocado, que no hay dogma que no se haya desmoronado ante el embate de los nuevos hechos.} -- Ernesto Sábato
	\end{flushright}
	\chapter*{Published and submitted content} 
	
	During the development of this thesis the author has published, as first author, the following articles in peer-reviewed journals. For each publication, his name is highlighted in bold and underlined font. Publications [I] and [II] contain the main results of this doctoral work and both are completely included in this dissertation. Chapters \ref{chap:Monoenergetic} and \ref{chap:MONKES} rely heavily on [I] and chapter \ref{chap:MONKES_applications} on [II]. At the beginning of each of these chapters, the connection between sections of the chapter and the corresponding publication will be clearly specified. 
	
	\subsection*{Publications in peer-reviewed journals}
	\begin{enumerate}[label={[\Roman*]}]
		
		\item \textbf{\underline{F. J. Escoto}}, J. L. Velasco, I. Calvo, M. Landreman, and F. I. Parra. \qmarks{MONKES: a fast neoclassical code for the evaluation of monoenergetic transport coefficients in stellarators}. \textit{Nuclear Fusion}, 64(7):076030, 2024.
		\href{https://iopscience.iop.org/article/10.1088/1741-4326/ad3fc9}{URL}.		
		
		\item \textbf{\underline{F. J. Escoto}}, J. L. Velasco, I. Calvo, E. Sánchez.
		Evaluation of neoclassical transport in nearly quasi-isodynamic stellarator magnetic fields using MONKES. \textit{Nuclear Fusion},
		65(3):036017, 2025. \href{https://iopscience.iop.org/article/10.1088/1741-4326/ada7e0}{URL}.
		
	\end{enumerate}

	\chapter*{Other research merits} 
	\subsection*{Publications in peer-reviewed journals}
	Publication [i] corresponds to a work in which the author of this dissertation has been actively involved and his name is highlighted in bold and underlined font. Publications [ii-iv] correspond to experimental campaigns in which the author of this dissertation has been included in alphabetical order as part of the experiment team.  
	\begin{enumerate}[label={[\roman*]}]		
		
		\item J. L. Velasco, I. Calvo, \textbf{\underline{F. J. Escoto}}, E. Sánchez, H. Thienpondt,
		and F.I. Parra. Piecewise omnigenous stellarators. \textit{Phys. Rev. Lett.},
		133:185101, 2024.
		\href{https://journals.aps.org/prl/abstract/10.1103/PhysRevLett.133.185101}{URL}.

		\item O. Grulke et al. \qmarks{Overview of the first Wendelstein 7-X long pulse campaign with fully water-cooled plasma facing components}. \textit{Nuclear Fusion}, 64 (11), 112002, 2024.
		\href{https://iopscience.iop.org/article/10.1088/1741-4326/ad2f4d}{URL}.
		
		\item T. S. Pedersen et al. \qmarks{Experimental confirmation of efficient island divertor operation and successful neoclassical transport optimization in Wendelstein 7-X}. \textit{Nuclear Fusion}, 62 (4), 042022, 2022.
		\href{https://iopscience.iop.org/article/10.1088/1741-4326/ac2cf5}{URL}.
		
		\item C. Hidalgo et al. \qmarks{Overview of the TJ-II stellarator research programme towards model validation in fusion plasmas}. \textit{Nuclear Fusion}, 62 (4), 042025, 2022.
		\href{https://iopscience.iop.org/article/10.1088/1741-4326/ac2ca1}{URL}.
	\end{enumerate}

	\subsection*{Poster contributions}
	\begin{enumerate}[label={[\roman*]}]  
		
		\item  \textbf{\underline{F. J. Escoto}}, J. L. Velasco, I. Calvo and M. Antoine. \qmarks{Towards a fast and accurate calculation of the bootstrap current in low collisionality stellarator plasmas}. Poster contribution to the 23$^\text{rd}$ Internationational Stellarator Heliotron Workshop (ISHW) in Warszaw, June 2022.
		
		\item  \textbf{\underline{F. J. Escoto}}, J. L. Velasco, I. Calvo, M. Landreman and F. I. Parra. \qmarks{Fast evaluation of the bootstrap current in stellarators}. Poster contribution to the 20$^\text{th}$ European Fusion Theory Conference (EFTC) in Padova, October 2023. Awarded with the \qmarks{Prize for the best poster presentation by a young researcher}. \href{https://www.researchgate.net/publication/380932499_Fast_evaluation_of_the_bootstrap_current_in_stellarators}{URL}.

		\item  \textbf{\underline{F. J. Escoto}}, J. L. Velasco, I. Calvo, M. Landreman, and F. I. Parra. \qmarks{MONKES: a fast neoclassical code for the evaluation of monoenergetic transport coefficients in stellarators}. Poster contribution to the Princeton Plasma Physics Laboratory/Simons Foundation
		Graduate Summer School
		July 29-August 2, 2024. \href{https://drive.google.com/file/d/1EtgyvPQZ2gwuaPYspRvJp8dXSqyY44KG/view}{URL}.
	\end{enumerate}

	\subsection*{Invited Talks}
	\begin{enumerate}[label={[\roman*]}]  %
		\item \qmarks{MONKES: a fast neoclassical code for direct optimization of the bootstrap current}. \textbf{\underline{F. J. Escoto}}, J. L. Velasco, I. Calvo, M. Landreman and F. I. Parra. Simons-CIEMAT Joint Meeting on Stellarator Turbulence Optimization, September 2023 Madrid. \href{https://sites.fusion.ciemat.es/stellarator-optimization-madrid-2023/files/2023/09/Escoto_Simons_CIEMAT_Meeting_September_2023.pdf}{URL}.

		\item \qmarks{MONKES: a neoclassical code for fast evaluation of the bootstrap current and stellarator optimization}. \textbf{\underline{F. J. Escoto}}, J. L. Velasco, I. Calvo, M. Landreman, F. I. Parra and E. Sánchez. Joint Varenna-Lausanne International Workshop on Theory of fusion plasmas. Varenna, October 2024. {Link to the list of invited speakers}. \href{https://varenna-lausanne.epfl.ch/Varenna2024/Speakers.pdf}{URL}.
	\end{enumerate}

%
%
	
	
	\hypersetup{linkcolor=black}
	\tableofcontents 
	\listoffigures 
	\listoftables 
	
	\clearpage
	\endgroup  
	
	\pagenumbering{arabic} 

	\pagestyle{fancy}
	\chapter{Introduction}
	\label{sec:Introduction} 
	
	\begin{figure}
		\centering
		\tikzsetnextfilename{Tokamak_sketch}
		\begin{subfigure}[t]{0.45\textwidth}
			\centering		
			\includegraphics{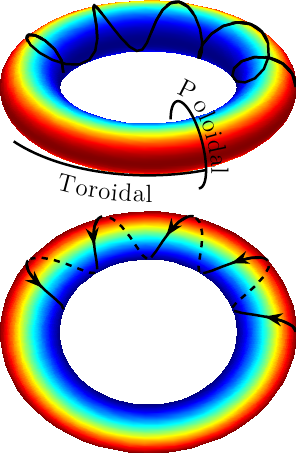}			
			\caption{}
			\label{fig:Tokamak_sketch}
		\end{subfigure}
		\tikzsetnextfilename{Stellarator_sketch}
		\begin{subfigure}[t]{0.45\textwidth}	
			\centering		
			\includegraphics{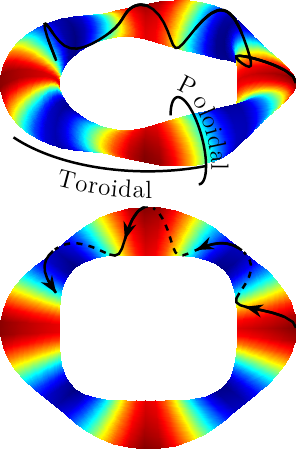}
			\caption{}
			\label{fig:Stellarator_sketch}
		\end{subfigure}
		
		\caption{Sketch of the flux surface shape of (a) a tokamak (b) a $4-$period stellarator. In black lines, the magnetic field $\vb*{B}$ tangent to the flux surface and in colors the magnetic field strength $B$.}
		
		\label{fig:Tokamak_Stellarator_sketch}
	\end{figure}
	
	Nuclear fusion reactions, the combination of two or more nuclei to form new ones, power stars and have produced most of the known elements by stellar nucleosynthesis. In addition, nuclear fusion is a promising mechanism for generating energy on Earth. When the mass of the reactant nuclei exceeds that of the products, energy is liberated due to Einstein's mass-energy equivalence. In order to overcome the electric repulsion between the nuclei, the reactants must possess a sufficiently large kinetic energy (i.e. temperature). Seemingly, the easiest fusion reaction to achieve on Earth is produced between deuterium and tritium, two isotopes of hydrogen, which may be fused to produce helium. For this reaction to take place, the hydrogen isotopes must be at a temperature of around $13.6$ keV, which corresponds to $\num{1.58e8}$ K. At this high temperature the hydrogen no longer behaves as a regular gas but as a \textit{hot plasma}. In order to produce electricity employing nuclear fusion power, the plasma (the fuel for feeding the fusion reactions) must be \textit{confined}.

\begin{figure}
		\tikzsetnextfilename{Larmor_motion}		
		\centering	
		\includegraphics{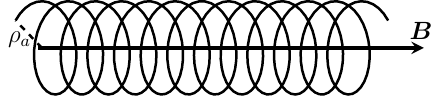}	
%
%
%
%
%
%
%
		\caption{Sketch of the Larmor gyration of charged particles around the magnetic field. In a strongly magnetized plasma  the Larmor radius $\rho_a$ is much smaller than the typical length of variation of the magnetic field $L\sim B/|\nabla \vb*{B}|$.}
		\label{fig:Larmor_motion}
	\end{figure}

	One way of confining the plasma is by employing a magnetic field, in which case one speaks of \textit{magnetic confinement}. The most promising devices for magnetically confining fusion plasmas are \textit{tokamaks} and \textit{stellarators}. As sketched in figure \ref{fig:Tokamak_Stellarator_sketch}, the magnetic field $\vb*{B}$ for both tokamaks and stellarators is such that its lines of force generate \textit{nested} \textit{toroidal} surfaces which are commonly known as \textit{flux surfaces} (a more precise definition is given in section \ref{sec:Force_balance}). Roughly speaking, the key idea is to force charged particles to follow magnetic field lines while they gyrate around them due to Larmor motion (see figure \ref{fig:Larmor_motion}). The outermost flux surface is called \textit{last closed flux surface} and the innermost \qmarks{surface} degenerates to a closed curve, to which $\vb*{B}$ is tangent, known as \textit{magnetic axis}. Thus, particles would ideally be confined in a toroidal volume so that fusion reactions can take place. However, in a toroidal magnetic field, charged particles not only follow its lines of force and gyrate around them. In addition, particles experience a secular \textit{drift} which has an outwards \textit{radial} (i.e. perpendicular to flux surfaces) component. For this reason, in order to confine particles, it is required that magnetic field lines wrap helically around the flux surface. That is, the lines of force of $\vb*{B}$ must rotate both in the toroidal (the long way around the torus) and the poloidal (the short way around the torus) directions as shown in figure \ref{fig:Tokamak_Stellarator_sketch}. This property of the magnetic field is commonly known as \textit{rotational transform}. The rationale behind this is that if particles visit the whole flux surface while following field lines, the radial drift averages out to zero. For this to happen, the rotational transform must be such that a single magnetic field line densely covers the whole flux surface \textit{ergodically} without ever closing itself. Flux surfaces in which the rotational transform has this property are called \textit{ergodic} and those without it are known as \textit{rational} (these denominations will be defined more precisely in section \ref{sec:Force_balance}). Due to the inhomogeneity of the magnetic field, the speed at which electrons and ions circulate along field lines varies spatially. Particles whose motion along field lines does not reverse direction are called \textit{passing} and are well confined by this mechanism. However, for particles which reverse the direction of their motion along field lines, called \textit{trapped} particles, rotational transform by itself does not guarantee that the radial drift averages out to zero. Extra conditions are required for having vanishing orbit-averaged radial drift of all types of particles. 
	
	The main difference between tokamaks and stellarators is that for a tokamak, both the magnetic field strength $B := |\vb*{B}|$ and the shape of the flux surface are \textit{axisymmetric}. This means that, in a tokamak, the shape of the flux surface and the value of $B$ do not depend on the position along the toroidal direction, as sketched in figure \ref{fig:Tokamak_sketch}. Stellarator magnetic fields are \textit{three dimensional}, the flux surface and magnetic field strength $B$ do not necessarily display any obvious continuous symmetry, as shown in the sketch of figure \ref{fig:Stellarator_sketch}\footnote{Stellarators can be composed of several identical sectors, called \textit{field periods}, and thus possess a discrete symmetry. The number of field periods $\Nfp$ induces an $\Nfp-$fold rotation symmetry on the magnetic field. For example, the stellarator flux surface sketched in figure \ref{fig:Stellarator_sketch} is composed of four field periods $\Nfp=4$.}. This difference between tokamaks and stellarators has an important consequence regarding confinement of particles. In particular, the \textit{orbit-averaged} motion of isolated charged particles is qualitatively different in tokamaks and a generic stellarator. The term \qmarks{orbit-average} refers to the time average defined by the motion of charged particles along magnetic field lines. In a tokamak, thanks to axisymmetry, the radial drift of trapped particles averages out to zero. In general, stellarators do not share this property. There is a class of stellarators called \textit{omnigenous} \cite{CaryPhysRevLett,Cary1997OmnigenityAQ} for which, as in axisymmetric fields, trapped particles experience a zero orbit-averaged radial drift (more precise definitions of orbit-average and omnigenity will be given in section \ref{subsec:Guiding_center_motion}). As a matter of fact, axisymmetry is a special case of omnigenity. Therefore, in order to have good confinement properties, it is usual to try to design stellarators to be close to omnigenity.
	
	
	The good confinement that axisymmetry grants comes at the expense of complicating tokamak operation. In a tokamak, due to axisymmetry, the rotational transform of the magnetic field is produced employing an inductive electric current flowing through the plasma. The presence of a large plasma current makes tokamaks susceptible to current instabilities, which can endanger their operation. On the other hand, stellarator magnetic fields can be generated entirely by means of external coils. Thanks to this, stellarators can avoid current-induced instabilities and facilitate steady state operation. In addition, through a process of \textit{optimization}, the coils and flux surface shape can be designed so that stellarator magnetic fields are approximately omnigenous. Thus, in order to be candidates for fusion power plants, stellarators can and must be optimized to be as close to omnigenity as possible. However, in plasmas confined by a toroidal magnetic field, \textit{transport processes} cause, among other things, the loss of particles and energy in the device.
	
	This dissertation is concerned with the theoretical description of a type of transport processes that occur in plasmas confined by a three dimensional stellarator magnetic field. Specifically, the work developed during this thesis focuses on \textit{neoclassical transport} in stellarators. Neoclassical transport is a theoretical description of the transport processes produced by 
	\textit{Coulomb collisions} between charged particles in a plasma confined by a \textit{strong} toroidal magnetic field. The term \qmarks{Coulomb collisions} refers to the binary collisions between charged particles produced by the Coulomb force. What is meant by \qmarks{strong} will be stated more precisely in the next paragraph. It is worth mentioning that, in a magnetically confined plasma, collisions and magnetic geometry are not the only cause of transport processes. Plasma microfluctuations on the scale of the Larmor radius give rise to \textit{turbulent} transport, which can produce losses of energy and particles. Studying this type of transport is out of the scope of this dissertation. 
	
	Magnetically confined plasmas are typically \textit{strongly magnetized}. A species $a$ in a plasma is said to be strongly magnetized when its Larmor radius $\rho_a$, also called \textit{gyroradius}, is much smaller than the typical length scale $L\sim B/|\nabla \vb*{B}|$ in which the magnetic field varies (see figure \ref{fig:Larmor_motion}). In other words, a (species $a$ in a) plasma is said to be strongly magnetized when the normalized gyroradius is small $\rhostar{a}:=\rho_a /L \ll 1$. Equivalently, in a strongly magnetized plasma the frequency of gyration of the Larmor motion $\Omega_a:= e_a B /m_a$, known as \textit{gyrofrequency}, is much larger than the frequency associated to the typical length scale and speed of particles $\omega_a := \vth{a}/L$, i.e. $\omega_a/\Omega_a\ll 1$. Here, $e_a$ and $m_a$ are, respectively, the mass and charge of species $a$, $\vth{a}:=\sqrt{2T_a/m_a}$ its thermal velocity and $T_a$ its temperature in energy units. The equivalency between these two conditions can be checked by estimating the size of the Larmor radius $\rho_a\sim \vth{a}/\Omega_a$, which implies, $\omega_a/\Omega_a\sim\rhostar{a}$. Thus, there are (at least) two different timescales in a strongly magnetized plasma: a fast scale associated to the Larmor gyration and slower scales, maximally defined by $\omega_a$, which correspond to \textit{transport} processes in the plasma. Thanks to this scale separation, it is possible to simplify the theoretical description of transport processes in a strongly magnetized plasma. The general idea is that the fast scale associated to Larmor motion can be averaged out so that the resulting equations describe only the motion of \textit{guiding-centers} on the slower transport scales. In short, a guiding-center is the point around which a magnetized particle would rotate due to Larmor motion if the electric and magnetic field that the particle \qmarks{felt} at a particular position and instant of time were constant and homogeneous (a more precise definition is given in section \ref{sec:Drift_kinetics}). Neoclassical transport can be described by \textit{drift-kinetics} \cite{Hazeltine_1973}, a kinetic theory for guiding-centers. The main result of this theory is the \textit{drift-kinetic equation} (DKE). By solving the DKE, it is possible to calculate the neoclassical {radial} fluxes of particles and energy and the flow velocity of each species. The mismatch between the flow parallel to magnetic field lines of electrons and ions produces a net current in the plasma, called \textit{bootstrap current} \cite{BickertonConnorTaylorBootstrap1971}. The bootstrap current flows parallel to magnetic field lines and is produced by a combination of plasma density and temperature gradients and collisional interaction between charged particles. Due to Ampère's law, this current modifies the magnetic field and its impact on the magnetic configuration can be critical. For instance, if the device is designed to exhaust power from the plasma by means of a divertor relying on a specific structure of magnetic islands at the plasma edge, the effect of the bootstrap current can alter significantly this structure and endanger the divertor's viability.
	
	 The difference in the confinement of \textit{collisionless} particles between omnigenous and non omnigenous magnetic fields has a strong impact on the neoclassical losses of particles and energy. When in a tokamak particles collide with a small collision frequency $\nu$ (reactor-relevant fusion plasmas are weakly collisional close to the magnetic axis) the neoclassical radial losses of particles and heat scale proportionally with the collision frequency $\nu$, giving rise to the so called \qmarks{banana regime}. This regime sets a lower bound on the minimal levels of neoclassical losses achievable in a toroidal device. Therefore, neoclassical losses are not a major concern for tokamaks. On the other hand, for a generic stellarator, the combination of a non zero orbit-averaged radial drift and a small collision frequency ${\nu}$, produces neoclassical losses of particles and energy that scale as $1/\nu$. This stellarator-specific transport regime, known as \qmarks{$1/\nu$ regime}, has a deleterious impact on the confinement and makes a generic stellarator invalid as a candidate for a future fusion reactor. From the viewpoint of neoclassical losses of particles and energy, an exactly omnigenous stellarator would behave as a tokamak, exhibiting a banana regime instead of the $1/\nu$ regime. Thus, omnigenity not only guarantees the confinement of collisionless charged  particles but also reduced losses due to collisional effects. Two subclasses of omnigenous magnetic fields have been historically considered for optimizing stellarators: \textit{quasi-symmetric} (QS) and \textit{quasi-isodynamic} (QI). An attractive feature of QS magnetic fields is that their neoclassical transport properties are isomorphic to those in a tokamak \cite{Pytte_Isomorphic,Boozer_Isomorphic}. For QS magnetic fields, the bootstrap current produced by the plasma can be large. Examples of this subclass are the Helically Symmetric eXperiment (HSX) \cite{HSX}, the design of the National Compact Stellarator Experiment (NCSX) \cite{NCSX} or the Chinese First Quasi-Axisymmetric Stellarator (CFQS) \cite{Shigeyoshi_KINOSHITA2019}. A QI magnetic field is an omnigenous field in which the isolines of $B$ on a flux surface close poloidally. The magnetic field strength $B$ on the flux surface sketched in figure \ref{fig:Stellarator_sketch} corresponds to that of a QI stellarator. The combination of omnigenity with poloidally closed isolines of $B$ on a flux surface grants QI stellarators the additional property of producing zero bootstrap current \cite{Helander_2009}. The Wendelstein 7-X (W7-X) experiment was designed to be relatively close to QI and demonstrates that theory-based stellarator optimization can be applied to construct a device with much better, reactor-relevant, confinement properties than any previous stellarator \cite{Beidler2021}. Moreover, the bootstrap current produced in W7-X plasmas is smaller than in non-optimized machines \cite{Dinklage2018}. However, despite its success, there is still room for improvement. The two main configurations of W7-X, the KJM (or so-called \qmarks{high mirror}) and the EIM (also known as \qmarks{standard}) are not optimized for simultaneously having low levels of radial and parallel neoclassical transport \cite{Beidler_2011, Beidler2021}: while W7-X EIM has small radial transport, it has intolerably large bootstrap current. Conversely, W7-X KJM displays small bootstrap current but larger levels of radial transport. Consequently, optimization of QI stellarators is a very active branch of research and, recently, much effort has been put in pushing forward the design and construction of quasi-isodynamic stellarators \cite{Sanchez_2023,velasco2023robust,RJorge_2022,camachomata_plunk_jorge_2022,Goodman_2023}.

	Roughly speaking, optimizing stellarators consists on varying the magnetic configuration until it meets a given set of desiderata. This is achieved by modifying appropriately the input parameters that determine the equilibrium magnetic field (e.g. the shape of the outermost flux surface). Typically, at each iteration of the optimization process a large number ($\sim$$10^2$) of magnetic configurations are generated. Therefore, in order to neoclassically optimize magnetic fields, it is required to be able to evaluate \textit{fast} the neoclassical properties of each configuration. 
	Ideally, this evaluation should be done \textit{directly}. That is, solving the DKE for each generated configuration and computing the neoclassical transport quantities of interest to be optimized. However, the DKE presented in \cite{Hazeltine_1973} is very complicated to solve and, even simplified versions of it, must be solved \textit{numerically}. At the beginning of this thesis, there was not a code for stellarators which could calculate neoclassical transport within and across flux surfaces sufficiently fast for optimization purposes. A paradigmatic example is the {\DKES} code \cite{DKES1986,VanRij_1989}, which has been the workhorse for neoclassical calculations in stellarators for almost four decades. However, as will be shown in chapter \ref{chap:MONKES}, for reactor-relevant (low) collisionalities, {\DKES} calculations can be very slow. Recent developments allow \textit{direct optimization} of radial neoclassical transport. Based on previous derivations \cite{Calvo_2017,dherbemont2022}, the code {\texttt{KNOSOS}} \cite{KNOSOSJCP,KNOSOSJPP} solves very fast an orbit-averaged DKE that is accurate for low collisionality regimes. {\texttt{KNOSOS}} is included in the stellarator optimization suite \texttt{STELLOPT} \cite{STELLOPT}. However, the orbit-averaged equations solved by {\texttt{KNOSOS}} only describe radial transport at low collisionalities.

	Due to this computational limitation and the requirement of fast neoclassical evaluation, neoclassical properties are typically addressed \textit{indirectly}. Omnigenity \cite{CaryPhysRevLett,Cary1997OmnigenityAQ} imposes several restrictions to the isolines of the magnetic field strength $B$ on a flux surface. In an omnigenous stellarator, the isolines of $B$ must close poloidally, toroidally or helically around the torus. In addition, the values of $B$ at its relative extrema along field lines are also constrained \cite{CaryPhysRevLett, Cary1997OmnigenityAQ, Parra_2015}. These restrictions can be employed for optimizing stellarators indirectly. For instance, one can tailor the variation of the magnetic field strength $B$ on the flux surface so that it nearly fulfils omnigenity: the isolines of $B$ can be forced to close in the desired manner (poloidally, toroidally or helically) and the variance of the extrema of $B$ along field lines can be minimized. A different indirect approach relies on figures of merit, which are easy to calculate, and that vanish in an exactly omnigenous configuration. For the $1/\nu$ regime, the code {\texttt{NEO}} \cite{Nemov1999EvaluationO1} computes the effective ripple $\epseff$, which encapsulates the dependence of radial neoclassical transport on the magnetic configuration. Minimizing $\epseff$ has the effect of shifting the $1/\nu$ regime to smaller values of the collisionality $\hat{\nu}$. 
	
	For neoclassical transport within the flux surface, there exist long mean free path formulae for parallel flow and bootstrap current \cite{Shaing-Callen-1983,Nakajima_1989,helander_parra_newton_2017}. Although they can be computed very fast and capture some qualitative behaviour, these formulae are plagued with noise due to resonances in rational surfaces and, even with smoothing ad hoc techniques, they are not accurate \cite{Landreman_SelfConsistent}. This lack of accuracy limits their application for optimization purposes. During the optimization process, an accurate calculation of the bootstrap current is required to account for its effect (e.g. for optimizing QS stellarators) or to keep it sufficiently small (when optimizing for quasi-isodinamicity). Traditionally, QI stellarators have been neoclasically optimized keeping in mind the constraints to the topology of the isolines of $B$ established in \cite{CaryPhysRevLett,Cary1997OmnigenityAQ}. For instance, one could try to force the isolines of $B$ to close poloidally. Then, one \textit{trusts} that minimizing proxies for general omnigenity while ensuring that most of the isolines of $B$ close poloidally will minimize the bootstrap current. Remarkably, this strategy has proven to be successful in the past for designing QI stellarators with small levels of radial and parallel transport \cite{Sanchez_2023,Goodman_2023}. Despite this ultimate success, simply following this strategy has two main drawbacks. The first one is the imperfect correlation between proxies and the physical quantities that they represent, which may make the process inefficient. Additionally, this strategy precludes the possibility of finding non traditional optimized configurations. In other words, if there exist nearly omnigenous equilibria different from those defined in \cite{CaryPhysRevLett,Cary1997OmnigenityAQ}, they will hardly be found this way.

	Despite its importance, direct optimization of the bootstrap current was not practically feasible with the standard neoclassical codes available at the beginning of this thesis. An accurate calculation of the bootstrap current in reactor-relevant stellarator plasmas was too slow to be included in the optimization process. The only exception are stellarators which are sufficiently close to quasi-symmetry for which semianalytical tokamak formulae \cite{Redl_Bootstrap} are available \cite{Landreman_SelfConsistent}. Hence, the primary goal of this PhD thesis was to provide a numerical tool which allowed, among other things, direct optimization of the bootstrap current in general stellarator geometry. As the general DKE derived in \cite{Hazeltine_1973} is too complicated for expecting fast computations, a simpler but sufficiently accurate DKE corresponding to the \textit{monoenergetic approximation} is the one that will be solved. The monoenergetic approximation consists of a set of assumptions made to simplify the DKE  \cite{DKES1986,Landreman_Monoenergetic}. In particular, the DKE in which this dissertation focuses is the one presented in \cite{DKES1986}, which is solved by the standard neoclassical code {\DKES} \cite{DKES1986,VanRij_1989}. The main result of this thesis is {\MONKES} (MONoenergetic Kinetic Equation Solver), a new neoclassical code conceived to satisfy the necessity of fast and accurate calculations of the bootstrap current for stellarators and in particular for stellarator optimization. Specifically, {\MONKES} makes it possible to compute the monoenergetic coefficients $\Dij{ij}$ where $i,j\in\{1,2,3\}$ (their precise definition is given in section \ref{sec:DKE}). These nine coefficients encapsulate neoclassical transport across and within flux surfaces. The parallel flow of each species can be calculated in terms of the coefficients $\Dij{3j}$ \cite{Taguchi,Sugama-PENTA,Sugama2008,MaasbergMomentumCorrection}. In the absence of externally applied loop voltage, the bootstrap current is driven by the radial electric field and gradients of density and temperature. The so called bootstrap current coefficient $\Dij{31}$ is the one that relates the parallel flow to these gradients. The six remaining coefficients $\Dij{ij}$ for $i\in\{1,2\}$ allow to compute the flux of particles and heat across the flux surface. {\MONKES} also computes fast these radial transport coefficients. Apart from optimization, {\MONKES} can find many other applications. For instance, it can be used for the analysis of experimental discharges or also be included in predictive transport frameworks. Similarly to the code \texttt{KNOSOS}, which is included in the predictive transport frameworks {\texttt{TANGO}} \cite{Banon_Navarro_2023} and \texttt{TRINITY} \cite{Barnes_Trinity_2010}, {\MONKES} could be used for computing the ambipolar radial electric field and neoclassical fluxes of energy in high fidelity simulations. In addition, {\MONKES} fast calculations of the bootstrap current can be used to evolve the magnetic configuration in predictive transport frameworks self-consistently with the ambipolar profile of bootstrap current. 
	
	
	The next chapter introduces some fundamental concepts related to magnetically confined plasmas and neoclassical transport that are required for understanding the work carried out during this thesis. The purpose is to provide the non expert reader with the minimal notions to understand the physical description associated to the DKE that {\MONKES} solves and how it is framed in the \qmarks{big picture} of magnetically confined fusion plasmas. It is important to clarify that all the contents from chapter \ref{chap:Fundamentals} are well-known in the fusion and plasma community and that the works on which it is based are not part of any of the publications of the author of this dissertation. A reader familiar with the area of neoclassical transport in magnetically confined fusion plasmas might want to skip chapter \ref{chap:Fundamentals} and go directly to chapter \ref{chap:Monoenergetic}.

	\chapter{Fundamentals of toroidal plasma confinement and neoclassical transport}\label{chap:Fundamentals} 
	
	In order to describe transport processes in a plasma two different viewpoints can be adopted. A very detailed description of plasma behaviour is given by a \textit{kinetic} treatment. Macroscopically, a plasma can be described as a \textit{fluid}. Theoretical understanding of plasma behaviour rests in appropriately combining these two different viewpoints. Due to the complexity and typical intractability of the general equations corresponding to both approaches, kinetic and fluid equations are simplified to focus on specific plasma processes. In section \ref{sec:Kinetic_fluid_description}, these two perspectives for describing a plasma are briefly introduced in a general manner. Additionally, the fluid equations for a plasma consisting of electrons and singly charged ions will be simplified employing the \textit{single fluid} approximation. In section \ref{sec:Force_balance}, the \textit{force balance} equation (\ref{eq:Ideal_MHD_Force_balance}) required for having a plasma in equilibrium is presented and briefly discussed. In this section, the force balance equation will be derived by simplifying the single fluid equations as in \cite{Freidberg_MHD_1982}. Additionally, how the force balance relation can be derived from kinetic arguments for plasmas with more than two species will be commented. However, its kinetic derivation will not be explained until the next section. This equation sets the minimal requirement that a magnetic field $\vb*{B}$ has to satisfy for confining the plasma. In particular, it dictates how the equilibrium magnetic field and electric current flowing through the plasma must be in order to withstand a finite pressure gradient within the plasma. Incidentally, many basic concepts of toroidal plasma confinement (e.g. flux surfaces) will be defined. Finally, the inadequacy of the force balance equation for describing neoclassical transport will also be illustrated. In section \ref{sec:Drift_kinetics}, \textit{guiding-center motion} and \textit{drift-kinetics}, the kinetic theory of guiding-centers, are briefly introduced. The basic assumptions and orderings will be listed and the two methods for deriving the DKE will be briefly reviewed. In section \ref{subsec:Guiding_center_motion}, the equations for {guiding-center motion} will be presented along with the guiding-center Lagrangian \cite{littlejohn_1983}. In section \ref{subsec:Recursive_DKE}, the general workflow of the recursive procedure introduced in \cite{Hazeltine_1973} for deriving the DKE as an asymptotic expansion in $\rhostar{a}$ will be reviewed. An important and instructive application of the recursive procedure presented in \cite{Hazeltine_1973} is the derivation of the force balance equation (\ref{eq:MHD_Momentum_balance_steady_state_species_a}) from section \ref{sec:Force_balance} employing kinetic arguments. In particular, the force balance equation can be derived as the fluid equation associated to a magnetized plasma close to thermodynamic equilibrium. Thus, from the kinetic point of view, neoclassical transport would arise from small deviations of the plasma from thermodynamic equilibrium. Finally, at the end of section \ref{subsec:Recursive_DKE}, the DKE (\ref{eq:DKE_Hazeltine_Hirshman_non_adiabatic}) that describes neoclassical transport in stellarator plasmas near thermodynamic equilibrium will be presented. 
	
	\section{Kinetic and fluid description of a plasma}\label{sec:Kinetic_fluid_description}
	As mentioned at the beginning of this chapter, the kinetic viewpoint provides a very detailed description of a plasma. The kinetic description of a plasma is given by the Fokker-Planck equation 
	\begin{align} 
		\pdv{F_a}{t}
		+
		\vb*{v}
		\cdot \nabla F_a
		+ 
		\frac{e_a}{m_a}\left( \vb*{E} + \vb*{v}\times \vb*{B} \right)
		\cdot \nabla_{\vb*{v}} F_a
		=
		\sum_{b}
		C_{ab} 
		\left(
		F_a,F_b
		\right)  
		.
		\label{eq:Kinetic_equation_Introduction}
	\end{align} 
	Here, $\vb*{r}$ and $\vb*{v}$ are respectively, the position and velocity of a particle, $t$ is the time, $F_a(\vb*{r},\vb*{v},t)$ is the distribution function for species $a$, $\vb*{E}(\vb*{r},t)$ is the electric field and $C_{ab}\left(F_a,F_b\right) $ is the bilinear Fokker-Planck collision operator between species $a$ and $b$ (its explicit expression and conservation properties are given in appendix \ref{sec:Fokker_Planck_operator}).
	
	A self-consistent evolution of the electromagnetic field and the plasma requires the
	Fokker-Planck equation (\ref{eq:Kinetic_equation_Introduction}) to be accompanied by Maxwell's equations
	\begin{align}
		& \nabla \times \vb*{E} = -\pdv{\vb*{B}}{t}
		, 
		\label{eq:Curl_E}
		\\
		& \nabla \cdot \vb*{B} = 0
		, 
		\label{eq:Divergence_free_B}
		\\
		& \nabla \times \vb*{B} = \mu_0 \vb*{J} +\frac{1}{c^2}\pdv{\vb*{E}}{t}
		,
		\label{eq:Ampere_law}
		\\
		& \nabla \cdot \vb*{E} = \frac{\rho_{\text{c}}}{\varepsilon_0}
		,
		\label{eq:Gauss_law}
	\end{align}
	where $\varepsilon_0$ is the vacuum permittivity, $c$ is the speed of light and $\mu_0=c^{-2} \varepsilon_0^{-1}$ is the vacuum permeability. Here, the electric current $\vb*{J}$ and the charge density $\rho_{\text{c}}$ have been introduced, respectively, in Ampère's and Gauss' laws. These two macroscopic quantities couple Maxwell's equations to the kinetic Fokker-Planck equation. Specifically, they are related to the distribution functions of the different species via
	\begin{align}
		\vb*{J} := \sum_{a} e_a n_a \vb*{V}_a, 
		\label{eq:Electric_current_flow}
		\\
		\rho_{\text{c}}  := \sum_{a} e_a n_a, 
	\end{align}
	where 
	\begin{align}
		n_a(\vb*{r},t) 
		:=
		\int
		F_a(\vb*{r},\vb*{v},t)
		\dd[3]{\vb*{v}}
	\end{align}
	is the particle number density and 
	\begin{align}
		\vb*{V}_a 
		:=
		\vaverage{a}{\vb*{v}}
	\end{align}
	is the flow velocity of species $a$. Here, the notation 
	\begin{align}
		\vaverage{a}{Q}(\vb*{r},t)
		:=
		\frac{1}{n_a(\vb*{r},t)}
		\vmoment{
			Q(\vb*{r},\vb*{v},t) 
			F_a(\vb*{r},\vb*{v},t)
		}
	\end{align} 
	for the macroscopic observable associated to a quantity $Q$ has been employed. Recall that given a vector potential $\vb*{A}$, the magnetic field is written $\vb*{B}=\nabla\times\vb*{A}$ and, in order to satisfy (\ref{eq:Curl_E}), the electric field has to be such that
	\begin{align}
		\vb*{E}= 
		- 
		\nabla \varphi 
		- 
		\pdv{\vb*{A}}{t}
		,
	\end{align}
	where $\varphi$ is the electrostatic potential. 
	
	The kinetic description given by the Fokker-Planck and Maxwell's equations is of little practical interest as it is an extremely complicated set of equations. In part, this is due to the disparate scales described by this model ranging from microscopic to macroscopic. For instance, in order to describe the motion of electrons, solving the characteristics of Fokker-Planck equation (\ref{eq:Kinetic_equation_Introduction}) would require describing the gyromotion of electrons. Numerically integrating the equations of motion for describing the gyromotion would require to take a time step of order $\Delta t\sim\Omega_{\text{e}}^{-1} $. On the other hand, the motion along field lines of electrons would be much slower, taking place in timescales of order $\omega_{\text{e}}^{-1}\gg \Omega_{\text{e}}^{-1}$. Thus, only after at least $\approx  \Omega_{\text{e}}  /\omega_{\text{e}}  \gg 1$ temporal steps of size $\Delta t$ the motion along field lines could be described. This is clearly a computationally expensive and inefficient approach. A less detailed description is given by the fluid perspective. Treating the plasma as a fluid allows to describe its motion in terms of a few macroscopic observables such as the density $n_a$, pressure $p_a$ and flow velocity $\vb*{V}_a $. From the moments $\vmoment{\text{Eq. }(\ref{eq:Kinetic_equation_Introduction})}$, $\vmoment{m_a\vb*{v} \text{ Eq. }(\ref{eq:Kinetic_equation_Introduction})}$ and  $\vmoment{m_a{v}^2 \text{ Eq. }(\ref{eq:Kinetic_equation_Introduction})/2}$ of the kinetic equation (\ref{eq:Kinetic_equation_Introduction}) (and some algebra explained in appendix \ref{sec:Appendix_fluid_equations}), the macroscopic fluid equations corresponding, respectively, to mass, momentum and energy conservation are obtained
	\begin{align}
		& 
		\dv{n_a}{t}
		+
		n_a 		
		\nabla\cdot
		\vb*{V}_a
		=
		0
		\label{eq:Fluid_mass_conservation}
		,
		\\
		& 
		n_a m_a \dv{\vb*{V}_a}{t}  
		+
		\nabla p_a
		+
		\nabla\cdot\vb*{\Pi}_a
		=
		\sum_{b}
		\vb*{F}_{ab}
		+
		e_a n_a\left(
		\vb*{E}
		+
		\vb*{V}_a
		\times
		\vb*{B}
		\right),
		\label{eq:Fluid_momentum_conservation_pressure}
		\\
		&
		\frac{3}{2}
		\dv{p_a}{t}
		+
		\frac{5}{2}
		p_a
		\nabla
		\cdot
		\vb*{V}_a
		+
		\vb*{\Pi}_a
		:
		\nabla \vb*{V}_a
		+
		\nabla \cdot 
		\vb*{h}_a
		=
		\sum_{b}
		W_{ab}
		.
		\label{eq:Fluid_energy_conservation}
	\end{align}
	Here, $v:=|\vb*{v}|$ is the speed, the material derivative $\dv*{t}=\pdv*{t} + \vb*{V}_a\cdot\nabla$ is taken along $\vb*{V}_a$ and the double contraction is defined for two dyads of vectors as $\vb*{a}_1\vb*{a}_2:\vb*{a}_3\vb*{a}_4 := (\vb*{a}_1\cdot\vb*{a}_4) (\vb*{a}_2\cdot\vb*{a}_3)$. In order to precisely define all the quantities in equations (\ref{eq:Fluid_momentum_conservation_pressure}) and (\ref{eq:Fluid_energy_conservation}), the velocity of particles relative to the fluid motion frame has to be introduced
	\begin{align}
		\vb*{w}_a:= \vb*{v} - \vb*{V}_a,
		\label{eq:Velocity_fluid_frame}
	\end{align}
	where $w_a:=|\vb*{w}_a|$ and note that, by definition, $\vaverage{a}{\vb*{w}_a}=0$. In terms of this variable the \textit{scalar pressure} is defined as 
	\begin{align}
		p_a := 
		\frac{1}{3}
		n_a m_a
		\vaverage{a}{ {w}_a^2 } =
		\frac{1}{3} \tr(\mathbb{P}_a)  ,
		\label{eq:Scalar_pressure_definition}
	\end{align}
	which constitutes the \textit{isotropic} piece of the \textit{pressure tensor}
	\begin{align}
		\mathbb{P}_a := n_a m_a \vaverage{a}{\vb*{w}_a\vb*{w}_a}
		=
		p_a I + \vb*{\Pi}_a
		,
		\label{eq:Pressure_tensor_definition}
	\end{align}
	where $I$ is the identity tensor and the anisotropic piece of $\mathbb{P}_a$ has been denoted by $\vb*{\Pi}_a$. 
	In the momentum conservation equation (\ref{eq:Fluid_momentum_conservation_pressure})	
	\begin{align}
		\vb*{F}_{ab}
		=
		\vmoment{m_a\vb*{v} C_{ab}(F_a,F_b)}
		,
	\end{align}
	is the friction force due to collisions. In the energy conservation equation (\ref{eq:Fluid_energy_conservation})
	\begin{align}
		\vb*{h}_a
		:=
		\frac{1}{2} 
		n_a m_a
		\vaverage{a}{ 
			w_a^2 
			\vb*{w}_a
		}
		,
		\label{eq:Heat_Flux_random_definition}
	\end{align}
	is the heat flux due to random motion and
	\begin{align}
		W_{ab}
		:=
		\frac{1}{2}
		m_a
		\vmoment{w_a^2 C_{ab}(F_a,F_b)}
		,
	\end{align}
	is the collisional exchange of kinetic energy due to random motion. The temperature of each species can be defined as usual from the density and scalar pressure
	\begin{align}
		T_a := p_a / n_a.
	\end{align} 
	Equations (\ref{eq:Fluid_mass_conservation})-(\ref{eq:Fluid_energy_conservation}) can be solved for $(n_a,\vb*{V}_a,p_a)$ when a \textit{closure} of the system is provided. Note that, collisions aside, each moment of equation (\ref{eq:Kinetic_equation_Introduction}) introduces an unknown variable which is a higher-order moment. From the zeroth order moment (\ref{eq:Fluid_mass_conservation}), the flow velocity $\vb*{V}_a$ appears. The first order moment (\ref{eq:Fluid_momentum_conservation_pressure}) provides an equation for $\vb*{V}_a$ but introduces the pressure tensor $\mathbb{P}_a$. Energy conservation (\ref{eq:Fluid_energy_conservation}) provides an equation for the scalar pressure $p_a$ but introduces $\vb*{h}_a$. Moreover, the moments of the collision operator introduce $\vb*{F}_{ab}$ and $W_{ab}$ which are, in principle, unknowns. Hence, for closing the system of equations for each species, it is required to give constitutive relations for $\{\vb*{\Pi}_a, \vb*{F}_{ab}, \vb*{h}_a, W_{ab}\}$. A rigorous closure would require, at least implicitly, to solve in some manner for the distribution function. Thus, the fluid approach does not seem an improvement compared to solving Fokker-Planck and Maxwell's equations. Nevertheless, the fluid description can be simplified. Thanks to the fact that typical fusion plasmas are mostly composed of hydrogen isotopes, a simpler set of approximate equations can be derived. These simplified equations will allow us to obtain, in the next section, an equation for describing plasma equilibrium without the need of solving any kinetic equation. 
	
	When the plasma is composed of electrons and singly charged ions, it is possible to obtain a \textit{single fluid} expression for the momentum equation as the result of two asymptotic limits \cite{Freidberg_MHD_1982}. The resulting momentum equation for ions will be simplified in section \ref{sec:Force_balance} to obtain an equation for the equilibrium magnetic field required for confining the plasma. From the momentum equation for electrons, a generalized Ohm's law for the plasma is obtained. The first asymptotic limit is carried over Maxwell's equations and given by $\varepsilon_0 \rightarrow 0$ while $\mu_0 $ is finite, which is equivalent to taking $c\rightarrow\infty$. Thus, the displacement current $c^{-2} \pdv*{\vb*{E}}{t}$ is neglected in Ampère's law (\ref{eq:Ampere_law}). The consequence of this limit over Gauss law (\ref{eq:Gauss_law}) is called the \qmarks{quasineutrality} approximation
	\begin{align}
		\rho_{\text{c}}
		=
		\sum_{a} n_a e_a \simeq 0.
		\label{eq:Charge_neutral_approximation}
	\end{align}
	
	For a pure plasma composed of singly charged ions and electrons, $e_{\text{i}}=-e_{\text{e}}=e$ and the \qmarks{quasineutrality} approximation (\ref{eq:Charge_neutral_approximation}) implies $n_{\text{i}}\simeq n_{\text{e}}$. Here, the subscripts \qmarks{i} and \qmarks{e} stand, respectively, for ions and electrons. Under these assumptions, the notation $n=n_{\text{i}}= n_{\text{e}}$ and $m = m_{\text{i}} + m_{\text{e}}$ is employed. The second asymptotic limit is neglecting the electrons inertia, i.e. taking $m_{\text{e}} \rightarrow 0$. Neglecting the electrons inertia amounts to say that electrons respond infinitely fast to any change in the plasma. As typically they respond much faster than ions, this limit is a reasonable approximation. In this double asymptotic limit, the equation describing the evolution of the flow velocity of the center of mass 
	\begin{align}
		m\vb*{V} 
		:= 
		m_{\text{i}}\vb*{V}_{\text{i}} 
		+  
		m_{\text{e}}\vb*{V}_{\text{e}}
		\sim 
		m_{\text{i}}\vb*{V}_{\text{i}} .
	\end{align}
	becomes to lowest order in $m_{\text{e}}/m_{\text{i}}$ (further details in section \ref{sec:Appendix_single_fluid}) 
	\begin{align}
		n m 
		\dv{ \vb*{V} }{t}
		+
		\nabla p 
		& =
		\vb*{J}\times\vb*{B}
		-
		\nabla
		\cdot
		\left(
		\vb*{\Pi}_{\text{i}}
		+
		\vb*{\Pi}_{\text{e}}
		\right),
		\label{eq:Single_fluid_momentum_equation}
	\end{align}
	where in equation (\ref{eq:Single_fluid_momentum_equation}) the material derivative of $\dv*{t}=\pdv*{t} + \vb*{V}\cdot\nabla$ is taken along $\vb*{V}$ and $p$ is the total (scalar) pressure in the plasma 
	\begin{align}
		p := \sum_{a} p_a.
		\label{eq:Total_pressure}
	\end{align} 
	On the other hand, the momentum equation for the electrons becomes the generalized Ohm's law
	\begin{align}  
		en
		\left(
		\vb*{E}
		+
		\vb*{V}
		\times
		\vb*{B}
		-
		\eta \vb*{J}
		\right)
		& =   
		\vb*{J}
		\times
		\vb*{B}
		- 
		\nabla p_{\text{e}} 
		-
		\nabla
		\cdot 
		\vb*{\Pi}_{\text{e}}
		+
		\widetilde{\vb*{F}}_{\text{e} \text{i}} .
		\label{eq:Single_fluid_Ohms_law}
	\end{align}
	Here, the friction force $ \vb*{F}_{\text{e} \text{i}} $ has been split in a piece proportional to the plasma current and a deviation $\widetilde{\vb*{F}}_{\text{e} \text{i}}$. Specifically,
	\begin{align}
		\vb*{F}_{\text{e} \text{i}} 
		=
		ne \eta \vb*{J}
		+
		\widetilde{\vb*{F}}_{\text{e} \text{i}} 
		,
	\end{align}
	where $\eta={\nu} m_\text{e}/(n e^2)$ is the plasma resistivity. When the right-hand side of equation (\ref{eq:Single_fluid_Ohms_law}) is neglected, the standard Ohm's law for plasmas is obtained
	\begin{align}   
		\vb*{E}
		+
		\vb*{V}
		\times
		\vb*{B}  
		& = 
		\eta \vb*{J}
		.
		\label{eq:Standard_fluid_Ohms_law}
	\end{align}

	\section{Force balance for plasma confinement}\label{sec:Force_balance}
	In steady state (i.e. $\pdv*{t}=0$), in the absence of center of mass flow (i.e. $\vb*{V}= 0$) and plasma pressure anisotropy (i.e. $\vb*{\Pi}_a= 0$), the single fluid momentum equation (\ref{eq:Single_fluid_momentum_equation}) becomes the \textit{force balance} equation
	\begin{align}
		\vb*{J}  \times\vb*{B} & = \nabla p,
		\label{eq:Ideal_MHD_Force_balance}
	\end{align}
	which is accompanied by Ampère's law (\ref{eq:Ampere_law}) in the limit $\varepsilon_0 \rightarrow 0$. Namely,
	\begin{align}
		\nabla\times\vb*{B} & = \mu_0 \vb*{J},
		\label{eq:Ideal_MHD_Amperes_law}
	\end{align}
	subject to the constraint that $\vb*{B}$ must be divergence-free (\ref{eq:Divergence_free_B}). For time-independent $\vb*{B}$, the electric field is electrostatic
	\begin{align}
		\vb*{E} = - \nabla \varphi,
		\label{eq:Electrostatic_field}
	\end{align}
	in order to satisfy induction equation (\ref{eq:Curl_E}) in steady state.

	Equations (\ref{eq:Ideal_MHD_Force_balance}) and (\ref{eq:Ideal_MHD_Amperes_law}) constitute the \textit{ideal magnetohydrodynamic equilibrium equations} and are the basis for magnetic confinement. An immediate consequence of (\ref{eq:Ideal_MHD_Force_balance}) is that both $\vb*{B}$ and $\vb*{J}$ are tangent to surfaces of constant pressure $p$. Namely,
	\begin{align}
		\vb*{B} \cdot \nabla p = \vb*{J} \cdot \nabla p = 0.
	\end{align}

	\begin{figure}
		\centering
		\tikzsetnextfilename{Nested_flux_surfaces}
		\includegraphics{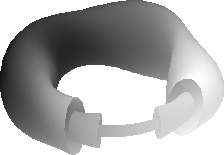}
%
%
%
%
%
%
%
%
%
		\caption{Sketch of a magnetic field with a structure of nested flux surfaces.}
	\end{figure}

\begin{figure}
		\centering
		\tikzsetnextfilename{toroidal-flux}
		\begin{subfigure}[]{0.45\textwidth}
			\includegraphics{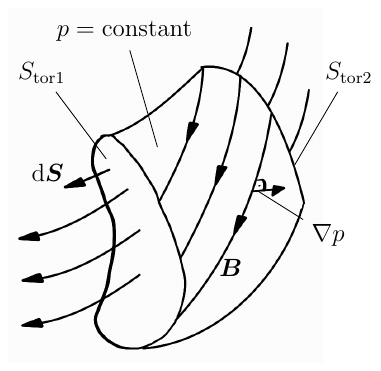}
%
%
%
%
%
%
%
%
			\caption{}
			\label{fig:Toroidal_flux}
		\end{subfigure}
		\tikzsetnextfilename{magnetic-coordinates}
		\begin{subfigure}[]{0.45\textwidth}
			\includegraphics{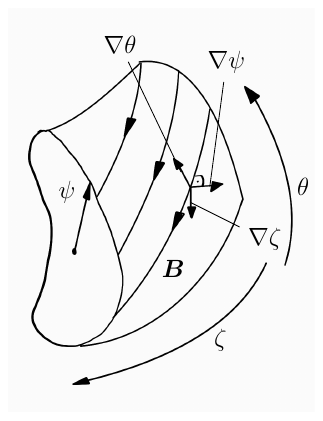}
%
%
%
%
%
%
%
%
			\caption{}
			\label{fig:Magnetic_coordinates}
		\end{subfigure}
		\caption{Sketch of a toroidal section of a flux surface. (a) The flux of $\vb*{B}$ across any toroidal section delimited by a surface of constant pressure $p$ is the same, regardless of the toroidal section. (b) Right-handed set of flux coordinates.}
		\label{fig:Toroidal_section}
	\end{figure}

	In order to magnetically confine a plasma it seems natural to require that all charged particles experience a finite magnetic field. In other words, it is desirable that the magnetic field never vanishes. As by the force balance equation (\ref{eq:Ideal_MHD_Force_balance}) $\vb*{B}$ is tangent to the surfaces of constant $p$, the requirement $\vb*{B}\ne 0$ imposes a strong condition on the topology of the isosurfaces of $p$. Note that, provided that $p$ is smooth and that $\nabla p \ne 0$, the regions of constant $p$ define (at least locally) a 2-dimensional smooth manifold embedded in $\mathbb{R}^3$ \cite{Lee00}, i.e. a smooth surface. Thus, the Poincaré-Hopf theorem \cite{Milnor1965} implies that if $\vb*{B}$ is a non vanishing vector field tangent to surfaces of constant $p$, then, these surfaces must be topologically equivalent to a torus. Therefore, the idyllic scenario for magnetic confinement is that the surfaces of constant plasma pressure $p$ consist on a set of \textit{nested toroidal surfaces}. Then, the innermost \qmarks{surface} is just a closed curve (degenerate torus) called \textit{magnetic axis}. These surfaces of constant pressure are commonly known as \textit{flux surfaces}. Consider the volume $V_{\text{tor}}(p)$ delimited by a surface of constant pressure $p$ and two toroidal sections $S_{\text{tor}1}(p)$ and $S_{\text{tor}2}(p)$ as sketched in figure \ref{fig:Toroidal_flux}. Then, Stokes' theorem and the divergence-free condition (\ref{eq:Divergence_free_B}), imply\footnote{We take the convention of considering the differential surface element vector $\dd{\vb*{S}}$ to be pointing outwards of the enclosed volume (see figure \ref{fig:Toroidal_flux}).}
	\begin{align}
		\int_{V_{\text{tor}}(p)} \nabla\cdot\vb*{B} \dd[3]{\vb*{r}}
		=
		\int_{S_{\text{tor}1}(p)} \vb*{B}\cdot \dd{\vb*{S}}
		+
		\int_{S_{\text{tor}2}(p)} \vb*{B}\cdot \dd{\vb*{S}}
		=
		0
		.
		\label{eq:Stokes_theorem_toroidal_flux}
	\end{align} 
	Equation (\ref{eq:Stokes_theorem_toroidal_flux}) reveals that the magnitude of the magnetic flux across any two toroidal sections of a flux surface is the same. We denote by $2\pi \psi(p)$ to the magnitude of the toroidal flux through any toroidal section $S_{\text{tor}}(p)$. Namely,
	\begin{align}
		2\pi \psi(p) := 
		\left| 
		\int_{S_{\text{tor}}(p)} \vb*{B}\cdot \dd{\vb*{S}}
		\right|
		.
		\label{eq:Toroidal_flux_definition}
	\end{align}	
	Definition (\ref{eq:Toroidal_flux_definition}) matches those of references \cite{RB_White_Toroidally_Confined,Helander_2014} but alternative definitions which allow for negative values of $\psi$ can also be found in the literature.
	
	As $\vb*{B}$ does not have zeros and is assumed to be smooth, the magnetic field does not reverse direction from one flux surface to another. This means that $\psi$ increases monotonically when we move from a flux surface to its outer neighbouring surface. More precisely, $\dv*{\psi}{p}\ne 0$ in the region where $\nabla p\ne 0$, which means that we can use $\psi$ as a \textit{radial coordinate} to write $p=p(\psi)$. The coordinate $\psi$ is known as \textit{flux surface label} and the value of $\psi$ corresponding to the \textit{last closed flux surface} in the plasma region is denoted by $\psilcfs$. The value $\psi=0$ corresponds to the magnetic axis. A function which only depends spatially on $\psi$ is called a \textit{flux function}. Spatial coordinates which employ $\psi$ to parametrize the toroidal plasma region are known as \textit{flux coordinates.} In this dissertation, if any, we will always employ a right-handed set of flux coordinates.
	
	Let $\theta$ and $\zeta$ be, respectively, poloidal and toroidal angles which parametrize the flux surface labelled by $\psi$. For the moment, the only requirement to these angles is that $\theta$ and $\zeta$ increase by $2\pi$ when the torus is traversed, respectively, in the poloidal and toroidal directions. Naturally, it is always possible to define a right-handed set of flux coordinates $(\psi,\theta,\zeta)$, i.e. such that $\nabla\psi\cdot\nabla\theta\times\nabla\zeta > 0$. As represented in figure \ref{fig:Magnetic_coordinates}, definition (\ref{eq:Toroidal_flux_definition}) implies that $\nabla\psi$ always points outwards of the flux surface. Additionally, as it can be observed from figure \ref{fig:Magnetic_coordinates}, it is adopted the convention that $\zeta$ goes in the same direction that $\vb*{B}$, i.e. that $\vb*{B}\cdot\nabla\zeta > 0$. Thus, the direction in which $\theta$ increases can be chosen so that the coordinate system $(\psi,\theta,\zeta)$ is right-handed.

	An almost identical argument to the one given above for the toroidal flux, reveals that the magnitude of the \textit{poloidal flux} enclosed by a flux surface is the same regardless of the poloidal section considered. By poloidal section we mean a \qmarks{ribbon-like} surface connecting the magnetic axis to the flux surface and which can be defined by $\theta=\text{constant}$. Now consider the volume $V_{\text{pol}}(\psi)$ delimited by a surface of constant pressure $p(\psi)$ and two poloidal sections $S_{\text{pol1}}(\psi)$ and $S_{\text{pol2}}(\psi)$ defined, respectively, by $\theta=\theta_1$ and $\theta=\theta_2$ where $\theta_1$ and $\theta_2$ are two distinct fixed values of the poloidal angle. Applying Stokes' theorem and the divergence-free condition (\ref{eq:Divergence_free_B}) yields $\int_{V_{\text{pol}}} \nabla\cdot\vb*{B} \dd[3]{\vb*{r}} =\int_{S_{\text{pol}1}} \vb*{B}\cdot \dd{\vb*{S}} + \int_{S_{\text{pol}2}}\vb*{B}\cdot \dd{\vb*{S}} = 0$. Thus, similarly to the toroidal flux, we denote by $2\pi \chi(\psi)$ the magnetic flux through any poloidal section $S_{\text{pol}}(\psi)$. Namely,
	\begin{align}
		2\pi \chi(\psi) := 
		\int_{S_{\text{pol}}(\psi)} \vb*{B}\cdot \dd{\vb*{S}}
		.
		\label{eq:Poloidal_flux_definition}
	\end{align}
	Note that, unlike the toroidal flux defined by (\ref{eq:Toroidal_flux_definition}), depending on the sign of $\vb*{B}\cdot\nabla\theta$, the poloidal flux $2\pi\chi$ can take positive or negative values.
	
	When the magnetic field $\vb*{B}$ satisfying (\ref{eq:Ideal_MHD_Force_balance}) and (\ref{eq:Ideal_MHD_Amperes_law}) consists of nested flux surfaces it is possible to define the angles $\theta$ and $\zeta$ so that magnetic field lines are represented as straight lines in the $\theta-\zeta$ plane (further details in appendix \ref{sec:Appendix_Magnetic_coordinates}). Flux coordinates in which $\vb*{B}$ can be represented as a straight line are called \textit{magnetic coordinates}. Thus, we can use a (right-handed) coordinate system $(\psi,\theta,\zeta)\in[0,\psilcfs]\times[0,2\pi]\times[0,2\pi/\Nfp]$ in which the contravariant representation of the magnetic field reads	
	\begin{align}
		\vb*{B}
		=
		\nabla\psi
		\times 
		\nabla\theta
		-
		\iota
		\nabla\psi
		\times 
		\nabla\zeta
		=
		\frac{1}{\sqrt{g}}
		\left(
		\vb*{e}_\zeta
		+
		\iota
		\vb*{e}_\theta
		\right)
		,
		\label{eq:Magnetic_Field_contravaraint}
	\end{align}
	where $\sqrt{g}:=(\nabla\psi\cdot\nabla\theta\times\nabla\zeta)^{-1} >0$ is the Jacobian associated to the parametrization given by these spatial coordinates. Here, the \textit{rotational transform}
	\begin{align}
		\iota(\psi) :=
		\dv{\chi}{\psi},
		\label{eq:Rotational_transform_definition}
	\end{align}
	which measures how magnetic field lines wrap around the torus, has been introduced. The quantity $\iota$ is the number of poloidal transits per single toroidal transit of a magnetic field line. The topology of a magnetic field line depends dramatically on $\iota$ being rational or irrational. When $\iota\in\mathbb{Q}$, the rotational transform can always be expressed as $\iota = N/M$, where $M$ and $N$ are coprime integers. Thus, magnetic field lines close themselves after $M$ toroidal transits and are topologically equivalent to a circle. A flux surface in which $\iota\in\mathbb{Q}$ is called a \textit{rational flux surface}. On the other hand, when $\iota\in\mathbb{R}\backslash\mathbb{Q}$ magnetic field lines do not close themselves and a single line of force \textit{densely} fills a flux surface. Due to the fact that, unlike rational numbers, the irrational numbers are not a countable set, the most common situation is to have ergodic surfaces. In such scenario, the topology of magnetic field lines is more exotic, corresponding to an \textit{irrational winding on the torus} and the flux surface is called \textit{ergodic}. As mentioned in chapter \ref{sec:Introduction}, the most favourable situation for confining charged particles is to have a non zero irrational rotational transform. It is a central result from the theory of magnetically confined plasmas that, in order to produce a finite $\iota$, it is required either a toroidal current in the plasma, a non planar magnetic axis or deformed non axisymmetric flux surfaces \cite{Mercier_1964}. Tokamaks produce the rotational transform using a large toroidal plasma current. On the other hand, stellarators produce most of the rotational transform by geometric shaping of the flux surfaces. 
	
	The contravariant representation (\ref{eq:Magnetic_Field_contravaraint}) is not unique, there are infinitely many sets of magnetic coordinates (i.e. flux coordinates in which $\vb*{B}$ is straight). There is a useful set of magnetic coordinates commonly known as \textit{Boozer coordinates} \cite{Boozer_coordinates} in which the covariant representation of the magnetic field is particularly simple. Boozer coordinates are specially convenient for transport calculations and, in what follows, the coordinate system $(\psi,\theta,\zeta)$ will refer to the Boozer coordinate system. In Boozer coordinates, in addition to the simple contravariant representation (\ref{eq:Magnetic_Field_contravaraint}), $\vb*{B}$ can be written as
	\begin{align}
		\vb*{B}
		=
		B_\psi(\psi,\theta,\zeta)
		\nabla\psi
		+
		B_\theta(\psi)
		\nabla\theta
		+
		B_\zeta(\psi)
		\nabla\zeta,
		\label{eq:Magnetic_Field_convaraint}
	\end{align}
	where it can be proven that $2\pi B_\theta/\mu_0$ and $2\pi B_\zeta/\mu_0$ are, respectively, the toroidal and poloidal electric currents enclosed by the flux surface. Dotting the covariant and contravariant representations of $\vb*{B}$ yields that the Jacobian in Boozer coordinates satisfies
	\begin{align}
		\sqrt{g}(\psi,\theta,\zeta)= \frac{B_\zeta(\psi) + \iota(\psi) B_\theta(\psi)}{B^2(\psi,\theta,\zeta)}.
	\end{align}
	
	Another useful set of magnetic coordinates are \textit{Clebsch coordinates} $(\psi,\alpha,l)$. Here, $l$ is the length along magnetic field lines and 
	\begin{align}
		\alpha := \theta- \iota \zeta,
		\label{eq:Clebsch_angle}
	\end{align} 
	is the \textit{Clebsch angle}. Note that in these coordinates the contravariant representation (\ref{eq:Magnetic_Field_contravaraint}) of the magnetic field becomes 
	\begin{align}
		\vb*{B} = \nabla\psi\times\nabla\alpha. 
		\label{eq:Magnetic_field_Clebsch}
	\end{align}	
	Thus, each magnetic field line is determined by a fixed value of $\psi$ and $\alpha$.

		The force balance equation (\ref{eq:Ideal_MHD_Force_balance}) can also be derived in strongly magnetized plasmas employing kinetic arguments. When each species is in \textit{radially local} thermodynamic equilibrium, that is, when the distribution function for each species is close to a Maxwellian at each flux surface
		\begin{align}
			f_{\text{M}a}(\psi, v) :=   n_a(\psi)  \pi^{-3/2}  {v_{\text{t}a}^{-3}(\psi)}  \exp(-\frac{v^2}{v_{\text{t}a}^2(\psi)}),
			\label{eq:Radially_local_Maxwellian}
		\end{align}
		the leading-order force balance relation is obtained (further details in section \ref{sec:Drift_kinetics}) \cite{hazeltine1992plasma,helander_parra_newton_2017}
		\begin{align}
			n_a e_a ( \vb*{E}_0 & + \vb*{V}_a \times \vb*{B} ) = \nabla p_a
			,
			\label{eq:MHD_Momentum_balance_steady_state_species_a}
		\end{align}	
		where the electric field $\vb*{E}_0 =  E_\psi(\psi) \nabla\psi$ is perpendicular to the flux surface. In section \ref{sec:Drift_kinetics}, it will be proven that (\ref{eq:MHD_Momentum_balance_steady_state_species_a}) is the momentum balance equation that, to lowest order in an asymptotic expansion in $\rhostar{a}$, the plasma flow has to satisfy. Note that, in the radially local Maxwellian (\ref{eq:Radially_local_Maxwellian}), the lowest order density $n_a$ and temperature $T_a$ (and therefore $p_a$) are flux functions. Summing (\ref{eq:MHD_Momentum_balance_steady_state_species_a}) over all species and taking into account definition (\ref{eq:Electric_current_flow}) gives the force balance equation (\ref{eq:Ideal_MHD_Force_balance}). The term containing the electric field $\vb*{E}_0$ in (\ref{eq:MHD_Momentum_balance_steady_state_species_a}) is eliminated by employing the \qmarks{quasineutrality} approximation (\ref{eq:Charge_neutral_approximation}). Recall that this approximation comes from the asymptotic limit $c\rightarrow\infty$. Hence, the displacement current is neglected in Ampère's law (\ref{eq:Ampere_law}), which takes the form given by equation (\ref{eq:Ideal_MHD_Amperes_law}). It is important to remark that, although the force balance equation obtained from an asymptotic expansion in $\rhostar{a}$ and by simplifying the single fluid approximation is superficially the same, the kinetic derivation permits considering a plasma consisting of more than two species, as long as they are strongly magnetized. Moreover, the kinetic derivation does not require to neglect the electrons inertia (i.e. does not require an expansion in $m_{\text{e}}/m_{\text{i}}\ll 1$).%

	In order to confine charged particles, the minimal requirement for the magnetic field $\vb*{B}$ is satisfying force balance (\ref{eq:Ideal_MHD_Force_balance}) and Ampère's law (\ref{eq:Ideal_MHD_Amperes_law}) while having a structure of nested flux surfaces with non zero rotational transform. However, the approximations employed to derive force balance are too crude to describe important phenomena in the plasma such as neoclassical transport. Two examples related to plasma flow across and within flux surfaces will be employed to emphasize the necessity and importance of drift-kinetics. For example, from (\ref{eq:MHD_Momentum_balance_steady_state_species_a}) it is immediate to note that $\vb*{V}_a\cdot\nabla\psi =0$, which would mean that there cannot be plasma flow across flux surfaces. Moreover, from force balance (\ref{eq:Ideal_MHD_Force_balance}) and Ampère's law (\ref{eq:Ideal_MHD_Amperes_law}) it is not possible to determine the value of the net parallel current carried by the plasma at each flux surface. Taking the cross product of $\vb*{B}$ with (\ref{eq:Ideal_MHD_Force_balance}) gives the piece of $\vb*{J}$ which is perpendicular to $\vb*{B}$ and the flux surface
	\begin{align}
		\vb*{J}_\perp = \frac{\vb*{B}\times\nabla p}{B^2}.
		\label{eq:Current_diamagnetic}
	\end{align}	
	Ampére's law (\ref{eq:Ideal_MHD_Amperes_law}) reveals that $\vb*{J}$ is divergence-free
	\begin{align}
		\nabla\cdot\vb*{J}=0. 
		\label{eq:Current_divergence_free}
	\end{align} 
	Thus, in principle, it is possible to calculate the piece of $\vb*{J}$ which is parallel to the magnetic field and ensures that (\ref{eq:Current_divergence_free}) is satisfied. Combining equations (\ref{eq:Current_diamagnetic}) and (\ref{eq:Current_divergence_free}) yields a \textit{magnetic differential equation} \cite{Newcomb1959MagneticDE} (further details in appendix \ref{sec:FSA_MDE})
	\begin{align}
		\vb*{B}\cdot\nabla 
		\left(
		\frac{J_\parallel}{B} 
		\right)
		=
		-
		\nabla \cdot \vb*{J}_\perp,
		\label{eq:MDE_Pfirsch_Schluter}
	\end{align}
	which can be solved for $J_\parallel := \vb*{J}\cdot\vb*{B}/B$. However, ${J_\parallel}/{B} $ is defined up to a free function which is constant along magnetic field lines. When the flux surface considered is ergodic, a single magnetic field line densely traces out a flux surface and the free function becomes a flux function\footnote{In rational flux surfaces the free function is not (in general) constant on flux surfaces. Moreover, the differential operator $\vb*{B}\cdot\nabla$ allows for solutions with singularities. For the sake of simplicity in exposition, this complication will be ignored here.}. The portion  $J_\parallel^{\text{PS}}$ of the parallel current which ensures $\nabla\cdot\vb*{J}=0$ is known as \textit{Pfirsch-Schlüter current}. The integration constant is commonly fixed by requiring that the Pfirsch-Schlüter current does not produce a net current over the flux surface, i.e. is fixed by setting 
	\begin{align}
		\mean*{J_\parallel^{\text{PS}} B} =0.
	\end{align}
	The symbol $\mean*{...}$ stands for the \textit{flux surface average} operation. Denoting by $V(\psi)$ the volume enclosed by the flux surface labelled by $\psi$, the flux surface average of a function $f$ can be defined as the limit
	\begin{align}
		\mean*{f}
		:=
		\lim_{\delta \psi \rightarrow 0} 
		\dfrac{\int_{V(\psi+\delta\psi)} f \dd[3]{\vb*{r}}- \int_{V(\psi)} f\dd[3]{\vb*{r}}}
		{V(\psi+\delta\psi) - V(\psi)},
		\label{eq:FSA}
	\end{align}
	where $\dd[3]{\vb*{r}}$ is the spatial volume form. In appendix \ref{sec:FSA_MDE}, two well-known properties of the flux surface average are derived.
	
	Thus, the parallel current is of the form 
	\begin{align}
		J_\parallel B = J_\parallel^{\text{PS}} B + \mean*{\vb*{J}\cdot\vb*{B}},
	\end{align}
	where the net parallel current $\mean*{\vb*{J}\cdot\vb*{B}}$ on the flux surface is arbitrary and cannot be determined solely from (\ref{eq:Ideal_MHD_Force_balance}) and (\ref{eq:Ideal_MHD_Amperes_law}). Neoclassical transport provides a kinetic theory which makes possible to calculate the net parallel current $\mean{\vb*{J}\cdot\vb*{B}}$ and the radial losses of particles and energy by solving the DKE.

	In this dissertation only one particular kind of net parallel current originated by neoclassical mechanisms is considered: the \textit{bootstrap current}. In stellarators, the bootstrap current is produced by a combination of plasma density and temperature gradients and collisional interaction between charged particles. From the fluid perspective, the bootstrap current is the parallel current that arises as a result of deviations from the standard Ohm's law (\ref{eq:Standard_fluid_Ohms_law}) \cite{BickertonConnorTaylorBootstrap1971, Boozer_bootstrap}. In a pure plasma consisting of electrons and singly charged ions it is possible to fix the net parallel current by employing the momentum equation for electrons. For example, the standard Ohm's law (\ref{eq:Standard_fluid_Ohms_law}) implies $\mean*{\vb*{J}\cdot \vb*{B}}= \mean*{\vb*{E}\cdot \vb*{B}}/\eta$. In stellarator transport theory, it is generally (and typically safely) assumed that $\vb*{E}$ is electrostatic (\ref{eq:Electrostatic_field}). It is well known that, for any differentiable (therefore single valued) function on the torus $f$, $\mean*{\vb*{B}\cdot \nabla f}=0$ \cite{Newcomb1959MagneticDE} (further details in appendix \ref{sec:FSA_MDE}). Hence, as the electrostatic potential $\varphi$ has to be differentiable on the flux surface $\mean*{\vb*{E}\cdot \vb*{B}} = \mean*{\vb*{B}\cdot \nabla \varphi}=0$ which implies $\mean*{\vb*{J}\cdot \vb*{B}}=0$. Thus, in a plasma in equilibrium (i.e. satisfying (\ref{eq:Ideal_MHD_Force_balance}) and (\ref{eq:Ideal_MHD_Amperes_law})) in which the standard Ohm's law is satisfied, the net parallel current is zero.	As equation (\ref{eq:Single_fluid_Ohms_law}) reveals, the deviations that cause a net parallel plasma current can originate from plasma pressure anisotropy and/or the portion of the friction force $\widetilde{\vb*{F}}_{\text{ei}}$ which is not proportional to the current.
	
	Assuming a structure of nested flux surfaces, equations (\ref{eq:Ideal_MHD_Force_balance}) and (\ref{eq:Ideal_MHD_Amperes_law}), subject to (\ref{eq:Divergence_free_B}), are solved by several numerical codes to produce stellarator magnetic configurations. The widespread code \texttt{VMEC} \cite{Hirshman_VMEC} employs a variational principle to solve these equations. Recently, a pseudospectral method to solve the magnetohydrodynamic equilibrium equations has been implemented in the \texttt{DESC} code \cite{DESC_equilibrium}. These codes compute the equilibrium magnetic field for prescribed profiles of pressure and currents. Thus, in practice, the calculation of the equilibrium magnetic field is uncoupled from solving the kinetic equation. Of course, each selection of the net parallel current which is undetermined from equations (\ref{eq:Ideal_MHD_Force_balance}) and (\ref{eq:Ideal_MHD_Amperes_law}) yields a different equilibrium magnetic field. Given a set of radial profiles for the plasma pressure and density of each species, there is a magnetic equilibrium which is consistent with the bootstrap current profile. One important application of the code {\MONKES} can be the self-consistent calculation of the equilibrium magnetic field with a bootstrap current profile. Codes for calculating equilibrium magnetic fields are crucial for stellarator \textit{optimization suites}. These suites vary the input parameters (e.g the shape of the last closed flux surface) that determine the magnetohydrodynamic equilibrium in order to find magnetic configurations with better confinement properties. The quality of the configuration is measured by a \textit{cost function} which is made as small as possible (further details in chapter \ref{chap:MONKES_applications}). The code \texttt{VMEC} is included in the optimization suites \texttt{STELLOPT} and \texttt{SIMSOPT} \cite{Landreman2021_SIMSOPT}. In parallel, the equilibrium code \texttt{DESC} has grown to become a stellarator optimization suite by itself \cite{Panici_Conlin_Dudt_Unalmis_Kolemen_2023, Conlin_Dudt_Panici_Kolemen_2023, Dudt_Conlin_Panici_Kolemen_2023}. Another important application of {\MONKES} is including neoclassical transport quantities within the cost function.
	
	\section{Drift-kinetics and neoclassical transport}
	\label{sec:Drift_kinetics}

	In the previous section, the derivation of the force balance equation by simplifying the plasma fluid equations was presented. As a consequence of the approximations that lead to equation (\ref{eq:Ideal_MHD_Force_balance}), plasma transport processes cannot be described by it. In this section, it will be reviewed how the kinetic treatment of a magnetically confined plasma can be simplified to study neoclassical transport phenomena. In particular, the drift-kinetic approximation and the DKE are briefly described. Mainly, there are two asymptotic methods for averaging out the fast Larmor motion and obtaining the DKE. For completeness, both of them will be reviewed. There exists a geometric approach due to Littlejohn \cite{littlejohn_1983, Parra_2011, Calvo_2012, Calvo_2013} relying on the machinery of \textit{phase-space Lagrangian and Hamiltonian} methods to uncouple the fast Larmor motion from the slower timescales. Applying this technique it is possible to obtain the equations which describe how guiding-centers move in the absence of collisions. The theory of \textit{guiding-center motion} provides a reduced dynamical description of the movement of particles by following guiding-centers instead of particles. Following this method, the Vlasov part of the DKE is obtained by employing the guiding-center motion equations as its characteristics. The general workflow of the Lagrangian approach will be illustrated in section \ref{subsec:Guiding_center_motion} for deriving the equations for {guiding-center motion}. Additionally, employing the equations for guiding-center motion will allow us to define more precisely the concept of omnigenity previously introduced in chapter \ref{sec:Introduction}. The second method is to obtain \textit{recursively} the DKE by working directly on the kinetic equation (\ref{eq:Kinetic_equation_Introduction}). This is the pioneer recursive method introduced by Hazeltine in \cite{Hazeltine_1973} and the one described in section \ref{subsec:Recursive_DKE}. In practice, both approaches provide equivalent versions of the DKE. However, employing the recursive approach will allow us to derive the force balance equation (\ref{eq:MHD_Momentum_balance_steady_state_species_a}) from kinetic arguments. After that, the DKE for treating situations near equilibrium will be presented in the coordinates that {\MONKES} employs. As a first step, we will recall and expand some of the orderings and assumptions mentioned at chapter \ref{sec:Introduction}.

	\begin{figure}[h]
		\centering
		\tikzsetnextfilename{guiding-center-coordinates}
		\includegraphics{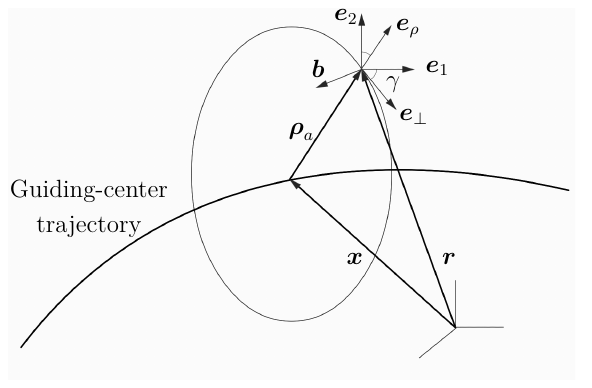}
%
%
%
%
%
%
%
%
%
%
%
%
		\caption{Sketch of the set of coordinates and frames of reference for describing guiding-center motion.}
		\label{fig:Drift_kinetics_coordinates}
	\end{figure}
	Stellarator plasmas are strongly magnetized, which means that there is a fast scale, given by the gyrofrequency $\Omega_a$, associated to the rapid gyration of charged particles around magnetic field lines. The frequency of this fast motion, is typically much larger than the one defined by the slow timescale in which the plasma varies $\omega_a/\Omega_a \ll 1$. Equivalently, the gyroradius $\rho_a$  is much smaller than the typical length scale $L\sim |B/\nabla\vb*{B}|$ in which the magnetic field varies, i.e. $\rhostar{a}=\rho_a/L \sim \omega_a/\Omega_a\ll 1$. Thanks to this separation of scales it is possible to simplify the modelling of transport in strongly magnetized plasmas. The dynamics of isolated magnetized particles can be approximately described employing {guiding-center motion} theory \cite{littlejohn_1983, Cary-Hamiltonian, Parra_2011}.  Neoclassical transport processes in a magnetized plasma can be described using {drift-kinetics}, which provides a kinetic equation for the collisional interaction of guiding-centers. 
	
	In addition to these assumptions, the \textit{drift ordering} (also called \textit{low flow regime}) has to be satisfied by the electric field
	\begin{align}
		\vb*{E}
		\sim 
		\frac{m_a \vth{a}}{e_a } 
		\frac{\vth{a}}{L} 
		\sim 
		\rhostar{a} \vth{a} B.
		\label{eq:Littlejohns_low_flow_ordering}
	\end{align} 
	
	An important consequence of the drift ordering is that, at most, the magnetic field varies \textit{slowly} in time. When the estimate (\ref{eq:Littlejohns_low_flow_ordering}) holds, it follows from the induction equation (\ref{eq:Curl_E}) that
	\begin{align}
		\pdv{\vb*{B}}{t}
		\sim 
		\frac{\vb*{E}}{L}
		\sim 
		\frac{\rhostar{a}\vth{a} B}{L}
		\sim 
		\rhostar{a}^2 \Omega_a B.
		\label{eq:Drift_kinetics_slowly_varying_B}
	\end{align}
	The magnetic field varies in a slower timescale than that in which the plasma varies. Therefore, in neoclassical transport theory, a usual and safe assumption is to consider the magnetic field $\vb*{B}$ to be time-independent. However, as it is not required for deriving the DKE, the assumption $\pdv*{\vb*{A}}{t}=0$ will be applied \textit{after} the DKE is presented. Another important aspect to remark is that, for deriving the DKE the magnetic field does not need to be of any particular shape (e.g. consisting of nested flux surfaces) as long as the orderings given above are satisfied.

	In order to define what is a guiding-center the velocity is represented as	
	\begin{align}
		\vb*{v} = \vp \vb*{b}+ \vb*{v}_\perp,
	\end{align}
	where the unit vector tangent to magnetic field lines is denoted by $\vb*{b}:=\vb*{B}/B$, $\vp:= \vb*{v}\cdot\vb*{b} $ and 
	\begin{align}
		\vb*{v}_\perp 
		= 
		{v}_\perp \vb*{e}_\perp,
	\end{align}
	is the portion of the velocity perpendicular to $\vb*{b}$.
	
	Employing the perpendicular velocity of a particle at a position $\vb*{r}$ and instant of time $t$ it is possible to define its associated {guiding-center}. The first step is to define the vector 
	\begin{align} 
		\vb*{\rho}_a
		:=
		\frac{\vb*{b}\times\vb*{v}_\perp}{\Omega_a} 
		=
		\rho_a \vb*{e}_\rho,
	\end{align}
	where $\vb*{e}_\rho:=\vb*{b}\times\vb*{e}_\perp  $ and $\rho_{a}:=v_\perp/\Omega_a \sim \vth{a}/\Omega_a \sim (\omega_a/\Omega_a) L \ll L$. 
	Note that, for a constant and uniform magnetic field $\vb*{B}$ and in the absence of electric field\footnote{In the presence of a constant and homogeneous electric field, the perpendicular velocity would have an additional term $\vb*{E}\times\vb*{B}/B^2$. It is possible to define the guiding-center aswell in this situation introducing a perpendicular velocity $\vb*{w}_\perp:=\vb*{v}_\perp-\vb*{E}\times\vb*{B}/B^2$. This splitting is convenient for situations in which the term $\vb*{E}\times\vb*{B}/B^2 \sim \vth{a}$ can be very large. However as this is not the case for drift-kinetics, this nuance will be ignored. }, $\vb*{v}_\perp= -\Omega_a \vb*{b}\times\vb*{\rho}_a$ is the rotation velocity of a particle in a non rotating frame centered at the point $\vb*{x}$ around which the particle would gyrate. For non constant and non uniform electromagnetic fields, the vector $\vb*{x}$ determines the position of the guiding-center and, as sketched in figure \ref{fig:Drift_kinetics_coordinates}, it is defined as
	\begin{align}
		\vb*{x}:= \vb*{r} - \vb*{\rho}_a.
	\end{align}
	
	The non inertial orthonormal frame $\{\vb*{e}_\perp, \vb*{e}_\rho, \vb*{b}\}$ is rotating fast due to Larmor motion. It is a convenient and standard practice to use the orthonormal frame $\{\vb*{e}_1, \vb*{e}_2,\vb*{b}\}$ attached to the particle position as sketched in figure \ref{fig:Drift_kinetics_coordinates} but whose axes do not rotate. The rotation angle between the frames $\{\vb*{e}_\perp, \vb*{e}_\rho, \vb*{b}\}$ and $\{\vb*{e}_1, \vb*{e}_2,\vb*{b}\}$ is the \textit{gyroangle}
	\begin{align}
		\gamma:= \atan(\frac{\vb*{v}\cdot\vb*{e}_2}{\vb*{v}\cdot\vb*{e}_1})
		,
	\end{align}
	whose variation in time sets the fast scale. Naturally, these two frames are related to each other via a rotation 
	\begin{align}
		\vb*{e}_\perp
		& = 
		\cos\gamma \vb*{e}_1 
		-
		\sin\gamma \vb*{e}_2
		,
		\\
		\vb*{e}_\rho
		& = 
		\sin\gamma \vb*{e}_1 
		+
		\cos\gamma \vb*{e}_2
		.
	\end{align}
	
	The introduction of the gyroangle as a velocity coordinate is very useful for averaging the fast motion. It is convenient to introduce the \textit{gyroaverage} operation 
	\begin{align}
		\gyroav{f}
		:=
		\frac{1}{2\pi}
		\int_{0}^{2\pi}
		f \dd{\gamma},
	\end{align}
	for a function $f$. 

	\subsection{Guiding-center motion and omnigenity}\label{subsec:Guiding_center_motion} 
	Guiding-center motion can be derived using Hamiltonian perturbation theory in non canonical coordinates \cite{Littlejohn_1982_noncanonical, littlejohn_1983,Cary-Hamiltonian, Parra_2011}. The idea is to employ near identity transformations to find new phase-space coordinates in which the fast gyromotion is uncoupled from the slow motion. In order to benefit from the flexibility of working with a Lagrangian formalism, which does not require employing canonical coordinates, and the conservation of invariants associated to Hamiltonian formulation, the \textit{phase-space Lagrangian} formalism is employed. In the phase-space Lagrangian formalism the Lagrangian is regarded as a function of $(\vb*{r},\vb*{v},\dot{\vb*{r}},\dot{\vb*{v}},t)$ instead of $(\vb*{r},\dot{\vb*{r}},t)$. The exact phase-space Lagrangian for a charged particle is given by
	\begin{align}
		L_a(\vb*{r},\vb*{v},\dot{\vb*{r}},t)
		:=
		\left(
		e_a \vb*{A}(\vb*{r},t)
		+
		m_a 
		\vb*{v}
		\right)
		\cdot
		\dot{\vb*{r}}
		-
		H_a(\vb*{r},\vb*{v},t),
		\label{eq:Charged_particle_PS_Lagrangian}
	\end{align}
	where the Hamiltonian is
	\begin{align}
		H_a(\vb*{r},\vb*{v},t)
		:=
		\frac{m_a |\vb*{v}|^2}{2}
		+
		e_a \varphi(\vb*{r},t).
		\label{eq:Charged_particle_Hamiltonian}
	\end{align}
	The Euler-Lagrange equations associated to the phase-space Lagrangian are
	\begin{align}
		&\dv{t}
		\left(
		\nabla_{\dot{\vb*{r}}}
		L_a
		\right)
		-
		\nabla 
		L_a
		=0, 
		\label{eq:Euler_Lagrange_PS_r}
		\\
		& \dv{t}
		\left(
		\nabla_{\dot{\vb*{v}}}
		L_a
		\right)
		-
		\nabla_{\vb*{v}}
		L_a
		=0
		\label{eq:Euler_Lagrange_PS_v}
		.
	\end{align}
	Of course, when applying Euler-Lagrange equations (\ref{eq:Euler_Lagrange_PS_r}) and (\ref{eq:Euler_Lagrange_PS_v}) to the phase-space Lagrangian (\ref{eq:Charged_particle_PS_Lagrangian}), the equations of motion corresponding to the electric and Lorentz force
	\begin{align}
		\dv{\vb*{v}}{t}
		& =
		\frac{e_a}{m_a}
		\left(
		\vb*{E}
		+
		\vb*{v}
		\times\vb*{B}
		\right)
		,
		\label{eq:Euler_Lagrange_Charged_particle_v}
		\\
		\dv{\vb*{r}}{t}
		& = 
		\vb*{v}
		,
		\label{eq:Euler_Lagrange_Charged_particle_r}
	\end{align}
	are obtained. 
	
	As mentioned above, by employing near identity transformations, it is possible to eliminate order by order the dependence of the Lagrangian on $\gamma$. Thus, as for the order of interest the $\gamma$ coordinate is ignorable, Noether's theorem guarantees the existence of an adiabatic invariant of the movement. Then, the phase-space Lagrangian for guiding-center motion is obtained as the gyroaverage of the exact phase-space Lagrangian for a charged particle. The near identity transformation can be obtained employing an elementary result from analytical mechanics that says that $L_a$ and the modified Lagrangian 
	\begin{align}
		L_a' = L_a + \dv{S}{t},
	\end{align}
	yield the same equations of motion. Here, $S$ can be any differentiable phase-space function commonly called \textit{generating function}. Note that the fluctuating piece of the Lagrangian can always be written as 
	\begin{align}
		L_a' - \gyroav{L_a'} = \Omega_a \pdv{S}{\gamma}
		\sim 
		\dv{S}{t}
		+
		O(\rhostar{a}\Omega_a S).
	\end{align}
	Thus, by appropriately selecting $S$ order by order, the difference between $L_a'$ and $\gyroav{L_a'}$ which is responsible for the difference on their associated equations of motion can be made arbitrarily small in $\rhostar{a}$. In \cite{Parra_2011}, the calculation to second order in $\rhostar{a}$ for electrostatic, but otherwise general, electromagnetic fields is carried out and a recursive method to proceed to arbitrary higher order is provided. However, for high order approximations, the calculations can become prohibitively complicated for hand-made derivations and computationally challenging for computer-based ones \cite{Burby_automation_GCM}. Fortunately, for most practical applications, the calculation to first order is sufficient and this case will be the only one considered.

	Carrying out the procedure presented in \cite{littlejohn_1983} to first order in $\rhostar{a}$ employing as phase-space coordinates $(\vb*{x},\vp,v_\perp,\gamma)$, yields the gyroaveraged Lagrangian 
	\begin{align}
		\gyroav{L_a'}
		(\vb*{x},\vp,v_\perp,\dot{\vb*{x}},\dot{\gamma},t)
		& =
		e_a 
		\vb*{A}^*(\vb*{x},\vp,t)
		\cdot
		\dot{\vb*{x}}
		+
		\frac{m_a}{e_a}
		\frac{m_a v_\perp^2}{2B(\vb*{x},t)} 
		\dot{\gamma}
		\nonumber
		\\
		& -
		\frac{m_a \vp^2}{2}
		-
		\frac{m_a v_\perp^2}{2}
		- 
		e_a \varphi(\vb*{x},t).
	\end{align}
	As a consequence of $\partial{\gyroav{L_a'}}/\partial{\gamma}=0$, the Euler-Lagrange equation associated to the gyroangle $\dv*{t}\partial{\gyroav{L_a'}}/\partial{\dot{\gamma}} - \partial{\gyroav{L_a'}}/\partial{\gamma} = 0$ yields that the \textit{magnetic moment} 
	\begin{align}
		\mu_a &:= \frac{m_a v_\perp^2}{2B}
		,
	\end{align}
	is a constant of the motion described by the gyroaveraged Lagrangian $\gyroav{L_a'}$. 
	
	Due to the (adiabatic) invariance of $\mu_a$, it is natural to replace the coordinate $v_\perp$ in favour of $\mu_a$
	\begin{align}
		L_{a}^{\text{gc}}
		(\vb*{x},\vp,\mu_a,\dot{\vb*{x}},\dot{\gamma},t)
		& =
		e_a 
		\vb*{A}^*(\vb*{x},\vp,t)
		\cdot
		\dot{\vb*{x}}
		+
		\frac{m_a}{e_a}
		\mu_a
		\dot{\gamma}
		-
		H_{a}^{\text{gc}}
		(\vb*{x},\vp,\mu_a,t)
		,
		\label{eq:Littlejohns_Lagrangian}
	\end{align}	
	where the notation $L_{a}^{\text{gc}}:=\gyroav{L_a'}$,  $\vb*{A}^*(\vb*{x},\vp,t) = \vb*{A}(\vb*{x},t) + m_a \vp \vb*{b}(\vb*{x},t)/e_a$ has been employed and
	\begin{align}
		H_{a}^{\text{gc}}
		(\vb*{x},\vp,\mu_a,t)
		:=
		\frac{m_a \vp^2}{2}
		+
		\mu_a B(\vb*{x},t)
		+ 
		e_a \varphi(\vb*{x},t) 
		,
	\end{align}
	is the guiding-center Hamiltonian. 
	
	The Euler-Lagrange equations associated to Littlejohn's Lagrangian (\ref{eq:Littlejohns_Lagrangian}) are
	\begin{align}
		\dot{\vb*{x}}
		&
		=		 
		\frac{
			\vp\vb*{B}^*
			-
			\vb*{b}\times\vb*{E}^*
		}{B_\parallel^*}
		,
		\label{eq:GCM_x}
		\\
		\dot{\vp}
		&
		=
		\frac{e_a}{m_a}
		\frac{\vb*{E}^*\cdot\vb*{B}^*}{B_\parallel^*}
		,
		\label{eq:GCM_v_parallel}
		\\
		\dot{\mu_a}
		&
		=
		0,
		\label{eq:GCM_mu}
		\\
		\dot{\gamma}
		&
		=
		\frac{e_a B}{m_a},
		\label{eq:GCM_gamma}
	\end{align}
	where $\vb*{E}^*:=-\nabla\varphi^* - \pdv*{\vb*{A}^*}{t}=\vb*{E}- (m_a\vp/e_a )\pdv*{\vb*{b}}{t} - \mu_a \nabla B / e_a$, $\varphi^*:=\varphi + \mu_a B / e_a$,  $\vb*{B}^*:=\nabla\times\vb*{A}^* = \vb*{B} + m_a \vp \nabla\times\vb*{b} / e_a$,  $B_\parallel^*:=\vb*{B}^*\cdot\vb*{b} = B + m_a \vp \vb*{b}\cdot\nabla\times\vb*{b} / e_a $. Comparing equations (\ref{eq:GCM_x})-(\ref{eq:GCM_mu}) to (\ref{eq:Euler_Lagrange_Charged_particle_v}) and (\ref{eq:Euler_Lagrange_Charged_particle_r}), the dynamical reduction thanks to the guiding-center approach is apparent. The system of six ordinary differential equations has been reduced to a system of four equations for $\vb*{x}$ and $\vp$ which determine the motion of guiding-centers. Besides, the gyromotion is uncoupled from the motion of the guiding-center. Once the evolution in time of $\vb*{x}$ is determined, equation (\ref{eq:GCM_gamma}) can be integrated to evolve $\gamma$ in time.

	It is instructive to split $\dot{\vb*{x}}$ in its parallel and perpendicular components to the magnetic field. 
	\begin{align}
		\dot{\vb*{x}}
		=
		\vp \vb*{b}
		+
		\dot{\vb*{x}}_\perp,
		\label{eq:GCM_dot_x}
	\end{align}
	where
	\begin{align}
		\dot{\vb*{x}}_\perp
		:=
		\frac{\vb*{E}\times \vb*{b}}{B_\parallel^*}
		+ 
		\vb*{b}
		\times
		\left(
		\frac{\mu_a
			\nabla B
		}{B_\parallel^*}
		+
		\frac{m_a\vp^2}{e_a B_\parallel^*} 
		\vb*{\kappa} 
		+
		\frac{m_a\vp}{e_a B_\parallel^*}
		\pdv{\vb*{b}}{t}
		\right)
		,
		\label{eq:GCM_dot_x_perp}
	\end{align}
	and $\vb*{\kappa}:=\vb*{b}\cdot\nabla\vb*{b} = -\vb*{b}\times\nabla\times\vb*{b}$ is the curvature of magnetic field lines. 
	
	In the derivation of the Lagrangian (\ref{eq:Littlejohns_Lagrangian}) presented in \cite{littlejohn_1983}, it is assumed that the electric field is ordered as (\ref{eq:Littlejohns_low_flow_ordering}), which implies ${\vb*{E}\times \vb*{b}}/{B_\parallel^*} \sim 
	\rhostar{a} \vth{a}$. Hence,
	\begin{align}
		\dot{\vb*{x}}_\perp
		\sim 
		\rhostar{a} \vth{a}
		,
	\end{align}
	and $\dot{\vb*{x}}_\perp$ describes the \textit{slow} drift of the guiding-center across magnetic field lines. 
	
	Due to the fact that $B_\parallel^*\sim B (1+O(\rhostar{a}))$, the perpendicular velocity $\dot{\vb*{x}}_\perp$  (\ref{eq:GCM_dot_x_perp}) is equivalent to first order in $\rhostar{a} \vth{a}$ to $\dot{\vb*{x}}_\perp B_\parallel^*/B$, which is commonly known as the \textit{drift velocity}

	\begin{align}
		\vb*{v}_{\text{d} a}  
		& :=
		\frac{B_\parallel^*}{B}
		\dot{\vb*{x}}_\perp 
		=
		\vb*{v}_{\text{m} a} 
		+
		\vb*{v}_{\vb*{E}\times\vb*{B}} 
		\label{eq:GCM_drift_velocity}
	\end{align}
	where
	\begin{align}
		\vb*{v}_{\text{m} a}  
		:= 
		\frac{1}{\Omega_a}
		\vb*{b}
		\times
		\left(
		\vp^2 
		\vb*{\kappa}
		+
		\frac{\mu_a}{m_a} \nabla B 
		+
		\vp
		\pdv{\vb*{b}}{t} 
		\right),
		\label{eq:Magnetic_drift_GCM}
	\end{align}
	is the \textit{magnetic drift} and
	\begin{align} 
		\vb*{v}_{\vb*{E}\times\vb*{B}}
		:=
		\frac{\vb*{E}\times\vb*{B}}{B^2},
		\label{eq:True_ExB}
	\end{align}
	is the $\vb*{E}\times\vb*{B}$ \textit{drift}. 
	
	It is usual to define the \textit{first order guiding-center velocity} as
		\begin{align}
			\vb*{v}_{\text{gc}a}
			:=
			\vp \vb*{b}
			+
			\vb*{v}_{\text{d}a}
			,
			\label{eq:Guiding_center_velocity}
		\end{align}
		which is equivalent (to this order) to the expression (\ref{eq:GCM_dot_x}) for $\dot{\vb*{x}}$.

	For time-independent magnetic fields, which are typical in neoclassical transport theory (recall the ordering (\ref{eq:Drift_kinetics_slowly_varying_B})), it is convenient to use in addition to $\mu_a$, the total energy
	\begin{align}
		\epsilon_a := \frac{m_a v^2}{2} + e_a \varphi,
	\end{align} 
	as velocity coordinate. Deriving $\epsilon_a$ along the guiding-center trajectories (\ref{eq:GCM_x})-(\ref{eq:GCM_mu}) yields
	\begin{align}
		\dot{\epsilon}_{a}
		=
		\mu_a 
		\pdv{B}{t}
		+
		e_a 
		\pdv{\varphi}{t}
		-
		e_a 
		\pdv{\vb*{A}}{t}
		\cdot\dot{\vb*{x}}.
		\label{eq:GCM_dot_epsilon}
	\end{align}
	Hence, when $\pdv*{\vb*{A}}{t}=\pdv*{\varphi}{t}=0$, guiding-centers move preserving both $\mu_a$ and $\epsilon_a$.
	
	Employing $\mu_a$ and $\epsilon_a$ as velocity coordinates requires to use as well the sign of $\vp$ so that 
	\begin{align}
		\vp(\vb*{x},\mu_a,\epsilon_a,\sigma)
		=
		\sigma
		\sqrt{
			\frac{2}{m_a}
			\left(
			\epsilon_a
			-
			e_a \varphi(\vb*{x})
			-
			\mu_a B(\vb*{x})
			\right)
		},
		\label{eq:Parallel_velocity_GCM_mu_epsilon}
	\end{align}
	where $\sigma:=\vp/|\vp|=\pm 1$ is the sign of the parallel velocity. The simple expression (\ref{eq:Parallel_velocity_GCM_mu_epsilon}) permits to classify the trajectories of guiding centers. In short, if the guiding-center of a particle is such that its total energy satisfies $\epsilon_a > e_a \varphi + \mu_a B $ along its orbit, then the parallel velocity is never zero. On the other hand, if $\epsilon_a$ equals $e_a \varphi + \mu_a B $ at some point, the parallel velocity vanishes and at that point the guiding-center reverses its direction in its motion along field lines.

	\begin{figure}[h]
		\centering
		\tikzsetnextfilename{GCM_PhaseSpace}
		\includegraphics{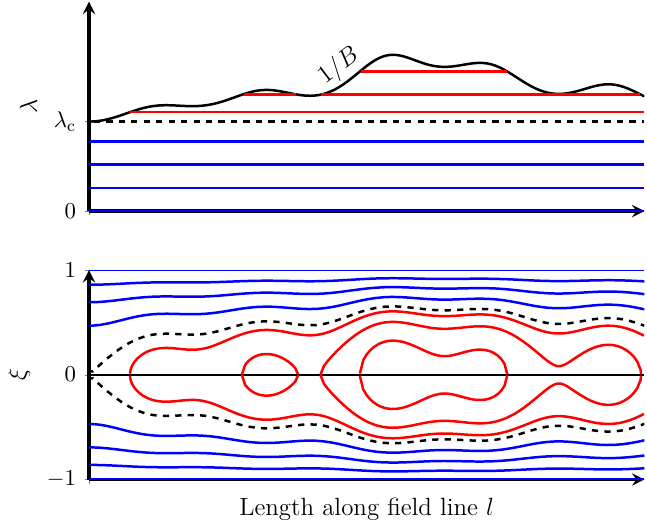}
		\caption{Isolines of $\lambda$ and phase-space classification between passing (blue) and trapped (red) trajectories. For $\lambda < \lambdac$, $\xi$ never vanishes and the trajectory extends along the whole field line. For $\lambda > \lambdac$, the isolines of $\lambda$ become loops in the $l-\xi$ plane. The intersections of these loops with the plane $\xi=0$ define the bounce points.}
		\label{fig:GCM_Phase_Space_sketch}
	\end{figure}

	In order to classify orbits for time-independent magnetic fields, it is convenient to replace the velocity coordinates $(\mu_a,\epsilon_a)$ by $(v,\lambda)$. As before, $v$ is the speed, and 
	\begin{align}
		\lambda(\vb*{x},\mu_a,\epsilon_a) := \frac{\mu_a}{\epsilon_a - e_a\varphi(\vb*{x})},
		\label{eq:GCM_normalized_magnetic_moment}
	\end{align}
	is the \textit{normalized magnetic moment} or also called \textit{pitch-angle coordinate}. Instead of employing an expression for $\vp$ in these coordinates it is more convenient to give an expression for the \textit{pitch-angle cosine} $\xi:=\vp/v\in[-1,1]$. In velocity coordinates $(v,\lambda,\sigma)$, $\xi$ is written as
	\begin{align}
		\xi(\vb*{x},\lambda,\sigma)
		=
		\sigma 
		\sqrt{1 - \lambda B(\vb*{x})
		}.
		\label{eq:Parallel_velocity_GCM_v_lambda}
	\end{align}
	
	A \textit{passing particle} is one for which $\xi$ never vanishes. For a \textit{trapped particle}, the parallel velocity changes its sign at points where $\lambda B=1$. Such points are called \textit{bounce points}. There is a threshold value of $\lambda$, called the \textit{passing-trapped boundary} $\lambdac:= 1/\Bmax$, which allows to distinguish between passing and trapped particles. Here, $\Bmax$ is the maximum value of $B$ on the flux surface. Employing $\lambdac$ orbits can be classified as follows
	\begin{equation} 
		\begin{aligned}
			\text{If }\lambda<\lambdac &\quad \Rightarrow \quad \text{Passing particle}. \\
			\text{If }\lambda>\lambdac &\quad \Rightarrow \quad \text{Trapped particle}.  
		\end{aligned}
		\label{eq:Orbit_classification}
	\end{equation}
	In figure \ref{fig:GCM_Phase_Space_sketch} a sketch of the division between passing and trapped particles is shown. Trajectories of passing particles are shown in blue and those of trapped particles in red. The boundary between passing and trapped particles is plotted with a black dashed line.

	In order to define the position of the bounce points, it is useful to employ {Clebsch coordinates} $(\psi,\alpha,l)$ (defined in section \ref{sec:Force_balance}). Thus, to lowest order in $\rhostar{a}$, magnetized particles move  keeping constant $\psi$ and $\alpha$ (i.e. along field lines.) Therefore, for fixed values of $\psi$, $\alpha$ and $\lambda$, the location of the bounce points is determined implicitly from condition
	\begin{align}
		\lambda B(\psi,\alpha,l_{\text{b}i}) = 1,
	\end{align}
	where $l_{\text{b}i}$ for $i\in\{1,2\}$ are the positions along a field line of the bounce points. 
	
	Note from figure \ref{fig:GCM_Phase_Space_sketch} that, for $\lambda>\lambdac$, the isolines of $\lambda$ can have multiple connected components. Each one of them corresponds to a particle trapped in a different \textit{well} with different bounce points $l_{\text{b}i}$. Denoting by $\Bmax^{\text{r}}$ the value of $B$ at a relative maxima on the flux surface, trapped orbits bifurcate into different wells at $\lambda=1/\Bmax^{\text{r}}$. Similarly to a pendulum, the curves $\xi(\vb*{x},1/\Bmax^{\text{r}},\sigma)$ act as a separatrix in phase-space whose equilibrium point is located at the bounce point where $B=\Bmax^{\text{r}}$. Trapped particles satisfying $\lambda\gtrsim\lambdac$ are called \textit{barely trapped}. Denoting by $\Bmin^{\text{r}}$ to the value of a relative minima of $B$ on the flux surface, those particles with $\lambda \sim 1/\Bmin^{\text{r}}$, are called \textit{deeply trapped}. This classification will be employed in sections \ref{subsec:Monoenergetic_lambda} and \ref{subsec:Contribution_lambda_MONKES}.

	Once passing and trapped particles have been defined, it is possible to make more precise why trapped particles are not always well confined. In addition, by giving an expression for the orbit-averaged radial drift for trapped particles, a better definition of omnigenity than the one given in chapter \ref{sec:Introduction} will be encountered. For time-independent $\vb*{B}$, the drift-velocity (\ref{eq:GCM_drift_velocity}) takes a particularly simple form when employing phase-space coordinates $(\vb*{x},\mu_a,\epsilon_a,\sigma)$	 
	\begin{align}
		\vb*{v}_{\text{d}a}(\vb*{x},\mu_a,\epsilon_a)
		=
		\frac{\vp}{\Omega_a}
		\nabla\times
		\left(\vp \vb*{b}\right)
		,
		\label{eq:Drift_velocity_mu_epsilon}
	\end{align}
	where $\vp$ is regarded as a function of $(\vb*{x},\mu_a,\epsilon_a,\sigma)$. Expression (\ref{eq:Drift_velocity_mu_epsilon}) for the drift velocity allows us to understand better why passing particles are always well confined as long as there is an irrational rotational transform. Conversely, it can be employed to shed light on why trapped particles are not always well confined in stellarators.

	Now, the \textit{orbit-average operation} is introduced for a function $f(\psi,\alpha,l,\mu_a,\epsilon_a,\sigma)$
	\begin{align}
		\orbav{f}
		:=
		\begin{dcases} 
			\frac{\mean*{Bf/\vp}}{\mean*{B/\vp}}
			,
			&
			\text{for passing particles}
			\\
			\sum_{\sigma}
			\frac{1}{t_{\text{b}}}
			\int_{l_{\text{b}1}}^{l_{\text{b}2}}
			\frac{f \dd{l}}{\vp}
			,
			&
			\text{for trapped particles.}
		\end{dcases}
		\label{eq:Orbit_average_definition}
	\end{align}
	where $t_{\text{b}}$ is the \textit{bounce time} which is set by requiring $\orbav{1}=1$. Note that the orbit-average operation corresponds to a time average along the motion parallel to field lines. For trapped particles, the orbit-average is a loop integral in phase-space. On the other hand, for passing particles in an ergodic surface, this movement extends to the whole flux surface. Thus, for passing particles, the orbit-average operation is written in terms of the flux surface average. 
	
	Expression (\ref{eq:Drift_velocity_mu_epsilon}) is useful for calculating the orbit-averaged radial drift that particles experience. For passing particles, the component of the drift velocity perpendicular to flux surfaces averages to zero
	\begin{align}
		\orbav{
			\vb*{v}_{\text{d}a}
			\cdot
			\nabla\psi
		}
		= 
		\frac{m_a}{e_a}
		\frac{\mean*{\nabla\psi \cdot \nabla\times
			\left(\vp \vb*{b}\right)} }{\mean*{B/\vp}}
		=0,
	\end{align}
	where property (\ref{eq:FSA_curl_perp}) has been used. On the other hand, for trapped particles it becomes the expression 
	\begin{align}
		\orbav{
			\vb*{v}_{\text{d}a}
			\cdot
			\nabla\psi
		}
		= 
		\frac{m_a}{e_a t_{\text{b}}}
		\pdv{J}{\alpha}
	\end{align}
	where the \textit{second adiabatic invariant}
	\begin{align}
		J(\psi,\alpha,\mu_a,\epsilon_a)
		:=
		2\int_{l_{\text{b}1}}^{l_{\text{b}2}}
		|\vp|(\psi,\alpha,l,\mu_a,\epsilon_a)
		\dd{l},
	\end{align}
	has been introduced. The reason why it is called adiabatic invariant is because, when orbit-averaged, trapped particles drift radially and precess poloidally preserving $J$. Indeed, denoting by $\Orbav{\dot{\psi}}:= \orbav{\vb*{v}_{\text{d}a}\cdot\nabla\psi}$
	and $\orbav{\dot{\alpha}}:= \orbav{\vb*{v}_{\text{d}a}\cdot\nabla\alpha}$ we have
	\begin{align}
		\Orbav{\dot{\psi}}
		&
		=
		\frac{m_a}{e_a t_{\text{b}}}
		\pdv{J}{\alpha}
		,
		\label{eq:Orbit_averaged_dot_psi}
		\\
		\orbav{\dot{\alpha}}
		,
		&
		=
		-
		\frac{m_a}{e_a t_{\text{b}}}
		\pdv{J}{\psi}
		\label{eq:Orbit_averaged_dot_alpha}
		.
	\end{align}
	Note that in (\ref{eq:Orbit_averaged_dot_psi}) and (\ref{eq:Orbit_averaged_dot_alpha}) $J$ acts as a time-independent Hamiltonian for describing the orbit-averaged motion of trapped particles, and therefore, is conserved. 
	
	The fact that trapped particles secularly drift preserving $J$ allows to give a more precise definition of omnigenity than the one introduced at chapter \ref{sec:Introduction}. A magnetic field is said to be omnigenous if the \textit{second adiabatic invariant does not vary along $\alpha$}. Hence, for an omnigenous magnetic field
	\begin{align}
		\pdv{J}{\alpha} = 0,
		\label{eq:Omnigenity_definition}
	\end{align}  
	wherever $J$ is defined for fixed $\psi$, $\mu_a$ and $\epsilon_a$. Thus, equation (\ref{eq:Orbit_averaged_dot_psi}) implies that if condition (\ref{eq:Omnigenity_definition}) holds, the radial drift of trapped particles averages out to zero. Equivalently, a magnetic field is said to be omnigenous if, for fixed $\psi$, $\mu_a$ and $\epsilon_{a}$, all the connected components of the region of the flux surface in which $J$ is constant close toroidally, poloidally or helically. A more common and less general definition of omnigenity is to define an omnigenous field to one in which \qmarks{$J$ is a flux function}. This definition, however, excludes less constrained omnigenous stellarators like those presented in \cite{Parra_2015}, which have more than one local minimum and maxima. In the magnetic fields presented in \cite{Parra_2015}, $J$ is a flux function \textit{within each well}. In an abuse of terminology, from now on, when it is said that for omnigenous stellarators \qmarks{$J$ is a flux function} it should be understood implicitly that \qmarks{$J$ is a flux function within each well}.

	\subsection{Recursive derivation of the drift-kinetic equation}\label{subsec:Recursive_DKE}
	In section \ref{subsec:Guiding_center_motion}, the  method for obtaining the equations describing guiding-center motion has been illustrated. Now the recursive method for obtaining the DKE presented in \cite{Hazeltine_1973} will be briefly explained. This asymptotic method exploits that in the kinetic equation (\ref{eq:Kinetic_equation_Introduction}), all terms are of order $\rhostar{a}\Omega_a F_a$ with the exception of that associated to Larmor motion. Namely, the largest term in Fokker-Planck equation  (\ref{eq:Kinetic_equation_Introduction}) is
	\begin{align} 
		\frac{e_a}{m_a}
		\vb*{v}
		\times
		\vb*{B} 
		\cdot 
		\nabla_{\vb*{v}}
		F_a
		\sim 
		\Omega_a 
		F_a, 
		\label{eq:Drift_kinetics_Larmor_motion_term_ordering}
	\end{align}
	and the remaining terms are ordered as	
	\begin{align}
		\pdv{F_a}{t}
		\sim 
		\omega_a F_a 
		\sim 
		\rhostar{a}
		\Omega_a F_a
		\label{eq:Drift_kinetics_temporal_derivative_ordering}
		,
		\\
		\sum_{b}
		C_{ab}(F_a,F_b)
		\sim 
		\nu^{a} F_a 
		\sim 
		\rhostar{a}
		\Omega_a 
		F_a,
		\label{eq:Drift_kinetics_collision_term_ordering}
		\\
		\vb*{v}\cdot\nabla F_a
		\sim 
		\omega_a F_a 
		\sim 
		\rhostar{a}
		\Omega_a F_a,
		\label{eq:Drift_kinetics_convective_term_ordering}
		\\
		\frac{e_a}{m_a}
		\vb*{E}_\parallel 
		\cdot 
		\nabla_{\vb*{v}}
		F_a 
		\lesssim
		\sum_{b}
		C_{ab}(F_a,F_b)
		\sim 
		\rhostar{a}
		\Omega_a 
		F_a,
		\label{eq:Drift_kinetics_parallel_E_term_ordering}
		\\
		\frac{e_a}{m_a}
		\vb*{E}_\perp 
		\cdot 
		\nabla_{\vb*{v}}
		F_a
		\sim 
		\frac{v_E}{\vth{a}}
		\Omega_a 
		F_a
		\sim 
		\rhostar{a}
		\Omega_a 
		F_a.
		\label{eq:Drift_kinetics_perpendicular_E_term_ordering}
	\end{align}
	The estimate (\ref{eq:Drift_kinetics_temporal_derivative_ordering}) implies that the timescale in which the plasma varies is much larger than the one associated to the Larmor motion. Similarly, (\ref{eq:Drift_kinetics_collision_term_ordering}) implies that the collision frequency is much smaller than the gyrofrequency $\nu^{a}\sim\omega_a\ll\Omega_a$. The ordering (\ref{eq:Drift_kinetics_convective_term_ordering}) associated to the convective term determines that the scale in which the plasma varies spatially is of order $L$, which is an specific assumption for studying neoclassical phenomena. For the ordering of the terms associated to the acceleration caused by the electric field (\ref{eq:Drift_kinetics_parallel_E_term_ordering}) and (\ref{eq:Drift_kinetics_perpendicular_E_term_ordering}), the splitting in its perpendicular and parallel components to $\vb*{b}$ has been used $\vb*{E}=\vb*{E}_\parallel + \vb*{E}_\perp$. It is important to remark that these two orderings can be derived from (\ref{eq:Littlejohns_low_flow_ordering}). However, it is instructive to consider them separately. The estimate (\ref{eq:Drift_kinetics_parallel_E_term_ordering}) is mandatory to treat situations near plasma equilibrium. The reason is that in the dynamics associated to the acceleration parallel to magnetic field lines only $\vb*{E}_\parallel$ and the collisions are involved. For treating situations near plasma equilibrium, the parallel acceleration due to $\vb*{E}_\parallel$ should be, at least, balanced by collisions. Thus, the situation ${e_a}/{m_a}
	\vb*{E}_\parallel 
	\cdot 
	\nabla_{\vb*{v}}
	F_a \ll \rhostar{a}\Omega_a F_a$ is an allowed limiting case by  (\ref{eq:Drift_kinetics_parallel_E_term_ordering}). Finally, for the ordering (\ref{eq:Drift_kinetics_perpendicular_E_term_ordering}), the perpendicular piece of the electric field is estimated from the $\vb*{E}\times\vb*{B}$ drift 	
	\begin{align}
		\vb*{v}_{\vb*{E}\times\vb*{B}} 
		\sim
		v_E
		:=
		\frac{|\vb*{E}_\perp|}{B}
		, 
	\end{align} 
	and the {drift ordering} (\ref{eq:Littlejohns_low_flow_ordering}) implies
	\begin{align}
		\frac{v_E}{\vth{a}} \sim \rhostar{a}.
	\end{align}

	With the above set of assumptions it is possible to effectively reduce the dimensionality of equation (\ref{eq:Kinetic_equation_Introduction}) from six to five. The complete derivation of the DKE, will not be given here. Instead, the general workflow of the recursion will be explained and how it can be used to derive the force balance equation from the previous section. For the complete derivation the reader can consult \cite{Hazeltine_1973}. A more detailed calculation to first order is given in \cite{hazeltine1992plasma} or in the low flow section of Lecture II from \cite{Parra_Collisionless}.

	In order to derive the DKE as in \cite{Hazeltine_1973}, it is convenient to recast the kinetic equation (\ref{eq:Kinetic_equation_Introduction}) as	
	\begin{align}
		\Omega_a
		\pdv{{F}_a}{\gamma}
		=
		-
		\left(
		\DD-\CC
		\right) 
		F_a
		,
		\label{eq:Kinetic_equation_gamma}
	\end{align}
	where the notations
	\begin{align}
		\DD F_a
		&:=
		\VV F_a
		-
		\Omega_a
		\pdv{F_a}{\gamma}
		\sim 
		\rhostar{a}\Omega_a
		F_a
		,
		\\
		\CC F_a
		&:=
		\sum_{b}
		C_{ab}(F_a,F_b)
		\sim 
		\rhostar{a}\Omega_a
		F_a,
	\end{align}
	and
	\begin{align}
		\VV
		& :=
		\pdv{t}
		+
		\vb*{v}
		\cdot		
		\nabla 
		+
		\frac{e_a}{m_a}
		\left(
		\vb*{E}+\vb*{v}\times\vb*{B}
		\right)
		\cdot		
		\nabla_{\vb*{v}},
	\end{align}
	for the Vlasov operator in the left-hand side of (\ref{eq:Kinetic_equation_Introduction}) have been employed.
	
	Now, the distribution function $F_a$ is splitted in its gyroaveraged and fluctuating pieces
	\begin{align}
		F_a = \overline{F}_a + \widetilde{F}_a,
	\end{align}
	where $\overline{F}_a := \gyroav{F_a}$ denotes the gyroaveraged piece and its fluctuating piece is small $\widetilde{F}_a \sim \rhostar{a} \overline{F}_a$.
	
	Imposing periodicity along $\gamma$ of $F_a$ in (\ref{eq:Kinetic_equation_gamma}) yields the solvability condition 
	\begin{align} 
		\gyroav{
			\left(\DD-\CC\right) 
			F_a
		} = 0.
	\end{align}
	Thus, the kinetic equation (\ref{eq:Kinetic_equation_gamma}) is equivalent to the system of equations 
	\begin{align}
		& \Omega_a
		\pdv{\widetilde{F}_a}{\gamma}
		=
		-
		\left(
		\DD-\CC
		\right) 
		\left(
		\overline{F}_a + \widetilde{F}_a
		\right)
		+
		\gyroav{		
			\left(
			\DD-\CC
			\right) 
			\left(
			\overline{F}_a + \widetilde{F}_a
			\right)
		},
		\label{eq:DKE_gyrophase_dependent}
		\\
		& \gyroav{
			\left(\DD-\CC\right)  
			\left(
			\overline{F}_a
			+
			\widetilde{F}_a
			\right) 
		} = 0.
		\label{eq:DKE_gyroaveraged}
	\end{align}
	The second term on the right-hand side of equation (\ref{eq:DKE_gyrophase_dependent}) is redundant due to (\ref{eq:DKE_gyroaveraged}). However, it is convenient to retain this term for approximating the solution to (\ref{eq:DKE_gyrophase_dependent}) perturbatively. Note that the right-hand side of (\ref{eq:DKE_gyrophase_dependent}) is of order $\rhostar{a}$ with respect to its left-hand side. This is a key aspect exploited by the recursive method presented in \cite{Hazeltine_1973}. This procedure consists on approximating $\widetilde{F}_a$ perturbatively from equation  (\ref{eq:DKE_gyrophase_dependent}) and then inserting this approximation in (\ref{eq:DKE_gyroaveraged}) to obtain the DKE. Specifically, the fluctuating piece is formally expanded as
	\begin{align}
		\widetilde{F}_a
		=
		\widetilde{F}_a^{(1)}
		+
		\widetilde{F}_a^{(2)}
		+
		\ldots,
		\label{eq:DKE_Fluctuating_piece_expansion}
	\end{align}
	where $\widetilde{F}_a^{(k)}\sim \rhostar{a}^k \overline{F}_a$. Inserting expansion (\ref{eq:DKE_Fluctuating_piece_expansion}) in (\ref{eq:DKE_gyrophase_dependent}) and grouping terms of the same order yields the sequence
	\begin{align}
		\Omega_a
		\pdv{\widetilde{F}_a^{(1)}}{\gamma}
		& =
		-
		\left(
		\DD-\CC
		\right)  
		\overline{F}_a  
		+
		\gyroav{		
			\left(
			\DD-\CC
			\right) 
			\overline{F}_a 
		}
		\label{eq:DKE_Fluctuating_step_1}
		,
		\\
		\Omega_a
		\pdv{\widetilde{F}_a^{(k+1)}}{\gamma}
		& =
		-
		\left(
		\DD-\CC
		\right)  
		\widetilde{F}_a^{(k)} 
		+
		\gyroav{		
			\left(
			\DD-\CC
			\right) 
			\widetilde{F}_a^{(k)} 
		}
		\label{eq:DKE_Fluctuating_step_k}
		,
	\end{align}
	for $k\ge 1$. The idea is to solve for $
	\widetilde{F}_a^{(k)} $ as a functional of $\overline{F}_a$ in a recursive manner. First, one would solve (\ref{eq:DKE_Fluctuating_step_1}) for $
	\widetilde{F}_a^{(1)} $. Then, using the functional form derived for $
	\widetilde{F}_a^{(1)} $ one can set $k=1$ in equation (\ref{eq:DKE_Fluctuating_step_k}) and solve it for $
	\widetilde{F}_a^{(2)} $ as a functional of $
	\overline{F}_a $ and proceed \textit{ad libitum}. Observe that equation (\ref{eq:DKE_Fluctuating_step_k}) can always be solved for $
	\widetilde{F}_a^{(k+1)} $ as the gyroaverage of the right-hand side is zero. The DKE is obtained by inserting the functional form of $\widetilde{F}_a= 
	\widetilde{F}_a^{(1)}
	+
	\widetilde{F}_a^{(2)}
	+
	\ldots$ to the desired order in equation (\ref{eq:DKE_gyroaveraged}). Fortunately, for the vast majority of applications, calculating the functional form of $
	\widetilde{F}_a^{(1)} $ is sufficient. In addition, for computing $ \widetilde{F}_a^{(1)} $, the term $\CC  
	\overline{F}_a -\gyroav{ \CC  	 	\overline{F}_a 
	}$ can be safely neglected in equation (\ref{eq:DKE_Fluctuating_step_1}).
	
	Employing velocity coordinates $(\mu_a,\epsilon_a,\gamma,\sigma)$ the functional form obtained for $\widetilde{F}_a^{(1)}$ is \cite{Hazeltine_1973, hazeltine1992plasma} 
	\begin{align}
		\widetilde{F}_a^{(1)}
		=
		-
		\vb*{\rho}_a
		\cdot
		\widetilde{\nabla}
		\overline{F}_a
		+
		g_a
		\pdv{\overline{F}_a}{\mu_a}
		.
		\label{eq:DKE_Fluctuating_piece_1}
	\end{align}
	Here,
	\begin{align}    	
		\widetilde{\nabla}
		:=
		\nabla 
		+
		e_a 
		\vb*{b}
		\times
		\vb*{v}_{\text{d}a}
		\pdv{\mu_a}
		-
		e_a
		\pdv{\vb*{A}}{t}
		\pdv{\epsilon_a}
	\end{align}
	and
	\begin{align}
		g_a
		:=
		\frac{\vp\mu_a}{\Omega_a}
		\left(
		\vb*{e}_\rho
		\vb*{v}_\perp
		:
		\nabla\vb*{b}
		-
		\frac{1}{2}
		\vb*{b}\cdot\nabla\times\vb*{b}
		\right).
	\end{align}
	Inserting the expression for $\widetilde{F}_a^{(1)}$ (\ref{eq:DKE_Fluctuating_piece_1}) in (\ref{eq:DKE_gyroaveraged}) (and lengthy algebra calculations explained in \cite{Hazeltine_1973,hazeltine1992plasma}) yields the \textit{first order DKE}
	\begin{align}
		\pdv{\overline{F}_a}{t}
		+
		(\vb*{v}_{\text{gc}a} + u_a \vb*{b})
		\cdot
		\nabla 
		\overline{F}_a
		+
		\dot{\mu}_a 
		\pdv{\overline{F}_a}{{\mu}_a}
		+
		\dot{\epsilon}_a 
		\pdv{\overline{F}_a}{{\epsilon}_a}
		& =
		\sum_{b}
		C_{ab}(\overline{F}_a,\overline{F}_b) 
		\label{eq:Drift_kinetics_DKE_mu_epsilon}
		\\
		& 
		+
		\sum_{b}
		\gyroav{
			C_{ab}(\widetilde{F}_a^{(1)},\widetilde{F}_b^{(1)})
		}
		,
		\nonumber
	\end{align}
	where  
	\begin{align}
		u_a(\vb*{x},\mu_a,t) 
		:=
		\frac{\mu_a }{e_a}
		\vb*{b}(\vb*{x},t)
		\cdot
		\nabla\times\vb*{b}(\vb*{x},t)
		,
		\label{eq:Baños_drift_mu_epsilon}
	\end{align}
	is the Baños parallel drift \cite{Baños_1967} and
	\begin{align}
		\dot{\mu}_a
		(\vb*{x},\mu_a,\epsilon_a, \sigma,t)  
		& 
		=
		\mu_a 
		\left[
		\frac{\vp}{\Omega_a}
		\nabla\cdot
		\left(
		\pdv{\vb*{b}}{t}
		\times 
		\vb*{b}
		\right)
		-
		\frac{\vb*{b}}{B}
		\cdot
		\pdv{\vb*{A}}{t}
		(\vb*{b}\cdot\nabla\times\vb*{b})
		\right]
		\nonumber
		\\
		& 
		+
		m_a 
		\vp 
		\vb*{b}
		\cdot
		\nabla 
		\left(
		\vp(\vb*{x},\mu_a,\epsilon_a, \sigma,t) 
		\frac{u_a(\vb*{x},\mu_a,t) }{B}
		\right)
		\label{eq:Drift_kinetics_dot_mu}
		,
		\\
		\dot{\epsilon}_a(\vb*{x},\mu_a,\epsilon_a, \sigma,t) 
		& 
		=	
		-
		\frac{m_a}{2} 
		\pdv{t} 	
		\left(\vp^2 (\vb*{x},\mu_a,\epsilon_a,t) \right)
		-
		{e_a} 
		\vb*{v}_{\text{gc}a}
		\cdot
		\pdv{\vb*{A}}{t}
		.
		\label{eq:Drift_kinetics_dot_epsilon}
	\end{align}
	Note that the Vlasov part of the DKE (\ref{eq:Drift_kinetics_DKE_mu_epsilon}) differs from the one that would be obtained employing the guiding-center equations (\ref{eq:GCM_x}), (\ref{eq:GCM_mu}) and (\ref{eq:GCM_dot_epsilon}). While the expression (\ref{eq:Drift_kinetics_dot_epsilon}) for $\dot{\epsilon}_a$  is equivalent to this order to the one obtained from the guiding-center Lagrangian (\ref{eq:GCM_dot_epsilon}), $\mu_a$ is not conserved. Besides, a correction $u_a \vb*{b}$ to the parallel velocity has appeared. As explained in section 4 of \cite{Parra_2011}, this correction can be made explicit by an appropriate selection of the generating function $S$ of section \ref{subsec:Guiding_center_motion}. These differences are, however, of more academic than practical interest as the refinements provided by the recursive procedure are rarely of importance. Nevertheless, the recursive procedure allows to prove that, to lowest order, collisions relax the plasma to a confined state which can be described by the force balance equation (\ref{eq:MHD_Momentum_balance_steady_state_species_a}). Typically, for calculating neoclassical transport, one is interested in the steady state towards which collisions relax the plasma. Therefore, the steady state version of the DKE (\ref{eq:Drift_kinetics_DKE_mu_epsilon}) is the one that will be considered from now on.

	 Employing the asymptotic expansion in $\rhostar{a}$, it is possible to prove that, within the volume enclosed by $\vb*{B}$, the lowest order gyroaveraged distribution function is given by a Maxwellian \cite{hazeltine1992plasma,Calvo_2012,Calvo_2013}. Expanding $\overline{F}_a$ as 
	\begin{align}
		\overline{F}_a = 
		\overline{F}_a^{(0)}
		+
		\overline{F}_a^{(1)}
		+
		\ldots
		,
	\end{align}
	where $\overline{F}_a^{(k+1)}\sim \rhostar{a}\overline{F}_a^{(k)}$, the lowest-order piece of the DKE (\ref{eq:Drift_kinetics_DKE_mu_epsilon}) is, in steady state,
	\begin{align}
		\vp
		\vb*{b}
		\cdot
		\nabla 
		\overline{F}_a^{(0)}
		=
		\sum_{b}
		C_{ab}\left(
		\overline{F}_a^{(0)},
		\overline{F}_b^{(0)}
		\right)
		.
		\label{eq:Drift_kinetics_lowest_order_DKE}
	\end{align}  
	In appendix \ref{sec:DKE_Maxwellian}, it is proven that the only solution to equation (\ref{eq:Drift_kinetics_lowest_order_DKE}) is a Maxwellian and that, when $\vb*{B}$ consists of nested flux surfaces, it is given by the radially local Maxwellian (\ref{eq:Radially_local_Maxwellian}) 
	\begin{align}
		\overline{F}_a^{(0)}
		(\psi,\epsilon_a)
		=
		f_{\text{M}a}
		(\psi,\epsilon_a)
		.
	\end{align}
	When the Fokker-Planck collision operator is employed, the temperature of all species is the same, i.e. $T_a=T_b$ for all species $a$ and $b$. A subsidiary expansion in the electron-to-ion mass ratio $m_{\text{e}}/m_{\text{i}}$ allows ions and electrons to have different temperatures. As discussed in appendix \ref{sec:DKE_Maxwellian}, replacing the Fokker-Planck operator by a simpler pitch-angle scattering collision operator in equation (\ref{eq:Drift_kinetics_lowest_order_DKE}), would allow $\overline{F}_a^{(0)}$ to have different temperature for each species. Regardless of which collision operator is employed, an important consequence of the solution to equation (\ref{eq:Drift_kinetics_lowest_order_DKE}) is that, to lowest order in $\rhostar{a}$, the electrostatic potential $\varphi$, the density $n_a$ and temperature $T_a$ are constant along field lines. Hence, when $\vb*{B}$ consists of nested flux surfaces, if $\varphi$ is split as
	\begin{align}
		\varphi(\vb*{x}) = \varphi_0(\psi) + \varphi_1(\vb*{x}),
		\label{eq:Electrostatic_potential_splitting}
	\end{align}
	the ordering $\varphi_1/\varphi_0\sim\rhostar{a}\ll 1$ holds.

	Thanks to the equation for the functional form of $\widetilde{F}_a^{(1)}$ (\ref{eq:DKE_Fluctuating_piece_1}) it is possible to derive the force balance equation for a plasma in radially local thermodynamic equilibrium (\ref{eq:MHD_Momentum_balance_steady_state_species_a}). As described above, when the magnetic field consists of nested flux surfaces, the gyroaveraged distribution function $\overline{F}_a$ is, to lowest order, given by the Maxwellian (\ref{eq:Radially_local_Maxwellian}). Hence, from (\ref{eq:DKE_Fluctuating_piece_1}) one obtains the  lowest order distribution function
	\begin{align}
		F_a^{(0)}
		=
		f_{\text{M}a}
		(\psi,\epsilon_a)
		-
		\vb*{\rho}_a
		\cdot\nabla 
		f_{\text{M}a}(\psi,\epsilon_a)
		,
		\label{eq:DKE_distribution_function_radially_local_equilibrium}
	\end{align}
	where the Maxwellian is regarded as a function of $(\psi,\epsilon_a)$ and thus the gradient $\nabla 
	f_{\text{M}a}$ is proportional to the so-called \textit{thermodynamic forces}
	\begin{align}
		\nabla 
		f_{\text{M}a}(\psi,\epsilon_a)
		=
		\left[
		\frac{\nabla n_a}{n_a}
		+
		\left(
		\frac{\epsilon_a-e_a\varphi_0}{T_a}
		-
		\frac{3}{2}
		\right)
		\frac{\nabla T_a}{T_a}
		+
		\frac{e_a\nabla \varphi_0}{T_a}
		\right] 
		f_{\text{M}a}(\psi,\epsilon_a).
	\end{align}
	Then, the flow velocity associated to the lowest order distribution function  (\ref{eq:DKE_distribution_function_radially_local_equilibrium}) is given by \cite{hazeltine1992plasma}
	\begin{align}
		n_a 
		\vb*{V}_{a}
		=
		-
		\vmoment{
			\vb*{v}_\perp\vb*{\rho}_a
			\cdot
			\nabla 
			f_{\text{M}a}
		}
		=
		\frac{\vb*{B}}{e_a B^2}	 	
		\times
		\left(
		\nabla p_a 
		+
		e_a n_a \nabla \varphi_0
		\right),
		\label{eq:LO_DKE_Macroscopic_flow}
	\end{align}
	which is precisely the one satisfying (\ref{eq:MHD_Momentum_balance_steady_state_species_a}). Note that the parallel flow associated to the distribution function (\ref{eq:DKE_distribution_function_radially_local_equilibrium}) is zero. It is important to emphasize that, without necessarily having a structure of nested flux surfaces, the lowest order flow velocity can still have the form given by (\ref{eq:LO_DKE_Macroscopic_flow}) with $\vb*{b}\cdot\nabla p_a= \vb*{b}\cdot\nabla n_a = \vb*{b}\cdot\nabla \varphi_0 = 0$. In appendix \ref{sec:DKE_Maxwellian}, it is proven that, as long as $\vb*{B}$ is tangent to a closed surface, the gyroaveraged, lowest order distribution function is a Maxwellian within the volume enclosed by the said surface. 
	
	Recall from section \ref{sec:Force_balance} that, when equation (\ref{eq:MHD_Momentum_balance_steady_state_species_a}) holds, flows across flux surfaces are not allowed. Similarly, from (\ref{eq:Ideal_MHD_Force_balance}) and (\ref{eq:Ideal_MHD_Amperes_law}), the net parallel current $\mean*{\vb*{J}\cdot\vb*{B}}$ is undetermined. Thus, in order to capture radial and parallel neoclassical flows, it is necessary to let the gyroaveraged piece of the distribution function to deviate from a Maxwellian. Importantly, deviations from (radially local) equilibrium allow for local radial currents $\vb*{J}\cdot\nabla\psi\ne0$. In order to be consistent with Ampère's law (\ref{eq:Ideal_MHD_Amperes_law}), these local currents must be such that each flux surface remains \textit{ambipolar}. Namely,
	\begin{align}
		\mean*{\vb*{J}\cdot\nabla\psi}=0,	 	
		\label{eq:Ambipolarity_condition}
	\end{align}
	must be satisfied\footnote{Here, Ampère's law (\ref{eq:Ideal_MHD_Amperes_law}) and property (\ref{eq:FSA_curl_perp}) have been employed to obtain $\mean*{\vb*{J}\cdot\nabla\psi}=\mean*{\nabla\times\vb*{B}\cdot\nabla\psi}/\mu_0  =0$.} regardless of whether or not force balance (\ref{eq:Ideal_MHD_Force_balance}) holds. Ambipolarity condition (\ref{eq:Ambipolarity_condition}) is important for determining the neoclassical radial electric field in stellarators. 
	
	Hence, in order to predict neoclassical phenomena, it is required to allow $\overline{F}_a$ to deviate from a Maxwellian. It has been proven above that, when the magnetic field is in equilibrium and consisting of flux surfaces, the lowest order (in $\rhostar{a}$) gyroaveraged distribution function is a Maxwellian. For this reason, it is a standard practice in neoclassical theory to expand \cite{DKES1986,Calvo_2013,Calvo_2017, Landreman_2014}
	\begin{align}
		\overline{F}_a 
		= 
		\left(
		1- \frac{e_a}{T_a}
		\varphi_1
		\right)
		f_{\text{M}a}
		+
		h_a 
		,
		\label{eq:Non_adiabatic_component_splitting}
	\end{align}
	where $h_a\sim\rhostar{a} f_{\text{M}a}$ is the non-adiabatic deviation of the distribution function from the radially local Maxwellian (\ref{eq:Radially_local_Maxwellian}). 
	
	Inserting splitting (\ref{eq:Non_adiabatic_component_splitting}) in (\ref{eq:Drift_kinetics_DKE_mu_epsilon}) and retaining only terms up to order $O(\rhostar{a}^2\Omega_a f_{\text{M}a})$ yields a DKE for $h_a$ to treat situations near equilibrium. As was mentioned at the beginning of this chapter, the drift ordering (\ref{eq:Littlejohns_low_flow_ordering}) implied
	that the magnetic field varies very slowly in time, according to estimate (\ref{eq:Drift_kinetics_slowly_varying_B}).
	Therefore, the magnetic field will be assumed
	to be static (i.e. $\pdv*{\vb*{A}}{t}=0$), which implies that the electric field is electrostatic
	(\ref{eq:Electrostatic_field}), (i.e. $\vb*{E} = -\nabla\varphi$). For numerical computations it is convenient to use coordinates whose domain is independent of the rest of variables. Hence, instead of writing this DKE in the original velocity coordinates $(\mu_a,\epsilon_a)$ in which it was derived, the magnitude of the velocity $v:=|\vb*{v}|\in [0,\infty)$ and the pitch-angle cosine $\xi:=\vb*{v}\cdot\vb*{b}/|\vb*{v}| \in [-1,1]$ will be employed. Moreover, the monoenergetic DKE described in chapter \ref{chap:Monoenergetic} and solved by {\MONKES} is written in these coordinates. The algebra for expressing the (magnetostatic) Vlasov part of the DKE in coordinates $(\xi,v)$ is explained in appendix \ref{sec:Hazeltine_DKE}. As a result, the DKE obtained is \cite{DKES1986} 
	\begin{align}
		(\vb*{v}_{\text{gc} a}
		+
		u_a \vb*{b}
		) 
		\cdot
		\nabla
		h_a  
		+
		\dot{\xi}
		\pdv{h_a}{\xi}
		+
		\dot{v}
		\pdv{h_a}{v}
		=
		\sum_{b}
		C_{ab}^{\text{L}}(h_a,h_b)
		+
		S_a.
		\label{eq:DKE_Hazeltine_Hirshman_non_adiabatic}
	\end{align} 
	The coefficients in front of the derivatives along $\xi$ and $v$ of the DKE (\ref{eq:DKE_Hazeltine_Hirshman_non_adiabatic}) are, respectively, the functions
	\begin{align}
		\dot{\xi}(\vb*{x},\xi,v) 
		&
		:=  
		(1-\xi^2)  
		\frac{\vb*{F}_a(\vb*{x},\xi,v)\cdot \vb*{b}(\vb*{x})}{m_a v}
		+
		\xi(1-\xi^2)
		\frac{\nabla\times\vb*{b}(\vb*{x})}{\Omega_a(\vb*{x})}
		\cdot
		\frac{\vb*{F}_a(\vb*{x},\xi,v)}{m_a}
		\nonumber
		\\
		&
		-
		3
		\xi
		\frac{\vb*{F}_a(\vb*{x},\xi,v) \cdot\vb*{b}(\vb*{x})}{m_a v^2}
		u_a(\vb*{x},\xi,v)   	
		-  
		\xi 
		\vb*{b}(\vb*{x})\cdot\nabla
		u_a(\vb*{x},\xi,v) 
		\nonumber
		\\
		& 
		-
		\frac{1}{2}
		\xi(1-\xi^2)
		(\vb*{v}_{\text{gc}a}(\vb*{x},\xi,v)
		+
		u_a(\vb*{x},\xi,v) \vb*{b}(\vb*{x}))
		\cdot
		\nabla \ln B(\vb*{x})  
		\label{eq:dot_xi_xi_v}	 
	\end{align}
	and
	\begin{align}
		\dot{v}(\vb*{x},\xi,v) := 
		\frac{e_a}{m_a v}
		\vb*{E}(\vb*{x})\cdot(v\xi\vb*{b}(\vb*{x}) + u_a \vb*{b}(\vb*{x}) + 
		\vb*{v}_{\text{m} a}(\vb*{x},\xi,v)  )
		,
		\label{eq:dot_v_xi_v}
	\end{align}	 
	where $\vb*{F}_a(\vb*{x},\xi,v) := e_a \vb*{E}^*(\vb*{x},\xi,v)= e_a \vb*{E}(\vb*{x}) - m_a v^2(1-\xi^2)\nabla \ln B(\vb*{x}) /2$.

	On the right-hand side of the DKE (\ref{eq:DKE_Hazeltine_Hirshman_non_adiabatic}), the source term $S_a$ contains the action of the Vlasov operator on $h_a - \overline{F}_a$. The specific form of $S_a$ (containing terms up to order $\rhostar{a}^2\Omega_a f_{\text{M}a}$) will be given in the next chapter. In regard to collisions, the terms $C_{ab}(h_a,h_b)$ and $\mean{C_{ab}(\widetilde{F}_a^{(1)},\widetilde{F}_b^{(1)})}_\gamma$ have been safely neglected, retaining only the \textit{linearized Fokker-Planck collision operator} 
	\begin{align}
		C_{ab}^{\text{L}}(h_a,h_b)
		:=
		C_{ab}(h_a,f_{\text{M}b})
		+
		C_{ab}(f_{\text{M}a},h_b)
		.
	\end{align} 
	It is important to remark that the linearized collision operator satisfies the same conservation properties as the non linear Fokker-Planck collision operator. Hence, in the conservation properties (\ref{eq:Fokker_Planck_mass_conservation}), (\ref{eq:Fokker_Planck_momentum_conservation}) and (\ref{eq:Fokker_Planck_energy_conservation}), $C_{ab}$ may be replaced by $C_{ab}^{\text{L}}$.
	
	Summarizing, the DKE (\ref{eq:DKE_Hazeltine_Hirshman_non_adiabatic}) is a simpler equation than (\ref{eq:Kinetic_equation_Introduction}) which rigorously describes the deviation of a plasma from radially local thermodynamic equilibrium due to neoclassical mechanisms. However, note that this equation is five dimensional and non linear in the unknown $h_a$ through $\varphi_1$, which has to be determined from (\ref{eq:Gauss_law}). Recall that the main mission of this dissertation was to provide a \textit{fast} numerical tool for evaluating neoclassical transport in stellarators. From numerically solving an equation as complicated as (\ref{eq:DKE_Hazeltine_Hirshman_non_adiabatic}) fast calculations are hardly expected. Therefore, a simpler DKE which approximates well the DKE (\ref{eq:DKE_Hazeltine_Hirshman_non_adiabatic}) is better suited for this purpose. In the next chapter, several simplifications applied to equation  (\ref{eq:DKE_Hazeltine_Hirshman_non_adiabatic}) in order to make it more tractable will be explained. These simplifications are commonly known as the \textit{monoenergetic approximation} and lead to a three dimensional DKE which is the one that the code {\MONKES} solves.
	
	

	\chapter{The monoenergetic approximation to neoclassical transport} 
	\label{chap:Monoenergetic}
	\thispagestyle{empty}

	In this chapter, the equation that {\MONKES} solves and the neoclassical transport coefficients that it computes are discussed. {\MONKES} solves a kinetic equation corresponding to the so called \textit{monoenergetic approximation}. In section \ref{sec:Drift_kinetics} the DKE (\ref{eq:DKE_Hazeltine_Hirshman_non_adiabatic}) describing neoclassical transport processes in toroidal plasmas close to equilibrium was introduced. However, due to its high dimensionality (five independent variables), most of the neoclassical transport calculations do not consist on solving this equation. For instance, the code {\DKES} \cite{DKES1986}, which has been the (\textit{de facto}) standard code for neoclassical transport calculations in stellarators for more than three decades, solves a DKE corresponding to the \textit{monoenergetic approximation}. The monoenergetic approximation to neoclassical transport consists on a series of simplifications applied to the rigorously derived DKE (\ref{eq:DKE_Hazeltine_Hirshman_non_adiabatic}) from \cite{Hazeltine_1973} in order to approximate it with a simpler, yet sufficiently accurate, DKE. Some of these changes are ad hoc but nevertheless reasonable. In \cite{dherbemont2022} it is proven that, to lowest order in the (double) limit of low collisionality, large aspect ratio stellarators with mirror ratios close to unity, the radial fluxes of heat and particles predicted by the monoenergetic DKE coincide with those obtained by solving the rigorously obtained DKE. Moreover, the neoclassical flows and ambipolar radial electric field predicted by the monoenergetic approximation have been compared satisfactorily against experimental measurements (see e.g. \cite{Alonso_2022}). As mentioned in chapter \ref{sec:Introduction}, an important feature of the monoenergetic approximation is that it permits to encapsulate the dependence of neoclassical transport on the magnetic configuration in at most four \textit{monoenergetic transport coefficients} $\Dij{ij}$. In what follows, we will call the \qmarks{monoenergetic DKE} simply \qmarks{DKE} and unless explicitly stated, when we say \qmarks{DKE} we mean the \qmarks{monoenergetic DKE}. In section \ref{sec:Monoenergetic_assumptions} the assumptions and modifications employed by the monoenergetic approximation are briefly listed. In section \ref{sec:DKE} we describe the monoenergetic DKE that {\MONKES} solves and the transport coefficients that it computes. Additionally, in section \ref{sec:DKE_adjoint_properties} we will define an appropriate mathematical framework in which it is possible to prove some general properties of the DKE and transport coefficients. These properties are useful for obtaining a method for computing derivatives of the transport coefficients with respect to parameters upon which the DKE depends. After that, in section \ref{sec:Appendix_Onsager_symmetry} these properties will be used to prove that the monoenergetic transport coefficients satisfy Onsager reciprocal relations \cite{Onsager_1,Onsager_2} under two (non exclusive) circumstances. In section \ref{sec:Legendre_representation} the representation of the DKE and its solution in a Legendre basis is explained. Based on the particular structure that the monoenergetic DKE displays when expressed in a Legendre basis, an algorithm for solving it is provided in section \ref{subsec:Legendre_expansion}. This algorithm is the one that is employed by the code {\MONKES} (see chapter \ref{chap:MONKES} for details on its implementation). Finally, in section \ref{sec:Derivatives_monoenergetic} several methods for computing the derivatives of the transport coefficients with respect to parameters upon which the DKE depends are discussed. These derivatives are useful for gradient-based optimization methods. For instance, one might want to be interested in compute the derivative of the bootstrap current with respect to $\iota$ or the poloidal current $B_\zeta$ in order to modify these two parameters so that the bootstrap current is reduced. Excepting sections \ref{sec:Monoenergetic_assumptions}, \ref{sec:DKE_adjoint_properties}, \ref{sec:Derivatives_monoenergetic} and \ref{subsec:Monoenergetic_lambda}, this chapter is mostly based on article [I] from the \textbf{\qmarks{PUBLISHED AND SUBMITTED CONTENT}} section at the beginning of this dissertation.

	\section{Simplifications to the DKE in the monoenergetic approximation} \label{sec:Monoenergetic_assumptions} 
	The DKE (\ref{eq:DKE_Hazeltine_Hirshman_non_adiabatic}) is an approximation to (\ref{eq:Kinetic_equation_Introduction}) obtained rigorously within the formal ordering of drift-kinetics for treating situations in which the plasma is near thermodynamic equilibrium. By solving equation  (\ref{eq:DKE_Hazeltine_Hirshman_non_adiabatic}), all the macroscopic observable quantities associated to neoclassical phenomena could be computed. It is important to remember that the DKE (\ref{eq:DKE_Hazeltine_Hirshman_non_adiabatic}) assumed the existence of an equilibrium magnetic field (i.e. satisfying (\ref{eq:Ideal_MHD_Force_balance}) and (\ref{eq:Ideal_MHD_Amperes_law})) consisting of nested flux surfaces. However, even though the fast scale of rapid gyration has been eliminated, its dimensionality is still too large for expecting fast numerical computations of neoclassical transport. In \cite{DKES1986} several ad hoc simplifications are carried out to equation (\ref{eq:DKE_Hazeltine_Hirshman_non_adiabatic}) in order to obtain a more tractable, but sufficiently accurate, version of the DKE, which we call monoenergetic DKE. Some of the terms of order $\rhostar{a}^2\Omega_a f_{\text{M}a}$ and higher will be dropped from equation  (\ref{eq:DKE_Hazeltine_Hirshman_non_adiabatic}) to reduce its dimensionality. In this section, these simplifications are listed below and the explicit expression of the monoenergetic DKE will be given section \ref{sec:DKE}. 
	
	\begin{enumerate}[label=S\arabic*]
		\item In all operators acting on $h_a$ of equation (\ref{eq:DKE_Hazeltine_Hirshman_non_adiabatic}), the electric field is assumed to be perpendicular to the flux surface. That is, $\vb*{E}$ is replaced by $ \vb*{E}_0 = E_\psi(\psi) \nabla\psi $ on $\vb*{v}_{\text{gc}a}\cdot\nabla h_a$, $\dot{\xi}\pdv*{h_a}{\xi}$ and $\dot{v}\pdv*{h_a}{v}$. This simplification allows to eliminate the non linearity of the DKE when $E_\psi$ is considered as an input. Recall from splitting (\ref{eq:Electrostatic_potential_splitting}) that this approximation is consistent within the formal ordering of the asymptotic expansion in $\rhostar{a}$.

		\item In the expression (\ref{eq:dot_xi_xi_v}) for $\dot{\xi}(\vb*{x},\xi,v)$, all terms of order $\rhostar{a}\vth{a}/L\sim \rhostar{a}^2\Omega_a$ are neglected, i.e. those including $u_a$, $\vb*{v}_{\text{d}a}$ or $\Omega_a^{-1}$. Besides, as by simplification 1, $\vb*{E}\cdot\vb*{b}=0$, expression (\ref{eq:dot_xi_xi_v}) is replaced by $\dot{\xi}(\vb*{x},\xi,v) = -v(1-\xi^2)\vb*{b}\cdot\nabla\ln B/2 = v(1-\xi^2)\nabla\cdot\vb*{b}/2$.
		
		\item In the expression (\ref{eq:dot_v_xi_v}) for $\dot{v}(\vb*{x},\xi,v)$, the contribution $\vb*{E}\cdot\vb*{v}_{\text{m}a} \sim \rhostar{a}\vth{a}^2/L$ in $\dot{v}\pdv*{h_a}{v}$ is neglected and as it was assumed before that $\vb*{E}\cdot\vb*{b}=0$, expression (\ref{eq:dot_v_xi_v}) is replaced by $\dot{v}(\vb*{x},\xi,v)=0$.

		\item In the convective term $(\vb*{v}_{\text{gc} a}+u_a\vb*{b})\cdot\nabla h_a$ of equation (\ref{eq:DKE_Hazeltine_Hirshman_non_adiabatic}), only the parallel streaming term $v\xi\vb*{b}$ and the $\vb*{E}\times\vb*{B}$ drift are retained in $\vb*{v}_{\text{gc} a}$. Neglecting the magnetic drift $\vb*{v}_{\text{m}a}$ while retaining the $\vb*{E}\times\vb*{B}$ drift is not consistent as, according to the drift ordering, these two drifts are of the same order. Nevertheless, this simplification is justified for large aspect ratio stellarators \cite{dherbemont2022}. Due to the previous assumption of $\vb*{E}\cdot\vb*{b}=0$, there is no radial component of the $\vb*{E}\times\vb*{B}$ drift and thus the resulting DKE is \textit{radially local}, i.e. there are no spatial derivatives in the direction perpendicular to flux surfaces. In order to obtain a kinetic equation which satisfies Liouville's theorem and can be written in divergence form, the $\vb*{E}\times\vb*{B}$ drift (\ref{eq:True_ExB}) is replaced by the \textit{incompressible} $\vb*{E}\times\vb*{B}$ drift \cite{dherbemont2022} (its explicit expression will be given in section \ref{sec:DKE}).

		\item The linearized Fokker-Planck collision operator appearing in equation (\ref{eq:DKE_Hazeltine_Hirshman_non_adiabatic}) is approximated by the piece of it that describes pitch-angle scattering collisions (its explicit expression will be given in section \ref{sec:DKE}). This piece, called \textit{pitch-angle collision operator}, only contains derivatives along $\xi$ and as $\dot{v}=0$, in the resulting (monoenergetic) DKE $v$ appears as a parameter.

	\end{enumerate} 
	
	Summarizing, when these simplifications are applied, a DKE in which $\psi$ and $v$ appear as parameters is obtained.


	\section{Monoenergetic drift-kinetic equation and transport coefficients}	  
	\label{sec:DKE}

	After applying the modifications corresponding to the monoenergetic approximation listed in section \ref{sec:Monoenergetic_assumptions}, the DKE (\ref{eq:DKE_Hazeltine_Hirshman_non_adiabatic}) becomes \cite{DKES1986}
	\begin{align}
		(v \xi \vb*{b}  + \vb*{v}_E) \cdot \nabla h_a 
		+
		v\nabla \cdot \vb*{b} \frac{(1-\xi^2)}{2}  \pdv{h_a}{\xi}  
		- \nu^{a} \Lorentz h_a
		= S_a,
		\label{eq:DKE_Original}
	\end{align}
	where, as for the non monoenergetic DKE (\ref{eq:DKE_Hazeltine_Hirshman_non_adiabatic}), the velocity coordinates employed are the cosine of the pitch-angle $\xi := \vb*{v}\cdot\vb*{b}/|\vb*{v}|$ and the magnitude of the velocity $v:=|\vb*{v}|$. 
	
	Recall from splitting (\ref{eq:Non_adiabatic_component_splitting}) that, in equation (\ref{eq:DKE_Original}), $h_a$ is the non-adiabatic component of the deviation of the (gyroaveraged) distribution function from a local Maxwellian for a plasma species $a$ 
	\begin{align}
		f_{\text{M}a}(\psi, v) :=   n_a(\psi)  \pi^{-3/2}  {v_{\text{t}a}^{-3}(\psi)}  \exp(-\frac{v^2}{v_{\text{t}a}^2(\psi)}).
	\end{align}

	For the convective term in equation (\ref{eq:DKE_Original})
	\begin{align}
		\vb*{v}_E 
		:= 
		\frac{\vb*{E}_0\times\vb*{B}}{\mean*{B^2}} 
		= 
		- 
		\frac{E_\psi}{\mean*{B^2}}\vb*{B}\times \nabla\psi
		\label{eq:Incompressible_ExB_definition}
	\end{align}
	denotes the incompressible $\vb*{E}\times\vb*{B}$ drift approximation \cite{dherbemont2022} and $\vb*{E}_0 = E_\psi(\psi) \nabla \psi$ is the electrostatic piece of the electric field $\vb*{E}$ perpendicular to the flux surface. Note that when $\vb*{B}$ satisfies (as is the case) (\ref{eq:Ideal_MHD_Force_balance}) and (\ref{eq:Ideal_MHD_Amperes_law}), $\nabla\cdot\vb*{v}_E=0$.
	
	The Lorentz pitch-angle scattering operator has been denoted by $\Lorentz$, which in coordinates $(\xi,v)$ takes the form
	\begin{align}
		\Lorentz   := \frac{1}{2}  \pdv{\xi}\left( (1-\xi^2)\pdv{}{\xi} \right).
		\label{eq:Pitch_angle_scattering_operator}
	\end{align}
	In the collision operator, $\nu^a(v) =\sum_{b}\nu^{ab}(v)$ and
	\begin{align}
		\nu^{ab}(v) := 
		\frac{4 \pi n_b e_a^2 e_b^2}
		{m_a^2 v_{\text{t}a}^3}
		\ln\Lambda_{ab}
		\frac{ \erf(v/v_{\text{t}b}) - G(v/v_{\text{t}b})}{v^3/v_{\text{t}a}^3}
		\label{eq:Collision_frequency}
	\end{align}
	stands for the pitch-angle collision frequency between species $a$ and $b$. Here, $G(x)=\left[\erf(x) - (2x/\sqrt{\pi}) \exp(-x^2)\right]/(2x^2)$ is the Chandrasekhar function, $\erf(x)$ is the error function and $\ln\Lambda_{ab}$ is the Coulomb logarithm \cite{Helander_2005}.

	On the right-hand-side of equation (\ref{eq:DKE_Original}) 
	\begin{align}
		S_a 
		& :=  
		- \vb*{v}_{\text{m} a} \cdot \nabla \psi 
		\left(
		A_{1a} 
		+  \frac{v^2}{v_{\text{t}a}^2}
		A_{2a}
		\right)
		f_{\text{M}a}
		+ 
		\frac{B}{B_0} v \xi A_{3a}f_{\text{M}a}
		\label{eq:DKE_Original_Source}
	\end{align}
	is the source term, 
	\begin{align}
		\vb*{v}_{\text{m} a}\cdot\nabla\psi
		=
		-\frac{Bv^2}{\Omega_a}
		\frac{1+\xi^2}{2B^3}
		\vb*{B}\times\nabla\psi \cdot \nabla B 
	\end{align}
	is the expression of the radial magnetic drift assuming ideal magnetohydrodynamic equilibrium (i.e. satisfying (\ref{eq:Ideal_MHD_Force_balance}) and  (\ref{eq:Ideal_MHD_Amperes_law})) and the flux-functions 
	\begin{align}
		A_{1a}(\psi) & := \dv{\ln n_a}{\psi} - \frac{3}{2} \dv{\ln T_a}{\psi} - \frac{e_a E_\psi}{T_a}, 
		\\
		A_{2a}(\psi) & := \dv{\ln T_a}{\psi} , 
		\\
		A_{3a}(\psi) & :=  \frac{e_a B_0}{T_a} \frac{\mean*{\vb*{E}\cdot\vb*{B}}}{\mean*{B^2}}
	\end{align}
	are the so-called thermodynamic forces.
	
	Mathematically speaking, there are still two additional conditions to completely determine the solution to equation (\ref{eq:DKE_Original}). First, equation (\ref{eq:DKE_Original}) must be solved imposing regularity conditions at $\xi =\pm 1$
	\begin{align}
		\eval{\left((1-\xi^2) \pdv{h_a}{\xi}\right)}_{\xi =\pm 1} = 0.
		\label{eq:Regularity_conditions}
	\end{align}
	Second, as the differential operator on the left-hand-side of equation (\ref{eq:DKE_Original}) has a non trivial kernel, the solution to equation (\ref{eq:DKE_Original}) is determined up to an additive function $g(\psi,v)$. This function is unimportant as it does not contribute to the neoclassical transport quantities of interest. Nevertheless, in order to have a unique solution to the DKE, it must be fixed by imposing an appropriate additional constraint. We will select this free function (for fixed $(\psi,v)$) by imposing
	\begin{align}
		\mean*{  \int_{-1}^{1} h_a \dd{\xi}  } = C,
		\label{eq:kernel_elimination_condition}
	\end{align}
	for some $C\in\mathbb{R}$. We will discuss this further in section \ref{subsec:Legendre_expansion}. 
	
	The DKE (\ref{eq:DKE_Original}) is the one solved by the standard neoclassical code {\DKES} \cite{DKES1986, VanRij_1989} using a variational principle. 
	Although the main feature of the code \texttt{SFINCS} \cite{Landreman_2014} is to solve a more complete non monoenergetic neoclassical DKE, it can also solve equation (\ref{eq:DKE_Original}). As it will be explained in chapter \ref{chap:MONKES}, this equation is also solved by the neoclassical code {\MONKES}, developed as part of this thesis project.
	
	Taking the moments $\{\vb*{v}_{\text{m} a} \cdot \nabla\psi,  (v^2/v_{\text{t}a}^2)\vb*{v}_{\text{m} a} \cdot \nabla\psi, v\xi B/B_0\}$ of $h_a$ and then the flux surface average yields, respectively, the radial particle flux, the radial heat flux and the parallel flow
	\begin{align}
		\mean*{\vb*{\Gamma}_a \cdot \nabla \psi} & := 
		\mean*{
			\int
			\vb*{v}_{\text{m} a} \cdot \nabla\psi	
			\ h_a
			\dd[3]{\vb*{v}}
		},
		\label{eq:Particle_flux_Original}
		\\
		\mean*{\frac{\vb*{Q}_a \cdot \nabla \psi}{T_a}} & := 
		\mean*{
			\int
			\frac{v^2}{v_{\text{t}a}^2}\vb*{v}_{\text{m} a} \cdot \nabla\psi	
			\ h_a
			\dd[3]{\vb*{v}}
		},
		\label{eq:Heat_flux_Original}
		\\
		\frac{\mean*{n_a \vb*{V}_{a} \cdot\vb*{B}}}{B_{0}} & :=
		\mean*{
			\frac{B}{B_0}
			\int
			v \xi 
			\ h_a
			\dd[3]{\vb*{v}}
		},
		\label{eq:Parallel_flow_Original}
	\end{align}
	where $\vb*{\Gamma}_a:= n_a \vb*{V}_a$ and $B_0(\psi)$ is a reference value for the magnetic field strength on the flux surface (its explicit definition is given in section \ref{chap:MONKES}).
	
	It is a common practice for linear drift-kinetic equations (e.g. \cite{DKES1986, Beidler_2011,Landreman_2014}) to apply superposition and split $h_a$ into several additive terms. As in the DKE (\ref{eq:DKE_Original}) there are no derivatives or integrals along $\psi$ nor $v$, it is convenient to use the splitting
	\begin{align}
		h_a 
		= 
		f_{\text{M}a}
		\left[
		\frac{B v}{\Omega_a} 
		\left(
		A_{1a} f_1 
		+ 
		A_{2a}  
		\frac{v^2}{v_{\text{t}a}^2}f_2
		\right)
		+
		A_{3a} f_3
		\right].
		\label{eq:Distribution_function_superposition}
	\end{align}
	The splitting is chosen so that the functions $\{f_j\}_{j=1}^{3}$ are solutions to
	\begin{align}
		\xi \vb*{b}  \cdot 
		\nabla f_j
		& +
		\nabla \cdot \vb*{b} \frac{(1-\xi^2)}{2}  \pdv{f_j}{\xi}  
		- 
		\frac{\widehat{E}_\psi}{\mean*{B^2}}
		\vb*{B}\times \nabla\psi\cdot \nabla f_j
		\label{eq:DKE}
		- \hat{\nu}\Lorentz f_j
		=  s_j, \quad 
	\end{align}
	for $j=1,2,3$, where $\hat{\nu} := \nu(v) / v$ and $\widehat{E}_\psi := {E}_\psi/v$. The source terms are defined as
	\begin{align}
		s_1 := - \vb*{v}_{\text{m} a} \cdot \nabla\psi \frac{\Omega_a}{B v^2},
		\quad
		s_2 :=  s_1, 
		\quad
		s_3 := \xi \frac{B}{B_0}.
		\label{eq:DKE_Sources}
	\end{align} 
	Note that each source $s_j$ corresponds to one of the three thermodynamic forces on the right-hand side of definition (\ref{eq:DKE_Original_Source}).
	
	The relation between $h_a$ and $f_j$ given by equation (\ref{eq:Distribution_function_superposition}) is such that the transport quantities (\ref{eq:Particle_flux_Original}), (\ref{eq:Heat_flux_Original}) and (\ref{eq:Parallel_flow_Original}) can be written in terms of four transport coefficients which, for fixed $(\hat{\nu}, \widehat{E}_\psi)$, depend only on the magnetic configuration. As $\dv*{\hat{\nu}}{v}$ never vanishes, the dependence of $f_j$ on the velocity $v$ can be parametrized by its dependence on $\hat{\nu}$. Thus, for fixed $(\hat{\nu}, \widehat{E}_\psi)$, equation (\ref{eq:DKE}) is completely determined by the magnetic configuration. Hence, its unique solutions $f_j$ that satisfy conditions (\ref{eq:Regularity_conditions}) and (\ref{eq:kernel_elimination_condition}) are also completely determined by the magnetic configuration. 
	
	Using splitting (\ref{eq:Distribution_function_superposition}) we can write the transport quantities (\ref{eq:Particle_flux_Original}), (\ref{eq:Heat_flux_Original}) and (\ref{eq:Parallel_flow_Original}) in terms of the Onsager matrix
	\begin{align}
		&\Matrix{c}
		{
			\mean*{\vb*{\Gamma}_a \cdot \nabla \psi} \\
			\mean*{ \dfrac{\vb*{Q}_a \cdot \nabla \psi}{T_a} }     \\ 
			\dfrac{\mean*{n_a \vb*{V}_{a} \cdot\vb*{B}}}{B_0}
		}
		=
		\Matrix{ccc}
		{
			L_{11a} & L_{12a}  & L_{13a} \\
			L_{21a} & L_{22a}  & L_{23a} \\
			L_{31a} & L_{32a}  & L_{33a} 
		}
		\Matrix{c}
		{ 
			A_{1a} \\
			A_{2a} \\
			A_{3a} 
		}.
	\end{align}
	Here, we have defined the thermal transport coefficients as 
	\begin{align}
		L_{ija} :=    
		\int_{0}^{\infty}
		2\pi v^2
		f_{\text{M}a} 
		w_i w_j 
		D_{ija} 
		\dd{v}, \ \ %
		\label{eq:Thermal_transport_coefficients}
	\end{align}
	where $w_1=w_3=1$, $w_2=v^2/v_{\text{t}a}^2$ and we have used that $\int g\dd[3]{\vb*{v}} = 2\pi \int_{0}^{\infty}\int_{-1}^{1} g v^2 \dd{\xi}\dd{v}$ for any integrable function $g(\xi,v)$. The quantities $D_{ija}$ are defined as
	\begin{align}
		D_{ija} & :=C_{ija} \Dij{ij},
	\end{align} 
	where
	\begin{align}
		C_{ija} & :=-\frac{B^2v^3}{\Omega_a^2}, &\quad i,j \in\{1,2\},
		\\
		C_{i3a} & :=  
		- \frac{B v^2}{\Omega_a} , &\quad i \in\{1,2\},
		\\
		C_{3ja} & := \frac{B v^2}{\Omega_a} = -C_{j3a} , &\quad j \in\{1,2\},
		\\
		C_{33a} & := v  , &
	\end{align}
	are species-dependent factors and  
	\begin{align}
		\widehat{D}_{ij}(\psi,v) := \mean*{ \int_{-1}^{1}  s_i f_j   \dd{\xi} }, \quad i,j\in\{1,2,3\}
		\label{eq:Monoenergetic_geometric_coefficients}
	\end{align} 
	are the monoenergetic geometric coefficients. Note that (unlike $D_{ija}$) the monoenergetic geometric coefficients $\widehat{D}_{ij}$ do not depend on the species for fixed $\hat{\nu}$ (however the correspondent value of $v$ associated to each $\hat{\nu}$ varies between species) and depend only on the magnetic geometry. In general, four independent monoenergetic geometric coefficients can be obtained by solving (\ref{eq:DKE}): $\widehat{D}_{11}$, $\widehat{D}_{13}$, $\widehat{D}_{31}$ and $\widehat{D}_{33}$. However, when the magnetic field possesses stellarator symmetry \cite{DEWAR1998275} or there is no radial electric field, Onsager symmetry implies $\widehat{D}_{13} = -\widehat{D}_{31}$ \cite{VanRij_1989} making only three of them independent (for further details see section \ref{sec:Appendix_Onsager_symmetry}). Note that when the monoenergetic transport coefficients satisfy Onsager symmetry, the Onsager matrix is symmetric, i.e. $L_{ija}=L_{jia}$. Hence, obtaining the transport coefficients for all species requires to solve (\ref{eq:DKE}) for two different source terms $s_1$ and $s_3$. The algorithm for solving equation (\ref{eq:DKE}) is described in section \ref{chap:MONKES}. 
	
	Finally, we briefly comment on the validity of the coefficients provided by equation (\ref{eq:DKE}) for the calculation of the bootstrap current. The pitch-angle scattering collision operator used in equation (\ref{eq:DKE_Original}) lacks parallel momentum conservation. Besides, the pitch-angle scattering operator is not adequate for calculating parallel flow of electrons, which is a quantity required to compute the bootstrap current. Hence, in principle, the parallel transport directly predicted by equation (\ref{eq:DKE_Original}) is not correct. Fortunately, there exist techniques  \cite{Taguchi,Sugama-PENTA,Sugama2008,MaasbergMomentumCorrection} to calculate the radial and parallel transport associated to more accurate momentum-conserving collision operators by just solving the simplified DKE (\ref{eq:DKE}). This has been done successfully in the past by the code \texttt{PENTA} \cite{Sugama-PENTA, Spong-PENTA}, using the results of {\DKES}. Nevertheless, the momentum-restoring technique is not needed for minimizing the bootstrap current. In the method presented in section V of \cite{MaasbergMomentumCorrection}, when there is no net parallel inductive electric field (i.e. $A_{3a}=0$), the parallel flow with the correct collision operator for any species vanishes when two integrals in $v$ of $\widehat{D}_{31}$ vanish. Thus, minimizing $\widehat{D}_{31}$ translates in a minimization of the parallel flows of all species involved in the bootstrap current calculation, and therefore of this current.  
	
	\subsection{Adjoint properties of the drift-kinetic equation}\label{sec:DKE_adjoint_properties}
	
	In this section, some mathematical properties of the DKE (\ref{eq:DKE}) will be reviewed. These properties are important for deriving Onsager symmetry relations of the transport coefficients (which will be done in section \ref{sec:Appendix_Onsager_symmetry}). Additionally, these properties will allow us to derive an adjoint method for computing derivatives of the transport coefficients with respect to parameters upon which the solution to the DKE (\ref{eq:DKE}) depends (which will be done in section \ref{sec:Derivatives_monoenergetic}). For each fixed value of the collisionality $\hat{\nu}$ and radial electric field $\Epsi$, the left-hand side of the DKE (\ref{eq:DKE}) can be interpreted as a linear operator which, given a magnetic field $\vb*{B}$ and a flux surface defined from the isosurfaces of $\psi$, acts on a smooth function $f_j$ to produce another smooth function $s_j$ (the coefficients are smooth). Thus, it is a linear operator from the space of smooth functions defined in $\mathcal{M}:=\mathbb{T}\times[-1,1]$, where $\mathbb{T}$ is the surface of the topological torus given by an isosurface of $\psi$. We denote this space of functions by $\Fsmooth$. We can rewrite the DKE (\ref{eq:DKE}) in a compact manner by defining a linear Vlasov operator
	\begin{align}
		\VV f:= \nabla\cdot( \xi \vb*{b} f + \widehat{\vb*{v}}_E f) +
		\pdv{\xi}\left(\frac{1}{2}(1-\xi^2)\nabla\cdot\vb*{b} f \right),
		\label{eq:Vlasov_operator_MONKES_equation}
	\end{align}
	containing the collisionless trajectories so that the DKE (\ref{eq:DKE}) can be written as
	\begin{align}
		\left(
		\VV - \hat{\nu}\Lorentz
		\right) f_j = s_j.
		\label{eq:DKE_compact}
	\end{align}
	Here, we have denoted $\widehat{\vb*{v}}_E:= \vb*{v}_E / v$ and have used the property $\nabla\cdot \vb*{v}_E = 0$ to write $\VV$ in a divergence form.

	It is useful to endow $\Fsmooth$ with an inner product
	\begin{align}
		\mean*{f,g}  := \mean*{\int_{-1}^{1} fg \dd{\xi}}.
		\label{eq:Monoenergetic_inner_product}
	\end{align} 
	In terms of the inner product, we can rewrite the monoenergetic transport coefficients (\ref{eq:Monoenergetic_geometric_coefficients}) as 
	\begin{align}
		\Dij{ij} = \mean*{s_i,f_j} .
		\label{eq:Monoenergetic_geometric_coefficients_inner_product}
	\end{align}

	It is well known that the operators $\VV$ and $\Lorentz$ satisfy the symmetry properties \cite{DKES1986}
	\begin{align}
		\mean*{\VV f,g}  = - \mean*{f,\VV g} ,
		\label{eq:Vlasov_skew_self_adjoint}
		\\ 
		\mean*{\Lorentz f,g}  =  \mean*{f,\Lorentz g} 
		\label{eq:Lorentz_self_adjoint}
	\end{align}
	reflecting that $\Lorentz$ and $\VV$ are, respectively symmetric (self-adjoint) and antisymmetric (skew-self-adjoint) with respect to this inner product. For obtaining the identity (\ref{eq:Vlasov_skew_self_adjoint}), we have used $\vb*{b}\cdot\nabla\psi = \vb*{v}_E\cdot\nabla\psi=0$ and property (\ref{eq:FSA_divergence}). For the symmetry of $\Lorentz$ see appendix \ref{sec:Appendix_Legendre}.
	
	Thus, identities (\ref{eq:Vlasov_skew_self_adjoint}) and (\ref{eq:Lorentz_self_adjoint}) imply that the adjoint of the differential operator $\VV-\hat{\nu}\Lorentz$ at the left-hand side of the DKE is given by
	\begin{align}
		\left(
		\VV - \hat{\nu}\Lorentz
		\right)^\dagger
		=
		-\VV - \hat{\nu}\Lorentz,
	\end{align}
	where the superscript $\dagger$ is used to indicate the adjoint of a linear operator. 
	
	It is useful to consider the solution to the adjoint problem for the three sources of (\ref{eq:DKE_compact})
	\begin{align}
		\left(
		\VV - \hat{\nu}\Lorentz
		\right)^\dagger f_i^\dagger = s_i.
		\label{eq:Adjoint_DKE_compact}
	\end{align}
	If we define the \textit{adjoint monoenergetic coefficients} $\Dij{ij}^\dagger$ as the monoenergetic coefficients given by the solution to the adjoint problem (\ref{eq:Adjoint_DKE_compact})
	\begin{align}
		\Dij{ij}^\dagger := \mean*{s_i,f_j^\dagger} ,
		\label{eq:Adjoint_monoenergetic_geometric_coefficients_inner_product}
	\end{align} 
	we can obtain the identity
	\begin{align}
		\Dij{ij}  = \Dij{ji}^\dagger,
		\label{eq:Monoenergetic_adjoint_relation}
	\end{align}	
	by projecting equation (\ref{eq:Adjoint_DKE_compact}) along the solution $f_j$ to (\ref{eq:DKE_compact}) and using the definition of adjoint. 
	
	Similarly, we can define the \textit{adjoint thermal transport coefficients} $L_{ija}^\dagger$ replacing $D_{ija}$ by $D_{ija}^\dagger:=C_{ija}\Dij{ij}^\dagger$ in (\ref{eq:Thermal_transport_coefficients}). Now, we can integrate (\ref{eq:Monoenergetic_adjoint_relation}) along $v$, weighted as in (\ref{eq:Thermal_transport_coefficients}), to obtain
	\begin{align*}
		L_{ija} & = L_{jia}^\dagger, & \quad i,j\in\{1,2\},
		\\
		L_{i3a} & =- L_{3ia}^\dagger, &\quad i\in\{1,2\},
		\\
		L_{3ja} & =- L_{j3a}^\dagger, &\quad j\in\{1,2\},
		\\
		L_{33a} & = L_{33a}^\dagger,		 &
	\end{align*}
	which can be written in a compact manner as
	\begin{align}
		L_{ija} & = (-1)^{\delta_{3i}+\delta_{3j}}
		L_{jia}^\dagger, & \quad i,j\in\{1,2,3\}
		\label{eq:Thermal_coefficients_adjoint_relation},
	\end{align}
	where $\delta_{ij}$ is the delta Kronecker symbol.
	
	Whenever Onsager symmetry is fulfilled, $L_{ija}=L_{jia}$ and the relation (\ref{eq:Thermal_coefficients_adjoint_relation}) becomes 
	\begin{align}
		L_{ija} & = (-1)^{\delta_{3i}+\delta_{3j}}
		L_{ija}^\dagger, & \quad i,j\in\{1,2,3\}
		\label{eq:Thermal_coefficients_adjoint_relation_Onsager}.
	\end{align}
	Relation (\ref{eq:Thermal_coefficients_adjoint_relation_Onsager}) reflects that, when the Onsager matrix is symmetric and there is no externally applied loop voltage ($A_{3a}=0$), the neoclassical fluxes and flows predicted by the DKE and its adjoint version are closely related. For fixed plasma profiles (i.e. fixed $A_{1a}$ and $A_{2a}$) radial neoclassical transport is identical ($L_{ija}=L_{ija}^\dagger$ for $i,j\in{1,2}$) and the parallel flow of each species is the opposite ($L_{3ja}=-L_{3ja}^\dagger$ for $j\in{1,2}$).

	\subsection{Onsager symmetry of the transport coefficients} 
	\label{sec:Appendix_Onsager_symmetry}
	In this section, it will be proven that the monoenergetic coefficients $\widehat{D}_{ij}$ defined by (\ref{eq:Monoenergetic_geometric_coefficients}) satisfy Onsager symmetry relations \cite{Onsager_1,Onsager_2} whenever there is no electric field $E_\psi=0$ or the magnetic field possesses stellarator symmetry. For this, we will prove a more general result involving linear equations defined in some domain (phase-space) $\mathcal{S}$. Suppose we have a space $\mathcal{F_S}$ of functions from $\mathcal{S}$ to $\mathbb{R}$ with inner product $\mean*{\cdot,\cdot}_{\mathcal{S}}$ and a set of linear equations
	\begin{align}
		\mathcal{V} f_j - \mathcal{C}f_j = s_j,
		\label{eq:Onsager_symmetry_DKE}
	\end{align}
	for $j=1,2\ldots, N_{\text{e}} $ where $s_j\in\mathcal{F_S}$ and the linear operators $\mathcal{C}$ and $\mathcal{V}$ are respectively symmetric and antisymmetric with respect to $\mean*{\cdot,\cdot}_{\mathcal{S}}$. Namely,
	\begin{align}
		\mean*{\mathcal{C}f,g}_{\mathcal{S}} & = \mean*{f,\mathcal{C}g}_{\mathcal{S}}, 
		\label{eq:Onsager_symmetry_Collision_Symmetry}\\
		\mean*{\mathcal{V}f,g}_{\mathcal{S}} & = -\mean*{f,\mathcal{V}g}_{\mathcal{S}}.
		\label{eq:Onsager_symmetry_Vlasov_Skew_Symmetry}
	\end{align}

	Now, we define the scalars 
	\begin{align}
		\mathcal{D}_{ij} := \mean*{ s_i, f_j}_{\mathcal{S}}
		,
		\label{eq:Onsager_symmetry_coefficients}
	\end{align}
	for $i,j=1,2\ldots, N_{\text{e}} $.
	
	Additionally, we define a property $\mathcal{P}$ to be a map which associates to each $f\in\mathcal{F_S}$ a function $\mathcal{P} f \in\mathcal{F_S}$ and is idempotent\footnote{This means that, for all $f\in\mathcal{F_S}$, $\mathcal{P} \mathcal{P} f=f$.}. Any function $f\in\mathcal{F_S}$ can be splitted in its even $f^+$ and odd $f^-$ portions with respect to the property $\mathcal{P}$ as follows
	\begin{align}
		f^\pm : = 
		\frac{1}{2}
		\left(
		f \pm \mathcal{P} f
		\right),
	\end{align}
	satisfying $\mathcal{P} f^\pm = \pm f^\pm$. Without loss of generality, we assume that $N^+\le N_{\text{e}}$ sources $s_j$ in (\ref{eq:Onsager_symmetry_DKE}) are even with respect to $\mathcal{P}$ and the remaining $N^- := N_{\text{e}}- N^+$ sources are odd. 
	
	The coefficients $\mathcal{D}_{ij}$ satisfy Onsager symmetry relations if three (sufficient) conditions are satisfied. 
	\begin{enumerate}
		\item Even and odd functions are mutually orthogonal $\mean*{f^\pm, g^\mp}_\mathcal{S}=0$. This implies that
		\begin{align}
			\mean*{f,g}_{\mathcal{S}} = \mean*{f^+,g^+}_{\mathcal{S}} + \mean*{f^-,g^-}_{\mathcal{S}}.
			\label{eq:Onsager_symmetry_orthogonality_Even_Odd}
		\end{align}
		
		\item The operator $\mathcal{C}$ is even with respect to property $\mathcal{P}$. Explicitly,
		\begin{align}
			(\mathcal{C} f)^\pm & = \mathcal{C} f^\pm.
			\label{eq:Onsager_symmetry_Collisions_Even}
		\end{align}
		
		\item The operator $\mathcal{V}$ is odd with respect to property $\mathcal{P}$. Explicitly,
		\begin{align}
			(\mathcal{V} f)^\pm & = \mathcal{V} f^\mp.
			\label{eq:Onsager_symmetry_Vlasov_Odd}
		\end{align}
	\end{enumerate}
	When conditions (\ref{eq:Onsager_symmetry_orthogonality_Even_Odd}), (\ref{eq:Onsager_symmetry_Collisions_Even}) and (\ref{eq:Onsager_symmetry_Vlasov_Odd}) are satisfied we have the following Onsager symmetry relations.
	\begin{itemize}
		\item For fixed $i$ and $j$, if $s_i$ and $s_j$ are both even, $\mathcal{D}_{ij} = \mathcal{D}_{ji}$. The proof is as follows 
		\begin{align*}
			\mathcal{D}_{ij} & = \mean*{s_i^+, f_j^+}_{\mathcal{S}} \\
			& 
			= \mean*{\mathcal{V} f_i^-, f_j^+}_{\mathcal{S}}
			- \mean*{\mathcal{C} f_i^+, f_j^+}_{\mathcal{S}} \\
			& 
			= - \mean*{f_i^-, \mathcal{V} f_j^+}_{\mathcal{S}}
			- \mean*{\mathcal{C} f_i^+, f_j^+}_{\mathcal{S}} \\
			& 
			= - \mean*{f_i^-, \mathcal{C} f_j^-}_{\mathcal{S}}
			- \mean*{\mathcal{C} f_i^+, f_j^+}_{\mathcal{S}}\\
			& 
			= - \mean*{f_i, \mathcal{C} f_j}_{\mathcal{S}}.
		\end{align*}
		As in the last equality, due to (\ref{eq:Onsager_symmetry_Collision_Symmetry}), the roles of $i$ and $j$ are interchangeable, we have that $\mathcal{D}_{ij} = \mathcal{D}_{ji}$.
		
		\item For fixed $i$ and $j$, if $s_i$ and $s_j$ are both odd, $\mathcal{D}_{ij} = \mathcal{D}_{ji}$. The proof is as follows 
		\begin{align*}
			\mathcal{D}_{ij} & = \mean*{s_i^-, f_j^-}_{\mathcal{S}} \\
			& 
			= \mean*{\mathcal{V} f_i^+, f_j^-}_{\mathcal{S}}
			- \mean*{\mathcal{C} f_i^-, f_j^-}_{\mathcal{S}} \\
			& 
			= - \mean*{f_i^+, \mathcal{V} f_j^-}_{\mathcal{S}}
			- \mean*{\mathcal{C} f_i^-, f_j^-}_{\mathcal{S}} \\
			& 
			= - \mean*{f_i^+, \mathcal{C} f_j^+}_{\mathcal{S}}
			- \mean*{\mathcal{C} f_i^-, f_j^-}_{\mathcal{S}} \\
			& 
			= - \mean*{f_i, \mathcal{C} f_j}_{\mathcal{S}}.
		\end{align*}
		As in the last equality, due to (\ref{eq:Onsager_symmetry_Collision_Symmetry}), the roles of $i$ and $j$ are interchangeable, we have that $\mathcal{D}_{ij} = \mathcal{D}_{ji}$.
		
		\item For fixed $i$ and $j$, if $s_i$ is even and $s_j$ is odd, $\mathcal{D}_{ij} = -\mathcal{D}_{ji}$. The proof is as follows 
		\begin{align*}
			\mathcal{D}_{ij} & = \mean*{s_i^+, f_j^+}_{\mathcal{S}} \\
			& 
			= \mean*{\mathcal{V} f_i^-, f_j^+}_{\mathcal{S}}
			- \mean*{\mathcal{C} f_i^+, f_j^+}_{\mathcal{S}} \\
			& 
			= \mean*{\mathcal{V} f_i^-, f_j^+}_{\mathcal{S}}
			- \mean*{f_i^+, \mathcal{C}  f_j^+}_{\mathcal{S}} \\
			& 
			= \mean*{\mathcal{V} f_i^-, f_j^+}_{\mathcal{S}}
			- \mean*{f_i^+, \mathcal{V}  f_j^-}_{\mathcal{S}} \\
			& 
			= \mean*{\mathcal{V} f_i, f_j}_{\mathcal{S}}.
		\end{align*}
		As in the last equality, due to (\ref{eq:Onsager_symmetry_Vlasov_Skew_Symmetry}), interchanging the roles of $i$ and $j$ switches signs, we have that $\mathcal{D}_{ij} = -\mathcal{D}_{ji}$.
		
	\end{itemize}
	
	With the three sufficient conditions (\ref{eq:Onsager_symmetry_orthogonality_Even_Odd}), (\ref{eq:Onsager_symmetry_Collisions_Even}) and (\ref{eq:Onsager_symmetry_Vlasov_Odd}) we can prove that the transport coefficients obtained from solving equation (\ref{eq:DKE}) satisfy Onsager symmetry for zero radial electric field and for stellarator-symmetric devices. In this case, the phase-space is $\mathcal{S}=\MM$. Note that the DKE (\ref{eq:DKE}) can be readily written in the form of (\ref{eq:Onsager_symmetry_DKE}) by setting
	the Vlasov and collision operators to match those of equation (\ref{eq:DKE_compact}). Namely,
	\begin{align}
		\mathcal{V} & := \xi \vb*{b}\cdot \nabla + \nabla\cdot\vb*{b} \frac{1-\xi^2}{2}\pdv{\xi}
		-  \frac{\widehat{E}_\psi}{\mean*{B^2}} \vb*{B}\times \nabla \psi \cdot \nabla,
		\\ 
		\mathcal{C} & := \hat{\nu} \Lorentz,
	\end{align}
	and the inner product to be the one given in (\ref{eq:Monoenergetic_inner_product})
	\begin{align}
		\mean*{f,g}_{\mathcal{S}}= \mean*{f,g}  = \mean*{\int_{-1}^{1}fg\dd{\xi}}.
		\label{eq:Onsager_Inner_Product}
	\end{align}
	
	With these definitions, we can check from identities (\ref{eq:Lorentz_self_adjoint}) and (\ref{eq:Vlasov_skew_self_adjoint}) that properties (\ref{eq:Onsager_symmetry_Collision_Symmetry}) and (\ref{eq:Onsager_symmetry_Vlasov_Skew_Symmetry}) are satisfied and $\mathcal{D}_{ij} = \widehat{D}_{ij} $. It is interesting to remark that the antisymmetry property (\ref{eq:Onsager_symmetry_Vlasov_Skew_Symmetry}) of $\mathcal{V}$ implies that the diagonal monoenergetic coefficients $\widehat{D}_{ii} $ are always positive. Note first that (\ref{eq:Onsager_symmetry_Vlasov_Skew_Symmetry}) implies $\mean*{f, \mathcal{V} f}_{\mathcal{S}} = 0$ for any $f\in \mathcal{F}_{\mathcal{S}}$. This implies that $\widehat{D}_{ii} = - \mean*{f_i, \hat{\nu}\Lorentz f_i}_{\mathcal{S}} $ and, as $\Lorentz$ is a negative operator (its eigenvalues are all negative or zero, see appendix \ref{sec:Appendix_Legendre}), $\widehat{D}_{ii} \ge 0$. Also note that properties (\ref{eq:Onsager_symmetry_Collision_Symmetry}) and (\ref{eq:Onsager_symmetry_Vlasov_Skew_Symmetry}) imply that $\mean*{\hat{\nu}\mathcal{L}f_j,1}_{\mathcal{S}} = 0$ and $\mean*{\mathcal{V}f_j,1}_{\mathcal{S}} = 0$. Thus, if the source term $s_j$ of the DKE (\ref{eq:DKE}) belongs to the image of the operator $\VV-\hat{\nu}\Lorentz$ on the left-hand side of the DKE (\ref{eq:DKE}), it is constrained by $\mean*{s_j,1}_{\mathcal{S}} = 0$.
	
	Now we distinguish the two cases for which the monoenergetic coefficients $\widehat{D}_{ij} $ satisfy Onsager symmetry relations. Apart from the velocity coordinate $\xi$, we will use Boozer angles $( {\theta},\zeta)$.
	\begin{enumerate}
		\item If $E_\psi=0$, the property is defined as 
		\begin{align}
			\mathcal{P} f( {\theta},\zeta,\xi) = f( {\theta},\zeta,-\xi).
		\end{align}
		It is straightforward to check that for this property, conditions (\ref{eq:Onsager_symmetry_orthogonality_Even_Odd}), (\ref{eq:Onsager_symmetry_Collisions_Even}) and (\ref{eq:Onsager_symmetry_Vlasov_Odd}) are satisfied. Also, $s_1= s_1^+$, $s_2 = s_2^+$ and $s_3 = s_3^-$. Hence, we have $\widehat{D}_{12}=\widehat{D}_{21}$, $\widehat{D}_{13}=-\widehat{D}_{31}$ and $\widehat{D}_{23}=-\widehat{D}_{32}$.

		\item When $E_\psi$ is not necessarily zero, we define the property $\mathcal{P}$ as the one that defines stellarator symmetry \cite{DEWAR1998275}
		\begin{align}
			\mathcal{P} f( {\theta},\zeta,\xi) = f( - {\theta}, - \zeta,\xi)
		\end{align}
		and we have assumed without loss of generality that the planes of symmetry are $\theta=0$ and $ \zeta =0$. Thus, when the magnetic field is stellarator-symmetric $B=B^+$. In this case, using (\ref{eq:FSA_Boozer}), (\ref{eq:Parallel_streaming_spatial_operator}) and (\ref{eq:ExB_spatial_operator}) it is straightforward to check\footnote{Note that derivatives along $\theta$ and $\zeta$ switch parities with respect to the stellarator symmetry property, i.e. $\pdv*{f^\pm}{\theta} = (\pdv*{f^\pm}{\theta} )^\mp$ and $\pdv*{f^\pm}{\zeta} = (\pdv*{f^\pm}{\zeta} )^\mp$. Also, as for stellarator-symmetric fields, $\sqrt{g}=\sqrt{g}^+$ the flux surface average satisfies $\mean*{f^-}=0$.} that conditions (\ref{eq:Onsager_symmetry_orthogonality_Even_Odd}), (\ref{eq:Onsager_symmetry_Collisions_Even}) and (\ref{eq:Onsager_symmetry_Vlasov_Odd}) are satisfied. Besides, $s_1=s_1^-$, $s_2 = s_2^-$ and $s_3 = s_3^+$. Hence, we have $\widehat{D}_{12}=\widehat{D}_{21}$, $\widehat{D}_{13}=-\widehat{D}_{31}$ and $\widehat{D}_{23}=-\widehat{D}_{32}$. 
	\end{enumerate}
	
	Note that for equation (\ref{eq:DKE}), the Onsager symmetry relation $\widehat{D}_{12}=\widehat{D}_{21}$ is trivial as $s_1=s_2$, which implies $f_1=f_2$ and thus $\widehat{D}_{12}=\widehat{D}_{21} = \widehat{D}_{11}= \widehat{D}_{22} $, $\widehat{D}_{31} = \widehat{D}_{32}$ and $\widehat{D}_{13} = \widehat{D}_{23}$. Nevertheless, if the definition of $s_1$ and $s_2$ was different, as long as their parity is the same, the relation $\widehat{D}_{12}=\widehat{D}_{21}$ would still hold.

	\section{Representation of the monoenergetic DKE in Legendre space}	
	\label{sec:Legendre_representation}
	In this section, an algorithm to solve the DKE (\ref{eq:DKE}) is presented. The algorithm, based on the tridiagonal representation of the DKE, emerges naturally when the velocity coordinate $\xi$ is discretized using a Legendre spectral method. We will present the algorithm in a formal way and describe some features of it. After that, in section \ref{subsec:Monoenergetic_lambda} we show how to use the solution to the DKE to elucidate which classes of particles contribute the most to the different monoenergetic coefficients.

	We will use (right-handed) Boozer coordinates\footnote{Even though we use Boozer coordinates, we want to stress out that the algorithm presented in subsection \ref{subsec:Legendre_expansion} is valid for any set of spatial coordinates in which $\psi$ labels flux surfaces and the two remaining coordinates parametrize the flux surface.} $(\psi,\theta,\zeta)\in[0,\psi_{\text{lcfs}}]\times[0,2\pi)\times[0,2\pi/\Nfp)$. The integer $\Nfp\ge 1$ denotes the number of toroidal periods of the device. The radial coordinate is selected so that $2\pi \psi$ is the toroidal flux of the magnetic field and $\theta$, $\zeta$ are respectively the poloidal and toroidal (in a single period) angles. As stated in section \ref{sec:Force_balance}, in these coordinates the magnetic field can be written as
	\begin{align}
		\vb*{B} & = \nabla\psi \times \nabla\theta - \iota(\psi) \nabla\psi \times \nabla\zeta 
		\nonumber\\
		& = B_\psi(\psi,\theta,\zeta) \nabla \psi + B_\theta(\psi) \nabla \theta + B_\zeta(\psi) \nabla \zeta,
		\label{eq:Magnetic_field_Boozer}
	\end{align}
	and the Jacobian of the transformation reads 
	\begin{align}
		\sqrt{g}(\psi,\theta,\zeta) 
		:=( 
		\nabla\psi \times \nabla \theta \cdot \nabla\zeta  
		)^{-1} 
		= 
		\frac{B_\zeta + \iota B_\theta}{B^2}.
		\label{eq:Jacobian_Boozer}
	\end{align} 
	The flux surface average operation (\ref{eq:FSA}) is written in Boozer angles as
	\begin{align}
		\mean*{f}
		=
		\left(\dv{V}{\psi}\right)^{-1}
		\oint\oint
		f
		\sqrt{g}
		\dd{\theta}\dd{\zeta}
		.
		\label{eq:FSA_Boozer}
	\end{align}
	
	We define the reference value for the magnetic field strength $B_0$ introduced in definition (\ref{eq:Parallel_flow_Original}) as the $(0,0)$ Fourier mode of the magnetic field strength. Namely, 
	\begin{align}
		B_0(\psi) := \frac{\Nfp}{4\pi^2 } 
		\oint\oint
		B(\psi,\theta,\zeta)
		\dd{\theta}\dd{\zeta}.
	\end{align}
	
	Using (\ref{eq:Magnetic_field_Boozer}) and (\ref{eq:Jacobian_Boozer}), the spatial differential operators present in the DKE (\ref{eq:DKE}) can be expressed in these coordinates as
	\begin{align}
		\vb*{b} \cdot \nabla & = 
		\frac{B}{B_\zeta + \iota B_\theta}
		\left(
		\iota \pdv{\theta}
		+ 
		\pdv{\zeta} 
		\right), \label{eq:Parallel_streaming_spatial_operator}
		\\
		\vb*{B}\times\nabla\psi \cdot \nabla & = 
		\frac{B^2}{B_\zeta + \iota B_\theta}
		\left(
		B_\zeta \pdv{\theta}
		-
		B_\theta \pdv{\zeta}
		\right). \label{eq:ExB_spatial_operator}
	\end{align}

	In order to ease the notation, in this section we will drop, when possible, the subscript $j$ that labels every different source term of the DKE (\ref{eq:DKE}). Also, as $\psi$ and $v$ act as mere parameters, we will omit their dependence and functions of these two variables will be referred to as constants.

	The algorithm is based on the approximate representation of the distribution function $f$ by a truncated Legendre series. We will search for approximate solutions to equation (\ref{eq:DKE}) of the form
	\begin{align}
		f(\theta,\zeta,\xi) = \sum_{k=0}^{N_\xi} f^{(k)}(\theta,\zeta) P_k(\xi), \label{eq:Legendre_expansion}
	\end{align} 
	where $f^{(k)} = \mean*{f,P_k}_\Lorentz / \mean*{P_k,P_k}_\Lorentz$ is the $k-$th Legendre mode of $f(\theta,\zeta,\xi)$ (see appendix \ref{sec:Appendix_Legendre}) and $N_\xi$ is an integer greater or equal to 1. As mentioned in appendix \ref{sec:Appendix_Legendre}, the expansion in Legendre polynomials (\ref{eq:Legendre_expansion}) ensures that the regularity conditions (\ref{eq:Regularity_conditions}) are satisfied. Of course, in general, the exact solution to equation (\ref{eq:DKE}) does not have a finite Legendre spectrum, but taking $N_\xi$ sufficiently high in expansion (\ref{eq:Legendre_expansion}) yields an approximate solution to the desired degree of accuracy (in infinite precision arithmetic).

	In appendix \ref{sec:Appendix_Legendre} we derive explicitly the projection of each term of the DKE (\ref{eq:DKE}) onto the Legendre basis when the representation (\ref{eq:Legendre_expansion}) is used. When doing so, we obtain that the Legendre modes of the DKE have the tridiagonal representation  
	\begin{align}
		L_k f^{(k-1)} + D_k f^{(k)} + U_k f^{(k+1)} = s^{(k)},  
		\label{eq:DKE_Legendre_expansion}
	\end{align}
	for $k=0,1,\ldots ,N_\xi$, where we have defined for convenience $f^{(-1)}:=0$ and from expansion (\ref{eq:Legendre_expansion}) it is clear that $f^{(N_\xi+1)}=0$. Analogously to (\ref{eq:Legendre_expansion}) the source term is expanded as $s=\sum_{k=0}^{N_\xi} s^{(k)} P_k$. For the sources given by (\ref{eq:DKE_Sources}) this expansion is exact when $N_\xi\ge2$ as $s_j^{(k)}=0$ for $k\ge 3$. The spatial differential operators read 
	\begin{align}
		L_k & = 
		\frac{k}{2k-1} 
		\left(
		\vb*{b} \cdot \nabla 
		+
		\frac{k-1}{2}
		\vb*{b}\cdot\nabla \ln B
		\right), \label{eq:DKE_Legendre_expansion_Lower}
		\\ 
		D_k & = - 
		\frac{\widehat{E}_\psi}{\mean*{B^2}}
		\vb*{B}\times \nabla\psi  \cdot \nabla 
		+  
		\frac{k(k+1)}{2}
		\hat{\nu} , \label{eq:DKE_Legendre_expansion_Diagonal}
		\\
		U_k & =  
		\frac{k+1}{2k+3} 
		\left(
		\vb*{b} \cdot \nabla 
		-
		\frac{k+2}{2}
		\vb*{b}\cdot\nabla \ln B
		\right). \label{eq:DKE_Legendre_expansion_Upper}
	\end{align}
	Thanks to its tridiagonal structure, the system of equations (\ref{eq:DKE_Legendre_expansion}) can be inverted using the standard Gaussian elimination algorithm for block tridiagonal matrices.

	Before introducing the algorithm we will explain how to fix the free constant of the solution to equation (\ref{eq:DKE_Legendre_expansion}) so that it can be inverted. Note that the aforementioned kernel of the DKE translates in the fact that $f^{(0)}$ is not completely determined from equation (\ref{eq:DKE_Legendre_expansion}). To prove this, we inspect the modes $k=0$ and $k=1$ of equation (\ref{eq:DKE_Legendre_expansion}), which are the ones that involve $f^{(0)}$. From expression (\ref{eq:ExB_spatial_operator}) we can deduce that the term $D_0 f^{(0)} + U_0 f^{(1)} $ is invariant if we add to $f^{(0)}$ any function of $B_\theta \theta + B_\zeta  \zeta$. For $\widehat{E}_\psi\ne 0$, functions of $B_\theta \theta + B_\zeta  \zeta$ lie on the kernel of $\vb*{B}\times\nabla \psi \cdot \nabla$ and for $\widehat{E}_\psi = 0$, $D_0$ is identically zero. Besides, the term $L_1 f^{(0)} + D_1 f^{(1)} + U_1 f^{(2)}$ remains invariant if we add to $f^{(0)} $ any function of $\theta-\iota\zeta$ (the kernel of $L_1=\vb*{b}\cdot\nabla$ consists of these functions). For ergodic flux surfaces, the only continuous functions on the torus that belong to the kernel of $L_1$ are constants. Thus, equation (\ref{eq:DKE_Legendre_expansion}) is unaltered when we add to $f^{(0)}$ any constant (a function that belongs simultaneously to the kernels of $\vb*{B}\times\nabla \psi \cdot \nabla$ and $\vb*{b}\cdot\nabla$). A constraint equivalent to condition (\ref{eq:kernel_elimination_condition}) is to fix the value of the $0-$th Legendre mode of the distribution function at a single point of the flux surface. For example,
	\begin{align}
		f^{(0)}(0,0)=0, \label{eq:kernel_elimination_condition_Legendre}
	\end{align}
	which implicitly fixes the value of the constant $C$ in (\ref{eq:kernel_elimination_condition}).
	With this condition, equation (\ref{eq:DKE_Legendre_expansion}) has a unique solution and its left-hand-side can be inverted  to solve for $f^{(k)}$ in two scenarios: when the flux surface is ergodic and in rational surfaces when $\widehat{E}_\psi\ne0$ (further details on its invertibility are given in appendix \ref{sec:Appendix_Invertibility}). Note that, as expansion (\ref{eq:Legendre_expansion}) is finite and representation (\ref{eq:DKE_Legendre_expansion}) is non diagonal, the functions $f^{(k)}$ obtained from inverting (\ref{eq:DKE_Legendre_expansion}) constrained by (\ref{eq:kernel_elimination_condition_Legendre}) are approximations to the first $N_\xi+1$ Legendre modes of the exact solution to (\ref{eq:DKE}) satisfying (\ref{eq:kernel_elimination_condition}) (further details at the end of appendix \ref{sec:Appendix_Legendre}).

	\subsection{Block TriDiagonal (BTD) solution to the DKE}\label{subsec:Legendre_expansion}
	In this section, the algorithm on which the new neoclassical code {\MONKES} is based will be presented. In particular, we will describe the algorithm for solving the truncated DKE (\ref{eq:DKE_Legendre_expansion}) which, for the sake of clarity, we repeat here  
	\begin{align}
		L_k f^{(k-1)} + D_k f^{(k)} + U_k f^{(k+1)} = s^{(k)}.  
		\tag{\ref{eq:DKE_Legendre_expansion}}
	\end{align}
	Equation (\ref{eq:DKE_Legendre_expansion}) possesses a Block TriDiagonal (BTD) structure in which each \qmarks{block} is a spatial differential operator. The algorithm for solving the BTD equation (\ref{eq:DKE_Legendre_expansion}) is a straightforward generalization of the LU factorization method for BTD matrices \cite{zhang05,Demmel1995BlockLU} and consists of two steps. 
	\begin{enumerate}
		\item \textbf{Forward elimination}
	\end{enumerate} 	
	Starting from $\Delta_{N_\xi} = D_{N_\xi}$ and $\sigma^{(N_\xi)} = s^{(N_\xi)}$ we can obtain recursively the operators
	\begin{align}
		\Delta_k = D_k - U_{k} \Delta_{k+1}^{-1} L_{k+1}, 
		\label{eq:Schur_complements}
	\end{align} 
	and the sources
	\begin{align}
		\sigma^{(k)} = s^{(k)} - U_k \Delta_{k+1}^{-1}    \sigma^{(k+1)},
		\label{eq:Forward_elimination_sources}
	\end{align}
	for $k=N_\xi-1, N_\xi-2, \ldots, 0$ (in this order). Equations (\ref{eq:Schur_complements}) and (\ref{eq:Forward_elimination_sources}) define the forward elimination. With this procedure we can transform equation (\ref{eq:DKE_Legendre_expansion}) to the equivalent system
	\begin{align}
		L_{k} f^{(k-1)} + \Delta_{k} f^{(k)} = \sigma^{(k)},
		\label{eq:DKE_Forward_elimination}
	\end{align}
	for $k=0,1, \ldots, N_\xi$. Note that this process corresponds to perform formal Gaussian elimination over 
	\begin{align}
		\Matrix{ccc|c}
		{
			L_{k} & D_{k} & U_k  & s^{(k)} \\
			0& L_{k+1} & \Delta_{k+1}  & \sigma^{(k+1)} \\
		}
		,
		\label{eq:DKE_Forward_Elimination}
	\end{align}
	to eliminate $U_k$ in the first row.
	
	\begin{enumerate}[resume]
		\item \textbf{Backward substitution}
	\end{enumerate}
	Once we have the system of equations in the form (\ref{eq:DKE_Forward_elimination}) it is immediate to solve recursively
	\begin{align}
		f^{(k)} = 
		\Delta_k^{-1}
		\left( 
		\sigma^{(k)} -  L_{k} f^{(k-1)} 
		\right), 	\label{eq:DKE_Backward_Substitution}
	\end{align}
	for $k=0,1,...,N_\xi$ (in this order). Here, $\Delta_0^{-1} \sigma^{(0)}$ denotes the unique solution to $\Delta_0 f^{(0)} = \sigma^{(0)} $ that satisfies (\ref{eq:kernel_elimination_condition_Legendre}). As $L_1= \vb*{b}\cdot \nabla$, using expression (\ref{eq:Parallel_streaming_spatial_operator}), it is clear from equation (\ref{eq:DKE_Backward_Substitution}) that the integration constant does not affect the value of $f^{(1)}$.
	
	We can apply this algorithm to solve equation (\ref{eq:DKE}) for $f_1$, $f_2$ and $f_3$ in order to compute approximations to the transport coefficients. In terms of the Legendre modes of $f_1$, $f_2$ and $f_3$, the monoenergetic geometric coefficients from definition (\ref{eq:Monoenergetic_geometric_coefficients}) read
	\begin{align}
		\widehat{D}_{11} & = 2\mean*{s_1^{(0)} f_1^{(0)}} + \frac{2}{5}\mean*{s_1^{(2)} f_1^{(2)}}, 
		\label{eq:Gamma_11_Legendre}\\ 
		\widehat{D}_{31} & = \frac{2}{3} \mean*{\frac{B}{B_0} f_1^{(1)}},\label{eq:Gamma_31_Legendre}\\ 
		\widehat{D}_{13} & = 2\mean*{s_1^{(0)} f_3^{(0)}} + \frac{2}{5}\mean*{s_1^{(2)} f_3^{(2)}}, \label{eq:Gamma_13_Legendre}\\ 
		\widehat{D}_{33} & =\frac{2}{3} \mean*{\frac{B}{B_0} f_3^{(1)}}, \label{eq:Gamma_33_Legendre}
	\end{align}
	where $3s_1^{(0)} /2= 3s_1^{(2)} = \vb*{B}\times\nabla\psi \cdot \nabla B / B^3$. Note  from expressions (\ref{eq:Gamma_11_Legendre}), (\ref{eq:Gamma_31_Legendre}), (\ref{eq:Gamma_13_Legendre}) and (\ref{eq:Gamma_33_Legendre}) that, in order to compute the monoenergetic geometric coefficients $\widehat{D}_{ij}$, we only need to calculate the Legendre modes $k=0,1,2$ of the solution and we can stop the backward substitution (\ref{eq:DKE_Backward_Substitution}) at $k=2$. This algorithm has been implemented in the code {\MONKES} and its implementation will be explained in chapter \ref{chap:MONKES}.
	
	\subsection{Contribution of different classes of particles to the monoenergetic coefficients}\label{subsec:Monoenergetic_lambda}
	Guiding-center motion equations (\ref{eq:Guiding_center_velocity}), (\ref{eq:dot_xi_xi_v}) and (\ref{eq:dot_v_xi_v}) reveal that, in the absence of collisions, guiding-centers move, to lowest order in $\rhostar{a}$, following magnetic field lines according to
	\begin{align}
		\dot{\vb*{x}} & = v\xi\vb*{b}, 
		\label{eq:Collisionless_trajectory_x}
		\\
		\dot{\xi} & = -\frac{(1-\xi^2)}{2} v\vb*{b}\cdot\nabla\ln B,
		\label{eq:Collisionless_trajectory_xi}
		\\
		\dot{v} & = 0,
		\label{eq:Collisionless_trajectory_v}
	\end{align}
	where it has been used that the lowest order portion of the electric field $\vb*{E}_0 = E_\psi(\psi) \nabla \psi$ is perpendicular to flux surfaces. It is immediate to check that guiding-centers whose motion is determined by (\ref{eq:Collisionless_trajectory_x})-(\ref{eq:Collisionless_trajectory_v}) preserve the normalized magnetic moment $\lambda$, which in coordinates $(\vb*{x},\xi,v)$ takes the form 
	\begin{align}
		\lambda(\vb*{x},\xi)=\frac{1- \xi^2}{B(\vb*{x})} \in [0,1/B].
	\end{align}
	In section \ref{subsec:Guiding_center_motion} it was shown that $\lambda$ allowed to classify different types of orbits. Recall that, according to classification (\ref{eq:Orbit_classification}), values of $\lambda$ smaller or greater than $\lambdac$ correspond, respectively, to passing and trapped particles. It was also stated that those particles with $\lambda \gtrsim \lambdac = 1/\Bmax $ were called barely trapped and those with $\lambda\sim 1/\Bmin^{\text{r}}$ were called deeply trapped. 
	
	It is natural to ask which classes of particles contribute the most to the radial neoclassical fluxes and to the parallel flow. This question can be answered by inspecting which classes of particles contribute the most to the monoenergetic coefficients at reactor-relevant collisionalities. With the solutions $\{ f_j\}_{j=1}^{3}$ to (\ref{eq:DKE}) it is possible to determine which particles contribute the most to the different monoenergetic coefficients.

	For the sake of clarity, we repeat the definition given for the monoenergetic coefficients in chapter \ref{sec:DKE}
	\begin{align}
		\widehat{D}_{ij} 
		:=
		\mean*
		{
			\int_{-1}^{1}
			s_i f_j 
			\dd{\xi}
		},
		\tag{\ref{eq:Monoenergetic_geometric_coefficients}}
	\end{align}	
	where $s_1 = s_2 = - \Omega_a \vb*{v}_{\text{m}a}\cdot\nabla\psi / Bv^2$ and $s_3=\xi B/ B_0$. Now we wonder, how the coefficients would be if in the integral of (\ref{eq:Monoenergetic_geometric_coefficients}) we only considered particles for which $(1-\xi^2)/B$ lies in the interval $[\lambda, B^{-1}]$ for a certain value of $\lambda$. This is equivalent to deactivating $f_j$ for particles with $\xi^2 > 1 - \lambda B $. In other words we substitute $f_j$ in (\ref{eq:Monoenergetic_geometric_coefficients}) by $f_j H(1-\lambda B -\xi^2)$ where $H(x)$ is a Heaviside function ($H(x\ge 0) = 1$ and $H(x<0)=0$). Thus, we obtain
	\begin{align}
		\widehat{d}_{ij}
		(\lambda) 
		&
		:=
		\mean*
		{
			\int_{-1}^{1}
			s_i f_j 
			H(1-\lambda B -\xi^2)
			\dd{\xi}
		}
		\nonumber\\
		& = 
		\mean*
		{
			H(1-\lambda B )
			\int_{-\sqrt{1-\lambda B }}^{\sqrt{1-\lambda B }}
			s_i f_j 
			\dd{\xi}
		}
		\nonumber\\
		&
		= 	
		\mean*{
			2
			H
			\left(
			1-\lambda B
			\right)
			\int_{0}^{\sqrt{1-\lambda B}}
			(s_i f_j)^+ 
			\dd{\xi}
		}.
		\label{eq:dij_definition}
	\end{align}
	for $\lambda\in[0,1/B_{\text{min}} ]$. Here, we denote by $g^+:=(g(\xi)+g(-\xi))/2$ to the even portion of a function $g$ with respect to $\xi$. Note that $\widehat{d}_{ij}(0)=\widehat{D}_{ij}$ and $\widehat{d}_{ij}(1/B_{\text{min}}) = 0$. Also note that $\widehat{d}_{ij}
	(\lambda_1) - \widehat{d}_{ij}
	(\lambda_2) $ is equivalent to \qmarks{activating} $f_j$ only for particles with $\lambda_1 \le \lambda \le \lambda_2$. Hence, $\widehat{d}_{ij}
	(\lambda_1) - \widehat{d}_{ij}
	(\lambda_2) $ measures the contribution of particles lying in $\lambda\in[\lambda_1,\lambda_2]$ to $\widehat{D}_{ij}$.

	Using the Legendre expansions for the solution $f_j = \sum_{k=0}^{N_\xi} f_j^{(k)} P_k(\xi)$ and sources $s_i = \sum_{k=0}^{2} s_i^{(k)} P_k(\xi)$, the function $\widehat{d}_{ij}$ can be rewritten as 
	\begin{align}
		\widehat{d}_{ij}(\lambda)
		& =
		\mean*{
			s_i^{(0)}
			H_j^{(0)}(\lambda,B)
		}
		+
		\mean*{
			s_i^{(1)}
			H_j^{(1)}(\lambda,B)
		}
		+
		\mean*{
			s_i^{(2)}
			H_j^{(2)}(\lambda,B)
		}
		,
		\label{eq:dij_definition_Legendre}
	\end{align}
	where we have defined
	\begin{align}
		H_j^{(0)}(\lambda,B)
		& :=
		\sum_{k\ge0}  
		f_j^{(2k)}
		I_{2k}^{(0)}
		\left(
		\sqrt{1-\lambda B}
		\right)		
		,
		\label{eq:Monoenergetic_lambda_H0} 
		\\
		H_j^{(1)}(\lambda,B)
		& :=
		\sum_{k\ge0}  
		f_j^{(2k+1)}
		I_{2k+1}^{(1)} 
		\left(
		\sqrt{1-\lambda B}
		\right) ,
		\label{eq:Monoenergetic_lambda_H1} 
		\\
		H_j^{(2)}(\lambda,B)
		& :=
		\sum_{k\ge1}  
		f_j^{(2k)}
		\frac{1}{2}
		\left[
		3
		I_{2k}^{(2)} 
		\left(
		\sqrt{1-\lambda B}
		\right)		
		-
		I_{2k}^{(0)} 
		\left(
		\sqrt{1-\lambda B}
		\right)
		\right] 
		.
		\label{eq:Monoenergetic_lambda_H2} 
	\end{align}
	
	The functions $\{H_j^{(k)}\}_{k=0}^{2}$ can be computed using the identities (for their proof see appendix \ref{sec:Appendix_Legendre})
	\begin{align}
		I_{2k}^{(0)}(x)
		& :=
		2H(x)
		\int_{0}^{x}
		P_{2k}(\xi) 
		\dd{\xi} 
		=
		\frac{2H(x)}{4k+1}
		\left(
		P_{2k+1}(x) - P_{2k-1}(x)
		\right)
		,
		\label{eq:I2k_0_identity}
		\\
		I_{2k+1}^{(1)}(x)
		&
		:=
		2H(x)
		\int_{0}^{x}
		\xi P_{2k+1}(\xi) 
		\dd{\xi}
		=
		\frac{(2k+2)}{(4k+3)}I_{2k+2}^{(0)}(x)
		+
		\frac{(2k+1)}{(4k+3)} I_{2k}^{(0)}(x) ,
		\label{eq:I2k_1_identity}
		\\
		I_{2k}^{(2)}(x)
		&
		:=
		2H(x)
		\int_{0}^{x}
		\xi^2 P_{2k}(\xi) 
		\dd{\xi}
		=
		\frac{1}{4k+1}
		(
		(2k+1) I_{2k+1}^{(1)}(x)
		+
		2k I_{2k-1}^{(1)}(x)
		),
		\label{eq:I2k_2_identity}
	\end{align}
	where $x\in\mathbb{R}$.
	
	Thus, once the functions $\{I_{2k}^{(0)}(x)\}_{0\le 2k\le N_\xi}$ are calculated, both $\{I_{2k+1}^{(1)}(x)\}_{1\le 2k+1\le N_\xi}$ and $\{I_{2k}^{(2)}(x)\}_{0\le 2k\le N_\xi}$ (in this order) can be obtained. With these functions we can evaluate $\widehat{d}_{ij}(\lambda)$ by simply setting $x=\sqrt{1-\lambda B}$. The calculation of the function $\widehat{d}_{ij}(\lambda)$ is implemented in the code {\MONKES}. In section \ref{subsec:Contribution_lambda_MONKES}, we will give some examples of how it is possible to employ the function $\dij{ij}(\lambda) $ that {\MONKES} computes to learn which classes of particles contribute the most to each monoenergetic coefficient.

	\section{Derivatives of the monoenergetic coefficients}\label{sec:Derivatives_monoenergetic}
	In this section, three methods for computing derivatives of the monoenergetic coefficients $\Dij{ij}$ will be described. Let $\eta$ be a parameter upon which the DKE (\ref{eq:DKE}) depends. For gradient-based optimization methods it is useful to compute the derivatives $\partial{\Dij{ij}}/\partial{\eta}$ of the monoenergetic coefficients. Deriving their definition (\ref{eq:Monoenergetic_geometric_coefficients_inner_product}) and using identity (\ref{eq:FSA_derivative_eta}) from appendix \ref{sec:Appendix_FSA_derivative}, we can express these derivatives as
	\begin{align} 
		\pdv{\Dij{ij}}{\eta}		
		=  
		\mean*{s_i , \pdv{f_j}{\eta}}
		+
		\mean*{\pdv{s_i }{\eta} , f_j}
		-
		2
		\mean*{
			\left(\pdv{\ln B}{\eta} -\mean*{\pdv{\ln B}{\eta} }\right)
			s_i, f_j
		}.
		\label{eq:Monoenergetic_derivatives_general}
	\end{align}
	Thus, the derivative $\partial{\Dij{ij}}/\partial{\eta}$ can be computed by computing three different inner products. The first two summands on the right-hand side of (\ref{eq:Monoenergetic_derivatives_general}) account, respectively, for the dependence on $\eta$ of the distribution function $f_j$ and the source term $s_i$. The latter term includes the dependence of the flux surface $2-$form $\sqrt{g} / (\dv*{V}{\psi})\dd{\theta}\dd{\zeta}$ on $\eta$. Note that the most complicated term to obtain is the one involving $\pdv*{f_j}{\eta}$. Naively, one could compute an approximation to it from its definition using first order finite differences
	\begin{align}
		\pdv{f_j}{\eta}
		:=
		\lim_{\Delta\eta\rightarrow0}
		\frac{ \Eval{f_j}_{\eta+\Delta\eta} - \Eval{f_j}_{\eta}}{\Delta\eta}
		\approx
		\frac{ \Eval{f_j}_{\eta+\Delta\eta} - \Eval{f_j}_{\eta}}{\Delta\eta},
		\label{eq:Distribution_function_derivative_eta}
	\end{align}
	for sufficiently small $\Delta\eta$. However, in order to approximate (\ref{eq:Distribution_function_derivative_eta}) using finite differences, it is required to know the solution to the DKE for at least two different values of the parameter: $\eta$ and $\eta+\Delta\eta$. At best, this would require solving the DKE twice for each different derivative of the transport coefficients. Fortunately, there are alternatives to this approach which, in most cases, are more efficient. 
	
	Given the linearity of the DKE with respect to its solution, it is possible to obtain a DKE whose solution is $\pdv*{f_j}{\eta}$. Deriving (\ref{eq:DKE_compact}) along $\eta$ yields
	\begin{align}
		\left(\VV-\hat{\nu}\Lorentz\right)\pdv{f_j}{\eta}
		=
		S_{j,\eta}
		,
		\label{eq:DKE_derivative_eta}
	\end{align}
	where 
	\begin{align}
		S_{j,\eta}
		:=
		\pdv{s_j}{\eta}
		-
		\VV_{\eta} f_j
		+
		\pdv{\hat{\nu}}{\eta}
		\Lorentz f_j,
	\end{align}
	is the resulting source term and the operator $\VV_{\eta}$ is the commutator between $\pdv*{\eta}$ and $\VV$, i.e.
	\begin{align}
		\VV_{\eta} 
		:= 
		\pdv{\eta} \VV 
		- 
		\VV \pdv{\eta}
		.
	\end{align}
	Equivalently, $\VV_{\eta}$ is the linear differential operator obtained from deriving the coefficients of $\VV$ along $\eta$ when $\VV$ is expressed in a certain set of coordinates.
	
	Thus, equation (\ref{eq:DKE_derivative_eta}) shows that the derivative $\pdv*{f_j}{\eta}$ satisfies almost the same DKE as $f_j$ but for a different source term. This equation provides all the required information for computing the term $\mean*{ s_i, \partial{f_j}/\partial{\eta} }$ in (\ref{eq:Monoenergetic_derivatives_general}). The \textit{direct method} to extract this information is to solve equation (\ref{eq:DKE_derivative_eta}) for $\pdv*{f_j}{\eta}$ applying the algorithm explained in section \ref{subsec:Legendre_expansion}. Alternatively, we can compute the projection $\mean*{ s_i, \partial{f_j}/\partial{\eta} }$ by using the solution $f_i^\dagger$ to the adjoint DKE (\ref{eq:Adjoint_DKE_compact}) without solving for or approximating $\pdv*{f_j}{\eta}$. This latter approach is known as an \textit{adjoint method}. In this section we will present formally how the term $\mean*{ s_i, \partial{f_j}/\partial{\eta} }$ can be computed by these two methods. In chapter \ref{chap:MONKES} we will revisit them and comment their computational aspects.
	
	\subsection{Direct method for computing derivatives}
	We can represent the DKE (\ref{eq:DKE_derivative_eta}) and its solution $\pdv*{f_j}{\eta}$ in a Legendre basis to obtain
	\begin{align}
		L_k 
		\pdv{f_j^{(k-1)}}{\eta}
		+
		D_k 
		\pdv{f_j^{(k)}}{\eta}
		+
		U_k 
		\pdv{f_j^{(k+1)}}{\eta}
		=
		S_{j,\eta}^{(k)}
		,
		\label{eq:DKE_derivative_eta_Legendre}
	\end{align}
	where 
	\begin{align}
		S_{j,\eta}^{(k)}
		:=
		\pdv{s_j^{(k)}}{\eta}
		-
		\left(\VV_{\eta} f_j\right)^{(k)}
		-
		\frac{k(k+1)}{2}
		\pdv{\hat{\nu}}{\eta}
		f_j^{(k)},
		\label{eq:DKE_derivatives_source}
	\end{align}
	is the source term,
	\begin{align}
		\left(\VV_{\eta} f_j\right)^{(k)}
		=
		L_{k,\eta} f_j^{(k-1)}
		+
		\left( D_{k,\eta} - \frac{k(k+1)}{2}
		\pdv{\hat{\nu}}{\eta} \right)f_j^{(k)}
		+
		U_{k,\eta} f_j^{(k-1)}
	\end{align}
	and $L_{k,\eta} $, $D_{k,\eta}$ and $U_{k,\eta}$ are, respectively, the commutators of $\pdv*{\eta}$ and $L_k$, $D_k$ and $U_k$. Namely, 
	\begin{align}
		L_{k,\eta} 
		& :=
		\pdv{\eta} L_k 
		-
		L_k \pdv{\eta}
		, 
		\\
		D_{k,\eta} 
		& :=
		\pdv{\eta} D_k 
		-
		D_k \pdv{\eta}
		,
		\\
		U_{k,\eta} 
		& :=
		\pdv{\eta} U_k 
		-
		U_k \pdv{\eta}
		,
	\end{align}
	are the linear differential operators obtained by deriving (respectively) the coefficients of $L_k$, $D_k$ and $U_k$ along $\eta$.

	Note that we can apply the BTD algorithm presented in section \ref{subsec:Legendre_expansion} to equation (\ref{eq:DKE_derivative_eta_Legendre}). In particular, in the forward elimination step we just have to substitute $s^{(k)}$ by $S_{j,\eta}^{(k)}$ in equation (\ref{eq:Forward_elimination_sources}) and $\sigma^{(k)}$ by $\Sigma^{(k)}_{j,\eta} $ in (\ref{eq:Forward_elimination_sources}) and (\ref{eq:DKE_Forward_elimination}). Here, we have defined $\Sigma^{(k)}_{j,\eta}$ as
	\begin{align}
		\Sigma^{(k)}_{j,\eta} 
		:=		
		S_{j,\eta}^{(k)}
		-
		U_k
		\Delta_{k+1}^{-1}
		\Sigma^{(k+1)}_{j,\eta},
		\label{eq:Forward_elimination_sources_derivatives}
	\end{align}
	which is obtained using the analogue recursion to (\ref{eq:Forward_elimination_sources}).

	The forward elimination procedure given by (\ref{eq:Schur_complements}) and (\ref{eq:Forward_elimination_sources_derivatives}) transforms (\ref{eq:DKE_derivative_eta_Legendre}) in the lower triangular system 
	\begin{align}
		\Delta_k \pdv{f_j^{(k)}}{\eta}
		+
		L_k \pdv{f_j^{(k-1)}}{\eta}
		=
		\Sigma^{(k)}_{j,\eta}. 
	\end{align}
	Thus, starting from $k=0$ we can solve for $\{ \partial{f_j^{(k)}}/\partial{\eta} \}_{k=0}^{N_\xi}$. However, as for computing the monoenergetic coefficients, we only need to compute the first three modes $\{\partial{f_j^{(k)}}/\partial{\eta}\}_{k=0}^{2}$ as they are the only modes that contribute to the term $\mean{s_i, \partial{f_j}/\partial{\eta}}$. In order to compute the sources $\{S^{(k)}_{j,\eta}\}_{k=0}^{N_\xi}$ and $\{\Sigma^{(k)}_{j,\eta}\}_{k=0}^{N_\xi}$ we do need, however, the full solution $\{f_j^{(k)}\}_{k=0}^{N_\xi}$ to (\ref{eq:DKE_Legendre_expansion}). Once we have computed $\{\partial{f_j^{(k)}}/\partial{\eta}\}_{k=0}^{2}$ we can calculate the term $\mean*{s_i, \partial{f_j }/\partial{\eta} }$ as
	\begin{align}
		\mean*{
			s_i, 
			\pdv{f_j }{\eta} 
		}
		=
		\sum_{k=0}^{2}
		\frac{2}{2k+1}
		\mean*{
			s_i^{(k)}
			\pdv{f_j^{(k)}}{\eta}
		}.
	\end{align}

	\subsection{Adjoint method for computing derivatives}
	A different approach for computing $\mean*{ s_i, \partial{f_j}/\partial{\eta} }$ employs the solution $f_i^\dagger$ to the adjoint DKE (\ref{eq:Adjoint_DKE_compact}). Projecting $\mean*{ f_i^\dagger, \text{Eq. (\ref{eq:DKE_derivative_eta})} }$, using the definition of adjoint and that $f_i^\dagger$ is the solution to the adjoint DKE (\ref{eq:Adjoint_DKE_compact}) we obtain
	\begin{align}
		\mean*{ s_i, \pdv{f_j}{\eta} } 
		=
		\mean*{ f_i^\dagger, \pdv{s_j}{\eta} }
		-
		\mean*{ f_i^\dagger, \VV_{\eta} f_j }
		+
		\pdv{\hat{\nu}}{\eta}
		\mean*{ f_i^\dagger,  \Lorentz f_j }.
		\label{eq:Adjoint_method_difficult_term}
	\end{align}
	Note that on the right-hand side of (\ref{eq:Adjoint_method_difficult_term}) there are no derivatives of $f_j$ nor $f_i^\dagger$. Thus, with the solutions to the DKE and its adjoint version we can readily compute the derivatives as 
	\begin{align} 
		\pdv{\Dij{ij}}{\eta}		
		& 
		=  
		\mean*{ f_i^\dagger, \pdv{s_j}{\eta} }
		-
		\mean*{ f_i^\dagger, \VV_{\eta} f_j }
		+
		\pdv{\hat{\nu}}{\eta}
		\mean*{ f_i^\dagger,  \Lorentz f_j }
		\nonumber
		\\
		& 
		+
		\mean*{\pdv{s_i }{\eta} , f_j}
		-
		2
		\mean*{
			\left(\pdv{\ln B}{\eta} -\mean*{\pdv{\ln B}{\eta} }\right)
			s_i, f_j
		}.
		\label{eq:Monoenergetic_derivatives_adjoint}
	\end{align}
	
	It is useful to express each inner product on the right-hand side of (\ref{eq:Monoenergetic_derivatives_adjoint}) in a Legendre basis. Suppose that we know the first $N_\xi+1$ Legendre modes of $f_j$ and $f_i^\dagger$, then
	\begin{align}
		\mean*{ f_i^\dagger, \pdv{s_j}{\eta} }
		& =
		\sum_{k=0}^{2}
		\frac{2}{2k+1}
		\mean*{
			(f_i^\dagger)^{(k)}
			\pdv{s_j ^{(k)}}{\eta}
		}
		,
		\\
		\mean*{ f_i^\dagger, \VV_{\eta} f_j }
		& =
		\sum_{k=0}^{N_\xi}
		\frac{2}{2k+1}
		\mean*{ (f_i^\dagger)^{(k)} (\VV_{\eta} f_j)^{(k)} }
		\label{eq:Adjoint_method_fi_dagger_Vfj}
		,
		\\
		\pdv{\hat{\nu}}{\eta}
		\mean*{ f_i^\dagger,  \Lorentz f_j }
		& =
		-
		\pdv{\hat{\nu}}{\eta}
		\sum_{k=1}^{N_\xi}
		\frac{k(k+1)}{2k+1}
		\mean*{(f_i^\dagger)^{(k)}
			f_j^{(k)}}
		,
		\label{eq:Adjoint_method_fi_dagger_Lfj}
		\\
		\mean*{\pdv{s_i }{\eta} , f_j}
		&
		=
		\sum_{k=0}^{2}
		\frac{2}{2k+1}
		\mean*{\pdv{s_i ^{(k)}}{\eta}
			f_j^{(k)}},
		\\
		2
		\mean*{
			\left(\pdv{\ln B}{\eta} -\mean*{\pdv{\ln B}{\eta} }\right)
			s_i, f_j
		}
		& =
		\sum_{k=0}^{2}
		\frac{4}{2k+1}
		\mean*
		{
			\pdv{\ln B}{\eta} 
			s_i^{(k)} f_j^{(k)}
		} 
		\\
		& 
		-
		\sum_{k=0}^{2}
		\frac{4}{2k+1}
		\mean*
		{
			\mean*{\pdv{\ln B}{\eta} } 
			s_i^{(k)} f_j^{(k)}
		}.
		\nonumber
	\end{align}
	As it was the case for the direct method, more than three Legendre modes of the solution are required. Specifically, the whole Legendre spectrum of $\{f_j^{(k)}\}_{k=0}^{N_\xi}$ and of the solution to the adjoint DKE $\{(f_i^\dagger)^{(k)}\}_{k=0}^{N_\xi}$ are needed for computing derivatives with the adjoint method.
	
	\chapter{BTD algorithm implementation: the {\MONKES} code}\label{chap:MONKES}
	\thispagestyle{empty}
	
	In chapter \ref{chap:Monoenergetic} (specifically in section \ref{subsec:Legendre_expansion}), an algorithm for formally solving the DKE in a Legendre basis (\ref{eq:DKE_Legendre_expansion}) has been proposed. In this chapter we present how this algorithm has been implemented in the new neoclassical code {\MONKES} to numerically solve equation (\ref{eq:DKE_Legendre_expansion}). The chapter is organized as follows. In section \ref{subsec:Algorithm_Implementation}, we describe the spatial discretization and the implementation of the algorithm in the code {\MONKES}. In section \ref{subsec:Convergence}, we carry out a convergence study to determine the required resolution in the spatial and velocity coordinates to correctly calculate the monoenergetic coefficients. In section \ref{subsec:Performance}, we evaluate the performance of {\MONKES} and compare it against that of {\DKES}. In section \ref{sec:Results_benchmark}, we benchmark the monoenergetic coefficients computed by {\MONKES} against those calculated by the codes {\DKES} and {\SFINCS}. Finally, in section \ref{sec:Other_capabilities_MONKES} we describe other capabilities of {\MONKES} apart from the computation of monoenergetic coefficients. Most of this chapter is based on article [I] from the \textbf{\qmarks{PUBLISHED AND SUBMITTED CONTENT}} section at the beginning of this dissertation. Specifically, sections \ref{subsec:Algorithm_Implementation}, \ref{subsec:Convergence}, \ref{subsec:Performance} and \ref{sec:Results_benchmark}.

	\section{Spatial discretization and implementation of the BTD algorithm}\label{subsec:Algorithm_Implementation}
	The algorithm described in section \ref{subsec:Legendre_expansion} allows, in principle, to compute the exact solution to the truncated DKE (\ref{eq:DKE_Legendre_expansion}) which is an approximate solution to the DKE (\ref{eq:DKE}). However, to our knowledge, it is not possible to give an exact expression for the operator $\Delta_k^{-1}$ except for $k=N_\xi \ge 1$ (see appendix \ref{sec:Appendix_Invertibility}). Instead, we are forced to compute an approximate solution to (\ref{eq:DKE_Legendre_expansion}).
	In order to obtain an approximate solution to equation (\ref{eq:DKE_Legendre_expansion}) we assume that each $f^{(k)}$ has a finite Fourier spectrum so that it can be expressed as
	\begin{align}	
		f^{(k)}(\theta,\zeta)
		& 
		=
		\vb*{I}(\theta,\zeta)
		\cdot
		\vb*{f}^{(k)},
		\label{eq:Fourier_expansion_f_k}
	\end{align}
	where the Fourier interpolant row vector map $\vb*{I}(\theta,\zeta)$ is defined at appendix \ref{sec:Appendix_Fourier} and the column vector $\vb*{f}^{(k)}\in\mathbb{R}^{N_{\text{fs}}}$
	contains $f^{(k)}$ evaluated at the equispaced grid points
	\begin{align}
		\theta_i & = 2\pi i / N_\theta, \quad & i=0,1,\ldots, N_\theta-1,  \label{eq:Theta_grid}
		\\ 
		\zeta_j & = 2\pi j / (N_\zeta \Nfp), \quad & j=0,1,\ldots, N_\zeta-1. \label{eq:Zeta_grid}
	\end{align}
	Here, $N_{\text{fs}}:=N_\theta N_\zeta$ is the number of points in which we discretize the flux surface being $N_\theta$ and $N_\zeta$ respectively the number of points in which we divide the domains of $\theta$ and $\zeta$. In general, the solution to equation (\ref{eq:DKE_Legendre_expansion}) has an infinite Fourier spectrum and cannot exactly be written as (\ref{eq:Fourier_expansion_f_k}) but, taking sufficiently large values of $N_\theta$ and $N_\zeta$, we can approximate the solution to equation (\ref{eq:DKE_Legendre_expansion}) to arbitrary degree of accuracy (in infinite precision arithmetic). As explained in appendix \ref{sec:Appendix_Fourier}, introducing the Fourier interpolant (\ref{eq:Fourier_expansion_f_k}) in equation (\ref{eq:DKE_Legendre_expansion}) and then evaluating the result at the grid points provides a system of $N_{\text{fs}}\times(N_\xi+1)$ equations which can be solved for $\{\vb*{f}^{(k)}\}_{k=0}^{N_\xi}$. This system of equations is obtained by substituting the operators $L_k$, $D_k$, $U_k$ in equation (\ref{eq:DKE_Legendre_expansion}) by the $N_{\text{fs}}\times N_{\text{fs}}$ matrices $\vb*{L}_k$, $\vb*{D}_k$, $\vb*{U}_k$, defined in appendix \ref{sec:Appendix_Fourier}. Thus, we discretize (\ref{eq:DKE_Legendre_expansion}) as %
	\begin{align}
		\vb*{L}_k  \vb*{f}^{(k-1)} + \vb*{D}_k  \vb*{f}^{(k)} + \vb*{U}_k   \vb*{f}^{(k+1)} = \vb*{s}^{(k)},   \label{eq:DKE_Legendre_expansion_Fourier_collocation}
	\end{align}
	for $k=0,1\ldots, N_\xi$ where $\vb*{s}^{(k)}\in\mathbb{R}^{N_{\text{fs}}}$
	contains $s^{(k)}$ evaluated at the equispaced grid points. This system has a block tridiagonal structure and the algorithm presented in subsection \ref{subsec:Legendre_expansion} can be applied. We just have to replace in equations (\ref{eq:Schur_complements}), (\ref{eq:Forward_elimination_sources}) and (\ref{eq:DKE_Backward_Substitution}) the operators and functions by their respective matrix and vector analogues, which we denote by boldface letters. 
	
	The matrix approximation to the forward elimination procedure given by equations (\ref{eq:Schur_complements}) and (\ref{eq:Forward_elimination_sources}) reads
	\begin{align}
		\vb*{\Delta}_k & = \vb*{D}_k - \vb*{U}_{k} \vb*{\Delta}_{k+1}^{-1} \vb*{L}_{k+1}, 
		\label{eq:Schur_complements_matrix}
		\\
		\vb*{\sigma}^{(k)} & = \vb*{s}^{(k)} - \vb*{U}_{k}  \vb*{\Delta}_{k+1}^{-1}    \vb*{\sigma}^{(k+1)},
		\label{eq:Forward_elimination_sources_matrix}
	\end{align}
	for $k=N_\xi-1, N_\xi-2, \ldots, 0$ (in this order). Thus, starting from $\vb*{\Delta}_{N_\xi}=\vb*{D}_{N_\xi}$ and $\vb*{\sigma}^{(N_\xi)}=\vb*{s}^{(N_\xi)}$, all the matrices $\vb*{\Delta}_k$ and the vectors $\vb*{\sigma}^{(k)}$ are defined from equations (\ref{eq:Schur_complements_matrix}) and (\ref{eq:Forward_elimination_sources_matrix}). Obtaining the matrix $\vb*{\Delta}_k$ directly from equation (\ref{eq:Schur_complements_matrix}) requires to invert $\vb*{\Delta}_{k+1}$, perform two matrix multiplications and a subtraction of matrices. The inversion using LU factorization and each matrix multiplication require $O(N_{\text{fs}}^3)$ operations so it is desirable to reduce the number of matrix multiplications as much as possible. We can reduce the number of matrix multiplications in determining $\vb*{\Delta}_{k}$ to one if instead of computing $\vb*{\Delta}_{k+1}^{-1}$ we solve the matrix system of equations
	\begin{align}
		\vb*{\Delta}_{k+1} \vb*{X}_{k+1} = \vb*{L}_{k+1},
		\label{eq:Forward_elimination_X}  
	\end{align}
	for $\vb*{X}_{k+1}$ and then obtain 
	\begin{align}
		\vb*{\Delta}_k = \vb*{D}_k - \vb*{U}_{k}\vb*{X}_{k+1}, 
		\label{eq:Schur_complements_Fourier_collocation}
	\end{align}
	for $k=N_\xi-1, N_\xi-2, \ldots, 0$. Thus, obtaining $\vb*{\Delta}_k$ requires $O(N_{\text{fs}}^3)$ operations for solving equation (\ref{eq:Forward_elimination_X}) (using LU factorization) and also $O(N_{\text{fs}}^3)$ operations for applying (\ref{eq:Schur_complements_Fourier_collocation}). In order to compute the monoenergetic coefficients, the backward substitution step requires solving equation (\ref{eq:DKE_Forward_elimination}) for $k=0,1$ and $2$. Therefore, for $k\le 1$, it is convenient to store $\vb*{\Delta}_{k+1}$ in the factorized LU form obtained when equation (\ref{eq:Forward_elimination_X}) was solved for $\vb*{X}_{k+1}$. The matrix $\vb*{\Delta}_{0}$ will be factorized later, during the backward substitution step.
	
	Similarly to what is done to obtain $\vb*{\Delta}_k$, to compute $\vb*{\sigma}^{(k)}$ we first solve 
	\begin{align}
		\vb*{\Delta}_{k+1} \vb*{y} = \vb*{\sigma}^{(k+1)}
		\label{eq:Forward_elimination_matrix_y_system}
	\end{align}
	for $\vb*{y}$ and then evaluate
	\begin{align}
		\vb*{\sigma}^{(k)} & = \vb*{s}^{(k)} - \vb*{U}_{k}  \vb*{y},
		\label{eq:Forward_elimination_sources_matrix_y}
	\end{align}
	for $k\ge 0$. Recall that none of the source terms $s_1$, $s_2$ and $s_3$ defined by (\ref{eq:DKE_Sources}) have Legendre modes greater than 2. Specifically, equation (\ref{eq:Forward_elimination_sources_matrix}) implies $\vb*{\sigma}_1^{(k)}, \vb*{\sigma}_3^{(k-1)} = 0$ for $k\ge 3$ and also $\vb*{\sigma}_1^{(2)} = \vb*{s}_1^{(2)}$, $\vb*{\sigma}_3^{(1)} = \vb*{s}_3^{(1)}$. Thus, we only have to solve equation (\ref{eq:Forward_elimination_matrix_y_system}) and apply (\ref{eq:Forward_elimination_sources_matrix_y}) to obtain $\{\vb*{\sigma}_1^{(k)}\}_{k=0}^{1}$ and $\vb*{\sigma}_3^{(0)}$. As $\{\vb*{\Delta}_{k+1}\}_{k=0}^{1}$ are already LU factorized, solving equation (\ref{eq:Forward_elimination_matrix_y_system}) and then applying (\ref{eq:Forward_elimination_sources_matrix_y}) requires $O(N_{\text{fs}}^2)$ operations and its contribution to the arithmetic complexity of the algorithm is subdominant with respect to the $O(N_{\text{fs}}^3)$ operations required to compute $\vb*{\Delta}_k$.
	
	For the backward substitution, we first note that solving the matrix version of equation (\ref{eq:DKE_Forward_elimination}) to obtain $\vb*{f}^{(0)}$ requires $O(N_{\text{fs}}^3)$ operations, as $\vb*{\Delta}_0$ has not been LU factorized during the forward elimination. On the other hand, obtaining the remaining modes  $\{\vb*{f}^{(k)}\}_{k=1}^{2}$, requires $O(N_{\text{fs}}^2)$ operations. As the resolution of the matrix system of equations (\ref{eq:Forward_elimination_X}) and the matrix multiplication in (\ref{eq:Schur_complements_Fourier_collocation}) must be done $N_\xi$ times, solving equation (\ref{eq:DKE_Legendre_expansion_Fourier_collocation}) by this method requires $O(N_\xi N_{\text{fs}}^3)$ operations.
	
	In what concerns to memory resources, as we are only interested in the Legendre modes $0$, $1$ and $2$, it is not necessary to store in memory all the matrices $\vb*{L}_k$, $\vb*{D}_k$, $\vb*{U}_k$ and $\vb*{\Delta}_k$. Instead, we store solely $\vb*{L}_k$, $\vb*{U}_k$ and $\vb*{\Delta}_k$ (in LU form) for $k=0,1,2$. For the intermediate steps we just need to use some auxiliary matrices $\vb*{L}$, $\vb*{D}$, $\vb*{U}$, $\vb*{\Delta}$ and $\vb*{X}$ of size $N_{\text{fs}}$. This makes the amount of memory required by {\MONKES} independent of $N_\xi$, being of order $N_{\text{fs}}^2$.
	\begin{algorithm}
		\caption{Block tridiagonal solution algorithm implemented in {\MONKES}.}\label{alg:MONKES_BTD}		
		\textbf{1. Forward elimination:}
		\begin{algorithmic}
			\State{
				$\vb*{L}\gets \vb*{L}_{N_\xi}$} \Comment{Starting value for $\vb*{L}$}
			\State{$\vb*{\Delta}\gets \vb*{D}_{N_\xi}$} \Comment{Starting value for $\vb*{\Delta}$} 
			\State{Solve $\vb*{\Delta} \vb*{X}=\vb*{L}$} \Comment{Compute $\vb*{X}_{N_\xi}$ stored in $\vb*{X}$}
			
			\For{$k=N_\xi-1$ \textbf{to} $0$}
			\State{$\vb*{L}\gets \vb*{L}_{k}$} 
			\Comment{Construct $\vb*{L}_k$ stored in $\vb*{L}$}
			\State{$\vb*{D}\gets \vb*{D}_{k}$} 
			\Comment{Construct $\vb*{D}_k$ stored in $\vb*{D}$}
			\State{$\vb*{U}\gets \vb*{U}_{k}$} 
			\Comment{Construct $\vb*{U}_k$ stored in $\vb*{U}$}
			\State{$\vb*{\Delta}\gets \vb*{D} - \vb*{U} \vb*{X}$} \Comment{Construct $\vb*{\Delta}_k$ stored in $\vb*{\Delta}$}
			\State{\textbf{if} $k>0$:\, Solve $\vb*{\Delta} \vb*{X}=\vb*{L}$} \Comment{Compute $\vb*{X}_{k}$ stored \qquad \qquad\qquad\qquad\qquad in $\vb*{X}$ for next iteration}
			
			\If{$k\le2$}\Comment{Save required matrices }
			\State{\textbf{if} $k=0$: $\vb*{L}_{k}\gets \vb*{L}$  	  \Comment{Save $\{\vb*{L}_k\}_{k=1}^{2}$}} 
			\State{$\vb*{U}_{k}\gets \vb*{U}$ 	\Comment{Save $\{\vb*{U}_k\}_{k=0}^{2}$  } }
			\State{$\vb*{\Delta}_{k}\gets \vb*{\Delta}$}    \Comment{Save $\{\vb*{\Delta}_k\}_{k=0}^{2}$}
			\EndIf
			\EndFor
			
			\State{}
			
			\For{$k=1$ to $0$}
			\State{Solve $\vb*{\Delta}_{k+1}\vb*{y}_1 = \vb*{\sigma}_1^{(k+1)} $} 
			
			\State{\textbf{if} $k=0$: Solve $\vb*{\Delta}_{k+1}\vb*{y}_3 = \vb*{\sigma}_3^{(k+1)} $}
			
			\State{$\vb*{\sigma}_1^{(k)} \gets \vb*{s}_1^{(k)} - \vb*{U}_k \vb*{y}_1$} \Comment{Construct $\vb*{\sigma}_1^{(k)}$ }
			\State{\textbf{if} $k=0$:  $\vb*{\sigma}_3^{(0)} \gets - \vb*{U}_0 \vb*{y}_3$} \Comment{Construct $\vb*{\sigma}_3^{(0)}$ }
			
			\EndFor

			\State{}
		\end{algorithmic}
		
		\textbf{2. Backward substitution:}
		\begin{algorithmic}
			\State{Solve $\vb*{\Delta}_0 \vb*{f}^{(0)} = \vb*{\sigma}^{(0)}$}
			\For{$k=1$ \textbf{to} $2$}    	
			\State{
				Solve $\vb*{\Delta}_k\vb*{f}^{(k)} = 
				\vb*{\sigma}^{(k)} - \vb*{L}_{k} \vb*{f}^{(k-1)} $
			}
			\EndFor
		\end{algorithmic}
	\end{algorithm}
	To summarize, the pseudocode of the implementation of the algorithm in {\MONKES} is given in Algorithm \ref{alg:MONKES_BTD}. In the first loop from $k=N_\xi-1$ to $k=0$ we construct and save only the matrices $\{\vb*{L}_k,\vb*{U}_k, \vb*{\Delta}_k\}_{k=0}^{2}$. At this point the matrices $\{\vb*{\Delta}_k\}_{k=1}^{2}$ are factorized in LU form. In the second loop, the sources $\{\vb*{\sigma}_1^{(k)}\}_{k=0}^{1}$ and $\vb*{\sigma}_3^{(0)}$ are computed and saved for the backward substitution. Finally, the backward substitution step is applied. For solving $\vb*{\Delta}_0 \vb*{f}^{(0)} = \vb*{\sigma}^{(0)}$ we have to perform the LU factorization of $\vb*{\Delta}_0$ (just for one of the two source terms) and then solve for $\vb*{f}^{(0)}$. For the remaining modes, the LU factorizations of $\{\vb*{\Delta}_k\}_{k=1}^{2}$ are reused to solve for $\{\vb*{f}^{(k)}\}_{k=1}^{2}$.
	
	Once we have solved equation (\ref{eq:DKE_Legendre_expansion_Fourier_collocation}) for $\vb*{f}^{(0)}$, $\vb*{f}^{(1)}$ and $\vb*{f}^{(2)}$, the integrals of the flux surface average operation involved in the monoenergetic coefficients (\ref{eq:Gamma_11_Legendre}), (\ref{eq:Gamma_31_Legendre}), (\ref{eq:Gamma_13_Legendre}) and (\ref{eq:Gamma_33_Legendre}), are conveniently computed using the trapezoidal rule, which for periodic analytic functions has geometric convergence \cite{Trapezoidal}. In section \ref{sec:Results_benchmark}, we will see that despite the cubic scaling in $N_{\text{fs}}$ of the arithmetical complexity of the algorithm, it is possible to obtain fast and accurate calculations of the monoenergetic geometric coefficients at low collisionality (and in particular $\widehat{D}_{31}$) in a single core. The reason behind this is that in the asymptotic relation $O(N_{\text{fs}}^3)\sim C_{\text{alg}} N_{\text{fs}}^3$, the constant $C_{\text{alg}}$ is small enough to allow $N_{\text{fs}}$ to take a sufficiently high value to capture accurately the spatial dependence of the distribution function without increasing much the wall-clock time. 
	
	The algorithm is implemented in the new code {\MONKES}, written in Fortran language. The matrix inversions and multiplications are computed using the linear algebra library \texttt{LAPACK} \cite{lapack99}.

	\section{Convergence of monoenergetic coefficients at low collisionality}
	\label{subsec:Convergence}
	
	In low collisionality regimes, convection is dominant with respect to diffusion. As equation (\ref{eq:DKE}) is singularly perturbed with respect to $\hat{\nu}$, its solution possesses internal boundary layers in $\xi$. These boundary layers appear at the interfaces between different classes of trapped particles. At these regions of phase-space, collisions are no longer subdominant with respect to advection. Besides, at these regions, the poloidal $\vb*{E}\times\vb*{B}$ precession from equation (\ref{eq:DKE}) can produce the chaotic transition of collisionless particles from one class to another due to separatrix crossing mechanisms \cite{Cary_Separatrix_Crossing, dherbemont2022}. The existence of these localized regions with large $\xi$ gradients demands a high number of Legendre modes $N_\xi$, explaining the difficulty to obtain fast and accurate solutions to equation (\ref{eq:DKE}) at low collisionality.

	\begin{figure}[h]
		\centering
		\begin{subfigure}[t]{0.48\textwidth}
			\tikzsetnextfilename{Convergence-Legendre-W7X-EIM-s0200-Er-0-D11}
			\includegraphics{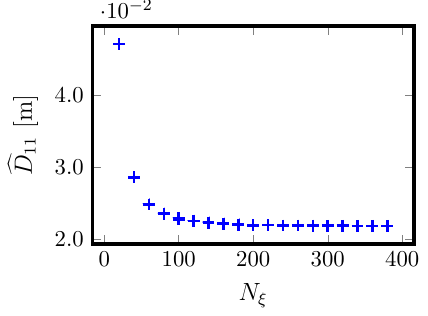}
			\caption{}
			\label{subfig:D11_Convergence_W7X_EIM_Er_0_Legendre}
		\end{subfigure}
		\begin{subfigure}[t]{0.48\textwidth}
			\tikzsetnextfilename{Convergence-Legendre-W7X-EIM-s0200-Er-0-D33}
			\includegraphics{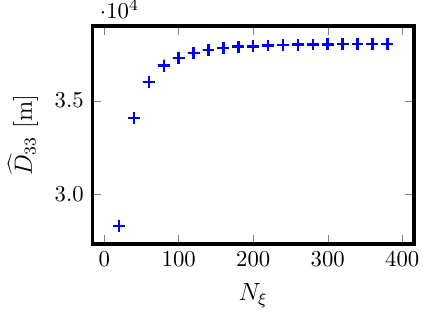}
			\caption{}
			\label{subfig:D33_Convergence_W7X_EIM_Er_0_Legendre}
		\end{subfigure}
		
		\begin{subfigure}[t]{0.48\textwidth}
			\tikzsetnextfilename{Convergence-Legendre-W7X-EIM-s0200-Er-0-D31-Detail}
			\includegraphics{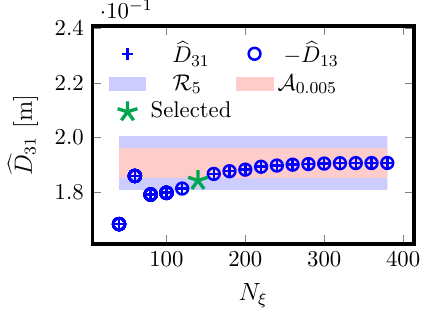}
			\caption{}
			\label{subfig:D31_convergence_Legendre_W7X_EIM_0200_Erho_0_Detail}
		\end{subfigure}
		\begin{subfigure}[t]{0.48\textwidth}
			\tikzsetnextfilename{Convergence-theta-zeta-W7X-EIM-s0200-Er-0-D31}
			\includegraphics{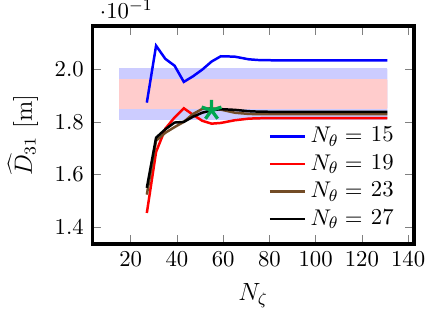}
            \caption{}
			\label{subfig:D31_convergence_theta_zeta_W7X_EIM_0200_Erho_0}
		\end{subfigure}
		\caption{Convergence of monoenergetic coefficients with the number of Legendre modes $N_\xi$ and convergence of $\Dij{31}$ with $N_\theta$ and $N_\zeta$ for the selected value of $N_\xi$ for W7X-EIM at the surface labelled by $\psi/\psi_{\text{lcfs}}=0.200$, for $\hat{\nu}=10^{-5}$ $\text{m}^{-1}$ and $\widehat{E}_r=0$ $\text{V}\cdot\text{s}/\text{m}^2$.}
		\label{fig:Convergence_W7X_EIM_Er_0}
	\end{figure}

	\begin{figure}[h]
		\centering
		\begin{subfigure}[t]{0.48\textwidth}
			\tikzsetnextfilename{Convergence-Legendre-W7X-EIM-s0200-Er-3e-4-D11}
			\includegraphics{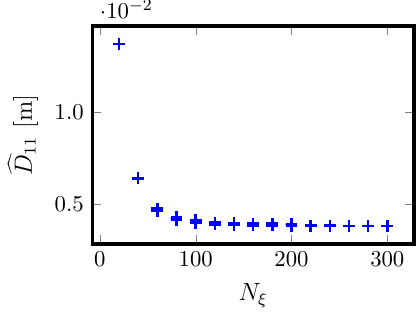}
			\caption{}
			\label{subfig:D11_convergence_Legendre_W7X_EIM_0200_Erho_3e-4}
		\end{subfigure}
		\begin{subfigure}[t]{0.48\textwidth}
			\tikzsetnextfilename{Convergence-Legendre-W7X-EIM-s0200-Er-3e-4-D33}
			\includegraphics{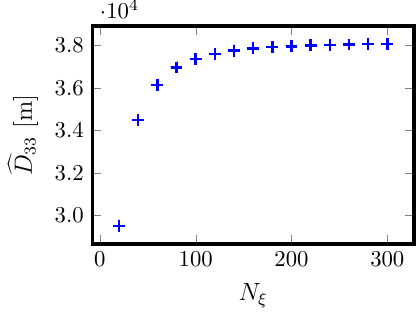}
			\caption{}
			\label{subfig:D33_convergence_Legendre_W7X_EIM_0200_Erho_3e-4}
		\end{subfigure}

		\begin{subfigure}[t]{0.48\textwidth}
			\tikzsetnextfilename{Convergence-Legendre-W7X-EIM-s0200-Er-3e-4-D31-Detail}
			\includegraphics{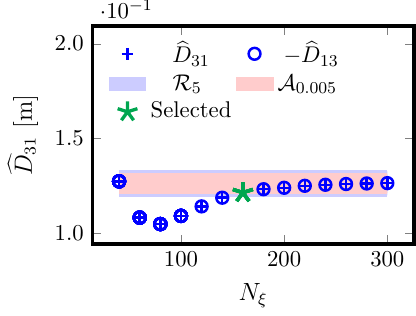}
			\caption{}
			\label{subfig:D31_convergence_Legendre_W7X_EIM_0200_Erho_3e-4_Detail}
		\end{subfigure}
		\begin{subfigure}[t]{0.48\textwidth}
			\tikzsetnextfilename{Convergence-theta-zeta-W7X-EIM-s0200-Er-3e-4-D31}
			\includegraphics{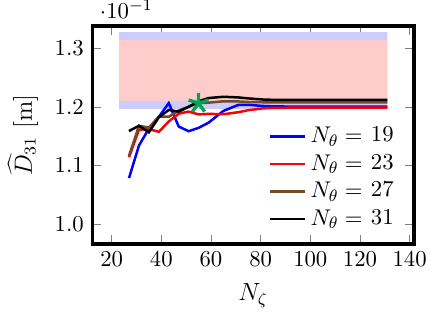}
			\caption{}
			\label{subfig:D31_convergence_theta_zeta_W7X_EIM_0200_Erho_3e-4_Detail}
		\end{subfigure}

		\caption{Convergence of monoenergetic coefficients with the number of Legendre modes $N_\xi$ and convergence of $\Dij{31}$ with $N_\theta$ and $N_\zeta$ for the selected value of $N_\xi$ for W7X-EIM at the surface labelled by $\psi/\psi_{\text{lcfs}}=0.200$, for $\hat{\nu}=10^{-5}$ $\text{m}^{-1}$ and $\widehat{E}_r=3\cdot 10^{-4}$ $\text{V}\cdot\text{s}/\text{m}^2$.}
		\label{fig:Convergence_W7X_EIM_Er_3e-4}
	\end{figure}
	
	In this subsection, we will select resolutions $N_\theta$, $N_\zeta$ and $N_\xi$ for which {\MONKES} provides accurate calculations of the monoenergetic coefficients in a wide range of collisionalities. For this, we will study how the monoenergetic coefficients computed by {\MONKES} converge with $N_\theta$, $N_\zeta$ and $N_\xi$ at low collisionality. From the point of view of numerical analysis, the need for large values of $N_\xi$ is due to the lack of diffusion along $\xi$ in equation (\ref{eq:DKE}). Hence, if {\MONKES} is capable of producing fast and accurate calculations at low collisionality, it will also produce fast and accurate calculations at higher collisionalities. 
	
	\begin{figure}[h]
		\centering
		\begin{subfigure}[t]{0.48\textwidth}
			\tikzsetnextfilename{Convergence-Legendre-W7X-KJM-s0204-Er-0-D11}
			\includegraphics{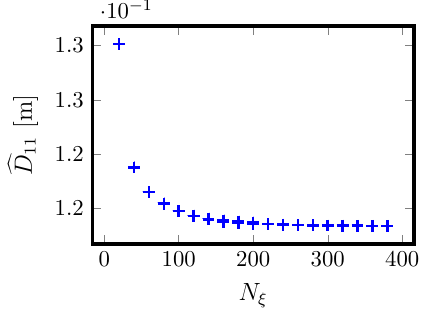}
			\caption{}
			\label{subfig:D11_convergence_Legendre_W7X_KJM_0204_Erho_0}
		\end{subfigure}
		\begin{subfigure}[t]{0.48\textwidth}
			\tikzsetnextfilename{Convergence-Legendre-W7X-KJM-s0204-Er-0-D33}
			\includegraphics{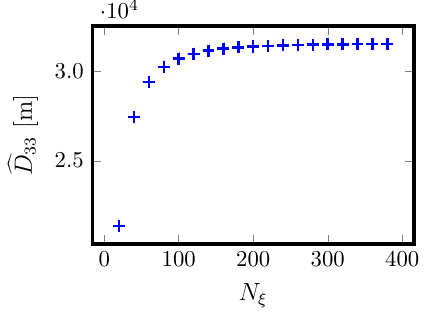}
			\caption{}
			\label{subfig:D33_convergence_Legendre_W7X_KJM_0204_Erho_0}
		\end{subfigure}

		\begin{subfigure}[t]{0.48\textwidth}
			\tikzsetnextfilename{Convergence-Legendre-W7X-KJM-s0204-Er-0-D31-Detail}
			\includegraphics{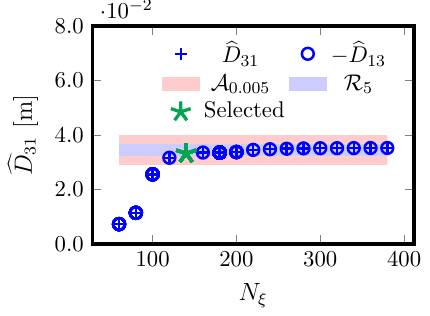} 
			\caption{}
			\label{subfig:D31_convergence_Legendre_W7X_KJM_0204_Erho_0_Detail}
		\end{subfigure}
		\begin{subfigure}[t]{0.48\textwidth}
			\tikzsetnextfilename{Convergence-theta-zeta-W7X-KJM-s0204-Er-0-D31}
			\includegraphics{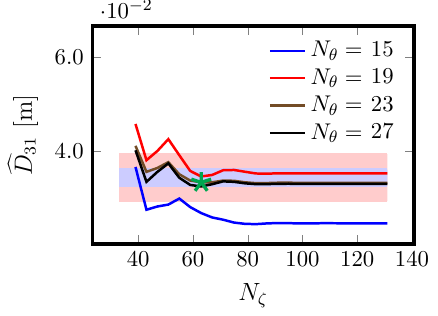}
			\caption{}
			\label{subfig:D31_convergence_theta_zeta_W7X_KJM_0204_Erho_0}
		\end{subfigure}
		\caption{Convergence of monoenergetic coefficients with the number of Legendre modes $N_\xi$ and convergence of $\Dij{31}$ with $N_\theta$ and $N_\zeta$ for the selected value of $N_\xi$ for W7X-KJM at the surface labelled by $\psi/\psi_{\text{lcfs}}=0.204$, for $\hat{\nu}=10^{-5}$ $\text{m}^{-1}$ and $\hat{E}_r=0$ $\text{V}\cdot\text{s}/\text{m}^2$.}
		\label{fig:Convergence_W7X_KJM_Er_0}
	\end{figure}
	
	For the convergence study, we select three different magnetic configurations at a single flux surface. Two of them correspond to configurations of W7-X: EIM and KJM. The third one corresponds to the new QI ``flat mirror'' \cite{velasco2023robust} configuration CIEMAT-QI \cite{Sanchez_2023}. The calculations are done for the $1/\nu$ (cases with $\widehat{E}_r=0$) and $\sqrt{\nu}$-$\nu$ regimes \cite{dherbemont2022} (cases with $\widehat{E}_r\ne 0$) at the low collisionality value $\hat{\nu}=10^{-5}$ $\text{m}^{-1}$. In table \ref{tab:Convergence_cases} the cases considered are listed, including their correspondent values of $\widehat{E}_r:= \widehat{E}_\psi\dv*{\psi}{r}$. We have denoted $r = r_{\text{lcfs}} \sqrt{\psi/\psi_{\text{lcfs}}}$ where $r_{\text{lcfs}}$ is the minor radius of the device\footnote{{\DKES} uses $r$ as radial coordinate instead of $\psi$. The quantities $\hat{\nu}$ and $\widehat{E}_r$ are denoted respectively \texttt{CMUL} and \texttt{EFIELD} in the code {\DKES}.}.
	\begin{table}[h]
		\centering
		\begin{tabular}{@{}lccc@{}}
			\toprule
			Configuration & $\psi/\psi_{\text{lcfs}}$ & $\hat{\nu}$ $[\text{m}^{-1}]$ & $\widehat{E}_r$  $[\text{V}\cdot\text{s}/\text{m}^2]$   \\ \midrule
			W7X-EIM       & 0.200                     & $10^{-5}$   & 0 \\
			W7X-EIM       & 0.200                     & $10^{-5}$   & $3\cdot10^{-4}$ \\
			W7X-KJM       & 0.204                     & $10^{-5}$   & 0 \\
			W7X-KJM       & 0.204                     & $10^{-5}$   & $3\cdot10^{-4}$ \\ 
			CIEMAT-QI     & 0.250                     & $10^{-5}$   & 0       \\
			CIEMAT-QI     & 0.250                     & $10^{-5}$   & $10^{-3}$       \\
			\bottomrule
		\end{tabular}
		\caption{Cases considered in the convergence study of monoenergetic coefficients and values of $(\hat{\nu},\widehat{E}_r)$.}
		\label{tab:Convergence_cases}
	\end{table}
	
	\begin{figure}[h]
		\centering
		\begin{subfigure}[t]{0.48\textwidth}
			\tikzsetnextfilename{Convergence-Legendre-W7X-KJM-s0204-Er-3e-4-D11}
			\includegraphics{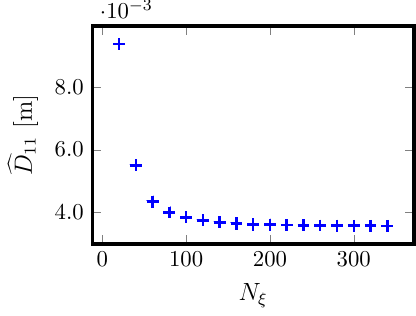}
			\caption{}
			\label{subfig:D11_convergence_Legendre_W7X_KJM_0204_Erho_3e-4}
		\end{subfigure}
		\begin{subfigure}[t]{0.48\textwidth}
			\tikzsetnextfilename{Convergence-Legendre-W7X-KJM-s0204-Er-3e-4-D33}
			\includegraphics{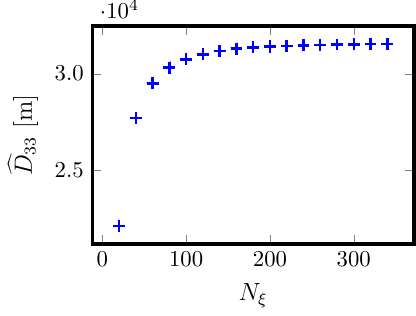}
			\caption{}
			\label{subfig:D33_convergence_Legendre_W7X_KJM_0204_Erho_3e-4}
		\end{subfigure}

		\begin{subfigure}[t]{0.48\textwidth}
			\tikzsetnextfilename{Convergence-Legendre-W7X-KJM-s0204-Er-3e-4-D31-Detail}
			\includegraphics{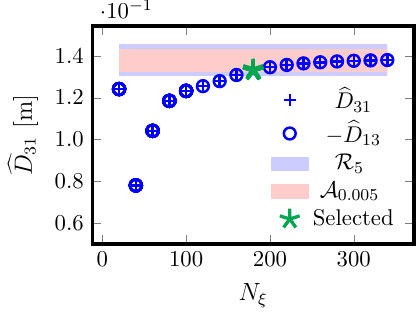}
			\caption{}
			\label{subfig:D31_convergence_Legendre_W7X_KJM_0204_Erho_3e-4_Detail}
		\end{subfigure}
		\begin{subfigure}[t]{0.48\textwidth}
			\tikzsetnextfilename{Convergence-theta-zeta-W7X-KJM-s0204-Er-3e-4-D31}
			\includegraphics{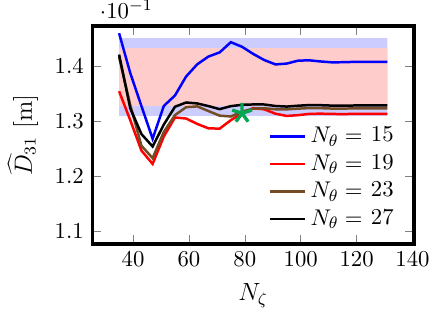}
			\caption{}
			\label{subfig:D31_convergence_theta_zeta_W7X_KJM_0204_Erho_3e-4_Detail}
		\end{subfigure}

		\caption{Convergence of monoenergetic coefficients with the number of Legendre modes $N_\xi$ and convergence of $\Dij{31}$ with $N_\theta$ and $N_\zeta$ for the selected value of $N_\xi$ for W7X-KJM at the surface labelled by $\psi/\psi_{\text{lcfs}}=0.204$, for $\hat{\nu}=10^{-5}$ $\text{m}^{-1}$ and $\widehat{E}_r=3\cdot 10^{-4}$ $\text{V}\cdot\text{s}/\text{m}^2$.}
		\label{fig:Convergence_W7X_KJM_Er_3e-4}
	\end{figure}
	
	In order to select the triplets $(N_\theta,N_\zeta, N_\xi)$ for sufficiently accurate calculations of $\widehat{D}_{31}$, we need to specify when we will consider that a computation has converged. For each case of table \ref{tab:Convergence_cases} we will proceed in the same manner. First, we plot the coefficients $\widehat{D}_{ij}$ as functions of the number of Legendre modes in a sufficiently wide interval. For each value of $N_\xi$, the selected spatial resolutions $N_\theta$ and $N_\zeta$ are large enough so that increasing them varies the monoenergetic coefficients in less than a 1\%. We will say that these calculations are ``spatially converged''. Since, typically, the most difficult coefficient to calculate is the bootstrap current coefficient, we will select the resolutions so that $\widehat{D}_{31}$ is accurately computed. From the curve of (spatially converged) $\widehat{D}_{31}$ as a function of $N_\xi$ we define our converged reference value, which we denote by $\widehat{D}_{31}^{\text{r}}$, as the converged calculation to three significant digits. From this converged reference value we will define two regions. A first region
	\begin{align}
		\mathcal{R}_{\epsilon}:=
		\left[
		(1-\epsilon/100)\widehat{D}_{31}^{\text{r}}, (1+\epsilon/100)\widehat{D}_{31}^{\text{r}} 
		\right]
	\end{align} 
	for calculations that deviate less than or equal to an $\epsilon$\% with respect to $\widehat{D}_{31}^{\text{r}}$. This interval will be used for selecting the resolutions through the following convergence criteria. We say that, for fixed $(N_\theta,N_\zeta,N_\xi)$ and $\epsilon$, a calculation $\widehat{D}_{31}\in\mathcal{R}_{\epsilon}$ is sufficiently converged if two conditions are satisfied 
	\begin{enumerate}
		\item Spatially converged calculations with $N_\xi'\ge N_\xi$ belong to $\mathcal{R}_{\epsilon}$.
		\item Increasing $N_\theta$ and $N_\zeta$ while keeping $N_\xi$ constant produces calculations which belong to $\mathcal{R}_{\epsilon}$.
	\end{enumerate}
	Condition (i) is used to select the number of Legendre modes $N_\xi$ and condition (ii) is used to select the values of $N_\theta$ and $N_\zeta$ once $N_\xi$ is fixed. 
	
	Additionally, we define a second interval 
	\begin{align}
		\mathcal{A}_{\epsilon}:=
		\left[
		\widehat{D}_{31}^{\text{r}}-\epsilon, \widehat{D}_{31}^{\text{r}}+\epsilon 
		\right]
	\end{align}
	to distinguish which calculations are at a distance smaller than or equal to $\epsilon$ from $\widehat{D}_{31}^{\text{r}}$. The reason to have two different regions is that for stellarators close to QI, the relative convergence criteria can become too demanding (the smaller $\widehat{D}_{31}^{\text{r}}$ is, the narrower $\mathcal{R}_{\epsilon}$ becomes). Nevertheless, for optimizing QI configurations, it is sufficient to ensure that $|\widehat{D}_{31}|$ is sufficiently small in absolute terms. If the absolute error is much smaller than a value of $|\widehat{D}_{31}|$ that can be considered sufficiently small, the calculation is converged for optimization purposes. We will use this interval for two reasons: first, to give a visual idea of how narrow $\mathcal{R}_{\epsilon}$ becomes. Second, to show that if $\mathcal{R}_{\epsilon}$ is very small, it is easier to satisfy an absolute criterion than a relative one.

	\begin{figure}[h]
		\centering
		\begin{subfigure}[t]{0.48\textwidth}
			\tikzsetnextfilename{Convergence-Legendre-CIEMAT-QI-s0250-Er-0-D11}
			\includegraphics{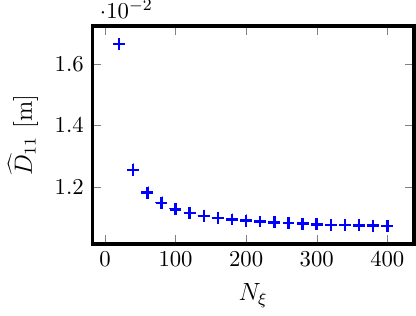}
			\caption{}
			\label{subfig:D11_convergence_Legendre_CIEMAT_QI_0250_Erho_0}
		\end{subfigure}
		\begin{subfigure}[t]{0.48\textwidth}
			\tikzsetnextfilename{Convergence-Legendre-CIEMAT-QI-s0250-Er-0-D33}
			\includegraphics{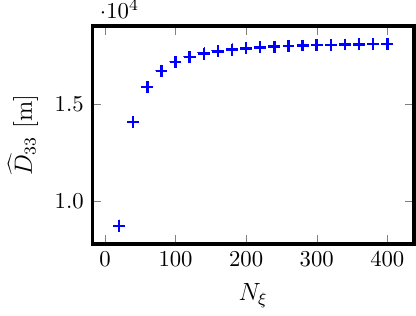}
			\caption{}
			\label{subfig:D33_convergence_Legendre_CIEMAT_QI_0250_Erho_0}
		\end{subfigure}

		\begin{subfigure}[t]{0.48\textwidth}
			\tikzsetnextfilename{Convergence-Legendre-CIEMAT-QI-s0250-Er-0-D31-Detail}
			\includegraphics{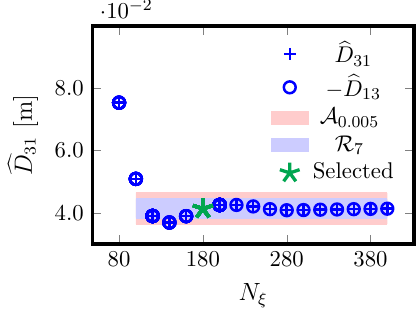}
			\caption{}
			\label{subfig:D31_convergence_Legendre_CIEMAT_QI_0250_Erho_0_Detail}
		\end{subfigure}
		\begin{subfigure}[t]{0.48\textwidth}
			\tikzsetnextfilename{Convergence-theta-zeta-CIEMAT-QI-s0250-Er-0-D31}
			\includegraphics{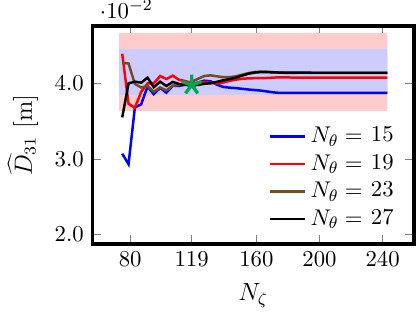}
			\caption{}
			\label{subfig:D31_convergence_theta_zeta_CIEMAT_QI_0250_Erho_0}
		\end{subfigure}

		\caption{Convergence of monoenergetic coefficients with the number of Legendre modes $N_\xi$ and convergence of $\Dij{31}$ with $N_\theta$ and $N_\zeta$ for the selected value of $N_\xi$ for CIEMAT-QI at the surface labelled by $\psi/\psi_{\text{lcfs}}=0.25$, for $\hat{\nu}=10^{-5}$ $\text{m}^{-1}$ and $\widehat{E}_r=0$ $\text{V}\cdot\text{s}/\text{m}^2$.}
		\label{fig:Convergence_CIEMAT_QI_Er_0}
	\end{figure}
	\begin{figure}[h]
		\centering
		\begin{subfigure}[t]{0.48\textwidth}
			\tikzsetnextfilename{Convergence-Legendre-CIEMAT-QI-s0250-Er-1e-3-D11}
			\includegraphics{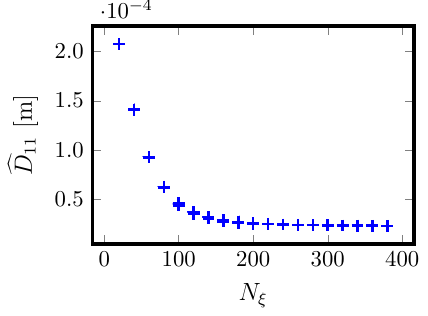}
			\caption{}
			\label{subfig:D11_convergence_Legendre_CIEMAT_QI_0250_Erho_1e-3}
		\end{subfigure}
		\begin{subfigure}[t]{0.48\textwidth}
			\tikzsetnextfilename{Convergence-Legendre-CIEMAT-QI-s0250-Er-1e-3-D33}
			\includegraphics{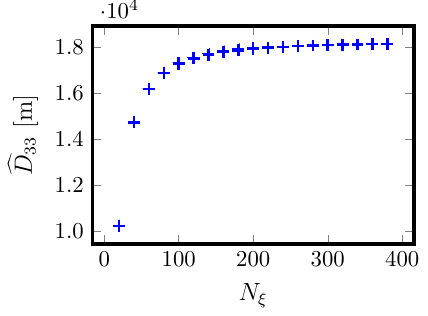}
			\caption{}
			\label{subfig:D33_convergence_Legendre_CIEMAT_QI_0250_Erho_1e-3}
		\end{subfigure}
		
		\begin{subfigure}[t]{0.48\textwidth}
			\tikzsetnextfilename{Convergence-Legendre-CIEMAT-QI-s0250-Er-1e-3-D31-Detail}
			\includegraphics{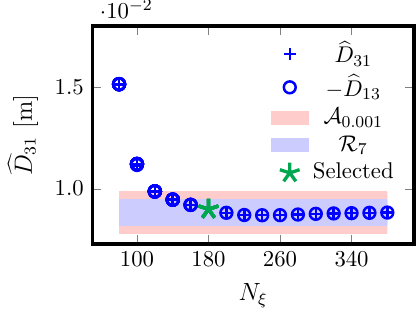}
			\caption{}
			\label{subfig:D31_convergence_Legendre_CIEMAT_QI_0250_Erho_1e-3_Detail}
		\end{subfigure}
		\begin{subfigure}[t]{0.48\textwidth}
			\tikzsetnextfilename{Convergence-theta-zeta-CIEMAT-QI-s0250-Er-1e-3-D31}
			\includegraphics{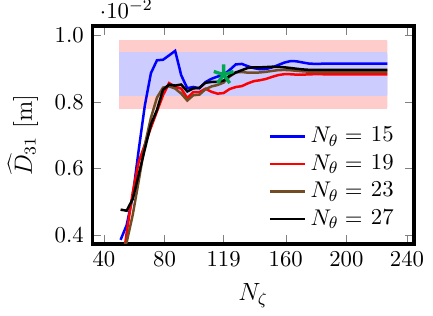}
			\caption{}
			\label{subfig:D31_convergence_theta_zeta_CIEMAT_QI_0250_Erho_1e-3}
		\end{subfigure}
		\caption{Convergence of monoenergetic coefficients with the number of Legendre modes $N_\xi$ and convergence of $\Dij{31}$ with $N_\theta$ and $N_\zeta$ for the selected value of $N_\xi$ for CIEMAT-QI at the surface labelled by $\psi/\psi_{\text{lcfs}}=0.25$, for $\hat{\nu}=10^{-5}$ $\text{m}^{-1}$ and $\widehat{E}_r=10^{-3}$ $\text{V}\cdot\text{s}/\text{m}^2$.}
		\label{fig:Convergence_CIEMAT_QI_Er_1e-3}
	\end{figure}

	Figure \ref{fig:Convergence_W7X_EIM_Er_0} shows the convergence of monoenergetic coefficients with the number of Legendre modes for W7-X EIM when $\widehat{E}_r=0$. From figures \ref{subfig:D11_Convergence_W7X_EIM_Er_0_Legendre} and \ref{subfig:D33_Convergence_W7X_EIM_Er_0_Legendre} we see that the radial transport ($\Dij{11}$) and parallel conductivity ($\Dij{33}$) coefficients converge monotonically with $N_\xi$. On the other hand, the bootstrap current coefficient is more difficult to converge as it can be seen on figure \ref{subfig:D31_convergence_Legendre_W7X_EIM_0200_Erho_0_Detail}. As a sanity check, the fulfilment of the Onsager symmetry relation $\widehat{D}_{31}= - \widehat{D}_{13}$ is included. The converged reference value $\widehat{D}_{31}^{\text{r}}$ is the spatially converged calculation for $N_\xi=380$. Defining a region of relative convergence of $\epsilon=5\%$, allows to select a resolution of $N_\xi=140$ Legendre modes to satisfy condition (i). The selection is indicated with a five-pointed green star. Note that for this case, an absolute deviation of $0.005$ m from $\widehat{D}_{31}^{\text{r}}$ is slightly more demanding than the relative deviation 
	condition. This absolute deviation is selected as the 5\% of $\widehat{D}_{31}\sim 0.1$ m, which can be considered a small value of $\widehat{D}_{31}$ (this value is typical of W7-X KJM). From figure \ref{subfig:D31_convergence_theta_zeta_W7X_EIM_0200_Erho_0} we choose the resolutions $(N_\theta,N_\zeta)=(23,55)$ to satisfy convergence condition (ii). 
	
	The case of W7-X EIM with $\widehat{E}_r\ne 0$ is shown in figure \ref{fig:Convergence_W7X_EIM_Er_3e-4}. We note from figure \ref{subfig:D31_convergence_Legendre_W7X_EIM_0200_Erho_3e-4_Detail} that obtaining sufficiently converged results for the region $\mathcal{R}_{5}$ is more difficult than in the case without radial electric field. For this case, the sizes of the intervals $\mathcal{A}_{0.005}$ and $\mathcal{R}_{5}$ are almost the same. This is in part due to the fact that the $\widehat{D}_{31}$ coefficient is smaller in absolute value and thus, the region $\mathcal{R}_{5}$ is narrower. We select $N_\xi=160$ to satisfy condition (i). The selection $(N_\theta,N_\zeta)=(27,55)$ satisfies condition (ii) as shown in figure \ref{subfig:D31_convergence_theta_zeta_W7X_EIM_0200_Erho_3e-4_Detail}. 
	
	The convergence curves for the case of W7-X KJM when $\widehat{E}_r=0$ are shown in figure \ref{fig:Convergence_W7X_KJM_Er_0}. Due to the smallness of $\widehat{D}_{31}^{\text{r}}$, the amplitude of the region $\mathcal{R}_{5}$ is much narrower than in the EIM case, being of order $10^{-3}$. It is so narrow that the absolute value region $\mathcal{A}_{0.005}$ contains the relative convergence region. It is shown in figure \ref{subfig:D31_convergence_Legendre_W7X_KJM_0204_Erho_0_Detail} that taking $N_\xi=140$ is sufficient to satisfy condition (i). According to the convergence curves plotted in figure \ref{subfig:D31_convergence_theta_zeta_W7X_KJM_0204_Erho_0}, selecting $(N_\theta,N_\zeta)=(23,63)$ ensures satisfying condition (ii). 
	
	The case of W7-X KJM for finite $\widehat{E}_r$ is shown in figure \ref{fig:Convergence_W7X_KJM_Er_3e-4}. The selection of $N_\xi=180$ Legendre modes, indicated in figure \ref{subfig:D31_convergence_Legendre_W7X_KJM_0204_Erho_3e-4_Detail}, satisfies convergence condition (i). As shown in figure \ref{subfig:D31_convergence_theta_zeta_W7X_KJM_0204_Erho_3e-4_Detail}, condition (ii) is satisfied by the selection $(N_\theta, N_\zeta)=(19,79)$. 
	
	The convergence of monoenergetic coefficients for CIEMAT-QI without $\widehat{E}_r$ is shown in figure \ref{fig:Convergence_CIEMAT_QI_Er_0}. Note that as in the W7-X KJM case at this regime, the region of absolute error $\mathcal{A}_{0.005}$ is bigger than the relative one. As the monoenergetic coefficients are smaller, we relax the relative convergence parameter to $\epsilon=7\%$. In figure \ref{subfig:D31_convergence_Legendre_CIEMAT_QI_0250_Erho_0_Detail} we see that the region of 7\% of deviation $\mathcal{R}_{7}$ is quite narrow and that selecting $N_\xi=180$ satisfies condition (i). To satisfy condition (ii), we choose the resolutions $(N_\theta,N_\zeta)=(15,119)$ as shown in figure \ref{subfig:D31_convergence_theta_zeta_CIEMAT_QI_0250_Erho_0}.

	Finally, the case of CIEMAT-QI with $\widehat{E}_r\ne 0$ is shown in figure \ref{fig:Convergence_CIEMAT_QI_Er_1e-3}. Looking at figure \ref{subfig:D31_convergence_Legendre_CIEMAT_QI_0250_Erho_1e-3_Detail} we can check that taking $N_\xi=180$ satisfies condition (i) for the region $\mathcal{R}_7$ of 7\% of deviation. In this case, the region of absolute error $\mathcal{A}_{0.001}$ is five times smaller than in the rest of cases and is still bigger than the relative error region. As shown in figure \ref{subfig:D31_convergence_theta_zeta_CIEMAT_QI_0250_Erho_1e-3}, the selection $(N_\theta,N_\zeta)=(15,119)$ satisfies condition (ii).

	\section{Code performance}
	\label{subsec:Performance}
	In this section we will compare {\MONKES} and {\DKES} performance in terms of the wall-clock time and describe {\MONKES} scaling properties. For the wall-clock time comparison, a convergence study (similar to the one explained in subsection \ref{subsec:Convergence}) is carried out for {\DKES} on appendix \ref{sec:Appendix_DKES_Bounds}. This convergence study is done to compare the wall-clock times between {\MONKES} and {\DKES} for the same level of relative convergence with respect to $\widehat{D}_{31}^{\text{r}}$. The comparison is displayed in table \ref{tab:DKES_MONKES_Comparison} along with the minimum number of Legendre modes for which the calculations of {\DKES} satisfy convergence condition (i). In all six cases, {\MONKES} is much faster than {\DKES} despite using more Legendre modes. Even for W7-X EIM, in which we have taken $N_\xi =40$ for {\DKES} calculations with finite $\widehat{E}_r$, {\MONKES} is $\sim 4$ times faster using almost four times the number of Legendre modes. For the W7-X EIM case without radial electric field, the speed-up is also of $ 4$. For the W7X-KJM configuration, {\MONKES} is $\sim 20$ times faster than {\DKES} without $\widehat{E}_r$ and $\sim 10$ times faster than {\DKES} when $\widehat{E}_r \ne 0$. In the case of CIEMAT-QI, {\MONKES} is more than $\sim 13$ times faster than {\DKES} without radial electric field. In the case with finite $\widehat{E}_r$, {\MONKES} calculations are around 64 times faster than {\DKES} ones. One calculation of {\MONKES} takes less than a minute and a half and the same calculation with {\DKES} requires waiting for almost an hour and a half. The disparity of wall-clock times reflects the superiority at low collisionality of the block tridiagonal algorithm used by {\MONKES} when compared to the iterative method used by {\DKES} to solve the variational principle. The conjugate gradient method used by {\DKES} converges slower (i.e. requires more iterations) when $\hat{\nu}$ decreases while the performance of the block tridiagonal method does not depend on $\hat{\nu}$. We point out that the wall-clock times for all the calculations shown are those from one of the partitions of CIEMAT's cluster XULA. Specifically, partition number 2 has been used, whose nodes run with Intel Xeon Gold 6254 cores at 3.10 GHz. 

	\begin{table}[h]
		\centering
		\begin{tabular}{ccccc}
			\toprule
			Case   & $N_\xi^{\DKES}$ & $N_\xi^{\MONKES}$  & $t_{\text{clock}}^{\DKES}$  & $t_{\text{clock}}^{\MONKES}$  \\ \midrule
			\makecell[c]{W7X-EIM   $\widehat{E}_r=0$}       &  80   &  140   &  90     s  &  22 s   
			\\
			\makecell[c]{W7X-EIM $\widehat{E}_r\ne 0$ }   &  40   &  160   &  172    s  &  35 s   
			\\ 
			\makecell[c]{W7X-KJM $\widehat{E}_r=0$ }      &  160  &  140   &  698    s  &  31 s   
			\\
			\makecell[c]{W7X-KJM $\widehat{E}_r\ne 0$ }   &  60   &  180   &  421    s  &  47 s   
			\\
			\makecell[c]{CIEMAT-QI $\widehat{E}_r=0$ }    &  160  &  180   &  1060   s  &  76 s  
			\\
			\makecell[c]{CIEMAT-QI $\widehat{E}_r\ne 0$ } &  160  &  180   &  4990   s  &  76 s   
			\\ 
			\bottomrule 
		\end{tabular}
		\caption{Comparison between the wall-clock time of {\DKES} and {\MONKES}.}
		\label{tab:DKES_MONKES_Comparison}
	\end{table}
	
	\begin{figure}[h]
		\centering
		\begin{subfigure}[t]{0.49\textwidth}
			\tikzsetnextfilename{MONKES-Scaling-Legendre}
			\includegraphics{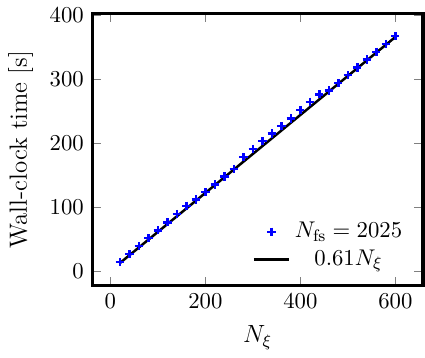} 
			\caption{}
			\label{subfig:MONKES_Scaling_Legendre}
		\end{subfigure}    
		\begin{subfigure}[t]{0.49\textwidth}
			\tikzsetnextfilename{MONKES-Scaling-Flux-Surface}
			\includegraphics{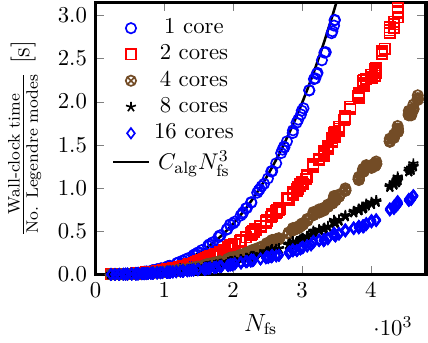}
			\caption{}
			\label{subfig:MONKES_Scaling_Nfs}
		\end{subfigure}
		\caption{Scaling of {\MONKES} wall-clock time. (a) Linear scaling with the number of Legendre modes for $N_{\text{fs}}=27\times 75 = 2025$ discretization points. (b) Cubic scaling with $N_{\text{fs}}$ for different number of cores used.}
		\label{fig:MONKES_Scaling}
	\end{figure}
	
	We next check that the arithmetic complexity of the algorithm described in section \ref{chap:MONKES} holds in practice. The scaling of {\MONKES} with the number of Legendre modes $N_\xi$ and the number of points in which the flux surface is discretized is shown in figure \ref{fig:MONKES_Scaling}. To demonstrate the linear scaling, the wall-clock time as a function of $N_\xi$ for $N_{\text{fs}}=2025$ points is represented in figure \ref{subfig:MONKES_Scaling_Legendre} and compared with the line of slope 0.61 seconds per Legendre mode. As can be seen in figure \ref{subfig:MONKES_Scaling_Nfs}, the wall-clock time (per Legendre mode) scales cubicly with the number of points $N_{\text{fs}}$ in which the flux surface is discretized. As it was mentioned at the end of section \ref{chap:MONKES}, the constant $C_{\text{alg}}$ in a single core is sufficiently small to give accurate calculations up to $\hat{\nu}\sim 10^{-5}$ $\text{m}^{-1}$. We have plotted in figure \ref{subfig:MONKES_Scaling_Nfs} the cubic fit $C_{\text{alg}}N_{\text{fs}}^3$, where $C_{\text{alg}}=0.61(1/2025)^3\sim 7\cdot 10^{-11}$ s. 
	
	As the {\texttt{LAPACK}} library is multithreaded and allows to parallelize the linear algebra operations through several cores, it is worth verifying the scaling of {\MONKES} when running in parallel. Additionally, for the resolutions selected in subsection \ref{subsec:Convergence}, we display in table \ref{tab:MONKES_Times_Multicore} the wall-clock time when running {\MONKES} using several cores in parallel. Note that for the W7-X cases, which require a smaller value of $N_{\text{fs}}$, the speed-up stalls at 8 cores. For CIEMAT-QI, that requires discretizing the flux surface on a finer mesh, this does not happen in the range of cores considered.

	\begin{table}[h]
		\centering
		\begin{tabular}{lccccc}
			\toprule
			\backslashbox{Case}{No. cores}   & 1  & 2  &  4 &  8 & 16\\ \midrule
			W7X-EIM $\widehat{E}_r=0$        & 22    & 13  &  8   &  5 &  5 \\
			W7X-EIM $\widehat{E}_r\ne 0$   & 40    & 20  & 12    & 8 &  6 \\ 
			W7X-KJM $\widehat{E}_r=0$        & 33    & 17  &  12 & 7 &   7   \\
			W7X-KJM $\widehat{E}_r\ne 0$   & 46    & 17  &  13 & 7 & 7  \\
			CIEMAT-QI $\widehat{E}_r=0$      & 78    & 45 & 29 & 21 & 16 \\
			CIEMAT-QI $\widehat{E}_r\ne 0$ & 78    & 45 & 29 & 21 & 16 \\\bottomrule
		\end{tabular}
		\caption{Wall-clock time of {\MONKES} in seconds for the triplets $(N_\theta,N_\zeta,N_\xi)$ selected to ensure convergence} when running in several cores.
		\label{tab:MONKES_Times_Multicore}
	\end{table}

	\section{Benchmark of transport coefficients}
	\label{sec:Results_benchmark}
	
	\tikzsetnextfilename{Benchmark-monoenergetic-D11}
	\begin{figure}[h]
		\includegraphics{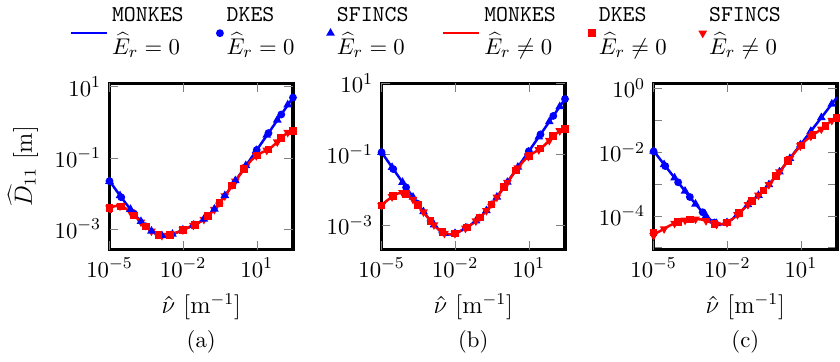}
		\caption{Calculation of the radial transport coefficient $\widehat{D}_{11}$ by \texttt{MONKES}, \texttt{DKES} and \texttt{SFINCS} for zero and finite $\widehat{E}_r$ for the three magnetic configurations considered. (a) W7X-EIM. (b) W7X-KJM. (c) CIEMAT-QI4.}
		\label{fig:Benchmark_monoenergetic_D11}
	\end{figure}

	Once we have chosen the resolutions $(N_\theta,N_\zeta,N_\xi)$ for each case, we need to verify that these selections indeed provide sufficiently accurate calculations of all the monoenergetic coefficients in the interval $\hat{\nu}\in[10^{-5},300]$ $\text{m}^{-1} $. It is instructive to recall what was mentioned at the beginning of subsection \ref{subsec:Convergence}: that the number of Legendre modes required for converged calculations of the monoenergetic coefficients decreases when $\hat{\nu}$ increases. Hence, the resolutions selected in subsection \ref{subsec:Convergence} also provide converged calculations for $\hat{\nu}\ge 10^{-5}$ $\text{m}^{-1}$. For instance, for the W7X-EIM case and collisionality $\hat{\nu}=10^{-4}$ $\text{m}^{-1}$, taking $N_\xi=20$ is sufficient to have calculations converged up to 5\% for zero and finite $\widehat{E}_r$. This means that for W7X-EIM the wall-clock times required by {\MONKES} calculations at $\hat{\nu}=10^{-4}$ $\text{m}^{-1}$ can be, at least, 7 times faster than for the case $\hat{\nu}=10^{-5}$ $\text{m}^{-1}$ shown in table \ref{tab:DKES_MONKES_Comparison}. In all cases, {\MONKES} calculations of the $\widehat{D}_{11}$ and $\widehat{D}_{31}$ coefficients will be benchmarked against converged calculations from {\DKES} (see appendix \ref{sec:Appendix_DKES_Bounds}) and from \texttt{SFINCS}\footnote{\texttt{SFINCS} calculations are converged up to 3\% in the three independent variables.}. The parallel conductivity coefficient will be benchmarked only against {\DKES}. 
	
	\tikzsetnextfilename{Benchmark-monoenergetic-D31} 
	\begin{figure}[h]
		\includegraphics{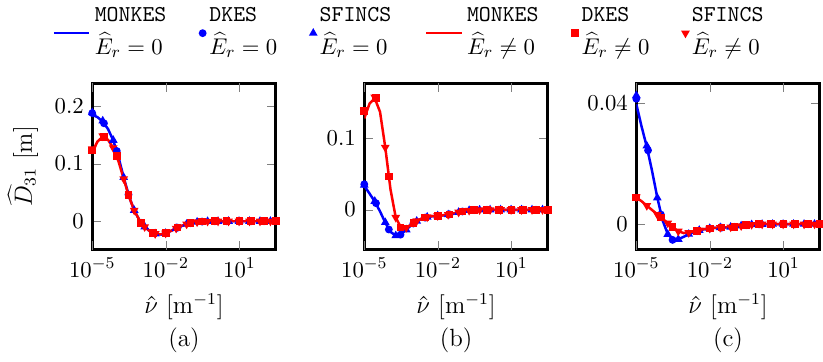}
		\caption{Calculation of the bootstrap current coefficient $\widehat{D}_{31}$ by \texttt{MONKES}, \texttt{DKES} and \texttt{SFINCS} for zero and finite $\widehat{E}_r$ for the three magnetic configurations considered. (a) W7X-EIM. (b) W7X-KJM. (c) CIEMAT-QI4.}
		\label{fig:Benchmark_monoenergetic_D31}
	\end{figure}

\begin{figure}[h]
		\includegraphics{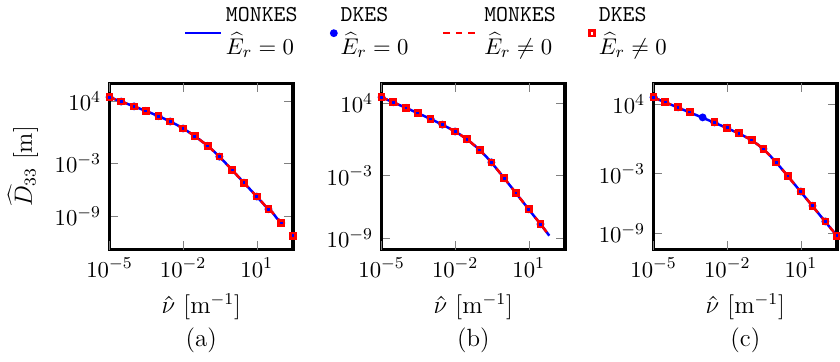}
		\caption{Calculation of the parallel conductivity coefficient $\widehat{D}_{33}$ by \texttt{MONKES} and \texttt{DKES} for zero and finite $\widehat{E}_r$ for the three magnetic configurations considered. (a) W7X-EIM. (b) W7X-KJM. (c) CIEMAT-QI4.}
		\label{fig:Benchmark_monoenergetic_D33}
	\end{figure}

	The benchmarking of the coefficient $\widehat{D}_{11}$ for the six different cases is shown in figure \ref{fig:Benchmark_monoenergetic_D11}. The result of the benchmark of the bootstrap current coefficient $\widehat{D}_{31}$ is shown in figure \ref{fig:Benchmark_monoenergetic_D31}. Finally, the parallel conductivity coefficient $\widehat{D}_{33}$ is benchmarked in figure \ref{fig:Benchmark_monoenergetic_D33}. Due to the weak effect of the radial electric field in the $\widehat{D}_{33}$ coefficient, the symbols for this plot have been changed. In all cases, the agreement between {\MONKES}, {\DKES} and \texttt{SFINCS} is almost perfect. Thus, we conclude that {\MONKES} calculations of the monoenergetic coefficients are not only fast, but also accurate. Additionally, we can evaluate the level of optimization of the three configurations considered by inspecting these plots. In figures \ref{fig:Benchmark_monoenergetic_D11}(a) and \ref{fig:Benchmark_monoenergetic_D11}(b) is shown that the W7X-EIM configuration has smaller radial transport coefficient than the W7X-KJM configuration. Figures \ref{fig:Benchmark_monoenergetic_D31}(a) and \ref{fig:Benchmark_monoenergetic_D31}(b) show that the smaller radial transport of the W7X-EIM  configuration comes at the expense of having larger bootstrap current coefficient. As shown in figures \ref{fig:Benchmark_monoenergetic_D11}(c) and \ref{fig:Benchmark_monoenergetic_D31}(c), the optimized stellarator CIEMAT-QI manages to achieve levels of radial transport similar or smaller than the W7X-EIM configuration and a bootstrap current coefficient as low as the W7X-KJM configuration.

	\tikzsetnextfilename{Benchmark-monoenergetic-D33}

	\section{Other capabilities of {\MONKES}}\label{sec:Other_capabilities_MONKES}
	Although the main purpose of {\MONKES} is the computation of monoenergetic coefficients $\Dij{ij}$, since the solution to the DKE (\ref{eq:DKE}) can also be computed, it is also possible to compute other quantities. In this section we will describe two non standard capabilities of {\MONKES}. The first one is the determination of the contribution of different classes of particles to the monoenergetic coefficients. A more practical capability of {\MONKES} for gradient-based optimization methods is the calculation of derivatives of the transport coefficients. The computational aspects of the methods for computing derivatives of the monoenergetic coefficients $\Dij{ij}$ described in section \ref{sec:Derivatives_monoenergetic} will be discussed. Additionally, some results of the adjoint method implemented in {\MONKES} will be shown.

	\subsection{Contribution of different classes of particles to the monoenergetic coefficients} \label{subsec:Contribution_lambda_MONKES}
	\begingroup
	\captionsetup[sub]{skip=-1.75pt, margin=100pt}
	\begin{figure}[h]
		\centering		         
		\tikzsetnextfilename{D11_lambda_W7X_KJM_Er_0} 
		\begin{subfigure}{0.42\textwidth}
			\includegraphics{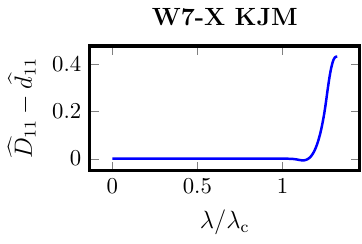} 
%
%
			\caption{}
			\label{subfig:D11_lambda_W7X_KJM_Er_0}
		\end{subfigure} 
		\tikzsetnextfilename{D11_lambda_W7X_KJM_Er_1e-3} 
		\begin{subfigure}{0.42\textwidth}
			\includegraphics{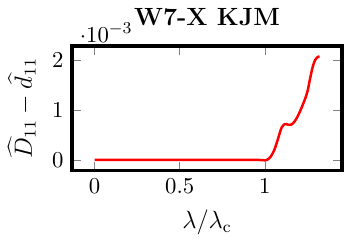} 
			\caption{}
			\label{subfig:D11_lambda_W7X_KJM_Er_1e-3}
		\end{subfigure} 

		\tikzsetnextfilename{D11_lambda_CIEMAT_QI_Er_0} 
		\begin{subfigure}{0.42\textwidth} 
			\includegraphics{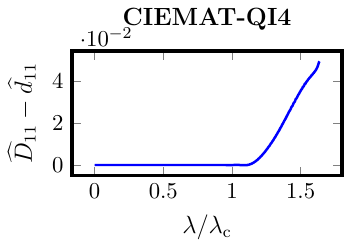}  
%
%
			\caption{}
			\label{subfig:D11_lambda_CIEMAT_QI_Er_0}
		\end{subfigure}
		\tikzsetnextfilename{D11_lambda_CIEMAT_QI_Er_1e-3}  
		\begin{subfigure}{0.42\textwidth}
			\includegraphics{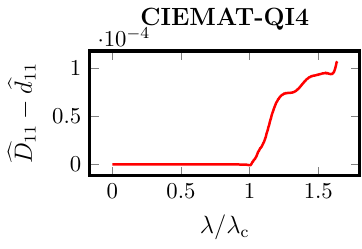}  
%
			\caption{}
			\label{subfig:D11_lambda_CIEMAT_QI_Er_1e-3}
		\end{subfigure}

		\tikzsetnextfilename{D11_lambda_QH_LP_Er_0}  
		\begin{subfigure}{0.42\textwidth}
			\includegraphics{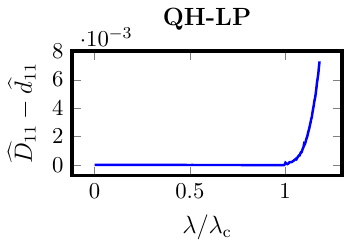}  
%
%
			\caption{}
			\label{subfig:D11_lambda_QH_LP_Er_0}
		\end{subfigure}
		\tikzsetnextfilename{D11_lambda_QH_LP_Er_1e-3}  
		\begin{subfigure}{0.42\textwidth}
			\includegraphics{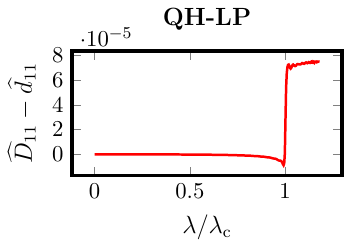}  
%
			\caption{}
			\label{subfig:D11_lambda_QH_LP_Er_1e-3}
		\end{subfigure}
		\label{fig:Monoenergetic_lambda_D11}
		\caption{Contribution of different classes of particles to the $\Dij{11}$ coefficient for the three magnetic configurations considered. Results with $\Er=0$ are indicated in blue and those with $\Er\ne 0$ in red.}
	\end{figure}

	In section \ref{subsec:Monoenergetic_lambda}, a manner to compute the contribution of different classes of particles (i.e. particles in different ranges of $\lambda$) to the monoenergetic coefficients employing a function $\dij{ij}(\lambda)$ was presented. The method described in that section to compute $\dij{ij}(\lambda)$ has been implemented in {\MONKES}. In this section, several examples of how this function can be used will be discussed, in particular, to confirm some analytical results from the literature. In regard to radial transport, we will confirm well known phase-space dependencies about the $1/\nu$ regime \cite{Nemov1999EvaluationO1} and the $\sqrt{\nu}$ regime for stellarators close to omnigenity \cite{Calvo_2017} employing the function $\dij{11}(\lambda)$. The curves for $\dij{31}(\lambda)$ and $\dij{13}(\lambda)$ will also be discussed. For the evaluation we have selected three different magnetic configurations, the first two are the W7X-KJM and CIEMAT-QI configurations from sections \ref{subsec:Convergence}, \ref{subsec:Performance} and \ref{sec:Results_benchmark}. The third configuration selected is the precise quasihelically (QH) symmetric configuration from \cite{Landreman_PreciseQS} at $\psi/\psi_{\text{lcfs}}=0.25$. For the three magnetic configurations, we have selected the values of (low) collisionality $\hat{\nu}$ and radial electric field $\Er$ corresponding to the CIEMAT-QI4 case from table \ref{tab:Convergence_cases}.

	In \cite{Nemov1999EvaluationO1}, it is shown that in the $1/\nu$ regime all classes of trapped particles contribute significantly to the effective ripple $\epseff$. As in this regime $\Dij{11}\propto\epseff^{3/2}/\hat{\nu}$, this result should also be reflected in the $\dij{11}(\lambda)$ curve. In figures \ref{subfig:D11_lambda_W7X_KJM_Er_0}, \ref{subfig:D11_lambda_CIEMAT_QI_Er_0} and 
	\ref{subfig:D11_lambda_QH_LP_Er_0} the dependence of $\Dij{11}-\dij{11}$ with $\lambda$ is shown for the three magnetic configurations. We recall that the difference in the value between two values of $\dij{11}(\lambda_1)-\dij{11}(\lambda_2)$ with $\lambda_1\le\lambda_2$ indicates the contribution to the $\Dij{11}$ coefficient of those classes of particles lying in the interval $[\lambda_1,\lambda_2]$. Hence, from the curves shown in figures \ref{subfig:D11_lambda_W7X_KJM_Er_0}, \ref{subfig:D11_lambda_CIEMAT_QI_Er_0} and 
	\ref{subfig:D11_lambda_QH_LP_Er_0}, we can immediately see that passing particles (those with $\lambda/\lambdac<1$) do not contribute to neoclassical radial transport in the $1/\nu$ regime. This was to be expected as the $1/\nu$ regime is originated from a non zero orbit-averaged radial drift (which corresponds to trapped particles) and a finite level of low collisionality in the absence (or irrelevance) of a radial electric field. From the lack of sudden jumps in the $\Dij{11}-\dij{11}$ curve for $\lambda/\lambdac>1$ we can see that there is no dominant class of trapped particles. That is, all classes of trapped particles contribute in a similar manner to radial transport in the $1/\nu$ regime.
	 
	\begin{figure}[h]
		\centering	
		\tikzsetnextfilename{D31_lambda_W7X_KJM}  
		\begin{subfigure}{0.42\textwidth}
			\includegraphics{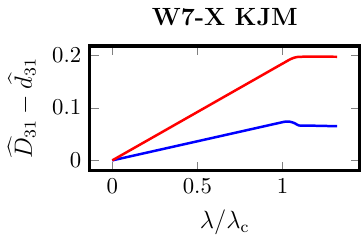}  
%
%
%
%
			\caption{}
			\label{subfig:D31_lambda_W7X_KJM}
		\end{subfigure}        
		\tikzsetnextfilename{D13_lambda_W7X_KJM}  
		\begin{subfigure}{0.42\textwidth}  
			\includegraphics{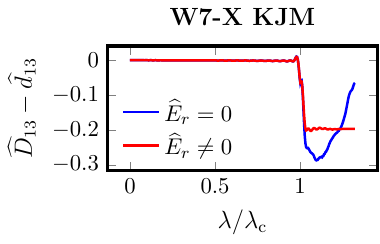}
%
%
%
			\caption{}
			\label{subfig:D13_lambda_W7X_KJM}
		\end{subfigure}

		\tikzsetnextfilename{D31_lambda_CIEMAT_QI} 
		\begin{subfigure}{0.42\textwidth} 
			\includegraphics{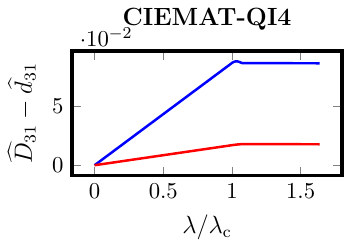}  
%
%
			\caption{}
			\label{subfig:D31_lambda_CIEMAT_QI}
		\end{subfigure}        
		\tikzsetnextfilename{D13_lambda_CIEMAT_QI}  
		\begin{subfigure}{0.42\textwidth}
			\includegraphics{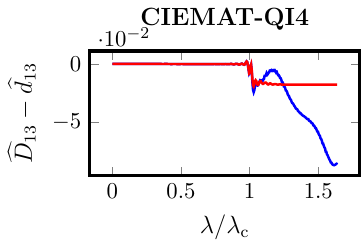}  
%
%
			\caption{}
			\label{subfig:D13_lambda_CIEMAT_QI}
		\end{subfigure}

		\tikzsetnextfilename{D31_lambda_QH_LP}  
		\begin{subfigure}{0.42\textwidth}  
			\includegraphics{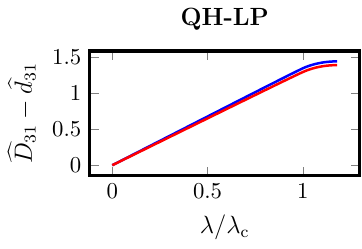}
%
%
			\caption{}
			\label{subfig:D31_lambda_QH_LP}
		\end{subfigure}        
		\tikzsetnextfilename{D13_lambda_QH_LP}
		\begin{subfigure}{0.42\textwidth}  
			\includegraphics{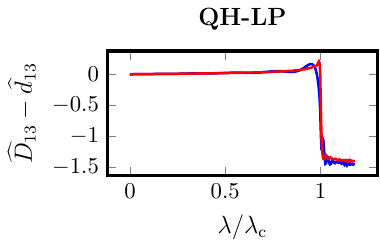}  
%
%
			\caption{}
			\label{subfig:D13_lambda_QH_LP}
		\end{subfigure}
		\caption{Contribution of different classes of particles to the $\Dij{31}$ and $\Dij{13}$ coefficients for the three magnetic configurations considered. Results with $\Er=0$ are indicated in blue and those with $\Er\ne 0$ in red.}
	\end{figure}
	\endgroup
	
	In nearly omnigenous stellarators, radial neoclassical transport in the $\sqrt{\nu}$ regime is dominated by charged particles in the boundary between passing and trapped particles \cite{Calvo_2017}. Employing the function $\dij{11}(\lambda)$ that {\MONKES} calculates, this analytical result and, to some extent, the proximity to omnigenity of each configuration can be visualized. In figures \ref{subfig:D11_lambda_W7X_KJM_Er_1e-3}, \ref{subfig:D11_lambda_CIEMAT_QI_Er_1e-3} and \ref{subfig:D11_lambda_QH_LP_Er_1e-3} the $\Dij{11}-\dij{11}$ curves for the case in the $\sqrt{\nu}$ regime are shown. As expected, only trapped particles contribute significantly to radial transport, and they do it in a different manner depending on the proximity to omnigenity of each configuration. Although W7-X KJM is relatively well optimized neoclassically, it is not very close to omnigenity. We can visualize this statement in figure \ref{subfig:D11_lambda_W7X_KJM_Er_1e-3}. Note that at $\lambda/\lambdac\sim 1$ there is a relatively sudden jump on the value of $\Dij{11}-\dij{11}$ but it does not represent even half of the total value of $\Dij{11}$. Instead, classes of particles which are more deeply trapped contribute to complete the total value of $\Dij{11}$. The flat mirror nearly QI configuration CIEMAT-QI4 is better optimized neoclassically than W7X-KJM and this is reflected on figure \ref{subfig:D11_lambda_CIEMAT_QI_Er_1e-3}. The sudden jump on the value of $\Dij{11}-\dij{11}$ at $\lambda/\lambdac\sim 1$ represents roughly $3/4$ of the total value of $\Dij{11}$. Still, more deeply trapped particles contribute significantly to radial transport, due to the imperfect optimization. From the three configurations shown, the precise QH configuration is the one that is closer to omnigenity. In figure \ref{subfig:D11_lambda_QH_LP_Er_1e-3} it can be seen that almost the total value of the $\Dij{11}$ is produced at $\lambda/\lambdac\sim 1$, in agreement with the analytical result from \cite{Calvo_2017}. Only a negligible contribution to radial transport is produced by more deeply trapped particles. Note from figures \ref{subfig:D11_lambda_QH_LP_Er_0} and \ref{subfig:D11_lambda_QH_LP_Er_1e-3} that for the QH configuration, the values of the $\Dij{11}$ with and without $\Er$ are almost identical. This is due to the fact that for the high degree of optimization of this configuration, the separation between the $1/\nu$ and $\sqrt{\nu}$ regimes appears at lower collisionalities than the one selected. 
	
	In the derivation of low collisionality formulas for the bootstrap current provided in \cite{Boozer_bootstrap} it is argued that, even though it is carried by passing particles, most of this current is produced by collisional exchange of momentum between passing and trapped particles. A similar explanation is also given in \cite{AGPeeters_2000}. The fact that the bootstrap current coefficient $\Dij{31}$ is dominated by passing particles can be observed in figures \ref{subfig:D31_lambda_W7X_KJM}, \ref{subfig:D31_lambda_CIEMAT_QI} and \ref{subfig:D31_lambda_QH_LP}. Note that $\Dij{31} - \dij{31}$ grows linearly in the passing region $\lambda/\lambdac < 1$ and becomes horizontal for trapped particles. In fact, the value of the $\Dij{31}$ coefficient is basically determined by the slope $\partial{\dij{31}}/\partial{\lambda}$ at $\lambda=\lambdac$. Note that there are no qualitative differences in the $\Dij{31} - \dij{31}$ curves between the cases $\Er = 0$ and $\Er\ne 0$. However, as will be argued in the following paragraph, for $\Er = 0$ the value of the slope of $\dij{31}$ at $\lambda=\lambdac$ is determined by the collisional interaction of trapped particles. In order to distinguish between the cases with zero and finite $\Er$, we will use Onsager symmetry $\Dij{13}=-\Dij{31}$ and inspect the curves $\Dij{13} - \dij{13}$ corresponding to the Ware pinch coefficient.
	
	Figures \ref{subfig:D13_lambda_W7X_KJM}, \ref{subfig:D13_lambda_CIEMAT_QI} and \ref{subfig:D13_lambda_QH_LP} reveal that only trapped particles contribute significantly to the Ware pinch coefficient $\Dij{13}$. Note that in the cases with $\Er\ne 0$, $\Dij{13}$ is completely determined by the region $\lambda/\lambdac\sim 1$, that is, by the boundary between passing and trapped particles. This is in agreement with \cite{Boozer_bootstrap, AGPeeters_2000} where it is claimed that the bootstrap current is dominated by collisional exchange of momentum between passing and trapped particles. However, with the exception of the precise QH, for the case with $\Er=0$ it can be seen from figures \ref{subfig:D13_lambda_W7X_KJM}, \ref{subfig:D13_lambda_CIEMAT_QI} and \ref{subfig:D13_lambda_QH_LP} that all classes of trapped particles contribute to the Ware pinch coefficient $\Dij{13}$, and by symmetry, to the bootstrap current coefficient $\Dij{31}=-\Dij{13}$. In this sense, the value of the slope $\partial{\dij{31}}/\partial{\lambda}$ at $\lambda=\lambdac$ for $\Er=0$ is determined by all classes of trapped particles. Finally, as a curiosity, we can see from \ref{subfig:D13_lambda_W7X_KJM} that the small value of $\Dij{31}$ for W7X-KJM is product of a cancellation between the contributions of barely and deeply trapped particles \cite{Beidler01011990, Beidler_2011}.

	
	
	To end this section, it is important to remark that this functionality of {\MONKES} is not as cheap in terms of memory as the computation of the monoenergetic coefficients $\Dij{ij}$. Note from expressions (\ref{eq:Monoenergetic_lambda_H0})-(\ref{eq:Monoenergetic_lambda_H2}) that in order to compute $\{H_j^{(k)}\}_{k=0}^{2}$ we need the full Legendre spectrum of the solution $\{f_j^{(k)}\}_{k=0}^{N_\xi}$. The amount of Legendre modes required for obtaining converged calculations of $\dij{ij}(\lambda)$ is of the same order of the resolution required to obtained converged calculations of the monoenergetic coefficients. Therefore, we need to slightly modify algorithm \ref{alg:MONKES_BTD} to store all the Schur complements $\{\vb*{\Delta}_{k}\}_{k=0}^{N_\xi}$. Thus, the memory requirements of this capability scale as $O(\Nfs^2 N_\xi)$.
	\FloatBarrier
	
	\subsection{Derivatives of the monoenergetic coefficients}	
	In section \ref{sec:Derivatives_monoenergetic}, three different approaches for computing the derivatives of the monoenergetic coefficients $\Dij{ij}$ with respect to a parameter $\eta$ upon which the DKE depends have been described. There, we just presented in a theoretical manner the different methods without paying particular attention to their computational aspects. In this section we will comment on the different advantages and drawbacks of each method taking into account their arithmetical complexity and memory requirements. Typically, one needs not only the derivative with respect to a single parameter but with respect to a set of them $\{\eta_m\}_{m=1}^{N_\eta}$. Hence, we denote by $\vb*{\eta}\in\mathbb{R}^{N_\eta}$ to a set of parameters with respect to which we want to differentiate the monoenergetic coefficients $\Dij{ij}$.
	
	The Finite Differences (FD) method of order $q$ for computing derivatives consists on approximating each $\partial{\Dij{ij}}/\partial{\eta_m}$ for $m=1,2,\ldots N_\eta$ using finite differences of order $q$, where $q \ge 1$ is an integer. Thus, for each $\eta_n$, it requires to compute $\sim q + 1$ values of the monoenergetic coefficients correspondent to the stencil of the FD method. As each solve using algorithm \ref{alg:MONKES_BTD} requires $O(\Nfs^3N_\xi)$ operations, the arithmetical complexity of this method scales as $O((N_\eta^q+1)\Nfs^3N_\xi)$ which, as will be shown, it is quite expensive compared with the other two methods. The only advantage of the FD method over the other two methods is that it only requires computing the monoenergetic coefficients $\Dij{ij}$ and not the full Legendre spectrum of the solution. Thus, its memory requirements are independent of $N_\xi$ and scale as $ O(\Nfs^2) $.
	
	The Direct Method (DM) requires to obtain the whole Legendre spectrum of the solution $\{f_j^{(k)}\}_{k=0}^{N_\xi}$ to the DKE in order to compute the source $S_{j,\eta}^{(k)}$ given by (\ref{eq:DKE_derivatives_source}). Hence, it requires to slightly modify the BTD algorithm \ref{alg:MONKES_BTD} to store all the Schur complements $\{\vb*{\Delta}_{k}\}_{k=0}^{N_\xi}$, not only the first three. Thus, its memory requirements scale as $ O(\Nfs^2 N_\xi) $. Obtaining and LU factorizing the Schur complements requires $O(\Nfs^3 N_\xi)$ operations. For calculating the derivatives with respect to a single $\eta_i$, only $O(\Nfs^2 N_\xi)$ operations are required, as the Schur complements are in LU form. Hence, the arithmetical complexity of the algorithm scales as $O(\Nfs^3 N_\xi) + O(N_\eta\Nfs^2 N_\xi)$.
	
	As the DM, the Adjoint Method (AM) requires to compute the whole Legendre spectrum of the solution $\{f_j^{(k)}\}_{k=0}^{N_\xi}$. Note respectively from (\ref{eq:Adjoint_method_fi_dagger_Vfj}) and (\ref{eq:Adjoint_method_fi_dagger_Lfj}) that, in order to compute $	\mean*{ f_i^\dagger, \VV_{\eta} f_j }$ or  $\mean*{ f_i^\dagger, \Lorentz f_j }$ the full spectrum of $f_j$ is required. Therefore, the memory requirements associated to this method also scale as $ O(\Nfs^2 N_\xi) $. As it requires to solve the DKE (\ref{eq:DKE_compact}) and its adjoint version (\ref{eq:Adjoint_DKE_compact}), if we use the BTD algorithm \ref{alg:MONKES_BTD} (with the slight modification to compute all the Legendre spectrum) its arithmetical complexity scales as
	$2 O(\Nfs^3 N_\xi)$. 
	
	In table \ref{tab:Summary_derivative_methods} we summarize the arithmetical complexity and memory requirements of each of the three methods. We can conclude that the FD method is only the best choice in the case in which we have very limited memory resources. That could be the case if one wanted to compute derivatives in a non dedicated core such as those of a personal computer. When comparing the DM and the AM, note that the DM might be slightly faster when $N_\eta < \Nfs $, which is typically the case at low collisionality. During the process of developing this thesis an algorithm for computing derivatives using the AM has been implemented in {\MONKES} but in future work the DM will also be considered. Although its arithmetical complexity is typically more favourable, it is not yet clear if the DM is a good alternative to the AM as the resolutions required to solve the DKE for $f_j$ might not be sufficient for solving for $\pdv*{f_j}{\eta_m}$. Recall that, for computing $\mean{s_i, \pdv*{f_j}{\eta}}$, the adjoint method only needs to compute $f_j$ and $f_j^\dagger$ and the resolutions required for computing $f_j^\dagger$ are quite similar. Another interesting comparison, left for future work, could be between the AM and/or DM with automatic differentiation, employed by the optimization suite \texttt{DESC} \cite{Dudt_Conlin_Panici_Kolemen_2023}.

	\begin{table}[]
		\centering
		\begin{tabular}{@{}ccc@{}}
			\toprule
			Method  & Arithmetical complexity & Memory requirements \\ 
			\midrule
			FD      &        $O((N_\eta^q+1) \Nfs^3 N_\xi)$                 &    $ O(\Nfs^2) $                 \\
			DM      &        $ O(\Nfs^3 N_\xi) + O(N_\eta\Nfs^2 N_\xi)$                 &         $O(\Nfs^2 N_\xi) $            \\
			AM &        $2O(\Nfs^3 N_\xi)$                 &       $O(\Nfs^2 N_\xi) $               \\ 
			\bottomrule
		\end{tabular}
		\caption{Summary of the arithmetical complexity and memory requirements of each of the methods for computing derivatives of the monoenergetic coefficients.}
		\label{tab:Summary_derivative_methods}
	\end{table}
	
	For the AM implemented in {\MONKES}, the parameter with respect to which the $\Dij{ij}$ are differentiated can be chosen among the set $\eta\in\{\{B_{mn}\}, \iota, B_\theta, B_\zeta, \hat{\nu}, \Er\}$. Here, $B_{mn}(\psi)$ are the stellarator-symmetric Fourier modes of the magnetic field strength in Boozer coordinates $(\theta,\zeta)$ at the flux surface labelled by $\psi$. Specifically, the magnetic field strength on a flux surface of an stellarator-symmetric device can be computed as
	\begin{align}
		B(\psi,\theta,\zeta)
		= 
		\sum_{m,n}
		B_{mn}(\psi)
		\cos(m\theta +n \Nfp \zeta).
	\end{align}
	\begin{figure}[h]
		\centering
		\tikzsetnextfilename{Dij_Derivatives_adjoint_Bmn_Er_0}
		\includegraphics{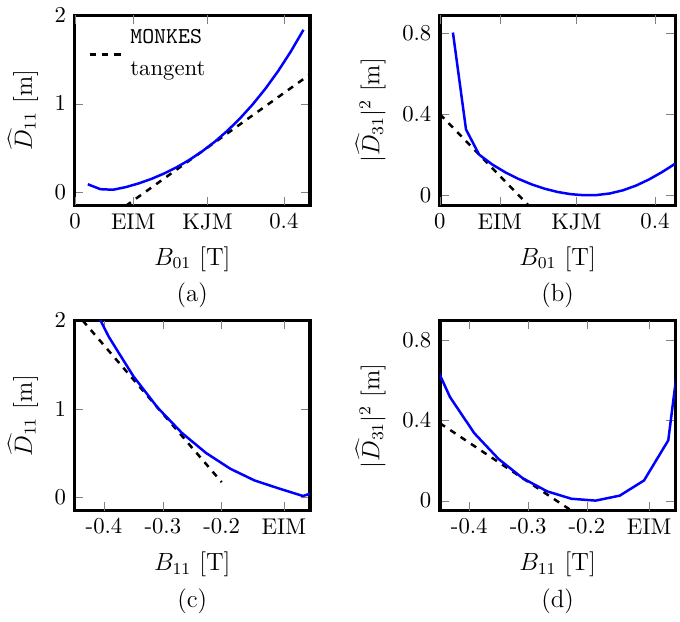}
		\caption{Dependence of the $\Dij{11}$ and $\Dij{31}$ coefficients with respect to the Fourier modes $B_{01}$ and $B_{11}$ of $B$ in Boozer coordinates and the tangents computed with the AM method implemented in {\MONKES}. The scans are carried out by varying $B_{01}$ and $B_{11}$ for W7X-EIM geometry at $\psi/\psi_{\text{lcfs}}=0.2$ keeping fixed the rest of Fourier modes $B_{mn}$.}
		\label{fig:Derivatives_Bmn_W7X_EIM}
	\end{figure}
	Note that the magnetic configuration of a particular flux surface is introduced in the DKE (\ref{eq:DKE}) via $\{\{B_{mn}\}, \iota, B_\theta, B_\zeta\}$. Thus, the usefulness of computing the derivatives of $\Dij{ij}$ with respect to these quantities for stellarator optimization is apparent. As an example of the capability of {\MONKES} for computing derivatives, we have carried out a scan for two Fourier modes taking as starting point the W7X-EIM geometry of sections \ref{subsec:Convergence} and \ref{sec:Results_benchmark}. We have selected the $B_{01}$ mode, related to the so called \qmarks{mirror term ratio} \cite{velasco2023robust} and the $B_{11}$ mode. For each case, the scan has been done by varying $B_{01}$ or $B_{11}$ while keeping the remaining Fourier modes constant (i.e. with the value correspondent to the W7-X EIM configuration). Then, we have evaluated the $\Dij{11}$ and $\Dij{31}$ coefficients and their derivatives at $\hat{\nu}=\num{e-5}$ and $\Er=0$ for each magnetic field of the scan. In figure \ref{fig:Derivatives_Bmn_W7X_EIM} the result of the evaluation is shown. In blue lines, the actual dependence of the $\Dij{11}$ and $\Dij{31}$ coefficients with the $B_{01}$ and $B_{11}$ modes is plotted. The value of the original values of the modes for the selected W7X-EIM configuration are indicated in the horizontal axis. In dashed black lines, the tangents that can be computed with the AM implemented in {\MONKES} are shown. For all cases, the tangent is accurately computed and the time required for computing both of these derivatives is below 2 minutes (and the time for computing derivatives of, say 100 Fourier modes would be the same). In figures \ref{fig:Derivatives_Bmn_W7X_EIM}(a) and \ref{fig:Derivatives_Bmn_W7X_EIM}(b) the dependence with the $B_{01}$ mode is illustrated. In the horizontal axis, the value of $B_{01}$ correspondent to the W7X-KJM configuration of sections \ref{subsec:Convergence} and \ref{sec:Results_benchmark} is indicated in figures \ref{fig:Derivatives_Bmn_W7X_EIM}(a) and \ref{fig:Derivatives_Bmn_W7X_EIM}(b). Note that, increasing the $B_{01}$ mode from EIM to the value corresponding to the KJM configuration increases the radial transport coefficient $\Dij{11}$ while diminishes the bootstrap current coefficient $\Dij{31}$. It can also be observed that the value of the $B_{01}$ mode correspondent to the EIM configuration is very close to the (local) optimum value for minimizing radial transport. Conversely, the value of the $B_{01}$ mode correspondent to the KJM configuration is close to the optimal one which minimizes $\Dij{31}$. The dependences of $\Dij{11}$ and $\Dij{31}$ with $B_{11}$ are shown, respectively, in figures \ref{fig:Derivatives_Bmn_W7X_EIM}(c) and \ref{fig:Derivatives_Bmn_W7X_EIM}(d). As for the $B_{01}$ mode, there is a trade off between radial and parallel transport. Increasing $B_{11}$ has the effect of decreasing $\Dij{11}$ at the expense of producing a larger value of $\Dij{31}$.

	\begin{figure}[h] 
		\centering
		\tikzsetnextfilename{Dij_Derivatives_adjoint_nu_Er}
		\includegraphics{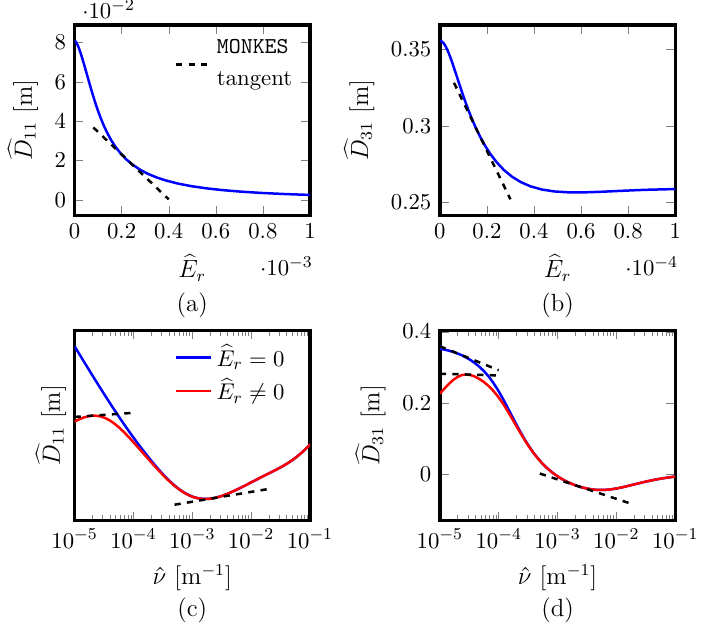}
		\caption{Dependence of the $\Dij{11}$ and $\Dij{31}$ coefficients with respect to the radial electric field at $\hat{\nu}=\num{e-5}$ $\text{m}^{-1}$ (top) and with respect to the collisionality (bottom) for $\widehat{E}_r=0$ and $\widehat{E}_r=\num{3e-4}$ and the tangents computed with the AM method implemented in {\MONKES} for W7X-EIM geometry at $\psi/\psi_{\text{lcfs}}=0.2$. $\widehat{E}_r$ is given in $\text{V}\cdot\text{m}^{-2}\cdot\text{s}$.}
		\label{fig:Derivatives_Er_nu_W7X_EIM}
	\end{figure}
	
	We end this section by also providing examples of how {\MONKES} can compute derivatives of $\Er$ and $\hat{\nu}$. Derivatives along $\Er$ can be useful for solving the ambipolar equation (\ref{eq:Ambipolarity_condition}). On the one hand, they can be useful for finding the solution to $\sum_{a} e_a \mean*{\vb*{\Gamma}_a\cdot\nabla\psi}=0$ employing a gradient-based method (e.g. a Newton-Rhapson). On the other hand, when multiple roots of ambipolarity occur, root-selection criteria such as the one presented in \cite{Hastings_1986_Bifurcation} require knowledge about the derivative  $\partial\mean*{\vb*{\Gamma}_a\cdot\nabla\psi}/\partial E_\psi$. For both scenarios, it is useful to compute the derivative $\partial\Dij{11}/\partial\Er$\footnote{Which is simply related to the derivative with respect to $E_\psi$ as $\partial\Dij{11}/\partial\Er=v\partial\Dij{11}/\partial E_\psi /\dv*{\psi}{r}$}. Derivatives along $\hat{\nu}$ might be useful for direct optimization purposes. For instance, one might want to minimize not only $\Dij{31}$ but also its derivative along $\hat{\nu}$. Ensuring flatness of the $\Dij{31}-\hat{\nu}$ curve guarantees that the plasma current does not strongly depend on the collisionality or, more generally, on the plasma scenario. In figures \ref{fig:Derivatives_Er_nu_W7X_EIM}(a) and 
	\ref{fig:Derivatives_Er_nu_W7X_EIM}(b), respectively, the dependences of $\Dij{11}$ and $\Dij{31}$ with $\Er$ are indicated with blue lines. Again, the tangents obtained by using the derivatives $\partial\Dij{ij}/\partial\Er$ provided by the AM implemented in {\MONKES} are represented with a black dashed line. In figures \ref{fig:Derivatives_Er_nu_W7X_EIM}(c) and 
	\ref{fig:Derivatives_Er_nu_W7X_EIM}(d), respectively, the dependences of $\Dij{11}$ and $\Dij{31}$ with $\hat{\nu}$ are shown with solid lines. The case with $\Er=0$ and the calculations with $\Er\ne 0$ are plotted, respectively, in blue and red colours. In both cases, the tangents computed with {\MONKES} are an excellent approximation to the tangent of the curve.

	\chapter{Evaluation of neoclassical transport for nearly QI magnetic fields using {\MONKES}}\label{chap:MONKES_applications}
	\thispagestyle{empty}

	In the previous chapters, the new code {\MONKES} and the theory behind it have been explained. Thanks to its speed, this new tool opens up, among other things, the possibility of \textit{direct neoclassical optimization} of stellarators. In particular, both radial transport and the bootstrap current can now be optimized directly in stellarators.

	In this chapter, the first two practical applications of the new neoclassical code {\MONKES}, which are connected to stellarator optimization, will be shown. In section \ref{sec:Correlations} {\MONKES} is employed to determine how efficient the indirect approach is to optimize QI magnetic fields and, in particular, to reduce the bootstrap current. In section \ref{sec:pwO_QI}, {\MONKES} is used to give the first steps in the exploration of the configuration space of the novel family of \textit{piecewise omnigenous} magnetic fields \cite{velasco2024piecewise} (their definition will be given in section \ref{sec:pwO_QI}). By approaching quasi-isodynamicity from piecewise omnigenity, we try to find regions of this configuration space with small levels of radial transport and bootstrap current. It is important to remark that the speed of {\MONKES} has been crucial to facilitate (if not to make possible) both applications. Most of this chapter is based on publication [II] from the \textbf{\qmarks{PUBLISHED AND SUBMITTED CONTENT}} section at the beginning of this dissertation.
	
	\section{Assessment of the efficiency of the indirect approach for optimizing QI magnetic fields}
	\label{sec:Correlations}
	\begingroup
	\captionsetup[sub]{skip=-1.75pt,aboveskip=-15pt, belowskip=5pt, margin=90pt}

	\begin{figure}[h]
		\centering
		\tikzsetnextfilename{D31_vs_D11_eps_0100_Er_0}
		\begin{subfigure}[t]{0.45\textwidth}	
			\includegraphics{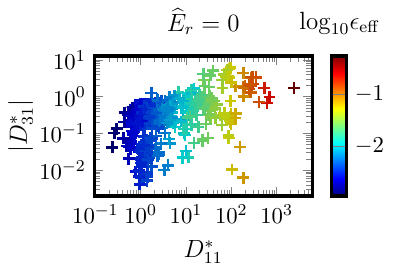}
			{}
			\caption{}
			\label{subfig:D31_vs_D11_eps_0100_Er_0}
		\end{subfigure} 
		\tikzsetnextfilename{D31_vs_D11_eps_0100_Er_1e-3}
		\begin{subfigure}[t]{0.45\textwidth}
			\includegraphics{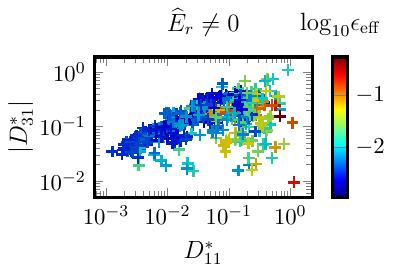}
			{}	
			\caption{}
			\label{subfig:D31_vs_D11_eps_0100_Er_1e-3}
		\end{subfigure} 
		
		\tikzsetnextfilename{D31_vs_D11_eps_0100_Er_0_restricted}
		\begin{subfigure}[t]{0.45\textwidth}	
			\includegraphics{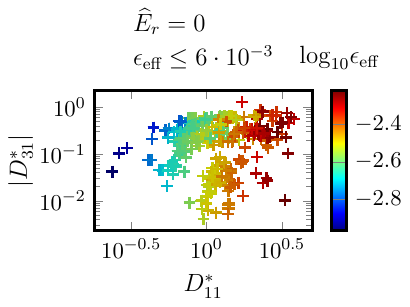}
			{}	
			\caption{}
			\label{subfig:D31_vs_D11_eps_0100_Er_0_restricted}
		\end{subfigure} 
		\tikzsetnextfilename{D31_vs_D11_eps_0100_Er_1e-3_restricted}
		\begin{subfigure}[t]{0.45\textwidth}	
			\includegraphics{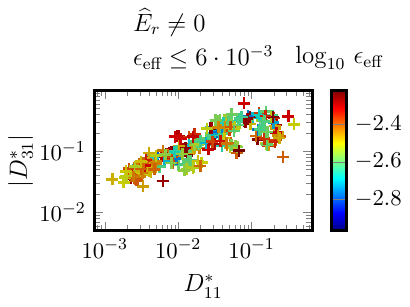}
			{}	
			\caption{}
			\label{subfig:D31_vs_D11_eps_0100_Er_1e-3_restricted}
		\end{subfigure}

		\tikzsetnextfilename{D31_vs_eps_0100_Er_0}
		\begin{subfigure}[t]{0.45\textwidth}
			\includegraphics{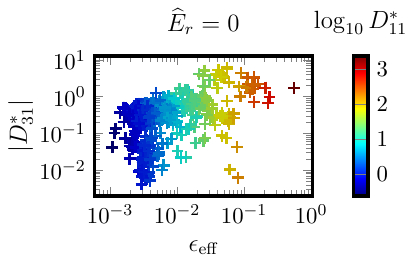}
			{}	
			\caption{}
			\label{subfig:D31_vs_eps_0100_Er_0}
		\end{subfigure} 
		\tikzsetnextfilename{D31_vs_eps_0100_Er_1e-3}
		\begin{subfigure}[t]{0.45\textwidth}	
			\includegraphics{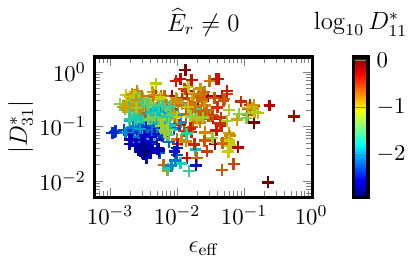}
			{}	
			\caption{}
			\label{subfig:D31_vs_eps_0100_Er_1e-3}
		\end{subfigure}      
		\caption{Relation of the radial transport $D_{11}^*$ and bootstrap current $D_{31}^*$ coefficients with $\epseff$.}
		\label{fig:Correlation_epseff}
	\end{figure} 
	\begin{figure}[h]
		\centering
		\tikzsetnextfilename{D31_vs_D11_KN_VBB_0100_Er_0}	
		\begin{subfigure}[t]{0.45\textwidth}	 		
			\includegraphics{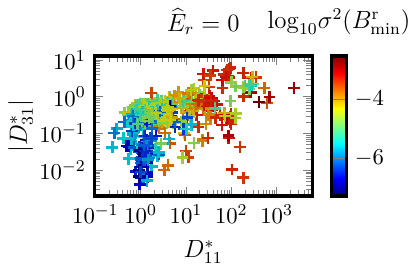}	
			{}
			\caption{}
			\label{subfig:D31_vs_D11_KN_VBB_0100_Er_0}
		\end{subfigure}     
		\tikzsetnextfilename{D31_vs_D11_KN_VBB_0100_Er_1e-3}	
		\begin{subfigure}[t]{0.45\textwidth}
			\includegraphics{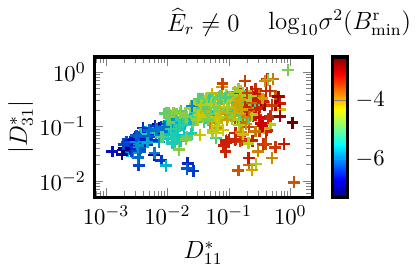}
			{}	
			\caption{}
			\label{subfig:D31_vs_D11_KN_VBB_0100_Er_1e-3}
		\end{subfigure} 

		\tikzsetnextfilename{D31_vs_KN_VBB_0100_Er_0}	
		\begin{subfigure}[t]{0.45\textwidth}	 	
			\includegraphics{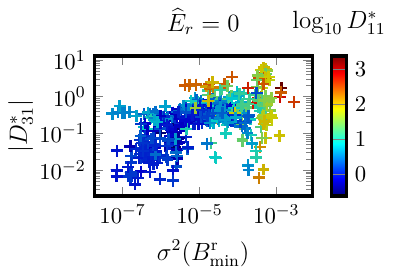}	
			{}	
			\vspace{0.022cm} 
			\caption{}
			\label{subfig:D31_vs_KN_VBB_0100_Er_0}
		\end{subfigure}     
		\tikzsetnextfilename{D31_vs_KN_VBB_0100_Er_1e-3}	
		\begin{subfigure}[t]{0.45\textwidth}	 	
			\includegraphics{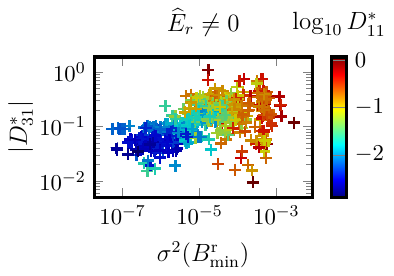}	
			{}	
			\caption{}
			\label{subfig:D31_vs_KN_VBB_0100_Er_1e-3}
		\end{subfigure} 

		\tikzsetnextfilename{D31_vs_KN_VBB_0100_Er_0_restricted}	
		\begin{subfigure}[t]{0.45\textwidth}	 		
			\includegraphics{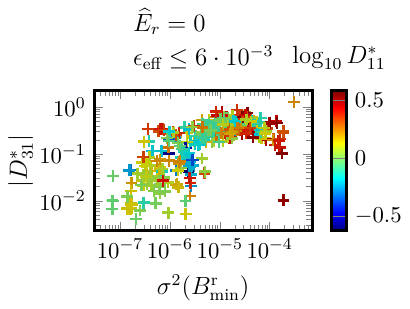}	
			{}
			\caption{}
			\label{subfig:D31_vs_KN_VBB_0100_Er_0_restricted}
		\end{subfigure}     
		\tikzsetnextfilename{D31_vs_KN_VBB_0100_Er_1e-3_restricted}
		\begin{subfigure}[t]{0.45\textwidth}	 			
			\includegraphics{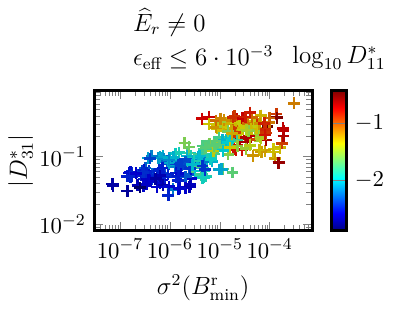}	
			{}
			\caption{}
			\label{subfig:D31_vs_KN_VBB_0100_Er_1e-3_restricted}
		\end{subfigure} 
		\caption{Relation of the radial transport $D_{11}^*$ and bootstrap current $D_{31}^*$ coefficients with $\sigma^2(\Bmin^{\text{r}})$.} 
		\label{fig:Correlation_VBB}
	\end{figure}

	\begin{figure}[h]
		
		\tikzsetnextfilename{D31_vs_D11_KN_VB0_0100_Er_0}	
		\begin{subfigure}[t]{0.45\textwidth}		
			\includegraphics{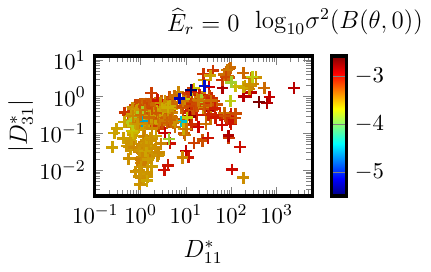}
			{}
			\caption{}
			\label{subfig:D31_vs_D11_KN_VB0_0100_Er_0}
		\end{subfigure}
		\tikzsetnextfilename{D31_vs_D11_KN_VB0_0100_Er_1e-3}	
		\begin{subfigure}[t]{0.45\textwidth}		
			\includegraphics{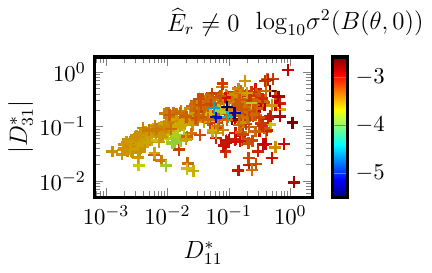}
			{}
			
			\caption{}
			\label{subfig:D31_vs_D11_KN_VB0_0100_Er_1e-3}
		\end{subfigure} 
		
		\tikzsetnextfilename{D31_vs_KN_VB0_0100_Er_0}	
		\begin{subfigure}[t]{0.45\textwidth}	
			\includegraphics{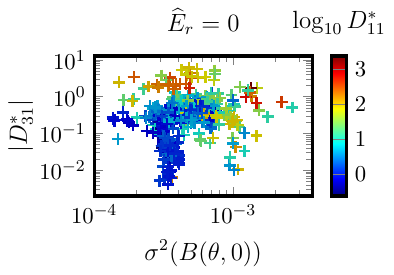}	
			\vspace{0.022cm} 
			\caption{}
			\label{subfig:D31_vs_KN_VB0_0100_Er_0}
		\end{subfigure}
		\tikzsetnextfilename{D31_vs_KN_VB0_0100_Er_1e-3}	
		\begin{subfigure}[t]{0.45\textwidth}		
			\includegraphics{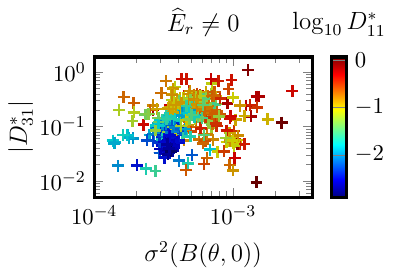}
			{}
			\caption{}
			\label{subfig:D31_vs_KN_VB0_0100_Er_1e-3}
		\end{subfigure}

		\tikzsetnextfilename{D31_vs_KN_VB0_0100_Er_0_restricted}
		\begin{subfigure}[t]{0.45\textwidth}		
			\includegraphics{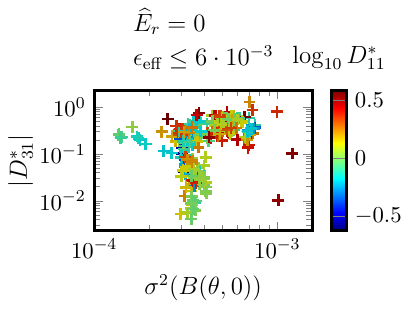}
			{}	
			\caption{}
			\label{subfig:D31_vs_KN_VB0_0100_Er_0_restricted}
		\end{subfigure}
		\tikzsetnextfilename{D31_vs_KN_VB0_0100_Er_1e-3_restricted}
		\begin{subfigure}[t]{0.45\textwidth}		
			\includegraphics{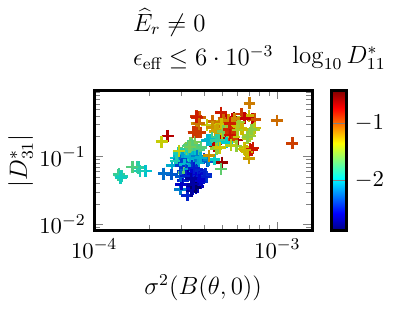}	
			{}
			\caption{}
			\label{subfig:D31_vs_KN_VB0_0100_Er_1e-3_restricted}
		\end{subfigure} 
		\caption{Relation of the radial transport $D_{11}^*$ and bootstrap current $D_{31}^*$ coefficients with $\sigma^2(B(\theta,0))$.}
		\label{fig:Correlation_VB0}
	\end{figure} 
	
	\begin{figure}[h]     
		\tikzsetnextfilename{D31_vs_D11_KN_GMC_0100_Er_0}	
		\begin{subfigure}[t]{0.45\textwidth}	 	
			\includegraphics{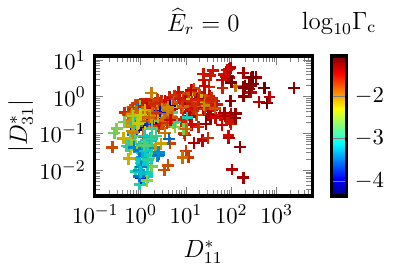}
			{}
			\caption{}
			\label{subfig:D31_vs_D11_KN_GMC_0100_Er_0}
		\end{subfigure}
		\tikzsetnextfilename{D31_vs_D11_KN_GMA_0100_Er_0}	
		\begin{subfigure}[t]{0.45\textwidth}	 	
			\includegraphics{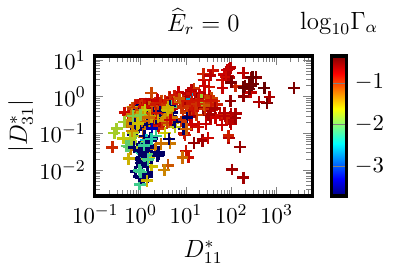}
			{}
			\caption{}
			\label{subfig:D31_vs_D11_KN_GMA_0100_Er_0}
		\end{subfigure}

		\tikzsetnextfilename{D31_vs_KN_GMC_0100_Er_0}	
		\begin{subfigure}[t]{0.45\textwidth}	
			\includegraphics{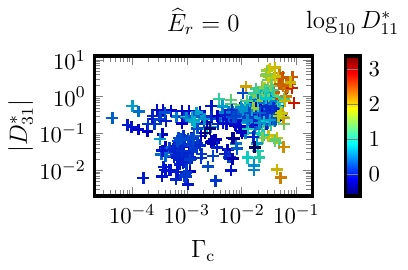}	 
			{}
			\caption{}
			\label{subfig:D31_vs_KN_GMC_0100_Er_0}
		\end{subfigure}
		\tikzsetnextfilename{D31_vs_KN_GMA_0100_Er_0}
		\begin{subfigure}[t]{0.45\textwidth}	 	
			\includegraphics{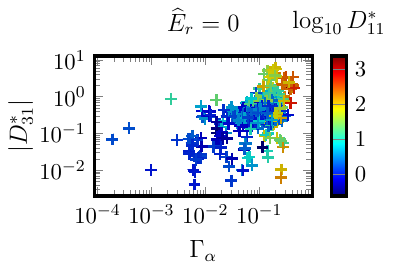}	
			\vspace{0.022cm} 
			\caption{}
			\label{subfig:D31_vs_KN_GMA_0100_Er_0}
		\end{subfigure}

		\tikzsetnextfilename{D31_vs_KN_GMC_0100_Er_0_restricted}	
		\begin{subfigure}[t]{0.45\textwidth}	 
			\includegraphics{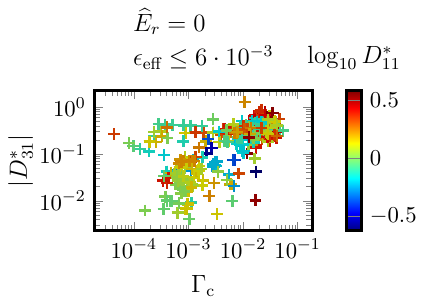}	
			{}
			\caption{}
			\label{subfig:D31_vs_KN_GMC_0100_Er_0_restricted}
		\end{subfigure}
		\tikzsetnextfilename{D31_vs_KN_GMA_0100_Er_0_restricted}
		\begin{subfigure}[t]{0.45\textwidth}	 	
			\includegraphics{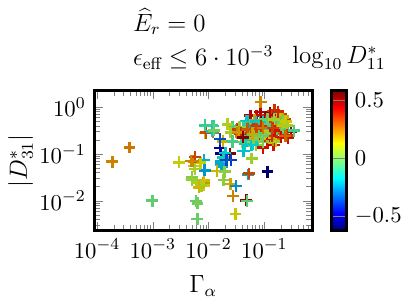}	
			{}
			\caption{}
			\label{subfig:D31_vs_KN_GMA_0100_Er_0_restricted}
		\end{subfigure}

		\caption{Relation of the radial transport $D_{11}^*$ and bootstrap current $D_{31}^*$ coefficients with $\GammaC$ and $\GammaAlpha$ for $\widehat{E}_r=0$.} 
		\label{fig:Correlation_GMC_GMA_Er_0}
	\end{figure}

	\begin{figure}[h]
		\centering
		\tikzsetnextfilename{D31_vs_D11_KN_GMC_0100_Er_1e-3}	
		\hfill
		\begin{subfigure}[t]{0.45\textwidth}	 	
			\includegraphics{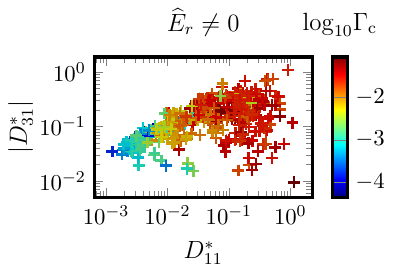}
			{}
			\caption{}
			\label{subfig:D31_vs_D11_KN_GMC_0100_Er_1e-3}
		\end{subfigure}
		\tikzsetnextfilename{D31_vs_D11_KN_GMA_0100_Er_1e-3}	
		\begin{subfigure}[t]{0.45\textwidth}	 
			\includegraphics{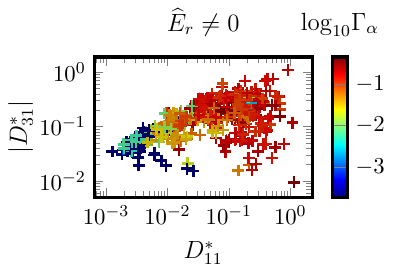}	
			{}
			\caption{}
			\label{subfig:D31_vs_D11_KN_GMA_0100_Er_1e-3}
		\end{subfigure} 
		
		\hfill
		\tikzsetnextfilename{D31_vs_KN_GMC_0100_Er_1e-3}
		\begin{subfigure}[t]{0.45\textwidth}	 	
			\includegraphics{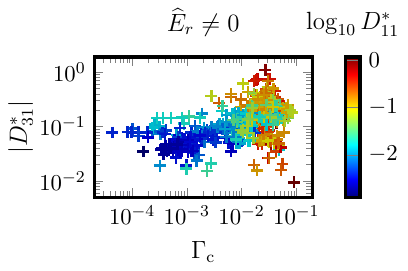}	
			{}
			\caption{}
			\label{subfig:D31_vs_KN_GMC_0100_Er_1e-3}
		\end{subfigure}
		\tikzsetnextfilename{D31_vs_KN_GMA_0100_Er_1e-3}	
		\begin{subfigure}[t]{0.45\textwidth}	 	
			\includegraphics{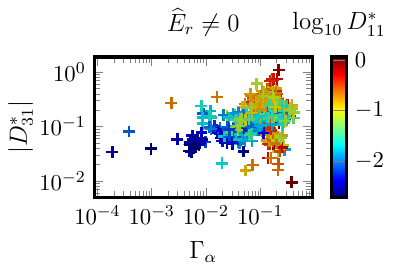}
			{}
			\caption{}
			\label{subfig:D31_vs_KN_GMA_0100_Er_1e-3}
		\end{subfigure}
		
		\hfill
		\tikzsetnextfilename{D31_vs_KN_GMC_0100_Er_1e-3_restricted}
		\begin{subfigure}[t]{0.45\textwidth}	 	
			\includegraphics{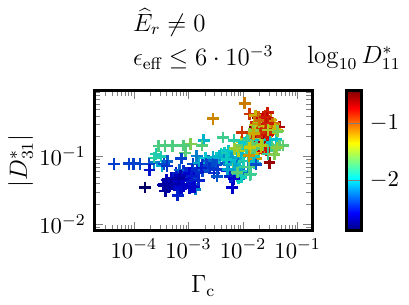}	
			{}
			\caption{}
			\label{subfig:D31_vs_KN_GMC_0100_Er_1e-3_restricted}
		\end{subfigure}
		\tikzsetnextfilename{D31_vs_KN_GMA_0100_Er_1e-3_restricted}		
		\begin{subfigure}[t]{0.45\textwidth}	 
			\includegraphics{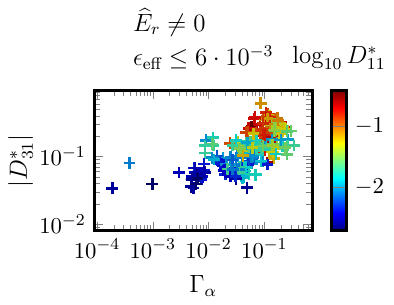}
			{}
			\caption{}
			\label{subfig:D31_vs_KN_GMA_0100_Er_1e-3_restricted}
		\end{subfigure}

		\caption{Relation of the radial transport $D_{11}^*$ and bootstrap current $D_{31}^*$ coefficients with $\GammaC$ and $\GammaAlpha$ for $\widehat{E}_r\ne 0$.} 
		\label{fig:Correlation_GMC_GMA_Er_1e-3}
	\end{figure}
	
	In neoclassical optimization, one typically pursues omnigenous configurations by minimizing a \textit{cost function} $\chi$. One manner to express this function is as a distance
	\begin{align}
		\chi^2 =
		\sum_{k}
		w_k^2 
		\left(
		\chi_k^{\text{target}}
		-
		\chi_k^{\text{eq}}
		\right)^2
		.
		\label{eq:Cost_function}
	\end{align}
	Here, $\chi_k$ stands for a specific proxy: a quantity that represents some property of the magnetic configuration. The value $\chi_k^{\text{target}}$ is the desired value for the aforementioned property and $\chi_k^{\text{eq}}$ is the actual value of $\chi_k$ for the magnetic configuration obtained by solving the magnetohydrodynamic equilibrium equation. For each value of $\chi$, the reciprocals of the scalars $w_k$ set an upper bound for the deviation  $|\chi_k^{\text{eq}} - \chi_k^{\text{target}}|\le |\chi/w_k|$. Hence, the weights $w_k$ set the relative importance of each proxy. Thus, a cost function is determined by a selection of proxies $\chi_k$, their target values $\chi_k^{\text{target}}$ and weights $w_k$.
	
	The selection of the proxies $\{\chi_k\}$ is meant to parametrize the type of stellarator that one wishes to obtain. For instance, in neoclassical optimization, the proxies $\{\chi_k\}$ should represent as well as possible the neoclassical properties of the magnetic configuration while being fast to calculate. In order to reduce the value of $\chi^2$, several quantities of the magnetic configuration called \textit{variables} are modified by an optimizer (e.g. the modes of the Fourier representation of the last closed flux surface). Selecting a single cost function $\chi^2$ is usually insufficient for satisfying all the criteria required for the magnetic configuration. Therefore, an optimization campaign consists on successive optimization steps until the obtained magnetic field satisfies the given desiderata. Each optimization step is defined by a different cost function $\chi^2$. That is, from one step to another, the proxy selection $\{\chi_k\}$, their target values $\chi_k^{\text{target}}$ and/or their relative importance (i.e. the values $\{w_k\}$) are varied. How to successfully change the cost function from one step to the next is a non straightforward process which, in most cases, requires some experience, intuition and luck.

	In order to neoclassically optimize QI configurations, the goal is to reduce not only $\Dij{11}$ but also $\Dij{31}$ as much as possible. Recall from section \ref{sec:DKE} that, $\Dij{11}$ and $\Dij{31}$ stand, respectively, for the radial transport and bootstrap current monoenergetic coefficients. As explained in section \ref{sec:DKE}, for fixed collisionality $\hat{\nu}$ and radial electric field ${E}_r$, the monoenergetic coefficients $\Dij{ij}$ encapsulate the dependence on the magnetic configuration of neoclassical transport in a given flux surface. As before the development of {\MONKES}, the inclusion of the $\Dij{31}$ coefficient in the optimization loop was practically impossible, the bootstrap current has traditionally been optimized indirectly. That is, some proxies which vanish for exactly QI configurations are used and then one \textit{hopes} that minimizing them will also minimize $|\Dij{31}|$. However, with this approach, one cannot guarantee that reducing the proxies will translate in a sufficient minimization of $|\Dij{31}|$. Moreover, the indirect approach does not allow to optimize taking into account the effect of the bootstrap current on the magnetic configuration and its neoclassical properties. For stellarators which are sufficiently close to quasi-symmetry \cite{Landreman_SelfConsistent}, optimization can be done in a self-consistent manner using analytical formulae for the bootstrap current in tokamaks \cite{Redl_Bootstrap} that are accurate and fast to compute.

	
	In \cite{Sanchez_2023}, a selection of new and standard proxies for quasi-isodynamicity is proposed, which allowed to obtain the \qmarks{flat-mirror} \cite{velasco2023robust} nearly QI configuration CIEMAT-QI4. In order to evaluate how efficient the optimization strategy was for minimizing neoclassical transport (and in particular the bootstrap current), we will use {\MONKES} to evaluate $\Dij{11}$ and $\Dij{31}$ for the database of magnetic configurations produced during the CIEMAT-QI4 optimization campaign. The efficiency of each proxy for indirect QI optimization will be assessed by investigating the correlation (or lack of it) between the proxy and $|\Dij{31}|$. It is important to remark that in the robust \qmarks{flat-mirror} strategy, many reactor-relevant properties are optimized simultaneously. The key idea is not to focus on being extremely close to QI and instead tailor the magnetic field so that particles drift tangentially to the flux surface. This trade-off facilitates to meet other reactor-relevant requirements that are not related to neoclassical transport e.g. magnetohydrodynamic stability. Therefore, this evaluation will clarify to what extent reducing these proxies translates into a reduction of $\Dij{31}$ when optimizing stellarators which are meant to be fusion reactor candidates. For the sake of clarity, we briefly review those proxies which vanish for exactly omnigenous and QI fields.
	
	For radial transport, the so-called effective ripple $\epseff$ \cite{Nemov1999EvaluationO1} encapsulates neoclassical losses of the bulk plasma in the $1/\nu$ regime. For fast ions, the proxies $\Gamma_{\text{c}}$ \cite{NemovGammaC} 
	\begin{align}
		\Gamma_{\text{c}}(s)
		=
		\frac{1}{\pi\sqrt{2}}
		\mean*{
			\int_{\Bmax^{-1}}^{\Bmin^{-1}}
			\left(
			\frac{
				\overline{\vb*{v}_{\text{m}a}\cdot\nabla s}
			}
			{
				\overline{\vb*{v}_{\text{m}a}\cdot\nabla\alpha}
			}
			\right)^2 
			\frac{B \dd{\lambda}}{\sqrt{1-\lambda B}} 
		}
		\label{eq:Gamma_c}
	\end{align}
	and its refinement $\Gamma_{\alpha}$ \cite{Velasco_2021} 
	\begin{align}
		\Gamma_{\alpha}(s)
		=
		\frac{e_a/m_a}{\pi\sqrt{2}} 
		\left\langle 
		\int_{\Bmax^{-1}}^{\Bmin^{-1}}
		H \left(
		(\alpha_{\text{out}}-\alpha)\overline{\vb*{v}_{\text{m}a}\cdot\nabla\alpha}
		\right) 
		\right.
		\nonumber
		\\
		\left.
		\times H \left(
		(\alpha-\alpha_{\text{in}})\overline{\vb*{v}_{\text{m}a}\cdot\nabla\alpha}
		\right)
		\frac{B \dd{\lambda}}{\sqrt{1-\lambda B}} 
		\right\rangle
		\label{eq:Gamma_alpha}
	\end{align}
	are used. Here, $s:=\psi/\psi_{\text{lcfs}}$ is the normalized flux surface label. Recall that $\vb*{v}_{\text{m}a}$ is the magnetic drift (\ref{eq:Magnetic_drift_GCM}) (for time-independent $\vb*{B}$), $\lambda$ is the so called \qmarks{pitch-angle coordinate} (\ref{eq:GCM_normalized_magnetic_moment}), $\alpha$ is the Clebsch poloidal angle (\ref{eq:Clebsch_angle}) which labels field lines and its values $\alpha_{\text{in}}$ and $\alpha_{\text{out}}$ are defined in \cite{Velasco_2021} (their specific definitions are not relevant for this dissertation).

	Several targets based on the shape of the isolines of $B$ in omnigenous configurations \cite{CaryPhysRevLett,Cary1997OmnigenityAQ} are also considered. In an omnigenous configuration (or for each well of those defined in \cite{Parra_2015}), all relative maxima and minima of $B$ have equal value. This implies that the variance of the relative maxima of $B$
	\begin{align}
		\sigma^2(\Bmax^{\text{r}}) 
		:=
		\frac{1}{  N_\theta}
		\sum_{i=0}^{N_\theta-1}
		\left(
		\dfrac{
			B_{\text{M}}(\theta_i)
			- 
			B_{\text{M}}^{\text{mean}} 
		}{B_{00}}
		\right)^2
		,
	\end{align}
	and the variance of the relative minima
	\begin{align}
		\sigma^2(\Bmin^{\text{r}}) 
		:=
		\frac{1}{  N_\theta}
		\sum_{i=0}^{N_\theta-1}
		\left(
		\dfrac{ 
			B_{\text{m}}(\theta_i)
			- 
			B_{\text{m}}^{\text{mean}}
		}{B_{00}}
		\right)^2
	\end{align} 
	vanish in a perfectly omnigenous configuration. Here, $B_{\text{M}}(\theta_i)=\max B(\theta_i,\zeta)$ and $B_{\text{m}}(\theta_i)=\min B(\theta_i,\zeta)$ for $0\le\zeta<2\pi/\Nfp$ are, respectively, the maximum and minimum values of $B$ in a poloidal equispaced grid $\theta_i =2\pi i/N_\theta$. The quantities $B_{\text{M}}^{\text{mean}} = \sum_{i=0}^{N_\theta-1} B_{\text{M}}(\theta_i)/N_\theta$ and $B_{\text{m}}^{\text{mean}} = \sum_{i=0}^{N_\theta-1} B_{\text{m}}(\theta_i)/N_\theta$ are, respectively, the mean values of $\{B_{\text{M}}(\theta_i)\}_{i=0}^{N_\theta-1}$ and $\{B_{\text{m}}(\theta_i)\}_{i=0}^{N_\theta-1}$.
	
	In a QI stellarator, stellarator symmetry \cite{DEWAR1998275} implies that the maximum value of $B$ in the flux surface must be attained at the beginning or the center of the field period along a curve that closes poloidally. Thus, stellarator symmetry implies that the isoline $B=\Bmax$ must match either the curve $\zeta=0$ or $\zeta=\pi/\Nfp$, where $\Nfp$ is the number of field periods of the device. However, redefining the beginning of the field period (i.e. mapping $\zeta\mapsto \zeta -\pi/\Nfp$) permits to agglutinate both cases in the case $\zeta=0$. Thus, specifically for obtaining (stellarator-symmetric) QI configurations, the variance of $B$ at $\zeta=0$ is considered
	\begin{align}
		\sigma^2(B(\theta,0)) 
		:=
		\frac{1}{N_\theta}
		\sum_{i=0}^{N_\theta-1}
		\left(
		\dfrac{
			B(\theta_i,0)
			-
			B_0^{\text{mean}}
		}{ B_{00} }
		\right)^2 
		,
	\end{align}  
	where $B_0^{\text{mean}} = \sum_{i=0}^{N_\theta-1} B(\theta_i,0)/N_\theta$. Note that for a perfectly QI stellarator-symmetric magnetic field $\sigma^2(B(\theta,0)) $ vanishes but, by itself, the nullity of $\sigma^2(B(\theta,0)) $ does not guarantee that the curve $\zeta=0$ coincides with the isoline $B=\Bmax$.

	For the neoclassical transport evaluation, we have
	selected a grid of 11 values of $\hat{\nu}$ in the low collisionality interval $\hat{\nu}\in[\num{e-5},\num{e-3}]$ $\text{m}^{-1}$ and two of the radial electric field $\widehat{E}_r \in \{0,\num{e-3}\}$ $\text{V}\cdot \text{s}/\text{m}^2$. Those cases with zero radial electric field are in the $1/\nu$ regime (typical of electrons) and those with finite $\widehat{E}_r$ are in the $\sqrt{\nu}$-$\nu$ regime \cite{dherbemont2022} (typical of bulk ions). For each pair $(\hat{\nu},\widehat{E}_r)$, we calculate the monoenergetic transport coefficients $\Dij{11}$ and $\Dij{31}$ of each configuration from the large database of 1165 configurations using {\MONKES}. In order to compare different magnetic configurations, we normalize the monoenergetic coefficients as in \cite{Beidler_2011} (further details in appendix \ref{sec:Appendix_monoenergetic_normalization}) and we denote them by $D_{ij}^*$. In Appendix \ref{sec:Appendix_Correlations_nu_scan}, we show that the conclusions
	regarding the efficiency of the proxies extracted from
	the results at $\hat{\nu}= \num{e-5}$ $\text{m}^{-1}$ are applicable to the whole interval $\hat{\nu}\in[\num{e-5},\num{e-3}]$ $\text{m}^{-1}$. Therefore, here we
	will only discuss the results for the lowest collisionality $\hat{\nu}= \num{e-5}$ $\text{m}^{-1}$. The rationale behind the selection of the values of $\hat{\nu}$ and $\widehat{E}_r$ is that, in order to minimize the bootstrap current in reactor-relevant scenarios, it is required to minimize $|D_{31}^*|$ at low collisionality with and without radial electric field. The value $\hat{\nu}= \num{e-5}$ $\text{m}^{-1}$ is usually a good estimate of the lowest collisionality that is important for computing the integrals of $D_{31}^*$ which yield the parallel flow of each species. 
	
	In figures \ref{subfig:D31_vs_D11_eps_0100_Er_0} and \ref{subfig:D31_vs_D11_eps_0100_Er_1e-3} the result of the neoclassical evaluation for the database of the CIEMAT-QI4 campaign is shown, along with the value of $\epseff$ in colours. Each point on the plane $D_{11}^*-|D_{31}^*|$ corresponds to a different configuration with a particular value of $\epseff$. Thus, configurations closer to being QI are located near the bottom left corner of these plots. Figure \ref{subfig:D31_vs_D11_eps_0100_Er_0} shows that, in the absence of radial electric field, configurations which were optimized for having small $D_{11}^*$ (equivalently $\epseff$) not necessarily had small bootstrap current coefficient. On the other hand, in the presence of a finite $\widehat{E}_r$, we can see from figure  \ref{subfig:D31_vs_D11_eps_0100_Er_1e-3} that minimizing radial transport entailed a minimization of $D_{31}^*$. In the complete database shown in figures \ref{subfig:D31_vs_D11_eps_0100_Er_0} and \ref{subfig:D31_vs_D11_eps_0100_Er_1e-3} there are many configurations corresponding to the initial stages of the optimization campaign and therefore, are not sufficiently optimized. As $\epseff$ is typically used as an indicator of overall radial neoclassical transport optimization, for inspecting potential correlations, it is useful to filter out non optimized configurations. When we restrict the database to those configurations with $ \epseff \lesssim \num{6e-3} $ for the case of $\widehat{E}_r=0$, the results shown in figure \ref{subfig:D31_vs_D11_eps_0100_Er_0_restricted} suggest a trade-off between $D_{11}^*$ and $|D_{31}^*|$. Conversely, note from figure  \ref{subfig:D31_vs_D11_eps_0100_Er_1e-3_restricted} that those configurations optimized to have $\epseff \lesssim \num{6e-3}$ cluster around a straight line of the $D_{11}^*-|D_{31}^*|$ plane. This clustering indicates a moderate correlation between $D_{11}^*$ and $|D_{31}^*|$ for sufficiently optimized configurations in the presence of a non zero radial electric field. The distribution of colours in figures \ref{subfig:D31_vs_D11_eps_0100_Er_0} and \ref{subfig:D31_vs_D11_eps_0100_Er_1e-3} also reveals that there is no correlation between $\epseff$ and $|D_{31}^*|$. This lack of correlation can be seen in more detail in figures \ref{subfig:D31_vs_eps_0100_Er_0} and \ref{subfig:D31_vs_eps_0100_Er_1e-3}, where the projection of the data onto the $|D_{31}^*|-\epseff$ plane is shown and the value of $D_{11}^*$ is represented in colours. Note that, for both values of $\widehat{E}_r$, those configurations that display simultaneously small levels of parallel and radial neoclassical transport are those with minimum $\epsilon_{\text{eff}}$. However, reducing $\epsilon_{\text{eff}}$ does not guarantee a reduction of the $D_{31}^*$ coefficient. For $\epseff\sim \num{2e-3}$ and $\widehat{E}_r=0$, we can see in figure \ref{subfig:D31_vs_eps_0100_Er_0} that the bootstrap current coefficient can range in an interval of almost three orders of magnitude, from $|D_{31}^*| \sim \num{e-3}$ to $|D_{31}^*| \sim 1$. For the case with finite $\widehat{E}_r$ shown in figure \ref{subfig:D31_vs_eps_0100_Er_1e-3}, the situation is similar but with a narrower interval of $|D_{31}^*|$. For $\epseff\sim \num{2e-3}$, the bootstrap current coefficient can change an order of magnitude, ranging between $|D_{31}^*| \sim \num{e-2}$ to $|D_{31}^*| \sim \num{e-1}$. This lack of correlation is unsurprising as reducing the effective ripple guarantees proximity to omnigenity, which is a necessary but not sufficient condition for quasi-isodynamicity. Finally, the fact that in the $1/\nu$ regime $D_{11}^* \propto \epseff^{3/2}/\hat{\nu}$ can be seen from figures \ref{subfig:D31_vs_D11_eps_0100_Er_0} and \ref{subfig:D31_vs_eps_0100_Er_0}. Note that the distribution of points and colour in figures \ref{subfig:D31_vs_D11_eps_0100_Er_0} and \ref{subfig:D31_vs_eps_0100_Er_0} is almost identical. Of course, this nearly perfect correlation is not preserved for the $\sqrt{\nu}$-$\nu$ regime, as shown in figures \ref{subfig:D31_vs_D11_eps_0100_Er_1e-3} and \ref{subfig:D31_vs_eps_0100_Er_1e-3}. This was expected as particle trajectories that cause the $\sqrt{\nu}$-$\nu$ flux are quite different from those that generate the $1/\nu$ flux.
	

	In figure \ref{fig:Correlation_VBB} the relation between $\sigma^2(\Bmin^{\text{r}})$ and the monoenergetic coefficients during the optimization campaign is shown. From figure \ref{subfig:D31_vs_D11_KN_VBB_0100_Er_0}, we can see that the smallest values of $\sigma^2(\Bmin^{\text{r}})$ cluster around the smallest values of $D_{11}^*$ and in the range of bootstrap current coefficient $\num{e-2}\lesssim D_{31}^*\lesssim \num{e-1}$. This suggests a slight correlation between $D_{31}^*$ and the variance $\sigma^2(\Bmin^{\text{r}})$. However, when inspecting this correlation in more detail in figure \ref{subfig:D31_vs_KN_VBB_0100_Er_0}, we can see that for very small values of $\sigma^2(\Bmin^{\text{r}})\lesssim 10^{-6}$, $D_{31}^*$ can vary almost two orders of magnitude, even if $D_{11}^*$ is also small. This simply indicates that it is possible to have a large deviation from quasi-isodynamicity even if $\sigma^2(\Bmin^{\text{r}})$ is close to zero. In figure \ref{subfig:D31_vs_KN_VBB_0100_Er_0_restricted} we have filtered out those configurations with $\epseff>\num{6e-3}$ and the slight correlation for sufficiently optimized configurations (in terms of the $\epseff$) is apparent, but far from ideal as $|D_{31}^*|$ can vary two orders of magnitude for $\sigma^2(\Bmin^{\text{r}})\sim\num{5e-7}$. The variability in the value of $|D_{31}^*|$ was expected, as this proxy is meant for approaching general omnigenity. As expected, there seems to be a trade-off between radial and parallel transport as the configurations with the smallest value of $|D_{31}^*|$ do not have the smallest values of $D_{11}^*$ aswell. For the case with radial electric field the correlation seems to be stronger. We can see in figure \ref{subfig:D31_vs_D11_KN_VBB_0100_Er_1e-3} that the smallest values of $\sigma^2(\Bmin^{\text{r}})$ are clustered very close to the left inferior corner in the $D_{11}^*-|D_{31}^*|$ plane. Indeed, for the smallest values of $\sigma^2(\Bmin^{\text{r}})$, we can see in figure \ref{subfig:D31_vs_KN_VBB_0100_Er_1e-3} that $|D_{31}^*|\lesssim \num{e-1}$. As shown in figure \ref{subfig:D31_vs_KN_VBB_0100_Er_1e-3_restricted}, the correlation is more evident for configurations optimized to have $\epseff\le\num{6e-3}$. The results suggest that for the case with finite radial electric field there is a moderate correlation between $\sigma^2(\Bmin^{\text{r}})$ and $|D_{31}^*|$. However, from the horizontal spread of the points shown in figure \ref{subfig:D31_vs_KN_VBB_0100_Er_1e-3_restricted}, we can conclude that minimizing $\sigma^2(\Bmin^{\text{r}})$ can be very inefficient for reducing $|D_{31}^*|$.

	In figure \ref{fig:Correlation_VB0} the relation between the monoenergetic coefficients and $\sigma^2(B(\theta,0))$ is shown. It is immediate to see from \ref{subfig:D31_vs_D11_KN_VB0_0100_Er_0} that, for $\widehat{E}_r=0$, there is no correlation between $\sigma^2(B(\theta,0))$ and the bootstrap current coefficient. Inspecting the lack of correlation in more detail in \ref{subfig:D31_vs_KN_VB0_0100_Er_0} we confirm that minimizing the variance from $\sigma^2(B(\theta,0))\sim \num{3e-4}$ to $\sigma^2(B(\theta,0))\sim\num{1e-4}$ can increase substantially the bootstrap current coefficient, even if $D_{11}^*$ is kept below 1. If we filter out configurations with $\epseff>\num{6e-3}$, as shown in figure \ref{subfig:D31_vs_KN_VB0_0100_Er_0_restricted}, this behaviour is confirmed and the results suggest that the simultaneous minimization of $\sigma^2(B(\theta,0))$ and $\epseff$ ($D_{11}^*$) is done at the expense of increasing $|D_{31}^*|$. Note from figure \ref{subfig:D31_vs_KN_VB0_0100_Er_0} that configurations with smaller levels of radial and parallel transport cluster at intermediate values of the variance $\sigma^2(B(\theta,0))$. This behaviour persists even for configurations with sufficiently optimized effective ripple, as shown in \ref{subfig:D31_vs_KN_VB0_0100_Er_0_restricted}. For the case with finite radial electric field, from figure \ref{subfig:D31_vs_D11_KN_VB0_0100_Er_1e-3}, we can see no appreciable correlation between $\sigma^2(B(\theta,0))$ and $|D_{31}^*|$. In figure \ref{subfig:D31_vs_KN_VB0_0100_Er_1e-3} we can see that configurations with small values of $D_{11}^*$ and $|D_{31}^*|$ cluster near the left of the plot, but still without strong correlation. When we filter configurations which are not sufficiently optimized in terms of $\epseff$, the results shown in figure \ref{subfig:D31_vs_KN_VB0_0100_Er_1e-3_restricted} suggest a mild correlation between $|D_{31}^*|$ and $\sigma^2(B(\theta,0))$ for $\widehat{E}_r\ne 0$. However, it is very far from ideal as for $\sigma^2(B(\theta,0))\sim\num{3e-4}$ the radial transport and bootstrap current coefficient can vary, respectively, two and one orders of magnitude. The inadequacy of $\sigma^2(B(\theta,0))$ for minimizing $|D_{31}^*|$ (even for configurations with small $\epseff$) is surprising as this is the only proxy specific for optimizing QI configurations and naively one would expect a better correlation. Finally, we point out that for the database considered, the proxies $\sigma^2(B(\theta,0))$ and $\sigma^2(\Bmax^{\text{r}}) $ are roughly equivalent and therefore we omit the results for the latter. This equivalency between the two proxies can be seen from figure \ref{fig:Correlation_VBM} in appendix \ref{sec:Appendix_VBM}, which is very similar to figure \ref{fig:Correlation_VB0}.
	

	Finally, we compare the relation of the monoenergetic coefficients with the fast ion proxies $\GammaC$ and $\GammaAlpha$. In figure \ref{fig:Correlation_GMC_GMA_Er_0}, the case for zero radial electric field is shown. Note from figure \ref{subfig:D31_vs_D11_KN_GMC_0100_Er_0} that configurations with the smallest values of $\GammaC$ do not cluster near the left inferior corner but on values $|D_{31}^*|\sim\num{3e-1}$. On the other hand, as figure \ref{subfig:D31_vs_D11_KN_GMA_0100_Er_0} shows, configurations with the smaller levels of parallel and radial transport also have the smallest values of $\GammaAlpha$. This difference suggests a slightly better correlation between $|D_{31}^*|$ and $\GammaAlpha$ than between $|D_{31}^*|$ and $\GammaC$. Inspecting this difference further, we can see in figure \ref{subfig:D31_vs_KN_GMC_0100_Er_0} that for $\GammaC$ there is an horizontal branch along which we can reduce $\GammaC$ but not $|D_{31}^*|$ and its value is not small ($|D_{31}^*|>10^{-1}$). As shown in figure \ref{subfig:D31_vs_KN_GMA_0100_Er_0}, this is not the case for $\GammaAlpha$ which seems to have a mild correlation with $|D_{31}^*|$. This difference in the behaviour persists even for configurations with low values of the effective ripple. From figure \ref{subfig:D31_vs_KN_GMC_0100_Er_0_restricted} we can see that the horizontal branch of $\GammaC$ is still present for configurations with low value of $\epseff$. Conversely, in figure \ref{subfig:D31_vs_KN_GMA_0100_Er_0_restricted} we can see that the correlation between $|D_{31}^*|$ and $\GammaAlpha$ is more pronounced for configurations with low $\epseff$. The case with finite radial electric field is shown in figure \ref{fig:Correlation_GMC_GMA_Er_1e-3}. For the case $\widehat{E}_r \ne 0$, the discussion is similar to the case without radial electric field. These numerical results suggest that in order to obtain a finite but small bootstrap current, it is more important to have contours of the second adiabatic invariant $J$ which close poloidally and do not deviate much from flux surfaces rather than exactly matching them. Specifically, for an approximately omnigenous configuration, reducing $\GammaC$ implies aligning all $J$ isosurfaces with flux surfaces. On the other hand, minimizing $\GammaAlpha$ entails an alignment of those constant $J$ surfaces which deviate the most from flux surfaces, but not all of them. A different (although non exclusive) possibility could be that, enforcing $J$ contours to be closed surfaces by minimizing $\GammaAlpha$ facilitates achieving the maximum$-J$ \cite{Rosenbluth_max_J,Helander_max_J} property to a sufficient degree of approximation. If so, the optimizer would be able to focus on minimizing other quantities, such as $\sigma^2(\Bmin^{\text{r}})$ to reduce $|D_{31}^*|$.

	%

	To summarize, in the light of these results, we can conclude that, although effective for obtaining nearly QI \qmarks{flat-mirror} configurations, the optimization strategy was not efficient for reducing the bootstrap current. The inefficiency of the indirect approach to minimize the $|D_{31}^*|$ coefficient is specially pronounced for the case without radial electric field. Thus, many intermediate configurations which are not sufficiently close from QI (in the sense of having too large $|D_{31}^*|$ or $D_{11}^*$) are produced during the optimization campaign. Apart from the imperfect correlation of the proxies used in indirect optimization, this inefficiency is probably enhanced by the multiple trade-offs that occur when many different requirements have to be met. It is reasonable to expect that a direct minimization of $|D_{31}^*|$, will be a much more efficient strategy for optimizing QI configurations.

	\endgroup
	\section{From piecewise omnigenity to quasi-isodynamicity}\label{sec:pwO_QI}
	\captionsetup[sub]{skip=-1.75pt, margin=50pt}
	Direct optimization has proven to be important not only for obtaining better magnetic configurations but also for finding new ideal stellarator designs. In \cite{Bindel_2023}, fast ion confinement was improved by including guiding-center trajectories in the optimization loop. Naturally, configurations with very small levels of fast ion losses were produced. When inspecting the isolines of $B$ of these configurations, the topology of constant $B$ contours within the flux surface differed significantly from what would be expected of an omnigenous configuration. Therefore, it came as a surprise that some of these configurations displayed also small values of $\epseff$. Inspired by this result, a new family of optimized stellarators denominated \textit{piecewise omnigenous} (pwO) \cite{velasco2024piecewise}, has emerged. 
	
	In order to define piecewise omnigenity, it is convenient to recall the definition of omnigenity given in section \ref{sec:Drift_kinetics}. As expressed in equations (\ref{eq:Orbit_averaged_dot_psi}) and (\ref{eq:Orbit_averaged_dot_alpha}), trapped particles drift preserving the second longitudinal adiabatic invariant $J$ in their bounce averaged movement \cite{dherbemont2022}. According to definition (\ref{eq:Omnigenity_definition}), in an omnigenous stellarator the isosurfaces of $J$ exactly match flux surfaces, i.e. for omnigenous stellarators $J$ is a flux function. Hence, the radial displacement that charged particles experience along their collisionless orbits averages to zero. In contrast, for a generic stellarator magnetic field, the isosurfaces of $J$ are \textit{transversal} to flux surfaces, which implies that trapped particles quickly drift out of the device. Requiring $J$ to be a flux function is what constrains the topology of the isolines of $B$ to close toroidally, poloidally or helically. 
	
	On the other hand, in a pwO field the second adiabatic invariant $J$ is a flux-function only \textit{piecewisely}, allowing jump discontinuities of $J$ on a flux surface along the poloidal direction. Among other things, this implies that the topology of the isolines of $B$ in a pwO field is not as limited as for an omnigenous stellarator. Imposing $J$ to be constant in a particular region of the flux surface constrains the isolines of $B$ in a similar way to the one presented in \cite{CaryPhysRevLett, Cary1997OmnigenityAQ} for omnigenous stellarators. Hence, as pwO magnetic fields have several regions in which $J$ is constant, the isolines are not necessarily forced to close poloidally, toroidally or helically. Those zero measure regions where $J$ can vary on the flux surface delimit different classes of trapped particles and therefore, transitions can occur when particles precess on the flux surface due to drifts or when particles collide. Remarkably, as indicated in \cite{velasco2024piecewise}, these transitions do not contribute to radial transport in the $1/\nu$ regime.

	In regard to parallel transport, the bootstrap current produced by piecewise omnigenous stellarators was still an unexplored area before this dissertation. For future stellarator designs, it could be very helpful to find pwO magnetic fields which have not only reduced $\epseff$ but also a small bootstrap current. In this section, we will employ {\MONKES} to investigate neoclasically nearly pwO magnetic fields that are approximately QI. The objective is to identify which levels of approximate piecewise omnigenity and quasi-isodynamicity are necessary to have low levels of radial and parallel neoclassical transport. Incidentally, we will demonstrate that {\MONKES} can be used to study neoclassical transport in stellarator configurations which are extremely complicated in terms of the Fourier spectra of $B$.


	A simple pwO magnetic field can be modelled using an exponential \cite{velasco2024piecewise}
	\begin{align}
		\widetilde{B}(\theta,\zeta)  
		& =
		\Bmin
		+
		(\Bmax-\Bmin)
		\exp(
		-
		\left(
		\frac{\zeta - \zeta_{\text{c}} }{w_\zeta}
		\right)^{2p}
		)
		\nonumber
		\\
		&
		\times
		\exp(
		-
		\left(
		\frac{\theta - \theta_{\text{c}} - t_\zeta(\zeta - \zeta_{\text{c}} )}{w_\theta}
		\right)^{2p}),
		\label{eq:Exponential_pwO}
	\end{align}
	along with the constraint to the rotational transform
	\begin{align}
		\iota =  -t_\zeta \frac{\Nfp w_\zeta}{\pi - \Nfp w_\zeta} ,
		\label{eq:iota_pwO}
	\end{align}
	becoming exactly pwO in the limit $p\rightarrow \infty$. When $p\rightarrow \infty$, the isolines $\Bmin<\widetilde{B}<\Bmax$ are compressed in a single parallelogram of center $(\theta_{\text{c}}, \zeta_{\text{c}})$ in the $(\theta,\zeta)$ plane. The four sides of this parallelogram are defined by the equations
	\begin{align}
		\theta - \theta_{\text{c}} & = \pm w_\theta + t_\zeta\left( \zeta - \zeta_{\text{c}}\right),
		\label{eq:pwO_tilted_sides}
		\\
		\zeta - \zeta_{\text{c}} & = \pm w_\zeta.
		\label{eq:pwO_vertical_sides}
	\end{align}
	Thus, the scalars $2w_\theta$ and $2w_\zeta<2\pi/\Nfp$ define, respectively, the poloidal and toroidal width of this parallelogram. The slope $t_\zeta$ defines the poloidal shear of the parallelogram, becoming a rectangle when $t_\zeta=0$. It is important to remark that the constraint (\ref{eq:iota_pwO}) guarantees that $J$ is a flux-function piecewisely for a field whose magnetic field strength is given by (\ref{eq:Exponential_pwO}) in the limit $p\rightarrow\infty$. We will make more precise this assertion later in this section.		
	 
	\begin{figure}[h]
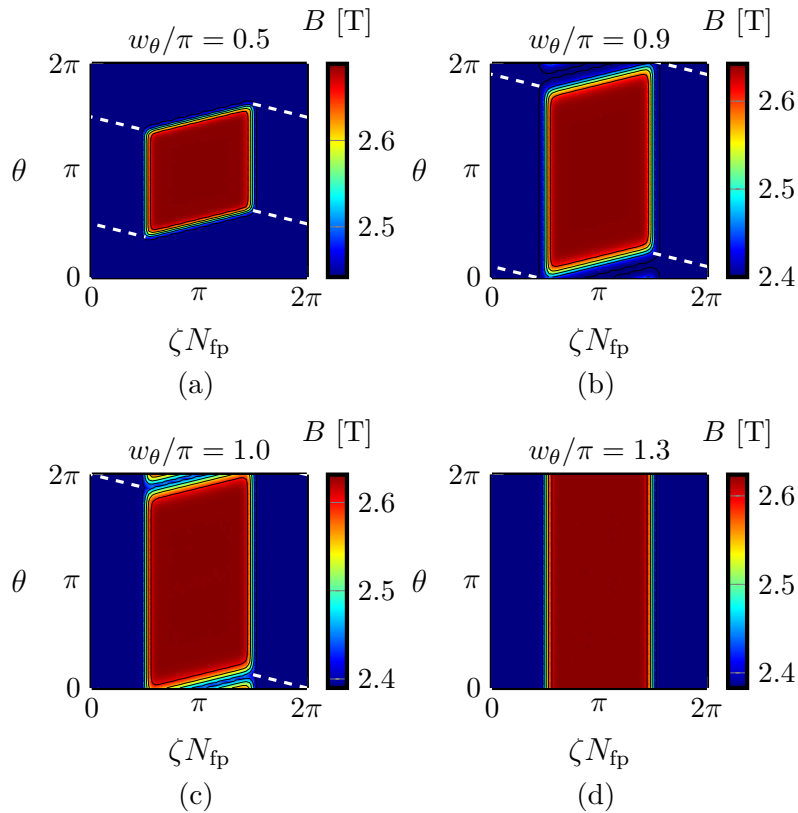

		\centering
		\foreach \p in {20}
		{ 
			\foreach \w in {0.5, 0.9, 1.0, 1.3}
			{%
				\tikzsetnextfilename{pwQI_B_\w_\p}
				\includepwOMagneticField{\p}{\w} 
			}  
			
		} 
		\caption{Magnetic field strength $B$ of an approximately pwO magnetic field ($p=10$) for several values of $w_\theta$.}
		\label{fig:Magnetic_field_strength_pwO_QI}
	\end{figure}

	\begin{figure}[h]
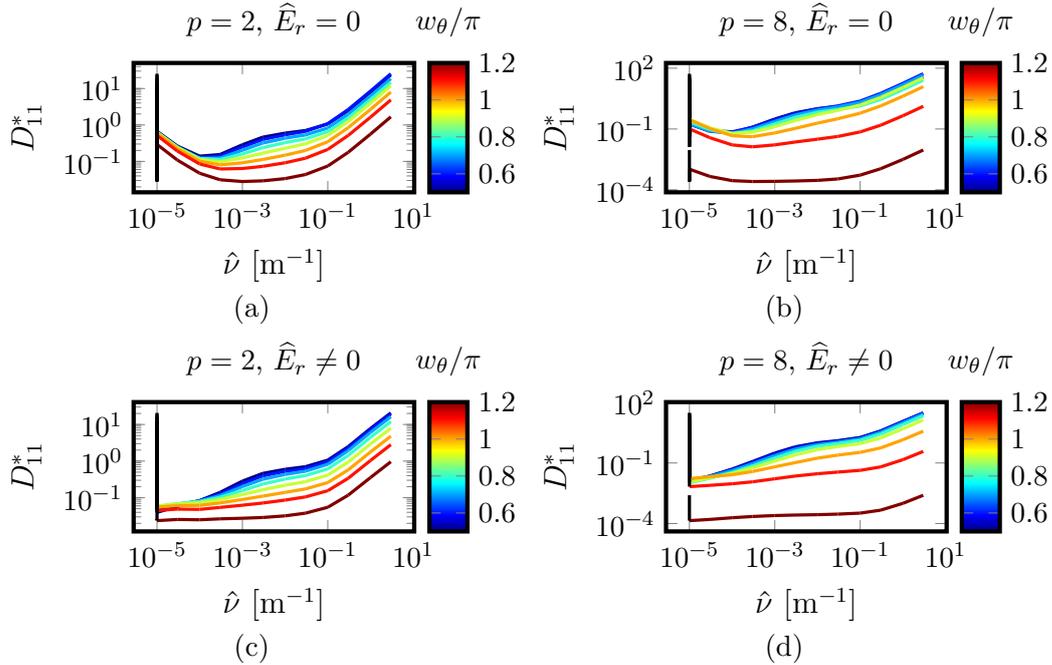
   
		\foreach \p in {4,16}
		{
			\tikzsetnextfilename{D11_vs_nu_pwQI_\p_Er_0}
			\begin{subfigure}[t]{0.45\textwidth}
				\includegraphics{D11_vs_nu_pwQI_\p_Er_0}
%
%
					\caption{}
					\label{subfig:D11_vs_nu_pwQI_\p_Er_0}
				\end{subfigure} 	     
			}	 
			
			\foreach \p in {4,16}
			{ 
				\tikzsetnextfilename{D11_vs_nu_pwQI_\p_Er_1e-3}
				\begin{subfigure}[t]{0.45\textwidth}
					\includegraphics{D11_vs_nu_pwQI_\p_Er_1e-3}
%
%
						\caption{}
						\label{subfig:D11_vs_nu_pwQI_\p_Er_1e-3}
					\end{subfigure}     	
				}
				
				\caption{Radial transport coefficient $D_{11}^*$ as a function of $\hat{\nu}$ and $w_\theta$ for $\widehat{E}_r = 0$ (top) and  $\widehat{E}_r \ne 0$ (bottom) for $p=2$ (left) and $p=8$ (right)}
				\label{fig:Dij_vs_nu_pwQI_Er_0}
			\end{figure}
			\begin{figure}[h]  
				\centering
				\begin{subfigure}[t]{0.45\textwidth}
					\tikzsetnextfilename{D11_vs_pi_factor_nu_1e-5_Er_0}
					\includegraphics{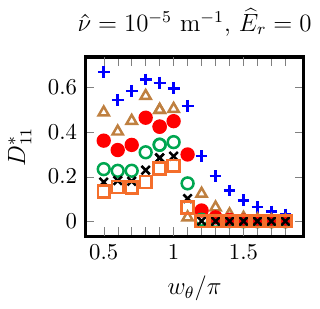}
					\caption{}
					\label{subfig:D11_vs_pi_factor_nu_1e-5}
				\end{subfigure}%
				\begin{subfigure}[t]{0.45\textwidth}		
					\tikzsetnextfilename{D11_vs_pi_factor_nu_1e-5_Er_1e-3}
					\includegraphics{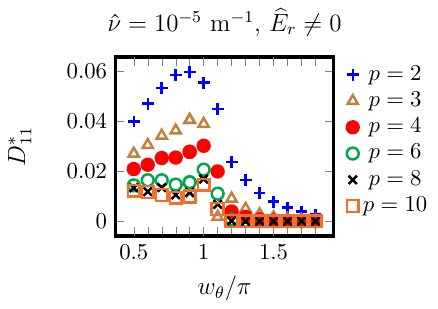}
					\caption{}
					\label{subfig:D11_vs_pi_factor_nu_1e-5_Er_1e-3}
				\end{subfigure}
				\caption{Radial transport coefficient $D_{11}^*$ as a function of $w_\theta$ and $p$ for $\hat{\nu}=\num{e-5}$ $\text{m}^{-1}$. (a) $\widehat{E}_r=0$ and (b) $\widehat{E}_r\ne0$. }
			\end{figure}	
			
			In order to identify regions of the pwO parameter space with small $D_{11}^*$ and $|D_{31}^*|$ we will evaluate neoclassically approximately pwO magnetic fields obtained from a scan in $w_\theta$ for several values of finite $p$. The idea is to start from a value of $w_\theta$ for which the configuration is nearly pwO and increase it until it becomes nearly QI. In figure \ref{fig:Magnetic_field_strength_pwO_QI}, we illustrate the scan in $w_\theta$ for a fixed value of $p=10$ using the magnetic field strength $B$ of an approximately pwO field constructed in the manner instructed in appendix \ref{sec:Appendix_pwO_B}. Note that in figure \ref{fig:Magnetic_field_strength_pwO_QI} we represent $B$ and not $\widetilde{B}$. We do this because the function $\widetilde{B}$ given by (\ref{eq:Exponential_pwO}) in the limit $p\rightarrow\infty$, by itself, can only define a stellarator-symmetric exactly pwO magnetic field strength $B$ for $w_\theta\le \pi-|t_\zeta| w_\zeta$. For a stellarator-symmetric exactly pwO field, at $w_\theta=\pi-|t_\zeta| w_\zeta$, two corners of the parallelogram are located at the poloidal positions $\theta=0$ and $\theta=2\pi$ and the remaining two somewhere in the interval $\theta \in (0,2\pi)$. An approximation to this situation is shown in figure \ref{subfig:pwOMagneticField_pow_20_wa0.9pi}. If we increase $w_\theta $ beyond this point, the parallelogram does not fit in the domain $\theta \in [0,2\pi]$ and imposing the constraint (\ref{eq:iota_pwO}) no longer guarantees exact piecewise omnigenity. In order to increase $w_\theta$ and maintain approximate pwO, the parallelogram must \qmarks{grow} in the way shown in figure \ref{subfig:pwOMagneticField_pow_20_wa1.0pi}. This behaviour cannot be obtained by simply increasing $w_\theta$ in the definition (\ref{eq:Exponential_pwO}). In addition, more complications arise when $p$ is finite. Nevertheless, as explained in appendix \ref{sec:Appendix_pwO_B}, we can circumvent these complications and use the exponential function from equation (\ref{eq:Exponential_pwO}) and the constraint (\ref{eq:iota_pwO}), to construct a stellarator-symmetric approximately pwO field for different values of $w_\theta$ and finite $p$, including $w_\theta > \pi - |t_\zeta| w_\zeta$. The approximately pwO magnetic field has been constructed so that it resembles that of a flux surface of Wendelstein 7-X KJM (further details in appendix \ref{sec:Appendix_pwO_B}). The parameters required for defining the magnetic field are listed in tables \ref{tab:pwO_parameters} and \ref{tab:pwO_parameters_KJM}. It is important to stress that this particular scan in $w_\theta$ and $p$ allows for the exploration of a \textit{very small} fraction of all the possible parametrizations of the configuration space of pwO fields.

			\begin{figure}[h]
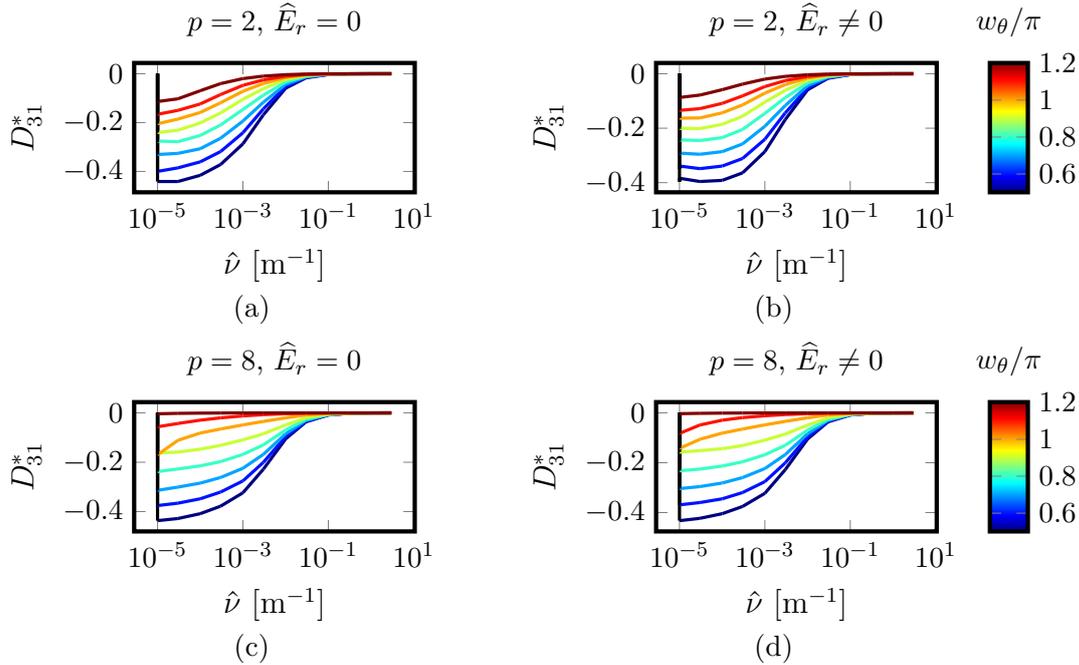
    
				\foreach \p in {4,16}
				{
					\tikzsetnextfilename{D31_vs_nu_pwQI_\p_Er_0} 
					\begin{subfigure}[t]{0.45\textwidth}
						\includegraphics{D31_vs_nu_pwQI_\p_Er_0} 
%
%
						\caption{}
						\label{subfig:D31_vs_nu_pwQI_\p_Er_0}
					\end{subfigure}
					\tikzsetnextfilename{D31_vs_nu_pwQI_\p_Er_1e-3} 
					\begin{subfigure}[t]{0.45\textwidth}
						\includegraphics{D31_vs_nu_pwQI_\p_Er_1e-3} 
%
%
						\caption{}
						\label{subfig:D31_vs_nu_pwQI_\p_Er_1e-3}
					\end{subfigure}
					
				}	 
				
				\caption{Bootstrap current coefficient $D_{31}^*$ as a function of $\hat{\nu}$ and $w_\theta$ for $\widehat{E}_r = 0$ (top) and  $\widehat{E}_r \ne 0$ (bottom) for $p=2$ (left) and $p=8$ (right)}
				\label{fig:Dij_vs_nu_pwQI_Er_1e-3}
			\end{figure}
			\begin{figure}[h]  
				\centering	%
				\begin{subfigure}[t]{0.45\textwidth}%
					\tikzsetnextfilename{D31_vs_pi_factor_nu_1e-5_Er_0}
					\includegraphics{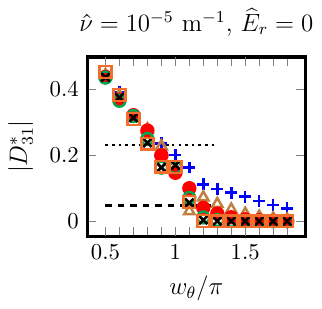}
					\caption{}
					\label{subfig:D31_vs_pi_factor_nu_1e-5}
				\end{subfigure} 
				\begin{subfigure}[t]{0.45\textwidth}		
					\tikzsetnextfilename{D31_vs_pi_factor_nu_1e-5_Er_1e-3}
					\includegraphics{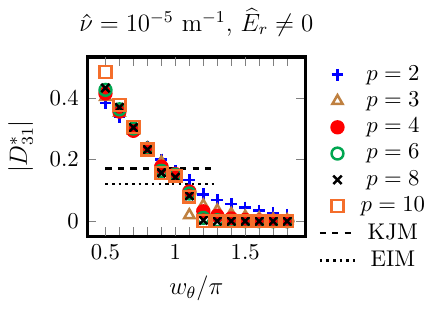}
					\caption{}
					\label{subfig:D31_vs_pi_factor_nu_1e-5_Er_1e-3}
				\end{subfigure} 
				\caption{Bootstrap current coefficient $D_{31}^*$ as a function of $w_\theta$ and $p$ for $\hat{\nu}=\num{e-5}$ $\text{m}^{-1}$.}
			\end{figure}
			We can use the approximately pwO fields represented in figure \ref{fig:Magnetic_field_strength_pwO_QI} to precise our previous comment about how $\iota$ guarantees that $J$ is a flux-function piecewisely in the limit $p\rightarrow\infty$. When $\iota$ is given by (\ref{eq:iota_pwO}), two field lines connect the four corners of the parallelogram. These field lines are indicated in figures \ref{subfig:pwOMagneticField_pow_20_wa0.5pi} and \ref{subfig:pwOMagneticField_pow_20_wa0.9pi} with a white dashed line and their intersections with the parallelogram define several regions in the $(\theta,\zeta)$ plane. As for each region all bounce points lie on two parallel segments of the parallelogram, the angular distance between bounce points does not depend on the field line chosen. Besides, as for each region $B$ is also constant, then $J$ must also be the same for any field line belonging to that region. Thus, the orbit-averaged drift that trapped particles experience at each region is zero. Across the two field lines that delimit different regions, the value of $J$ can change abruptly. For an exactly pwO field, these transitions do not contribute to radial transport in the $1/\nu$ regime (see \cite{velasco2024piecewise}). For our model, the benignancy of transitioning particles is guaranteed by the fact that in the limit $p\rightarrow\infty$ the isolines have pointy corners. Hence, in the limit $p\rightarrow \infty$ any field where $B$ and $\iota$ are appropriately defined by (\ref{eq:Exponential_pwO}) and (\ref{eq:iota_pwO}) (e.g. as in the manner explained in appendix \ref{sec:Appendix_pwO_B}) would have $\epseff=0$. From figures \ref{subfig:pwOMagneticField_pow_20_wa1.0pi} and \ref{subfig:pwOMagneticField_pow_20_wa1.3pi}, we can verify that increasing $w_\theta$ beyond $\pi$ forces the isolines of $B$ to close poloidally. As a consequence, at some point in the scan, the different classes of trapped particles disappear and $J$ becomes constant on the whole flux surface, making the resulting field quasi-poloidally symmetric (a particular case of QI). Note from figure \ref{subfig:pwOMagneticField_pow_20_wa1.0pi} that for $w_\theta=\pi$ the isoline $B=\Bmin$ is not poloidally closed due to the finiteness of $p$. In the limit $p\rightarrow\infty$, this isoline would close at precisely $w_\theta=\pi$. We recall that, by definition, the integer power $p$ represents the proximity to piecewise omnigenity of the model field. Similarly, $w_\theta$ controls closeness to quasi-isodynamicity of the configuration.

			A very attractive feature of pwO fields is that rough approximations to an exactly pwO field can have low levels of radial neoclassical transport \cite{velasco2024piecewise}. In particular, we will see that using the model given by equations (\ref{eq:Exponential_pwO}) and (\ref{eq:iota_pwO}) for $p=2$ (the lowest value of $p$ considered), a banana-like regime \cite{Landreman_PreciseQS} appears between the plateau and the deleterious $1/\nu$ regime. Thus, the reduction of radial neoclassical transport appears for values of $p$ for which the magnetic field varies in a scale compatible with rigorous neoclassical theory \cite{velasco2024piecewise}. For this reason, pwO magnetic fields are very promising as an ideal design goal for optimization. Therefore, in this section, we will explore the parameter space $(p,w_\theta)$ to identify portions of it which have simultaneously small levels of radial and parallel transport. In the light of what has been exposed we expect $\Dij{11}$ to decrease with increasing $p$ and, for each fixed $p$, $|\Dij{31}|$ to be a monotonically decreasing function of $w_\theta$. Besides, for sufficiently large $w_\theta$ we expect $|\Dij{31}|$ to be a monotonically decreasing function of $p$.

			In order to verify numerically our theoretical expectations, we have computed the monoenergetic coefficients $\Dij{11}$ and $\Dij{31}$ for collisionalities $\hat{\nu}\in[\num{e-5},3]$ $\text{m}^{-1}$ and radial electric field $\widehat{E}_r\in\{0,\num{e-3}\}$ $\text{V}\cdot\text{s} /\text{m}^{2}$. This scan in collisionality and radial electric field has been carried out for approximately pwO fields constructed as indicated in appendix \ref{sec:Appendix_pwO_B} for $p\in[2,10]$ and $w_\theta/\pi \in[0.5,1.9]$. In figure \ref{fig:Dij_vs_nu_pwQI_Er_0} the result of the scan in collisionality is shown for $\widehat{E}_r=0$ for $p=2$ and $p=8$. In colours, the value of $w_\theta/\pi$ for each case is displayed. As it was mentioned, we can see from the curve of $D_{11}^*$ plotted in figure \ref{subfig:D11_vs_nu_pwQI_4_Er_0} that even for $p=2$ a banana-like regime appears for $w_\theta/\pi \le 0.8 $. For higher values of $p$ the situation is similar, as shown in \ref{subfig:D11_vs_nu_pwQI_16_Er_0} for $p=8$. For $\hat{\nu}\gtrsim \num{e-4}$, increasing $w_\theta$ diminishes the value of $D_{11}^*$ for all values of $p$ considered. This happens even when $w_\theta$ is increased beyond $ 0.8\pi$ and the banana-like regime disappears. For large values of $w_\theta$ the width in $\hat{\nu}$ of the plateau region seems to increase. The apparent spread of the plateau region is due to the reduction of the $D^*_{11}$ in the low collisionality region where the $1/\nu$ regime would appear in a magnetic field far from omnigenity. The reduction of the value of $D_{11}^*$ in the plateau region with $w_\theta$ is due to the fact that for high values of $w_\theta$, the magnetic field becomes almost constant along $\theta$ \cite{EmiliaRodriguez1987}. For $\hat{\nu}\lesssim \num{e-4}$ there is, however, an increase of transport in the $1/\nu$ regime for $w_\theta\sim\pi$ as figure \ref{subfig:D11_vs_nu_pwQI_16_Er_0} reveals. This effect is seen in more detail in \ref{subfig:D11_vs_pi_factor_nu_1e-5} where the value of $D_{11}^*$ for $\hat{\nu}=\num{e-5}$ and $\widehat{E}_r=0$ is shown. We can see from this figure that $D_{11}^*(\hat{\nu}=\num{e-5})$ grows as it approaches $w_\theta=\pi$ and that this growth is ameliorated when $p$ is increased. The peak of radial transport when $w_\theta$ grows ($w_\theta<\pi$) may be caused by a combination of the finiteness of $p$ and the fact that orbits become shorter in the region between the tilted sides of the parallelogram defined by (\ref{eq:pwO_tilted_sides}) as $w_\theta$ grows. Due to the smaller value of $J$ in this narrow region, minimizing $\partial_{\alpha}{J}/J$ for these orbits requires a larger value of $p$ when $w_\theta$ grows. Nevertheless, this increase in radial transport is not larger than a factor of 2 for any of the cases considered. When there is a radial electric field, increasing $w_\theta$ also produces a flattening of the $D_{11}^*$ curve from plateau to low collisionality. Again, there is an increasing of the radial transport coefficient at low collisionality when $w_\theta\sim\pi$. This effect can be observed in more detail in figure \ref{subfig:D11_vs_pi_factor_nu_1e-5_Er_1e-3}. As for the $1/\nu$ regime, the growth in $D_{11}^*$ is less pronounced for the largest value of $p$ considered.

			In regard to the bootstrap current coefficient, excepting a few cases of small $p$ at low collisionality, the results for finite and zero $\widehat{E}_r$ are very similar due to the extreme proximity to omnigenity of the magnetic fields considered. For small values of $p$, the effect of increasing $w_\theta$ is to reduce the value of $|D_{31}^*|$ uniformly, as shown in figures \ref{subfig:D31_vs_nu_pwQI_4_Er_0} and \ref{subfig:D31_vs_nu_pwQI_4_Er_1e-3}. For higher values of $p\ge 4$, in the region $w_\theta \sim \pi$, the $|D_{31}^*|$ curve changes its convexity in the range of collisionalities considered. Thus, at the lowest collisionality, the value of $D_{31}^*$ for $w_\theta=\pi$ (orange curve) can be approximately equal to the one for $w_\theta=0.9\pi$ (lime curve). Nevertheless, as can be seen in figures \ref{subfig:D31_vs_pi_factor_nu_1e-5} and \ref{subfig:D31_vs_pi_factor_nu_1e-5_Er_1e-3}, the effect of increasing $w_\theta$ is to reduce $|D_{31}^*|$ at low collisionality. As expected, we can observe from comparing figures \ref{subfig:D31_vs_nu_pwQI_4_Er_0} and \ref{subfig:D31_vs_nu_pwQI_16_Er_0} or \ref{subfig:D31_vs_nu_pwQI_4_Er_1e-3} and \ref{subfig:D31_vs_nu_pwQI_16_Er_1e-3} that, for fixed $w_\theta>\pi$, the reduction in $|D_{31}^*|$ is  typically more pronounced for bigger values of $p$. We can see from figure \ref{subfig:D31_vs_pi_factor_nu_1e-5} that when $\widehat{E}_r=0$ and $w_\theta/\pi \ge 1.2$ and $p\ge4$, the value of $|D_{31}^*|$ is smaller than that of the KJM configuration (also without $\widehat{E}_r$). Another case with such small value of the bootstrap current coefficient is the case $p=3$ and $w_\theta/\pi=1.1$. These results suggest that it is possible to design stellarators that deviate from QI to approach pwO with small levels of both radial and parallel transport. However, in this first exploration, the results indicate that it is necessary to be close to QI to have small $|D_{31}^*|$. In order to have a $|D_{31}^*|$ value equal or smaller than that of the KJM configuration without $\widehat{E}_r$ we need at least $p=3$ and $w_\theta/\pi=1.1$ or $p=4$ and $w_\theta/\pi=1.2 $. By inspecting figures \ref{subfig:pwOMagneticField_Isolines_pow_6_wa1.1pi} and \ref{subfig:pwOMagneticField_Isolines_pow_8_wa1.2pi}, we can check that most of the isolines of $B$ for this case are poloidally closed. Interestingly, we can see from this figure that those isolines which do not close poloidally are located around $\Bmax$. This is in agreement with what the numerical results shown in section \ref{sec:Correlations} suggest. In section \ref{sec:Correlations} we verified that minimizing the proxy $\sigma^2(B(\theta,0))$ (equivalently $\sigma^2(\Bmax^{\text{r}})$) did not entail a reduction of $|D^*_{31}|$. We emphasize that this exploration is far from being exhaustive and the configuration space of pwO fields has to be investigated further in future work. Hence, this exploration does not rule out the existence of other types of pwO magnetic fields with very small bootstrap current. In a very recent work \cite{Calvo_pwO_zero_bootstrap} (posterior to the findings of this section), pwO fields with zero bootstrap current in the limit of low collisionality have been discovered and characterized using {\MONKES}. 

			\captionsetup[sub]{skip=-1.75pt, margin=40pt}
			\begin{figure}[h]
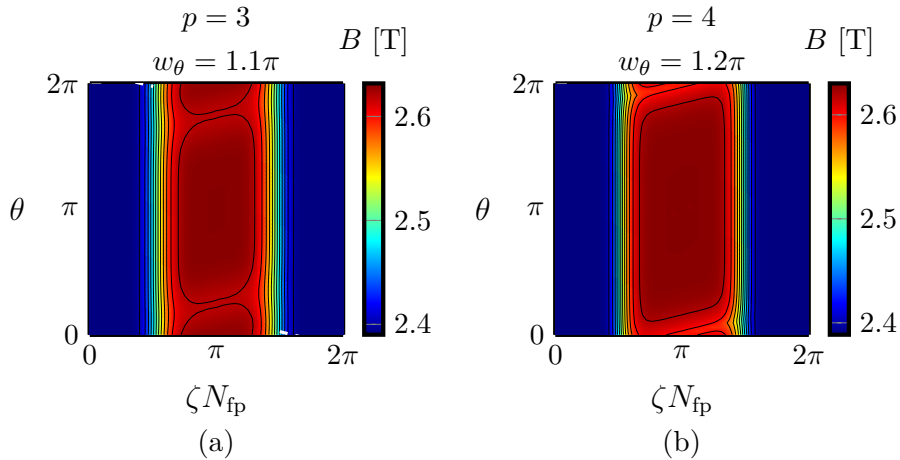
  
				\centering
				\foreach \w in {1.1}
				{ 
					\foreach \p in {6}
					{%
						\tikzsetnextfilename{pwQI_B_\w_\p_small_D31}
						\includepwOMagneticFieldIsolines{\p}{\w}
					}%
				} 
				\foreach \w in {1.2}
				{ 
					\foreach \p in {8}
					{%
						\tikzsetnextfilename{pwQI_B_\w_\p_small_D31}
						\includepwOMagneticFieldIsolines{\p}{\w}
					}%
				} 
				\caption{Magnetic field strength $B$ of those approximately pwO magnetic fields with smaller $D_{31}^*$ ($w_\theta=1.2\pi$) than the KJM configuration.}
				\label{fig:Magnetic_field_strength_pwO_QI_small_D31}
			\end{figure}

			Finally, we point out that an exactly pwO magnetic field cannot be represented with a Fourier series without suffering the Gibbs phenomenon (further details in appendix \ref{sec:Appendix_pwO_B}). This phenomenon is caused by the discontinuity of $B$ in the perimeter of the parallelogram. Even though we have considered only finite values of $p$, due to the large gradients of $B$ in the vicinity of the perimeter, approximating $\widetilde{B}$ with a Fourier series requires a large amount of modes $\{B_{mn}\}$ with big mode numbers $(m,n)$. This unusually broad spectrum of $B$ for high $p$ implies that, in order to solve the DKE (\ref{eq:DKE}) at low collisionality, the spatial and Legendre resolutions must be very large. For instance, for $p= 10$ and $\hat{\nu}=\num{e-5}$, calculating the monoenergetic coefficients using {\MONKES} required around $ 12000 $ discrete Fourier modes and $200$ Legendre modes. For this extremely (and unusually) large spatial resolution, the wall-clock time for computing the monoenergetic coefficients for each pair $(\hat{\nu},\widehat{E}_r)$ was of approximately 14 minutes while running using 30 cores of CIEMAT's cluster XULA. Hence, the investigation of pwO magnetic fields would have been much more difficult (if not practically impossible) without a fast neoclassical code like {\MONKES}.

			\chapter{Conclusions}  
			\label{chap:Conclusions}
			\thispagestyle{empty}
			In this thesis, an algorithm for computing fast and accurately the bootstrap current in reactor-relevant, low collisionality stellarator plasmas, which is based on the analytical properties of the monoenergetic DKE (\ref{eq:DKE}) has been provided. As a result of the implementation of this algorithm, a new fast neoclassical code named {\MONKES} has been developed. {\MONKES} is thus a natural successor of the widespread neoclassical code {\DKES}, which has been the workhorse for neoclassical transport calculations in stellarator plasmas for more than thirty years. {\MONKES} rapid computations make it possible to include bootstrap current calculations in numerical studies in which it was not possible before. In particular, this thesis opens up the possibility of systematic, direct optimization of the bootstrap current (and neoclassical transport) in general stellarator geometry. In addition, it also makes it possible to treat the effect of the bootstrap current self-consistently in predictive transport frameworks. The impact of {\MONKES} and its algorithm goes beyond its direct applications. Recently, the author of this dissertation has been collaborating to develop a Python version of {\MONKES} that uses the \texttt{JAX} library and which will be included in the stellarator optimization suite \texttt{DESC}. In addition, the \texttt{JAX} version of {\MONKES}, will be tested as a preconditioner for a new implementation of the code {\SFINCS}.
			
			In chapter \ref{chap:Fundamentals}, the fundamental concepts of toroidal plasma confinement and neoclassical transport in stellarators have been reviewed. The general kinetic and fluid descriptions of a plasma have been explained. The ideal magnetohydrodynamic equilibrium equations for a toroidal plasma have been derived as a simplified approximation of the fluid description. The assumptions and orderings of drift-kinetics have been listed and its main result, the DKE, has been presented. Incidentally, the Lagrangian approach to guiding-center motion and omnigenity have been described. The ideal magnetohydrodynamic equilibrium equations for a toroidal plasma have been derived as the fluid equations corresponding to a toroidal plasma in thermodynamical equilibrium. Finally, the DKE to treat situations near radially local equilibrium has been obtained.

			In chapter \ref{chap:Monoenergetic}, the monoenergetic approximation to neoclassical transport and the DKE corresponding to this approximation have been reviewed. In addition, some well known properties of the monoenergetic DKE and transport coefficients have been derived. An algorithm, based on the structure of the DKE in a Legendre basis, for solving the DKE at any finite collisionality has also been provided. Three different methods for obtaining derivatives of the monoenergetic transport coefficients have also been described and discussed from the theoretical point of view: the finite differences, the direct and the adjoint methods.
			
			In chapter \ref{chap:MONKES}, the implementation of the algorithm presented in chapter \ref{chap:Monoenergetic} in the new neoclassical code {\MONKES} has been detailed. By means of a convergence study and a thorough benchmark it has been shown that {\MONKES} is fast, accurate and memory efficient. Calculations of all the monoenergetic coefficients $\Dij{ij}$ at a reactor-relevant low collisionality, take approximately one minute of wall-clock time when running in a single core. Besides, the memory required for computing the monoenergetic coefficients $\Dij{ij}$ is sufficiently low so that calculations fit in a single core and can be carried out in a personal computer. In addition, it has been shown that, when multiple cores are available, {\MONKES} calculations can be even faster when running in parallel. Finally, other capabilities of {\MONKES} which can be useful but are not standard in neoclassical codes, like the computation of derivatives of the monoenergetic coefficients have been demonstrated. 
			
			In chapter \ref{chap:MONKES_applications}, we have shown two applications of {\MONKES} related to stellarator optimization which have been possible to be carried out during this thesis thanks to {\MONKES} speed of computation. Using as an example the large database of intermediate magnetic configurations that lead to the flat mirror nearly QI configuration CIEMAT-QI4, we have illustrated the inefficiency of the indirect approach to stellarator optimization for reducing the bootstrap current. After that, we have used {\MONKES} to identify a direction of the configuration space of QI configurations along which we can deviate without compromising the smallness of radial and parallel transport. In particular, we have shown that we can deviate from a purely QI magnetic field to approach piecewise omnigenity.

			\appendix 
			\chapter{The Fokker-Planck collision operator}\label{sec:Fokker_Planck_operator}%

			In this appendix, the explicit expression of the Fokker-Planck collision operator, its conservation properties and the $H-$theorem are reviewed. The Fokker-Planck operator can be written as \cite{Helander_2005, Parra_Collisional}
			\begin{align}
				C_{ab}\left( F_a, F_b \right)
				& 
				:= 
				- \nabla_{\vect{v}}
				\cdot \vect{j}_{ab}(F_a,F_b)
				,
				\label{eq:Fokker_Planck_conservation}
			\end{align}
			where 
			\begin{align}
				\vb*{j}_{ab}(F_a,F_b)
				& 
				:=
				-
				\gamma_{ab}
				\nabla_{\vect{v}}
				\cdot
				\left[
				\int
				W(\vect{v}-\vect{v}')
				\cdot
				\left(
				\frac{F_b}{m_a}
				\nabla_{\vect{v}}{F_a}
				-
				\frac{F_a}{m_b}
				\nabla_{\vect{v}'}{F_b}
				\right)
				\dd[3]{\vect{v}'}
				\right] 
				\label{eq:Fokker_Planck_velocity_flux}
			\end{align}
			is the flux in velocity space associated to the collision of species $a$ with species $b$, $\gamma_{ab} = \Frac{e_a^2 e_b^2 \ln\Lambda_{ab}}{8\pi\epsilon_0^2 m_a}$ and $W(\vect{x}) = (Ix^2 - \vect{x}\vect{x})/x^3$. Here, $\ln\Lambda_{ab}$ is the Coulomb logarithm. Note that in (\ref{eq:Fokker_Planck_velocity_flux}), $\vb*{j}_{ab}$ and $F_a$ depend on $\vb*{v}$ while $F_b$ depends on the dummy integration variable $\vb*{v}'$.

			\section{Conservation properties}\label{sec:Fokker_Planck_conservation_properties}
			The divergence form of (\ref{eq:Fokker_Planck_conservation}) permits to write (using Stokes' theorem on a sphere in the velocity space with radius $R_v \rightarrow \infty$) the integral identity for any function $\phi$
			\begin{align}
				\int \phi \ C_{ab}(F_a,F_b) \dd[3]{\vect{v}}
				=
				\int \nabla_{\vect{v}}{\phi} 
				\cdot 
				\vect{j}_{ab}(F_a,F_b) 
				\dd[3]{\vect{v}},
				\label{ec:Fokker_Planck_integral_identity}
			\end{align}
			which will be useful to obtain the conservation properties of $C_{ab}(F_a,F_b)$.
			
			The (species dependent) quantity $\phi$ is conserved by the collision operator if
			\begin{align}
				\sum_{a}\sum_{b}\int \phi_a \ C_{ab}(F_a,F_b) \dd[3]{\vect{v}}
				=
				0.
				\label{eq:Collision_Conserved_quantity_definition}
			\end{align}
			Note that, mass conservation is a consequence of evaluating (\ref{ec:Fokker_Planck_integral_identity}) for $\phi=m_a$ which is a stronger condition than (\ref{eq:Collision_Conserved_quantity_definition}). A property of the Fokker-Planck operator is that conservation of momentum and energy are satisfied pairwise by the collisions $a-b$ and $b-a$. This is a consequence of employing binary Coulomb collisions (which conserve these quantities pairwise) to construct the Fokker-Planck collision operator. 
			
			
			Indeed, using (\ref{ec:Fokker_Planck_integral_identity}) the contribution of the collisions $a-b$ and $b-a$ to (\ref{eq:Collision_Conserved_quantity_definition}) can be written as\footnote{For the second equality we have used the cyclic permutation of dummy variables $(\vect{v},\vect{v}')\mapsto(\vect{v}',\vect{v})$ in $\vect{j}_{ba}$. We do this to obtain in the definition of $\vect{j}_{ba}$ the variables associated to $a$ as functions of $\vect{v}$ and the associated to $b$ as functions of $\vect{v}'$ (which of course includes $\vect{j}_{ba}$). By doing this, we can factorize the term $\left( F_b m_a^{-1}	\nabla_{\vect{v}}{F_a} - {F_a}{m_b}^{-1}
				\nabla_{\vect{v}'}{F_b} \right)$.}
			\begin{align}
				\int \phi_a& \ C_{ab}(F_a,F_b) \dd[3]{\vect{v}}
				+
				\int \phi_b \ C_{ba}(F_b,F_a) \dd[3]{\vect{v}}
				\nonumber
				\\
				&
				=
				\int 
				\left(
				\nabla_{\vect{v}} {\phi_a}\cdot \vect{j}_{ab} 
				+
				\nabla_{\vect{v}}{\phi_b} \cdot \vect{j}_{ba}
				\right)
				\dd[3]{\vect{v}}
				\nonumber
				\\
				&
				=
				\int 
				\nabla_{\vect{v}}{\phi_a} \cdot \vect{j}_{ab}
				\dd[3]{\vect{v}}
				-
				\frac{m_a}{m_b}
				\int 
				\nabla_{\vect{v}'}{\phi_b}
				\cdot \vect{j}_{ba}(\vect{v}')
				\dd[3]{\vect{v}'}
				\nonumber
				\\
				& =
				\gamma_{ab}
				\iint
				\left(
				\nabla_{\vect{v}}{\phi_a}
				-
				\frac{m_a}{m_b}
				\nabla_{\vect{v}'}{\phi_b}
				\right) 
				\cdot 
				W(\vect{v}-\vect{v}')
				\cdot
				\left(
				\frac{F_b}{m_a}
				\nabla_{\vect{v}}{F_a}
				-
				\frac{F_a}{m_b}
				\nabla_{\vect{v}'}{F_b}
				\right)
				\dd[3]{\vect{v}'}
				\dd[3]{\vect{v}}.
				\label{eq:Conservation_properties_Fokker_Planck}
			\end{align}
			Conservation of mass, momentum and energy are obtained by introducing respectively $(\phi_a,\phi_b)\in\{ (m_a,m_b), (m_a\vect{v},m_b\vect{v}'), (m_av^2/2,m_b{v'}^2/2)  \}$ in (\ref{eq:Conservation_properties_Fokker_Planck}). For both mass and momentum conservation we have $\pdv*{\phi_a}{\vect{v}}-{m_a}{m_b}^{-1}\pdv*{\phi_b}{\vect{v}'} = 0$. Energy conservation comes from the fact that $\nabla_{\vect{v}}{\phi_a}-{m_a}{m_b}^{-1}\nabla_{\vect{v}'}{\phi_b} = m_a(\vect{v} - \vect{v}')$ is orthogonal to the image of $W(\vect{v}-\vect{v}')$. For these three cases, the integrand of (\ref{eq:Conservation_properties_Fokker_Planck}) is zero and therefore the quantities are conserved.
			
			Hence, definition (\ref{eq:Collision_Conserved_quantity_definition}) can be refined for binary collisions. The Fokker-Planck collision operator is said to \textit{conserve the (species dependent) quantity $\phi$} if 
			\begin{align}
				\int \phi_a \ C_{ab}(F_a,F_b) \dd[3]{\vect{v}}
				+
				\int \phi_b \ C_{ba}(F_b,F_a) \dd[3]{\vect{v}}
				=
				0.
				\label{eq:Conservation_properties_Fokker_Planck_pairwise}
			\end{align}
			In particular, the Fokker-Planck collision operator preserves mass, momentum and energy. These conservation properties can be expressed, respectively as
			\begin{align}
				&
				\int \ C_{ab}(F_a,F_b) \dd[3]{\vect{v}}
				=
				0
				,
				\label{eq:Fokker_Planck_mass_conservation}
				\\
				&
				\int m_a \vect{v}\ C_{ab}(F_a,F_b) \dd[3]{\vect{v}}
				+
				\int m_b \vect{v} \ C_{ba}(F_b,F_a) \dd[3]{\vect{v}}
				=
				0
				,
				\label{eq:Fokker_Planck_momentum_conservation}
				\\
				&
				\int \frac{m_a v^2}{2} \ C_{ab}(F_a,F_b) \dd[3]{\vect{v}}
				+
				\int \frac{m_b v^2}{2} \ C_{ba}(F_b,F_a) \dd[3]{\vect{v}}
				=
				0
				.
				\label{eq:Fokker_Planck_energy_conservation}
			\end{align}

			\section{$H$-theorem for the Fokker-Planck operator}\label{sec:H_theorem}
			The entropy associated to species $a$ is defined as
			\begin{align}
				S_a := -\int F_a \ln F_a \dd[3]{\vect{v}}.
				\label{eq:Fokker_Planck_entropy_a}
			\end{align}
			
			As $F_a$ evolves according to Fokker-Planck equation (\ref{eq:Kinetic_equation_Introduction}), the total derivative of $S_a$, known as \textit{entropy production}, satisfies
			\begin{align}
				\dv{S_a}{t}= \sum_{b}\dot{\sigma}_{ab},
				\label{eq:Fokker_Planck_entropy_production_a}
			\end{align} 
			where the entropy production associated to the Coulomb collisions described by the Fokker-Planck operator between species $a$ and $b$ has been defined as
			\begin{align}
				\dot{\sigma}_{ab}
				:=
				-\int  \ln F_a C_{ab}(F_a,F_b)\dd[3]{\vect{v}}.
				\label{eq:Fokker_Planck_entropy_production_ab}
			\end{align}
			
			The $H$-theorem states that the total entropy production due to collisions between species $a$ and $b$ satisfies the inequality
			\begin{align}
				\dot{\sigma}_{ab}
				+
				\dot{\sigma}_{ba}
				\geq 
				0,
				\label{eq:H-theorem}
			\end{align}
			which in particular implies
			\begin{align}
				\sum_a 
				\dv{S_{a} }{t}
				\geq 
				0.
				\label{eq:H-theorem_a}
			\end{align}

			In order to prove the $H$-theorem for the Fokker-Planck collision operator we use (\ref{eq:Conservation_properties_Fokker_Planck}) for $(\phi_a,\phi_b)=(\ln F_a, \ln F_b)$ to write the left-hand side of (\ref{eq:H-theorem}) as
			\begin{align}
				& \dot{\sigma}_{ab}
				+
				\dot{\sigma}_{ba}
				=
				\gamma_{ab}
				\iint
				\frac{F_a F_b}{m_a}
				\nonumber 
				\\
				&
				\times
				\left(
				\nabla_{\vect{v}}{\ln F_a}
				-
				\frac{m_a}{m_b}
				\nabla_{\vect{v}'}{\ln F_b}
				\right) 
				\cdot 
				W(\vect{v}-\vect{v}')
				\cdot
				\left(
				\nabla_{\vect{v}}{\ln F_a}
				-
				\frac{m_a}{m_b}
				\nabla_{\vect{v}'}{\ln F_b}
				\right) 
				\dd[3]{\vect{v}'}
				\dd[3]{\vect{v}}.
				\label{eq:H_theorem_proof}
			\end{align}
			The $H$-theorem is a consequence of the fact that $|\vect{v}-\vect{v}'|W(\vect{v}-\vect{v}')$ is a projection matrix (to the space orthogonal to $\vect{v}-\vect{v}'$) and therefore $W(\vect{v}-\vect{v}')$ is semidefinite positive, i.e. $\vect{a}\cdot W(\vect{v}-\vect{v}')\cdot\vect{a}\geq 0$ for any vector $\vect{a}$. Hence, as $\gamma_{ab}$ and the integrand in the right-hand side of (\ref{eq:H_theorem_proof}) is always positive or zero, so is the integral.  
			
			Moreover, the $H$-theorem also reveals which pair of functions $(F_a,F_b)$ lie in the kernel of $C_{ab}$. The functions $(F_a,F_b)$ for which equality holds in (\ref{eq:H-theorem}) also satisfy $C_{ab}(F_a,F_b)=0$. Equality in (\ref{eq:H-theorem}) implies that $\nabla_{\vect{v}}{ \ln F_a}-{m_a}{m_b}^{-1}\nabla_{\vect{v}'}{\ln F_b}$ is orthogonal to the image of $W(\vect{v}-\vect{v}')$. Inspecting (\ref{eq:Fokker_Planck_velocity_flux}) is easy to write the integrand of $\vect{j}_{ab}$ as proportional to $W(\vect{v}-\vect{v}')\cdot(\nabla_{\vect{v}}{ \ln F_a}-{m_a}{m_b}^{-1}\nabla_{\vect{v}'}{\ln F_b})$, which proves that equality holds in (\ref{eq:H-theorem}) if and only if $C_{ab}(F_a,F_b)=0$. As the only direction orthogonal to $W(\vect{v}-\vect{v}')$ is the one spanned by $\vect{v}-\vect{v}'$, imposing $\dot{\sigma}_{ab} + \dot{\sigma}_{ba}= 0$ is equivalent to demand that $F_a$ and $F_b$ are such that
			\begin{align}
				\frac{1}{m_a}
				\nabla_{\vect{v}}{ \ln F_a}
				-
				\frac{1}{m_b}
				\nabla_{\vect{v}'}{\ln F_b}
				=
				K (\vect{v}-\vect{v}'),
				\label{eq:H_theorem_orthogonality_condition}
			\end{align}
			for some constant\footnote{The fact that $K$ is constant is consequence of that $\ln F_a(\vect{v})$ and $\ln F_b(\vect{v}')$ are differentiable functions which depend respectively only on $\vect{v}$ and $\vect{v}'$. For the complete proof we refer the reader to section \ref{Appendix:K_constant}.} $K$. Hence, we have that
			\begin{align}
				\ln F_a = 
				K 
				\frac{m_a {\left(\vect{v}-\vect{V}\right)}^2}{2} 
				+ K_a , \qquad  
				\ln F_b = 
				K 
				\frac{m_b \left(\vect{v}'-\vect{V}\right)^2}{2} 
				+ K_b	,
				\label{eq:Logarithms_Maxwellians}
			\end{align}
			for two integration constants $K_a(\vect{r},t)$ and $K_b(\vect{r},t)$. The vector field $\vect{V}(\vect{r},t)$ is the mean flow velocity of species $a$ and $b$\footnote{The appearance of the integration constant $\vect{V}$ is a consequence of the Galilean invariance of (\ref{eq:H_theorem_orthogonality_condition}). That is, of its invariance with respect to the transformation $(\vect{v},\vect{v}')\mapsto(\vect{v}-\vect{V},\vect{v}'-\vect{V})$. Solving in the translated variables yields the solution (\ref{eq:Logarithms_Maxwellians}).}. As the distribution functions must be integrable in the whole velocity space, the factor $K$ must be negative and corresponds to the equilibrium temperature $ T = -1/K $ of both species. The constants $K_a$ and $K_b$ are fixed imposing that the zeroth order moment of $F_a$ and $F_b$ match, respectively, the densities $n_a$ and $n_b$. Hence, the only functions that satisfy $\dot{\sigma}_{ab}+\dot{\sigma}_{ba}=0$ and $C_{ab}(F_a,F_b)=C_{ba}(F_b,F_a)=0$ are the equilibrium Maxwellian distributions
			\begin{align}
				F_a
				=
				f_{\text{M}a} 
				& = 
				n_a \left(\frac{m_a}{2\pi T}\right)^{3/2}
				\exp(-\frac{m_a {\left(\vect{v}-\vect{V}\right)}^2}{2T} )
				\label{eq:H_theorem_Maxwellian_a}
				,
				\\
				F_b
				=
				f_{\text{M}b} 
				& = 
				n_b \left(\frac{m_b}{2\pi T}\right)^{3/2}
				\exp(-\frac{m_b {\left(\vect{v}'-\vect{V}\right)}^2}{2T} ).
				\label{eq:H_theorem_Maxwellian_b}
			\end{align}
			
			\section{Proof that $K$ is a constant}\label{Appendix:K_constant}
			In this section it is proven that in (\ref{eq:H_theorem_orthogonality_condition}) $K$ is a constant. In order to do this, consider the scenario that comes from relaxing (\ref{eq:H_theorem_orthogonality_condition}) by allowing $K$ be a function of $\vect{v}$ and $\vect{v}'$
			\begin{align}
				\nabla_{\vect{v}}{f} - \nabla_{\vect{v}'} {g}
				= 
				K(\vect{v},\vect{v}')(\vect{v}-\vect{v}')
				=
				\Delta_K\left|
				\nabla_{\vect{v}}{f}
				- \nabla_{\vect{v}'}{g}
				\right|
				\frac{\vect{v}-\vect{v}'}{|\vect{v}-\vect{v}'|},
				\label{eq:Appendix_K_Orthogonality_condition}
			\end{align}
			where $f=\ln F_a /m_a$ and $g=\ln F_b /m_b$ are differentiable and $K(\vect{v},\vect{v}')$ is a function which is sufficiently regular so that $K(\vect{v},\vect{v}')(\vect{v}-\vect{v}')$ is continuous. In the second equality from (\ref{eq:Appendix_K_Orthogonality_condition}), $\Delta_K = K/|K|=\pm1$ is the sign of $K$ and we have made explicit the fact that $\nabla_{\vect{v}}{f} - \nabla_{\vect{v}'}{g}$ is parallel to $\vect{v}-\vect{v}'$. As the right-hand side of (\ref{eq:Appendix_K_Orthogonality_condition}) is continuous and $(\vect{v}-\vect{v}')/|\vect{v}-\vect{v}'|$ is not defined at the limit $\vect{v}\rightarrow\vect{v}'$ we conclude that at this limit $\left|\nabla_{\vect{v}}{f} - \nabla_{\vect{v}'}{g}\right|\rightarrow 0$ and therefore for any $\vect{v}$
			\begin{align}
				\nabla_{\vect{v}}{f} = \eval{\nabla_{\vect{v}'}{g}}_{\vect{v}'=\vect{v}} 
				\label{eq:Appendix_H_Theorem_f_equal_g}.
			\end{align}

			Evaluating (\ref{eq:Appendix_K_Orthogonality_condition}) at $\vect{v}'=0$ and $\vect{v}=0$ yields, respectively,
			\begin{align}
				\nabla_{\vect{v}} {f}
				& =
				K(\vect{v},0)\vect{v}
				+  
				\eval{\nabla_{\vect{v}'} {g}}_{\vect{v}'=0},
				\label{eq:Appendix_H_Theorem_eval_v_prime}
				\\
				\nabla_{\vect{v}'}{g} 
				& = 
				K(0,\vect{v}')\vect{v}' + \eval{\nabla_{\vect{v}}{f}}_{\vect{v}=0} .
				\label{eq:Appendix_H_Theorem_eval_v}
			\end{align}
			
			Substracting (\ref{eq:Appendix_H_Theorem_eval_v_prime}) - (\ref{eq:Appendix_H_Theorem_eval_v}) and using (\ref{eq:Appendix_H_Theorem_f_equal_g}) evaluated at $\vect{v}=0$ and (\ref{eq:Appendix_K_Orthogonality_condition}) gives
			\begin{align}
				K(\vect{v},\vect{v}')(\vect{v}-\vect{v}')
				=
				K(\vect{v},0)\vect{v} 
				-
				K(0,\vect{v}')\vect{v}'.
				\label{eq:Appendix_H_Theorem_K_vv}
			\end{align}
			
			Finally, projecting (\ref{eq:Appendix_H_Theorem_K_vv}) on $W(\vect{v}')$ and $W(\vect{v})$ gives that for any pair $(\vect{v},\vect{v}')$
			\begin{align}
				K(\vect{v},\vect{v}') W(\vect{v}') \cdot \vect{v}
				& =
				K(\vect{v},0)W(\vect{v}') \cdot \vect{v} 
				,
				\label{eq:Appendix_H_Theorem_Wv_K_vv}
				\\
				K(\vect{v},\vect{v}') W(\vect{v}) \cdot \vect{v}'
				& =
				K(0,\vect{v}')W(\vect{v}) \cdot \vect{v}' 
				.
				\label{eq:Appendix_H_Theorem_Wv_K_vv_prime}
			\end{align}
			Identity (\ref{eq:Appendix_H_Theorem_Wv_K_vv}) implies $K(\vect{v},\vect{v}')=K(\vect{v},0)$ for all $\vb*{v}'$ and thus $K$ cannot depend on $\vect{v}'$. Similarly, (\ref{eq:Appendix_H_Theorem_Wv_K_vv_prime}) implies $K(\vect{v},\vect{v}')=K(0,\vect{v}')$ for all $\vb*{v}$ and thus $K$ cannot depend on $\vect{v}$. Hence, $K$ is a constant.

			\chapter{Fluid equations and the single fluid approximation}\label{sec:Appendix_fluid_equations} 
			In this appendix, the (lengthy) algebraic calculations that lead to equations (\ref{eq:Fluid_mass_conservation})-(\ref{eq:Fluid_energy_conservation}) from the velocity moments of (\ref{eq:Kinetic_equation_Introduction}) are explained. The first step is to employ the properties $\nabla\cdot\vb*{v}=\nabla_{\vb*{v}}\cdot(\vb*{E}+\vb*{v\times\vb*{B}})=0$ to write (\ref{eq:Kinetic_equation_Introduction}) in divergence form 
			\begin{align} 
				\pdv{F_a}{t}
				+
				\nabla\cdot
				\left(
				\vb*{v}  
				F_a
				\right)
				+ 
				\nabla_{\vb*{v}}
				\cdot 
				\left(
				\frac{e_a}{m_a}
				\left( \vb*{E} + \vb*{v}\times \vb*{B} \right)
				F_a
				\right)
				=
				\sum_{b}
				C_{ab} 
				\left(
				F_a,F_b
				\right)  
				.
				\label{eq:Kinetic_equation_Appendix}
			\end{align} 
			
			First, by taking the moment $\vmoment{\text{Eq. }(\ref{eq:Kinetic_equation_Appendix})}$ the equation
			\begin{align}    	
				& 
				\pdv{n_a}{t}
				+	
				\nabla\cdot
				\left(
				n_a 	
				\vb*{V}_a
				\right)
				=
				\dv{n_a}{t}
				+	
				n_a 	
				\nabla\cdot
				\vb*{V}_a
				=
				0
				\label{eq:Fluid_mass_conservation_Moments}
				,
			\end{align}
			is obtained, which is exactly (\ref{eq:Fluid_mass_conservation}). 
			
			From the first moment $\vmoment{m_a\vb*{v} \text{ Eq. (\ref{eq:Kinetic_equation_Appendix})}}$ the equation
			\begin{align} 
				& 
				\pdv{t}
				\left(n_a m_a \vb*{V}_a\right)
				+
				\nabla\cdot
				\left(m_a n_a \vaverage{a}{\vb*{v}\vb*{v}}\right)
				-
				e_a n_a\left(
				\vb*{E}
				+
				\vb*{V}_a
				\times
				\vb*{B}
				\right)
				=
				\sum_{b}
				\vb*{F}_{ab}
				,
				\label{eq:Fluid_momentum_conservation_Moments} 
			\end{align}
			is obtained. It is important to remark that conservation of momentum (\ref{eq:Fokker_Planck_momentum_conservation}) by the Fokker-Planck collision operator implies that the friction force satisfies Newton's third law
			\begin{align} 
				\vb*{F}_{ab}
				+
				\vb*{F}_{ba}
				=
				0
				,
				\label{eq:Friction_force_3rd_Newtons_law}
			\end{align}
			which, in particular, implies $\sum_{a}\sum_{b} \vb*{F}_{ab}=0$.

			Using splitting (\ref{eq:Velocity_fluid_frame}), the property $\vaverage{a}{\vb*{w}_a}=0$ and the definition of the pressure tensor (\ref{eq:Pressure_tensor_definition}), the second moment can be written as
			\begin{align}
				n_a m_a \vaverage{a}{\vb*{v}\vb*{v}} = \mathbb{P}_a + n_a m_a \vb*{V}_a\vb*{V}_a.
			\end{align}
			Then, employing the identity $ \nabla\cdot (n_a m_a \vb*{V}_a\vb*{V}_a) =  m_a \vb*{V}_a \nabla\cdot(n_a\vb*{V}_a ) + n_a m_a \vb*{V}_a \cdot\nabla\vb*{V}_a  $, equation (\ref{eq:Fluid_momentum_conservation_Moments}) becomes
			\begin{align} 
				n_a m_a 
				\underbrace{\left(
					\pdv{\vb*{V}_a}{t}
					+ 
					\vb*{V}_a \cdot\nabla\vb*{V}_a
					\right)  }_{\dv*{\vb*{V}_a}{t}}
				&  
				+
				\nabla\cdot\mathbb{P}_a 
				-
				e_a n_a\left(
				\vb*{E}
				+
				\vb*{V}_a
				\times
				\vb*{B}
				\right)
				\nonumber
				\\
				&
				+ 
				m_a \vb*{V}_a
				\underbrace{\left(
					\pdv{n_a}{t}
					+
					\nabla\cdot(n_a\vb*{V}_a )
					\right)}_{=0\text{ by (\ref{eq:Fluid_mass_conservation_Moments})}}
				=
				\sum_{b}
				\vb*{F}_{ab}
				.
				\label{eq:Fluid_momentum_conservation_with_continuity} 
			\end{align}
			Finally, imposing mass conservation (\ref{eq:Fluid_mass_conservation_Moments}) in (\ref{eq:Fluid_momentum_conservation_with_continuity}) and splitting the pressure tensor $\mathbb{P}_a=p_aI +\vb*{\Pi}_a$ yields (\ref{eq:Fluid_momentum_conservation_pressure}). Thus, (\ref{eq:Fluid_mass_conservation}) and (\ref{eq:Fluid_momentum_conservation_pressure}) are equivalent to (\ref{eq:Fluid_mass_conservation_Moments}) and (\ref{eq:Fluid_momentum_conservation_Moments}).
			
			Computing the second moment $\vmoment{m_a{v}^2 \text{ Eq. }(\ref{eq:Kinetic_equation_Appendix})/2}$ yields
			\begin{align}
				&
				\pdv{t}
				\left(
				n_a
				\frac{1}{2} m_a
				\vaverage{a}{v^2}
				\right)
				+
				\nabla
				\cdot
				\left(
				n_a
				\frac{1}{2} m_a
				\vaverage{a}{v^2 \vb*{v}}
				\right)
				-
				e_a n_a 
				\vb*{E}
				\cdot
				\vb*{V}_a
				=
				\sum_{b}
				W_{ab}^{\text{L}}
				,
				\label{eq:Fluid_energy_conservation_Moments}
			\end{align}
			
			where 
			\begin{align}
				W_{ab}^{\text{L}}
				:=
				\frac{1}{2}
				m_a
				\vmoment{v^2 C_{ab}(F_a,F_b)}
			\end{align}
			is the collisional exchange of kinetic energy in the \qmarks{laboratory} reference frame. Due to the fact that the collision operator preserves mass (\ref{eq:Fokker_Planck_mass_conservation}), the collisional exchange of kinetic energy in the laboratory frame $W_{ab}^{\text{L}}$ is related to the exchange due to random motion $W_{ab}$ via
			\begin{align} 
				W_{ab}^{\text{L}}
				=
				W_{ab}
				+
				\vb*{F}_{ab}
				\cdot
				\vb*{V}_a
				.
				\label{eq:Collisional_exchange_energy_laboratory_random_relation}
			\end{align}
			
			Using splitting (\ref{eq:Velocity_fluid_frame}), property $\vaverage{a}{\vb*{w}_a}=0$ and the scalar pressure definition (\ref{eq:Scalar_pressure_definition}), the partial temporal derivative of (\ref{eq:Fluid_energy_conservation_Moments}) can be rewritten as
			\begin{align} 
				\pdv{t}
				\left(
				n_a
				\frac{1}{2} m_a
				\vaverage{a}{v^2}
				\right)
				& =
				\pdv{t}
				\left(
				n_a
				\frac{1}{2} m_a
				\vaverage{a}{w_a^2}
				+ 
				n_a
				\frac{1}{2} m_a
				\vb*{V}_a
				\cdot 
				\vb*{V}_a 
				\right)
				\nonumber
				\\
				& =
				\frac{3}{2} 
				\pdv{p_a}{t}
				+ 
				\frac{1}{2} m_a
				\vb*{V}_a
				\cdot 
				\vb*{V}_a 
				\pdv{n_a}{t} 
				+  
				\vb*{V}_a
				\cdot 
				n_a m_a
				\pdv{\vb*{V}_a}{t}  .
				\label{eq:Fluid_energy_temporal_derivative_term}
			\end{align}
			If in addition, definition (\ref{eq:Heat_Flux_random_definition}) is employed, the divergence term in the energy conservation equation (\ref{eq:Fluid_energy_conservation_Moments}) can be expressed as
			\begin{align} 
				\nabla
				\cdot
				\left(
				n_a
				\frac{1}{2} m_a
				\vaverage{a}{v^2 \vb*{v}}
				\right)
				&
				= 
				\nabla
				\cdot
				\left(
				n_a
				\frac{1}{2} m_a
				\vaverage{a}{
					\left(
					w_a^2 
					+
					2
					\vb*{w}_a
					\cdot 
					\vb*{V}_a
					\right)    		
					\vb*{w}_a
				}
				\right)
				\nonumber
				\\
				&
				+ 
				\nabla
				\cdot
				\left(
				n_a
				\frac{1}{2} m_a
				\vaverage{a}{
					\left(
					w_a^2 
					+ 
					\vb*{V}_a
					\cdot 
					\vb*{V}_a
					\right)
					\vb*{V}_a
				} 
				\right)  	
				\nonumber	
				\\
				&
				= 
				\nabla
				\cdot
				\vb*{h}_a
				+ 
				\nabla
				\cdot
				\left(
				\mathbb{P}_a 
				\cdot 
				\vb*{V}_a
				\right)
				+
				\nabla\cdot
				\left(
				\frac{3}{2}p_a
				\vb*{V}_a
				\right) 
				+
				\nabla\cdot
				\left(
				\frac{3}{2}p_a
				\vb*{V}_a
				\right)
				\nonumber
				\\
				&
				+ 
				\frac{1}{2} m_a
				\vb*{V}_a
				\cdot 
				\vb*{V}_a
				\nabla
				\cdot
				\left(
				n_a
				\vb*{V}_a
				\right)  	 
				+ 
				\vb*{V}_a  
				\cdot 
				\left(
				m_a
				n_a
				\vb*{V}_a
				\cdot
				\nabla  
				\vb*{V}_a
				\right)
				\nonumber	 
				\\
				&
				= 
				\nabla
				\cdot
				\vb*{h}_a
				+  
				\vb*{\Pi}_a 
				: 
				\nabla
				\vb*{V}_a  
				+
				\vb*{V}_a\cdot
				\nabla
				\left(
				\frac{3}{2}p_a
				\right) 
				+
				\frac{5}{2}p_a
				\nabla\cdot 
				\vb*{V}_a 
				\nonumber
				\\
				&
				+ 
				\frac{1}{2} m_a
				\vb*{V}_a
				\cdot 
				\vb*{V}_a
				\nabla
				\cdot
				\left(
				n_a
				\vb*{V}_a
				\right)  	 
				+ 
				\vb*{V}_a  
				\cdot 
				\left(    	
				\nabla 
				\cdot
				\mathbb{P}_a 
				+
				m_a
				n_a
				\vb*{V}_a
				\cdot
				\nabla  
				\vb*{V}_a
				\right)	
				\label{eq:Fluid_energy_divergence_term}
			\end{align}
			
			Summing (\ref{eq:Fluid_energy_temporal_derivative_term}) and (\ref{eq:Fluid_energy_divergence_term}) yields
			\begin{align} 
				\pdv{t}
				\left(
				\frac{1}{2} 
				n_a m_a
				\vaverage{a}{v^2}
				\right)
				+ 
				\nabla
				\cdot
				\left(
				\frac{1}{2} 
				n_a m_a
				\vaverage{a}{v^2 \vb*{v}}
				\right)
				=
				\label{eq:Fluid_energy_temporal_derivative_and_divergence_term}
				\\
				\frac{3}{2} 
				\underbrace{\left(
					\pdv{p_a}{t}
					+
					\vb*{V}_a\cdot
					\nabla p_a
					\right) }_{
					\dv*{p_a}{t} }
				+
				\frac{5}{2}p_a
				\nabla\cdot 
				\vb*{V}_a 
				+
				\nabla
				\cdot
				\vb*{h}_a
				+  
				\vb*{\Pi}_a 
				: 
				\nabla
				\vb*{V}_a  
				\nonumber
				\\
				+  
				\vb*{V}_a
				\cdot
				\underbrace{
					\left( 
					n_a m_a
					\pdv{\vb*{V}_a}{t}
					+
					m_a
					n_a
					\vb*{V}_a
					\cdot
					\nabla  
					\vb*{V}_a
					+   	
					\nabla 
					\cdot
					\mathbb{P}_a 
					\right)}_{=\sum_b \vb*{F}_{ab} +e_a n_a (\vb*{E} + \vb*{V}_a\times\vb*{B})\text{ by (\ref{eq:Fluid_momentum_conservation_with_continuity})}
				}
				\nonumber
				\\
				+ 
				\frac{1}{2} m_a
				\vb*{V}_a
				\cdot 
				\vb*{V}_a 
				\underbrace{\left(
					\pdv{n_a}{t} 
					+
					\nabla
					\cdot
					\left(
					n_a
					\vb*{V}_a
					\right)  
					\right)}_{=0\text{ by (\ref{eq:Fluid_mass_conservation_Moments})}}
				\nonumber
			\end{align}
			Thus, inserting (\ref{eq:Fluid_energy_temporal_derivative_and_divergence_term}) in the energy equation (\ref{eq:Fluid_energy_conservation_Moments}) assuming mass (\ref{eq:Fluid_mass_conservation_Moments}) and momentum conservation  (\ref{eq:Fluid_momentum_conservation_with_continuity}) yields
			\begin{align}
				\frac{3}{2} 
				\dv{p_a}{t} 
				+
				\frac{5}{2}p_a
				\nabla\cdot 
				\vb*{V}_a 
				+
				\nabla
				\cdot
				\vb*{h}_a
				+  
				\vb*{\Pi}_a 
				: 
				\nabla
				\vb*{V}_a  
				=
				\sum_{b}
				\left(
				W_{ab}^{\text{L}}
				-
				\vb*{F}_{ab}
				\cdot
				\vb*{V}_a
				\right).
				\label{eq:Fluid_energy_conservation_appendix}
			\end{align}
			Finally, by taking into account relation (\ref{eq:Collisional_exchange_energy_laboratory_random_relation}) in (\ref{eq:Fluid_energy_conservation_appendix}), the energy conservation equation (\ref{eq:Fluid_energy_conservation}) is recovered.
			
			\section{Single fluid approximation}\label{sec:Appendix_single_fluid}
			
			In this section, the single fluid approximation will be applied. In particular, the equations (\ref{eq:Single_fluid_momentum_equation}) and (\ref{eq:Single_fluid_Ohms_law}) will be obtained.
			Summing the momentum equation (\ref{eq:Fluid_momentum_conservation_pressure}) for electrons and ions yields
			\begin{align}
				n m 
				\dv{ \vb*{V} }{t}
				+
				\nabla p 
				& =
				\vb*{J}\times\vb*{B}
				-
				\nabla
				\cdot
				\left(
				\vb*{\Pi}_{\text{i}}
				+
				\vb*{\Pi}_{\text{e}}
				\right)
				\label{eq:Fluid_momentum_conservation_center_of_mass_Appendix}
				\\
				&
				-
				n m 
				\frac{m_{\text{e}}}{m_{\text{i}}}
				\left[
				\left(
				\vb*{V}
				-
				\vb*{V}_{\text{e}}
				\right)	
				\cdot
				\nabla \vb*{V}
				+ 
				\left(
				\vb*{V}_{\text{e}}
				-
				\vb*{V}_{\text{i}}
				\right)	
				\cdot
				\nabla \vb*{V}_{\text{e}}
				\right]	 
				\nonumber
			\end{align}
			and in this case 
			\begin{align}
				\vb*{V}_{\text{e}}
				=  
				\vb*{V}
				-
				\frac{1}{en}
				\vb*{J}
				+
				\frac{m_{\text{e}}}{m_{\text{i}}}
				\left(
				\vb*{V}_{\text{e}}
				+ 
				\vb*{V}
				\right).
				\label{eq:Single_fluid_current}
			\end{align}
			Note that the contribution of the friction forces to the momentum equation (\ref{eq:Fluid_momentum_conservation_center_of_mass_Appendix}) vanishes due to momentum conservation (\ref{eq:Friction_force_3rd_Newtons_law}).

			Taking into account (\ref{eq:Single_fluid_current}), the momentum equation (\ref{eq:Fluid_momentum_conservation_pressure}) for electrons becomes
			\begin{align}  
				en
				\left(
				\vb*{E}
				+
				\vb*{V}
				\times
				\vb*{B}
				\right)
				& =  
				\vb*{J}
				\times
				\vb*{B}
				- 
				\nabla p_{\text{e}} 
				-
				\nabla
				\cdot 
				\vb*{\Pi}_{\text{e}}
				+
				\vb*{F}_{\text{e} \text{i}}
				\label{eq:Fluid_momentum_conservation_electrons_Appendix}
				\\
				& 
				-
				n m_{\text{e}} 
				\dv{ \vb*{V}_{\text{e}} }{t}
				-
				en
				\frac{m_{\text{e}}}{m_{\text{i}}}
				\left(
				\vb*{V}_{\text{e}}
				+ 
				\vb*{V}
				\right)
				\times
				\vb*{B}
				\nonumber
			\end{align}
			
			As stated in section \ref{sec:Kinetic_fluid_description}, the second asymptotic limit is neglecting the electrons inertia, i.e. taking $m_{\text{e}} \rightarrow 0$. Dropping all the terms proportional to $m_{\text{e}}$ in equations (\ref{eq:Fluid_momentum_conservation_center_of_mass_Appendix}) and (\ref{eq:Fluid_momentum_conservation_electrons_Appendix}) yields, respectively, equations (\ref{eq:Single_fluid_momentum_equation}) and (\ref{eq:Single_fluid_Ohms_law}).

			\chapter{Magnetic coordinates} \label{sec:Appendix_Magnetic_coordinates}
			In this chapter, the contravariant representation (\ref{eq:Magnetic_Field_contravaraint}) of a non vanishing magnetic field $\vb*{B}$ tangent to nested flux surfaces will be derived. Let $\vb*{A}'$ be a magnetic vector potential, i.e. $\vb*{B}=\nabla\times\vb*{A}'$. It is an elemental result from vector calculus that any other magnetic vector potential of the form
			\begin{align}
				\vb*{A}
				:=
				\vb*{A}'
				+
				\nabla F
			\end{align}
			yields the same magnetic field. Here, $F$ is any smooth function called magnetic gauge. 
			
			Let $\vartheta$ and $\zeta$ be, respectively, the poloidal and toroidal angles that parametrize a flux surface, labelled by the coordinate $\psi$ defined in (\ref{eq:Toroidal_flux_definition}). For the moment, these two angles are arbitrary. The only requirement is that $\vartheta$ and $\zeta$ increase monotonically their value in $2\pi$ when a complete turn around the torus is produced, respectively, along the poloidal and toroidal directions. It is possible to select the gauge $F$ so that $A_\psi =0$. This is accomplished by setting
			\begin{align}
				\pdv{F}{\psi}
				+
				A_\psi' = 0,
			\end{align}
			and therefore
			\begin{align}
				\vb*{A}
				= 
				A_\vartheta \nabla\theta
				+
				A_\zeta \nabla \zeta.
			\end{align}
			
			Thus, the magnetic field can be written as
			\begin{align}
				\vb*{B}
				& =
				\nabla 	A_\vartheta 
				\times
				\nabla\theta
				+
				\nabla 	A_\zeta 
				\times
				\nabla\zeta
				\nonumber 
				\\
				& = 
				\pdv{A_\vartheta}{\psi}  
				\nabla 	\psi 
				\times
				\nabla\theta
				+
				\left( 
				\pdv{A_\zeta}{\vartheta}
				-
				\pdv{A_\vartheta}{\zeta}
				\right)
				\nabla 	\vartheta 
				\times
				\nabla\zeta
				+
				\pdv{A_\zeta}{\psi}  
				\nabla \psi	 
				\times
				\nabla\zeta
				.
			\end{align}
			Hence, the magnetic field is perpendicular to flux surfaces (i.e. $\vb*{B}\cdot\nabla\psi=0$) when
			\begin{align} 
				\pdv{A_\zeta}{\vartheta}
				=
				\pdv{A_\vartheta}{\zeta}
				,
				\label{eq:B_perp_FS_requirement}
			\end{align}
			holds. As $A_\vartheta$ and $A_\zeta$ are single valued on the torus, condition (\ref{eq:B_perp_FS_requirement}) implies
			\begin{align}
				\pdv{\zeta}
				\oint{A_\vartheta}\dd{\vartheta}=0,
				\qquad
				\pdv{\vartheta}
				\oint{A_\zeta}\dd{\zeta}=0,
			\end{align}
			which entails
			\begin{align}
				A_\vartheta
				& =
				f(\psi)
				+
				\pdv{H}{\vartheta}
				,
				\\
				A_\zeta
				& =
				g(\psi)
				+
				\pdv{H}{\zeta},
			\end{align}
			for arbitrary differentiable functions $f(\psi)$, $g(\psi)$ and a periodic function $H$. Hence, 	
			\begin{align}
				\vb*{B}
				&=
				\dv{f}{\psi} 
				\nabla\psi
				\times
				\nabla\vartheta
				+		
				\dv{g}{\psi} 
				\nabla\psi
				\times
				\nabla\zeta
				+
				\nabla\psi
				\times
				\nabla
				\left(
				\dv{f}{\psi}
				h
				\right)
				\nonumber
				\\
				&
				=
				\dv{f}{\psi}
				\nabla\psi
				\times
				\nabla\left(\vartheta + h\right)
				+		
				\dv{g}{\psi}
				\nabla\psi
				\times
				\nabla\zeta
			\end{align}
			where $h=\pdv*{H}{\psi}/\dv*{f}{\psi}$. Note that $h$ can be absorbed in the definition of a new poloidal angle $\theta= \vartheta + h$. In coordinates $(\psi,\theta,\zeta)$ the magnetic field can be written as
			\begin{align}
				\vb*{B}
				&=
				\dv{f}{\psi}
				\nabla\psi
				\times
				\nabla\theta
				+		
				\dv{g}{\psi}
				\nabla\psi
				\times
				\nabla\zeta.
				\label{eq:Appendix_Contravariant_B_f_g}
			\end{align}
			
			Now the magnetic field has a straight contravariant representation (\ref{eq:Appendix_Contravariant_B_f_g}). However, the assumption $\dv*{f}{\psi}\ne 0$ has been done implicitly to define $h$. This assumption can be verified computing the magnetic flux through the toroidal section $S_{\text{tor}}$ given by constant $\zeta$ in coordinates $(\psi,\theta,\zeta)$. For this toroidal section the surface differential form is
			\begin{align}
				\dd{\vb*{S}}
				=
				\vb*{e}_\psi \times 
				\vb*{e}_\theta 
				\dd{\psi}
				\dd{\theta} 
				=
				\sqrt{g}
				\nabla\zeta 
				\dd{\psi}
				\dd{\theta}
			\end{align}
			and using (\ref{eq:Appendix_Contravariant_B_f_g}) the toroidal flux across the surface of constant $\zeta$ can be computed as
			\begin{align}
				\int_{S_{\text{tor}}}
				\vb*{B}
				\cdot
				\dd{\vb*{S}} 
				=
				\oint
				\int_{0}^{\psi}
				\dv{\psi'}
				f(\psi')
				\dd{\psi'}
				\dd{\theta}
				=
				2\pi 
				f(\psi)
				.
			\end{align}
			Thus, from definition (\ref{eq:Toroidal_flux_definition}) it is obtained $\left|f(\psi)\right|=\psi$, which implies $f=\pm\psi$. One can choose
			\begin{align}
				f(\psi)=\psi
				,
			\end{align}
			which implies $\dv*{f}{\psi}=1$.
			
			Similarly, the value of $-2\pi g(\psi)$ can be proven to be equal to the poloidal flux (\ref{eq:Poloidal_flux_definition}). For the poloidal section $S_{\text{pol}}$ given by constant $\theta$ the surface differential form is
			\begin{align}
				\dd{\vb*{S}}
				=
				\vb*{e}_\zeta \times 
				\vb*{e}_\psi 
				\dd{\zeta} 
				\dd{\psi}
				=
				\sqrt{g}
				\nabla\theta 
				\dd{\zeta} 
				\dd{\psi}
				.
			\end{align} 
			Using (\ref{eq:Appendix_Contravariant_B_f_g}), the poloidal flux  across the surface of constant $\theta$ can be computed as
			\begin{align}
				\int_{S_{\text{pol}}}
				\vb*{B}
				\cdot
				\dd{\vb*{S}} 
				=
				-
				\oint
				\int_{0}^{\psi}
				\dv{\psi'}
				g(\psi')
				\dd{\psi'}
				\dd{\zeta}
				=
				-2\pi g(\psi)
			\end{align}
			which, according to definition (\ref{eq:Poloidal_flux_definition}) implies 
			\begin{align}
				g(\psi)
				=
				-
				\chi(\psi).
			\end{align}
			
			Thus, representation (\ref{eq:Appendix_Contravariant_B_f_g}) becomes
			\begin{align}
				\vb*{B}
				=
				\nabla\psi
				\times 
				\nabla\theta
				-
				\dv{\chi}{\psi}
				\nabla\psi
				\times 
				\nabla\zeta,
			\end{align}
			which is exactly (\ref{eq:Magnetic_Field_contravaraint}) by definition (\ref{eq:Rotational_transform_definition}). Note that for the selection $f(\psi)=-\psi$, the poloidal angle could be redefined as $-\theta$ to obtain (\ref{eq:Magnetic_Field_contravaraint}). It is important to remark that as $h$ has not been specified, representation  (\ref{eq:Magnetic_Field_contravaraint}) is not unique. An appropriate selection of $h$ \cite{Boozer_coordinates,Helander_2014} leads to Boozer coordinates and the covariant representation (\ref{eq:Magnetic_Field_convaraint}). 
			
			%
			%
			%
			%
			
			\chapter{The flux surface average and magnetic differential equations} \label{sec:FSA_MDE}     
			In section \ref{sec:Force_balance}, the flux surface average has been introduced. In this appendix, several well-known properties of the flux surface average and its connection to magnetic differential equations are reviewed. For the sake of clarity, the definition of flux surface average is repeated here
			\begin{align}
				\mean*{f}
				:=
				\lim_{\delta \psi \rightarrow 0} 
				\dfrac{\int_{V(\psi+\delta\psi)} f \dd[3]{\vb*{r}}- \int_{V(\psi)} f\dd[3]{\vb*{r}}}
				{V(\psi+\delta\psi) - V(\psi)}.
				\tag{\ref{eq:FSA}}
			\end{align}
			
			By applying Stokes' theorem to the volume integrals in (\ref{eq:FSA}), the property
			\begin{align}
				\mean*{\nabla\cdot \vb*{F}}
				=
				\left(
				\dv{V}{\psi}\right)^{-1}
				\pdv{\psi}
				\left(
				\dv{V}{\psi}
				\mean*{
					\vb*{F}\cdot\nabla\psi
				}
				\right)
				\label{eq:FSA_divergence}
			\end{align}
			is obtained, which can be used to obtain another useful property
			\begin{align}
				\mean*{\nabla\psi
					\cdot
					\nabla\times \vb*{F} }
				=  
				\mean*{\nabla\cdot 
					(\vb*{F}			
					\times \nabla\psi) 
				}
				=
				0
				.
				\label{eq:FSA_curl_perp}
			\end{align}
			Property (\ref{eq:FSA_divergence}) is important for defining a solvability condition for magnetic differential equations in ergodic flux surfaces. 
			
			Magnetic differential equations are first order differential equations of the form
			\begin{align}
				\vb*{B}\cdot \nabla f
				=
				s
				,
				\label{eq:MDE}
			\end{align}
			for some source $s$. 
			
			When $\vb*{B}$ is tangent to a flux surface, due to the fact that $\vb*{B}$ is divergence-free (\ref{eq:Divergence_free_B}), there is a necessary condition for the solution $f$ to (\ref{eq:MDE}) to be single valued on the torus. Taking the flux surface average of the left-hand side of (\ref{eq:MDE}) while assuming that $f$ is smooth on the torus yields the identity
			\begin{align}
				\mean*{\vb*{B}\cdot\nabla f}
				=
				\mean*{\nabla\cdot (\vb*{B} f)} 
				=
				\left(\dv{V}{\psi}\right)^{-1}
				\pdv{\psi}
				\mean*{
					\dv{V}{\psi}
					f\vb*{B} \cdot\nabla\psi
				}
				=
				0
				,
				\label{eq:MDE_Annihilator}
			\end{align}
			where (\ref{eq:Divergence_free_B}) and property (\ref{eq:FSA_divergence}) have been used. Note from identity (\ref{eq:MDE_Annihilator}) that the flux surface average is the \textit{annihilator} of $\vb*{B}\cdot\nabla$ when regarded as an operator from the space of smooth functions on the torus to itself.
			
			Hence, a \textit{necessary} condition for the continuity (on the torus) of the solution to (\ref{eq:MDE}) is
			\begin{align}
				\mean*{s} = 0.
				\label{eq:MDE_Solvability_ergodic}
			\end{align}
			Solvability condition (\ref{eq:MDE_Solvability_ergodic}) is also \textit{sufficient} when the surface is ergodic. For rational surfaces, condition (\ref{eq:MDE_Solvability_ergodic}) is not enough to guarantee that $f$ is single valued. In order to make more precise this statement, it is convenient to employ Clebsch coordinates $(\alpha,l)$ where $\alpha$ is the Clebsch angle (\ref{eq:Clebsch_angle}) and $l$ is the length along magnetic field lines. In a rational surface, a magnetic field line labelled by $\alpha$ closes on itself after a length $L_{\text{c}}(\alpha)$.
			Dividing equation (\ref{eq:MDE}) by $B$ and then integrating in $l$ until the magnetic field line closes itself gives, for each fixed $\alpha$, the solvability condition
			\begin{align} 
				\int_{0}^{L_{\text{c}}(\alpha)} 
				\frac{s}{B} \dd{l}   	
				=
				0,
				\label{eq:MDE_Solvability_rational}
			\end{align} 
			where continuity of $f$, i.e. $f(\alpha,0)=f(\alpha,L_{\text{c}}(\alpha))$, has been imposed. Note that satisfying condition (\ref{eq:MDE_Solvability_rational}) implies fulfilling (\ref{eq:MDE_Solvability_ergodic}) as well.

			\chapter{Solution to the lowest order DKE}\label{sec:DKE_Maxwellian}
			In this appendix we prove that, assuming the existence of nested flux surfaces, the only solution to the lowest order DKE (\ref{eq:Drift_kinetics_lowest_order_DKE}) is the radially local Maxwellian (\ref{eq:Radially_local_Maxwellian}) with $T_a=T_b$ for all species $a$ and $b$. We will also comment on how employing a simpler pitch-angle collision operator allows for a Maxwellian with different temperatures between species. It is important to emphasize that the solution to the lowest order DKE can be a Maxwellian without necessarily assuming a structure of nested flux surfaces. We comment on this aspect at the end of the appendix. The lowest order DKE reads
			\begin{align}
				\vp
				\vb*{b}
				\cdot
				\nabla 
				\overline{F}_a^{(0)}
				=
				\sum_{b}
				C_{ab}\left(
				\overline{F}_a^{(0)},
				\overline{F}_b^{(0)}
				\right)
				.
				\tag{\ref{eq:Drift_kinetics_lowest_order_DKE}}
			\end{align}
			We will prove that the only solution to (\ref{eq:Drift_kinetics_lowest_order_DKE}) is the one that simultaneously makes zero both the left-hand and the right-hand side of equation (\ref{eq:Drift_kinetics_lowest_order_DKE}). In order to prove this, the only requirement is that $\vb*{B}$ is such that there is a closed surface that encloses the plasma and to which $\vb*{B}$ is tangent. A special case of this situation is when $\vb*{B}$ consists of nested flux surfaces.
			
			As a consequence of mass conservation of the Fokker-Planck collision operator (\ref{eq:Fokker_Planck_mass_conservation}), integrating (\ref{eq:Drift_kinetics_lowest_order_DKE}) in velocities yields
			\begin{align}
				\vmoment{		
					\vp\vb*{b}\cdot\nabla
					\overline{F}_a^{(0)}
					(\vb*{x},\mu_a,\epsilon_a,\sigma)
				}
				=
				0
				,
				\label{eq:LO_DKE_Property_1}
			\end{align}

			Multiplying (\ref{eq:Drift_kinetics_lowest_order_DKE}) by $\ln \overline{F}_a^{(0)}$ and integrating in velocities yields
			\begin{align}		
				\vmoment{
					\vp\vb*{b}\cdot\nabla
					\left(
					\overline{F}_a^{(0)}
					\ln 
					\overline{F}_a^{(0)}
					(\vb*{x},\mu_a,\epsilon_a,\sigma)
					\right)
				}
				=
				-
				\sum_{b}
				\dot{\sigma}_{ab}
				\label{eq:LO_DKE_logarithm_integral}
			\end{align}
			where $\dot{\sigma}_{ab}$ is the entropy production between species $a$ in $b$ defined in (\ref{eq:Fokker_Planck_entropy_production_ab}) and property (\ref{eq:LO_DKE_Property_1}) has been employed. 
			
			Employing the property\footnote{This property can be proven employing coordinates $(\vp,\mu_a)$ instead of $(\mu_a,\epsilon_a)$. Let $h(\vb*{x},\vp,\mu_a) = g(\vb*{x},\mu_a,\epsilon_a(\vb*{x},\vp,\mu_a),\sigma)$, then $\vp\vb*{b}\cdot\nabla g = \vb*{b}\cdot\nabla( \vp h)  - \pdv*{\vp}( (\mu_a \vb*{b}\cdot\nabla B + e_a \vb*{b}\cdot\nabla \varphi / m_a) h ) $. In coordinates, $(\vp,\mu_a)$ the velocity integral of any gyroaveraged function $f$ takes the form $\int f\dd[3]{\vb*{v}}= 2\pi B / m_a\int_{0}^{\infty} \int_{-\infty}^{\infty} f\dd{\vp}\dd{\mu_a}$. Thus, integrating $\vb*{b}\cdot\nabla( \vp h)  - \pdv*{\vp}( (\mu_a \vb*{b}\cdot\nabla B + e_a \vb*{b}\cdot\nabla \varphi / m_a) h ) $ in velocities and imposing that $h\rightarrow0$ when $\vp\rightarrow\pm\infty$ yields property (\ref{eq:LO_DKE_Property_2}).} 
			\begin{align}
				\vmoment{		
					\vp\vb*{b}\cdot\nabla
					g
					(\vb*{x},\mu_a,\epsilon_a,\sigma)
				}
				=
				\vb*{B} 
				\cdot
				\nabla
				\left(
				\vmoment{
					\frac{\vp}{B}
					g
				}
				\right),
				\label{eq:LO_DKE_Property_2}
			\end{align}
			for any function $g$, equation (\ref{eq:LO_DKE_logarithm_integral}) can be written as
			\begin{align}
				\vb*{B}
				\cdot\nabla 
				\left(
				\vmoment{
					\frac{\vp}{B}
					\overline{F}_a^{(0)}
					\ln 
					\overline{F}_a^{(0)}
				}
				\right)
				=
				-
				\sum_{b}
				\dot{\sigma}_{ab}
				.
				\label{eq:LO_DKE_Entropy_production}
			\end{align}
			
			Up to this point, no particular shape of the magnetic field $\vb*{B}$ has been assumed as long as the plasma is strongly magnetized. When the magnetic field consists of a structure of nested flux surfaces, the operator $\vb*{B}\cdot\nabla$, regarded as a linear map from the space of smooth functions on the torus to itself, possesses an annihilator, which is the flux surface average (see appendix \ref{sec:FSA_MDE}). Hence, taking the flux surface average of equation (\ref{eq:LO_DKE_Entropy_production}) yields
			\begin{align}  
				\sum_{b}
				\mean*{\dot{\sigma}_{ab}}
				=
				0
				.
				\label{eq:LO_DKE_FSA_entropy_production}
			\end{align}
			
			Summing equation (\ref{eq:LO_DKE_FSA_entropy_production}) over species yields 
			\begin{align}
				\sum_a
				\mean*{\dv{S_a}{t}}
				=
				0
				.
				\label{eq:LO_DKE_H_theorem}
			\end{align}
			
			Thus, the solution to the lowest order DKE  (\ref{eq:Drift_kinetics_lowest_order_DKE}) is such that the total entropy of the plasma at each flux surface does not increase. As, by virtue of the $H-$theorem (\ref{eq:H-theorem_a}) (see section \ref{sec:H_theorem}) $\dv*{S_a}{t}\ge 0$, equation (\ref{eq:LO_DKE_H_theorem}) implies $\dv*{S_a}{t}=0$. In section \ref{sec:H_theorem}, it is proven that $\dv*{S_a}{t}=0$ is only satisfied when the distribution function of all species is given by the isothermal Maxwellian (\ref{eq:H_theorem_Maxwellian_a}). Hence, the only non zero solution to the lowest order DKE (\ref{eq:Drift_kinetics_lowest_order_DKE}) is of the form
			\begin{align}
				\overline{F}_a^{(0)}
				(\vb*{x},\mu_a,\epsilon_a,\sigma)
				=
				\frac{n_a(\vb*{x})}{\pi^{3/2}\vth{a}(\vb*{x})}
				\exp(- \frac{(\vb*{v}(\vb*{x},\mu_a,\epsilon_a,\sigma)-V_\parallel^{(0)}(\vb*{x})\vb*{b}(\vb*{x}))^2}{\vth{a}^2(\vb*{x})})
				, 
				\label{eq:LO_DKE_Maxwellian}
			\end{align}
			where $V_\parallel^{(0)}$ is the lowest order parallel flow velocity and $T_a=T_b$ for all species $a$ and $b$. Note that, as $\overline{F}_a^{(0)}$ is gyrophase independent, the mean flow associated to the Maxwellian cannot be perpendicular to magnetic field lines. 
			
			As the Maxwellian distribution functions belong to the kernel of the collision operator, i.e. $C_{ab}\left(\overline{F}_a^{(0)},\overline{F}_b^{(0)}\right)=0$, the lowest order solution $\overline{F}_a^{(0)}$ must also satisfy $\vp\vb*{b}\cdot\nabla \overline{F}_a^{(0)} = 0$. Namely, 
			\begin{align}
				& \left[
				\vp\vb*{b}\cdot
				\frac{\nabla n_a}{n_a} 
				+
				\frac{m_a\vp(\vp-V_\parallel^{(0)}	)}{T_a}
				\vb*{b}\cdot\nabla V_\parallel
				-
				\frac{V_\parallel^{(0)}}{T_a} 
				\mu_a \vb*{b}\cdot\nabla B 
				+
				\frac{\vp-V_\parallel^{(0)}}{T_a} 
				e_a \vb*{b}\cdot\nabla \varphi_0
				\right]
				\overline{F}_a^{(0)} 
				\nonumber
				\\
				& 
				+
				\left[ 
				\left(
				\frac{\mu_a B}{T_a}
				+
				\frac{m_a \left(\vp - V_\parallel^{(0)} \right)^2}{2T_a}
				- 
				\frac{3}{2}
				\right)
				\vp\vb*{b}\cdot
				\frac{\nabla T_a}{T_a}
				\right]
				\overline{F}_a^{(0)}
				=
				0
				,
				\label{eq:LO_DKE_Constant_along_field_lines}
			\end{align}
			must be satisfied for all $\vp$ and $\mu_a$, which implies
			\begin{align}
				V_\parallel^{(0)}=0,
				\label{eq:LO_DKE_Parallel_flow}
				\\
				\vb*{b}\cdot\nabla \varphi_0
				=
				\vb*{b}\cdot\nabla n_a 
				=
				\vb*{b}\cdot\nabla T_a = 0.
				\label{eq:LO_DKE_Radial_profiles}
			\end{align}
			For ergodic flux surfaces, condition (\ref{eq:LO_DKE_Radial_profiles}) implies that, to lowest order, the electrostatic potential, the density and temperature are flux functions. As long as $\dv*{\iota}{\psi}= 0$ only at isolated values of $\psi$, continuity implies that these quantities are also flux functions at rational surfaces. Thus, when the magnetic field consists of nested flux surfaces, the only solution to the lowest order DKE (\ref{eq:Drift_kinetics_lowest_order_DKE}) is the Maxwellian (\ref{eq:Radially_local_Maxwellian}) with $T_a=T_b$ for all species $a$ and $b$. 
			
			Now we comment on how the lowest order solution associated to a pitch-angle scattering collision operator would allow for different temperatures. If for collisions, $\sum_{b}C_{ab}(\overline{F}_a^{(0)},\overline{F}_b^{(0)})$ is replaced in (\ref{eq:Drift_kinetics_lowest_order_DKE}) by $\nu^a \Lorentz \overline{F}_a^{(0)}$ where 
			\begin{align}
				\Lorentz f
				:=
				\frac{1}{2}v^3
				\nabla_{\vb*{v}}
				\cdot
				\left(
				W(\vb*{v})
				\cdot 
				\nabla_{\vb*{v}}
				f
				\right)
				,
				\label{eq:Lorentz_operator}
			\end{align}
			is the Lorentz operator, $\nu^a(v)$ is the collision frequency and, as in appendix \ref{sec:Fokker_Planck_operator}, $W(\vb*{x}) := (Ix^2 - \vect{x}\vect{x})/x^3$. The explicit definition of $\nu^a$ is given in equation (\ref{eq:Collision_frequency}) but is not relevant for the discussion here. Identically to the case with the Fokker-Planck collision operator, the solution to the lowest order DKE must lie simultaneously on the kernels of $\vp\vb*{b}\cdot\nabla$ and $\Lorentz$. Any (gyroaveraged) function of the form $f(\vb*{x},v)$ belongs to the kernel of $\Lorentz$. Thus, the Maxwellian (\ref{eq:LO_DKE_Maxwellian}) can satisfy $T_a\ne T_b$ and be a solution to $\vp\vb*{b}\cdot\nabla \overline{F}_a^{(0)} = \nu^a \Lorentz\overline{F}_a^{(0)} $.
			
			Finally, we comment on how one could obtain a superficially similar equation to (\ref{eq:MHD_Momentum_balance_steady_state_species_a}) but without assuming that $\vb*{B}$ consists of nested flux surfaces. Let $V_{\text{c}}$ be a control volume whose boundary is tangent to magnetic field lines. Then, integrating equation (\ref{eq:LO_DKE_Entropy_production}) in this region and applying Stokes' theorem yields $\sum_b \int_{V_{\text{c}}} \dot{\sigma}_{ab} \dd[3]{\vb*{r}}=0$ which implies $ \int_{V_{\text{c}}} \sum_a\dv*{S_a}{t} \dd[3]{\vb*{r}}=0$. Hence, the $H-$theorem (\ref{eq:H-theorem_a}) implies that $\dv*{S_a}{t}= 0$. Thus, in the region $V_{\text{c}}$, the gyroaveraged, lowest order distribution function $\overline{F}_a^{(0)}$ is the (isothermal) Maxwellian (\ref{eq:LO_DKE_Maxwellian}) also satisfying (\ref{eq:LO_DKE_Parallel_flow}) and (\ref{eq:LO_DKE_Radial_profiles}) but with the difference that $\vb*{B}$ is not necessarily tangent to a flux surface. Then, we can obtain the lowest order distribution function ${F}_a^{(0)}$ exactly in the same way as for the case in which nested flux surfaces where assumed, applying equation (\ref{eq:DKE_distribution_function_radially_local_equilibrium}) but substituting $\psi$ by $\vb*{x}$. Doing so, we obtain ${F}_a^{(0)} = \overline{F}_a^{(0)} - \vb*{\rho}_a \cdot\nabla \overline{F}_a^{(0)}$. Hence, the macroscopic flow associated to ${F}_a^{(0)}$ is superficially identical to the one given in equation (\ref{eq:LO_DKE_Macroscopic_flow}) with the difference that $\nabla p_a$ and $\nabla \varphi_0$ are perpendicular to $\vb*{B}$ without (necessarily) being perpendicular to any toroidal surface. Namely, one would obtain the lowest order flow velocity
			\begin{align}
				n_a \vb*{V}_a^{(0)}
				:=
				\vmoment{\vb*{v} F_a^{(0)}}
				= 
				\frac{\vb*{B}}{e_a B^2 }
				\times 
				\left(\nabla p_a +e_a n_a \nabla\varphi_0\right)
				.
				\label{eq_LO_DKE_Macroscopic_flow_force_balance}
			\end{align}
			Then, taking the cross product of $\vb*{B}$ with (\ref{eq_LO_DKE_Macroscopic_flow_force_balance}) and imposing (\ref{eq:LO_DKE_Radial_profiles}) implies that
			\begin{align}
				n_a e_a
				\left(
				-\nabla \varphi_0
				+
				\vb*{V}_a^{(0)}
				\times
				\vb*{B}
				\right)
				=
				\nabla p_a 
				.
				\label{eq:LO_DKE_Force_Balance_a}
			\end{align}
			must be satisfied in the volume $V_{\text{c}}$.	Summing (\ref{eq:LO_DKE_Force_Balance_a}) over all species, applying the quasineutrality approximation (\ref{eq:Charge_neutral_approximation}) and taking into account definitions (\ref{eq:Electric_current_flow}) and (\ref{eq:Total_pressure}) gives $\vb*{J}\times\vb*{B}=\nabla p$ in the volume $V_{\text{c}}$.

			\chapter{Magnetostatic Hazeltine's DKE in coordinates $(\xi,v)$}
			\label{sec:Hazeltine_DKE}
			In this appendix we carry out the change of variables in the DKE (\ref{eq:Drift_kinetics_DKE_mu_epsilon}) from \cite{Hazeltine_1973} to coordinates $(\vb*{x},\xi,v)$ in the electrostatic case (i.e. $\pdv*{\vb*{A}}{t}=0$). For notational convenience we drop the subscript $a$ indicating the species and denote by $f(\vb*{x},\mu,\epsilon)$ to the gyroaveraged distribution function. In steady state, the electrostatic limit of the DKE (\ref{eq:Drift_kinetics_DKE_mu_epsilon}) takes the form
			\begin{align} 
				(\vb*{v}_{\text{gc}} + u \vb*{b})
				\cdot 
				\nabla f 
				+
				\dot{\mu}
				\pdv{f}{\mu}
				=
				C(f),
				\label{eq:Hazeltine_DKE_mu_epsilon}
			\end{align}
			where, for ease of notation, we have denoted the collision operator by $C(f)$. Recall that
			\begin{align}
				\vb*{v}_{\text{gc}}(\vb*{x},\mu,\epsilon)
				& 
				=
				\vp(\vb*{x},\mu,\epsilon,\sigma)
				\vb*{b}(\vb*{x})
				+
				\vb*{v}_{\text{d}}(\vb*{x},\mu,\epsilon),
				\\
				\dot{\mu}(\vb*{x},\mu,\epsilon)
				&
				= 
				m\vp(\vb*{x},\mu,\epsilon,\sigma)
				\vb*{b}(\vb*{x})
				\cdot
				\nabla
				\left(
				\vp(\vb*{x},\mu,\epsilon,\sigma)
				\frac{u(\vb*{x},\mu,\epsilon)}{B(\vb*{x})}
				\right),
				\label{eq:dot_mu_mu_eps}
			\end{align}
			where $\vp(\vb*{x},\mu,\epsilon,\sigma)=\sigma\sqrt{2(\epsilon-\mu B(\vb*{x}) - e\varphi(\vb*{x}))/m}$ and $u(\vb*{x},\mu,\epsilon)
			=
			{\mu}
			\vb*{b}
			\cdot
			\nabla
			\times
			\vb*{b} / {e}$. It is convenient to rewrite the drift velocity as
			\begin{align} 
				\vb*{v}_{\text{d}}
				(\vb*{x},\mu,\epsilon)
				\nonumber &
				:=
				\frac{\vb*{F}(\vb*{x},\mu,\epsilon)\times\vb*{b}}{m \Omega}
				+
				\frac{\vp^2(\vb*{x},\mu,\epsilon)}{\Omega}
				\left(
				I
				-
				\vb*{b}
				\vb*{b}
				\right)
				\cdot
				\nabla
				\times
				\vb*{b}
				\\
				&
				=
				\frac{\vb*{F}(\vb*{x},\mu,\epsilon)\times\vb*{b}}{m \Omega}
				+
				\frac{\vp^2(\vb*{x},\mu,\epsilon)}{\Omega}
				\left( 
				\nabla
				\times
				\vb*{b}
				- 
				\frac{\Omega m}{\mu B} 
				u(\vb*{x},\mu,\epsilon)
				\vb*{b}
				\right)
				,
			\end{align}
			where $\vb*{F}(\vb*{x},\mu,\epsilon) 
			:=
			e\vb*{E}
			-
			m
			\mu 
			\nabla B$. 
			
			The goal is to express equation (\ref{eq:Hazeltine_DKE_mu_epsilon}) using coordinates $(\vb*{x},\xi,v)$ where $\vp(v,\xi):=v\xi$, $\mu(\vb*{x},\xi,v):=mv^2(1-\xi^2)/(2B(\vb*{x}))$ and $\epsilon(\vb*{x},v):=mv^2/2 + e\varphi(\vb*{x}) $. Given a function $g(\vb*{x},\xi,v):=f(\vb*{x}, \mu(\vb*{x},\xi,v), \epsilon(\vb*{x},v))$ we have the relations
			\begin{align}
				\nabla
				f(\vb*{x},\mu,\epsilon)
				& =
				\nabla 
				g(\vb*{x},\xi,v)
				+
				\frac{e\vb*{E}}{mv}
				\pdv{g(\vb*{x},\xi,v)}{v}
				\nonumber
				\\
				&
				+
				\frac{1-\xi^2}{\xi v^2}
				\left(
				\frac{\vb*{F}}{m}
				-
				\frac{v^2\xi^2}{2B}
				\nabla B
				\right)
				\pdv{g(\vb*{x},\xi,v)}{\xi}
				\label{eq:nabla_mu_eps_xi_v}
				\\
				\pdv{f(\vb*{x},\mu,\epsilon)}{\mu}
				& =
				-
				\frac{B}{mv^2 \xi}
				\pdv{g(\vb*{x},v,\xi)}{\xi}.
				\label{eq:pdv_mu_mu_eps_xi_v}
			\end{align}
			Thus, we transform the left-hand side of (\ref{eq:Hazeltine_DKE_mu_epsilon}) as
			\begin{align*}
				&
				(\vb*{v}_{\text{gc}} + u \vb*{b})
				\cdot 
				\nabla f(\vb*{x},\mu,\epsilon) 
				+
				\dot{\mu}(\vb*{x},\mu,\epsilon)
				\pdv{f(\vb*{x},\mu,\epsilon)}{\mu}
				\\ 
				&
				=
				(\vb*{v}_{\text{gc}} + u \vb*{b})
				\cdot
				\nabla 
				g(\vb*{x},\xi,v)
				+
				\frac{e\vb*{E}\cdot 
					(\vb*{v}_{\text{gc}} + u \vb*{b})}{mv}
				\pdv{g(\vb*{x},\xi,v)}{v}
				\\ 
				&
				+
				\left[
				\frac{1-\xi^2}{\xi v^2}
				\left(
				\frac{\vb*{F}}{m}
				-
				\frac{v^2\xi^2}{2B}
				\nabla B
				\right)
				\cdot 
				\vb*{v}(\vb*{x},\xi,v) 
				-
				\frac{B \dot{\mu}(\vb*{x},\xi,v)}{mv^2 \xi}
				\right]
				\pdv{g(\vb*{x},\xi,v)}{\xi}
				\\
				&
				=
				(\vb*{v}_{\text{gc}} + u \vb*{b})
				\cdot
				\nabla 
				g(\vb*{x},\xi,v)
				+
				\dot{v}
				(\vb*{x},\xi,v)
				\pdv{g(\vb*{x},v,\xi)}{v}
				+
				\dot{\xi}
				(\vb*{x},\xi,v)
				\pdv{g(\vb*{x},v,\xi)}{\xi}.
			\end{align*}
			
			Note that it is immediate to obtain 
			\begin{align}
				\dot{v}
				(\vb*{x},\xi,v)
				=
				\frac{e\vb*{E}(\vb*{x})\cdot 
					(\vb*{v}_{\text{gc}}(\vb*{x},\xi,v) + u(\vb*{x},\xi,v)\vb*{b}(\vb*{x}))
				}{mv},
			\end{align}
			which is expression (\ref{eq:dot_v_xi_v}).
			Obtaining equation (\ref{eq:dot_xi_xi_v}) requires some straightforward but lengthy algebra. First, we need to give an explicit expression for $\dot{\mu}(\vb*{x},\xi,v)B/(mv^2\xi)$. Applying identity (\ref{eq:nabla_mu_eps_xi_v}) to
			(\ref{eq:dot_mu_mu_eps}) gives
			\begin{align*}
				\dot{\mu}
				(\vb*{x},\xi,v)
				& =
				m\vp(\vb*{x},\mu,\epsilon)
				\vb*{b}
				\cdot
				\nabla
				\left(
				\vp(\vb*{x},\mu,\epsilon)
				\frac{u(\vb*{x},\mu,\epsilon)}{B}
				\right)
				\\
				& 
				=		
				\frac{mv^2\xi^2}{B}
				\vb*{b}
				\cdot
				\nabla
				u(\vb*{x},\mu,\epsilon)
				+
				mv\xi
				\vb*{b}
				\cdot
				\nabla 
				\vp(\vb*{x},\mu,\epsilon) 
				\frac{u(\vb*{x},\xi,v)}{B}
				\\
				&
				+
				mv^2\xi^2
				u(\vb*{x},\xi,v)
				\vb*{b}
				\cdot
				\nabla
				\left(
				\frac{1}{B}
				\right)
				\\
				& 
				=		
				\frac{mv^2\xi^2}{B}
				\left(
				\vb*{b}
				\cdot
				\nabla
				u(\vb*{x},\mu,\epsilon)
				-
				\vb*{b}
				\cdot
				\nabla
				\ln{B} 
				\
				u(\vb*{x},\xi,v)
				\right)
				+
					\vb*{F}\cdot\vb*{b}
				\frac{u(\vb*{x},\xi,v)}{B} .
			\end{align*}
			
			Hence, using that $\nabla
			u(\vb*{x},\mu,\epsilon) = \nabla
			u(\vb*{x},\xi,v) + \nabla
			\ln{B} \ u(\vb*{x},\xi,v) $ we obtain
			\begin{align} 
				\frac{\dot{\mu}
					(\vb*{x},\xi,v)B}{mv^2\xi}
				& = 
				\xi
				\vb*{b}
				\cdot
				\nabla
				u(\vb*{x},\xi,v) 
				+
				\frac{1}{\xi}
				\frac{\vb*{F}\cdot\vb*{b}}{mv^2}
				\frac{u(\vb*{x},\xi,v)}{B} .
			\end{align}
			
			Now, we operate on the term proportional to $\vb*{F}\cdot\vb*{v}$ to obtain
			\begin{align*} 
				\frac{1-\xi^2}{\xi v^2}
				&
				\frac{\vb*{F}}{m}
				\cdot 
				(\vb*{v}_{\text{gc}}
				+
				u\vb*{b})
				=
				\frac{1-\xi^2}{\xi v^2}
				\frac{\vb*{F}}{m}
				\cdot 
				\left(
				v\xi
				+
				u
				\vb*{b}
				+
				\vb*{v}_{\text{d}}
				\right)
				\\
				&
				= 
				(1-\xi^2) 
				\frac{\vb*{F}\cdot \vb*{b}}{mv} 		
				+		
				\frac{1-\xi^2}{\xi}
				\frac{\vb*{F} \cdot \vb*{b}}{m v^2} 
				u
				+
				\frac{1-\xi^2}{\xi }
				\frac{\vb*{F} \cdot \vb*{v}_{\text{d}}}{mv^2} 
				\\
				&
				=   
				(1-\xi^2) 
				\frac{\vb*{F}\cdot \vb*{b}}{mv}  
				+		 
				\frac{1-\xi^2}{\xi}
				\frac{\vb*{F} \cdot \vb*{b}}{m v^2} 
				u	 
				+      
				\xi(1-\xi^2)		
				\frac{\vb*{F} \cdot 
					\nabla
					\times
					\vb*{b}}{m\Omega} 	
				-  
				2 \xi
				u
				\frac{\vb*{F} \cdot 
					\vb*{b}}{mv^2}		
				.  
			\end{align*}
			
			Thus, we obtain
			\begin{align}
				\dot{\xi}(\vb*{x},\xi,v)
				& 
				:=
				\frac{1-\xi^2}{\xi v^2}
				\left(
				\frac{\vb*{F}(\vb*{x},\xi,v)}{m}
				-
				\frac{v^2\xi^2}{2B(\vb*{x})}
				\nabla B(\vb*{x})
				\right)
				\cdot  
				(\vb*{v}_{\text{gc}}(\vb*{x},\xi,v)
				+
				u(\vb*{x},\xi,v)\vb*{b}(\vb*{x}))\nonumber
				\\
				&
				-
				\frac{\dot{\mu}
					(\vb*{x},\xi,v)B(\vb*{x})}{v^2\xi}
				\nonumber
				\\
				&
				=  
				(1-\xi^2) 
				\frac{\vb*{F}(\vb*{x},\xi,v)\cdot \vb*{b}(\vb*{x})}{mv}
				+    
				\xi(1-\xi^2)		
				\frac{\vb*{F}(\vb*{x},\xi,v) \cdot 
					\nabla
					\times
					\vb*{b}(\vb*{x})
				}{m\Omega(\vb*{x})} 	   
				\\
				&
				- 
				3 \xi
				u(\vb*{x},\xi,v)
				\frac{\vb*{F}(\vb*{x},\xi,v) \cdot 
					\vb*{b}(\vb*{x})}{mv^2}	
				- 
				\xi
				\vb*{b}(\vb*{x})
				\cdot
				\nabla
				u(\vb*{x},\xi,v)  
				\nonumber	 
				\\
				&
				-
				\xi
				\frac{1-\xi^2}{2B}		 
				(\vb*{v}_{\text{gc}}(\vb*{x},\xi,v)
				+
				u(\vb*{x},\xi,v)\vb*{b}(\vb*{x}))\cdot\nabla B  
				,
			\end{align}
			which exactly matches expression (\ref{eq:dot_xi_xi_v}).

			\chapter{Legendre polynomials}
			\label{sec:Appendix_Legendre}
			
			Legendre polynomials are the eigenfunctions of the Sturm-Liouville problem in the interval $\xi\in[-1,1]$ defined by the differential equation%
			\begin{align}
				2\Lorentz P_k(\xi) = -k(k+1) P_k(\xi), \label{eq:Legendre_eigenvalues}
			\end{align}
			and regularity boundary conditions at $\xi = \pm 1 $
			\begin{align}
				\eval{(1-\xi^2)\dv{P_k}{\xi}}_{\xi = \pm 1} = 0,
			\end{align}
			where $k\ge 0$ is an integer. 
			
			As $\Lorentz$ has a discrete spectrum and is self-adjoint with respect to the inner product
			\begin{align}
				\mean*{f,g}_\Lorentz := \int_{-1}^{1} fg \dd{\xi} ,
			\end{align}
			in the space of functions that satisfy the regularity condition, $\{P_k\}_{k=0}^{\infty}$ is an orthogonal basis satisfying $\mean*{P_j,P_k}_\Lorentz = 2 \delta_{jk}/(2k+1)$. Hence, these polynomials satisfy the three-term recurrence formula
			\begin{align}
				(2k+1)\xi P_k(\xi) = (k+1)P_{k+1}(\xi) + k P_{k-1}(\xi),
				\label{eq:Legendre_Three_Term_Recurrence}
			\end{align}
			obtained by Gram-Schmidt orthogonalization. Starting from the initial values $P_0=1$ and $P_1=\xi$, the recurrence defines the rest of the Legendre polynomials. Additionally, they satisfy the differential identity
			\begin{align}
				(1-\xi^2)\dv{P_k}{\xi} = k P_{k-1}(\xi) - k \xi P_k(\xi).
				\label{eq:Legendre_Differential_Recurrence}
			\end{align}
			Identities (\ref{eq:Legendre_Three_Term_Recurrence}) and (\ref{eq:Legendre_Differential_Recurrence}) are useful to represent tridiagonally the left-hand side of equation (\ref{eq:DKE}) when we use the expansion (\ref{eq:Legendre_expansion}). The $k-$th Legendre mode of the term $\xi\vb*{b}\cdot \nabla f $ is expressed in terms of the modes $f^{(k-1)}$ and $f^{(k+1)}$ using (\ref{eq:Legendre_Three_Term_Recurrence})
			\begin{align}	
				\mean*{\xi \vb*{b}\cdot\nabla f, P_k}_\Lorentz
				=
				\frac{2}{2k+1}
				\left[
				\frac{k}{2k-1} 
				\vb*{b}\cdot\nabla f^{(k-1)} 
				+
				\frac{k+1}{2k+3} 
				\vb*{b}\cdot\nabla f^{(k+1)} 
				\right].
			\end{align}
			Combining both (\ref{eq:Legendre_Three_Term_Recurrence}) and (\ref{eq:Legendre_Differential_Recurrence}) allows to express the $k-$th Legendre mode of the mirror term $\nabla\cdot \vb*{b}((1-\xi^2)/2)\pdv*{f}{\xi}$ in terms of the modes $f^{(k-1)}$ and $f^{(k+1)}$ as
			\begin{align} 	
				& 
				\mean*{ 
					\frac{1}{2}(1-\xi^2)
					\nabla\cdot\vb*{b}  
					\pdv{f}{\xi}, P_k}_\Lorentz
				=	\\
				&\frac{\vb*{b}\cdot\nabla \ln B}{2k+1}
				\left[
				\frac{k(k-1)}{2k-1} 
				f^{(k-1)} 
				-
				\frac{(k+1)(k+2)}{2k+3}
				f^{(k+1)} 
				\right], \nonumber
			\end{align}
			where we have also used $\nabla\cdot\vb*{b}  = - \vb*{b}\cdot \nabla \ln B$. The term proportional to $\widehat{E}_\psi$ is diagonal in a Legendre representation
			\begin{align}
				\mean*{\frac{\widehat{E}_\psi}{\mean*{B^2}}
					\vb*{B}\times \nabla\psi \cdot\nabla f , P_k}_\Lorentz
				=
				\frac{2}{2k+1}
				\frac{\widehat{E}_\psi}{\mean*{B^2}}
				\vb*{B}\times \nabla\psi \cdot\nabla f^{(k)}.
				\nonumber
			\end{align}
			For the collision operator used in equation (\ref{eq:DKE}), as Legendre polynomials are eigenfunctions of the pitch-angle scattering operator, using (\ref{eq:Legendre_eigenvalues}) we obtain the diagonal representation 
			\begin{align}
				\mean*{\hat{\nu} \Lorentz f , P_k}_\Lorentz
				&
				=
				-\hat{\nu}
				\frac{k(k+1)}{2k+1}	
				f^{(k)}.
			\end{align}
			
			Now, we briefly comment on why the truncation error from (\ref{eq:Legendre_expansion}) implies that the solution to (\ref{eq:DKE_Legendre_expansion}) and (\ref{eq:kernel_elimination_condition_Legendre}) is an approximation of the Legendre spectrum of the exact solution to (\ref{eq:DKE}) satisfying (\ref{eq:kernel_elimination_condition}). For this, we will assume that the solution to (\ref{eq:DKE}) and (\ref{eq:kernel_elimination_condition}) is unique (which it is, see appendix \ref{sec:Appendix_Invertibility}). We denote this exact solution by $f_{\text{ex}}$ and its Legendre modes by $f^{(k)}_{\text{ex}}$. The Legendre modes $f^{(k)}_{\text{ex}}$ satisfy (\ref{eq:DKE_Legendre_expansion}) for all values of $k$, including $k>N_\xi$ and, in general, $f^{(N_\xi+1)}_{\text{ex}}\ne 0$. Denoting the error of the solution $f^{(k)}$ to (\ref{eq:DKE_Legendre_expansion}) and (\ref{eq:kernel_elimination_condition_Legendre}) by
			\begin{align}
				E^{(k)} : = f^{(k)}_{\text{ex}} - f^{(k)} ,
			\end{align}
			is easy to prove that
			\begin{align}
				L_k E^{(k-1)} + D_k E^{(k)} + U_k E^{(k+1)} = 0, 
				\label{eq:Error_DKE_Legendre_k}
			\end{align}
			for $k =0,1,\ldots, N_\xi-1$ and 
			\begin{align}
				L_{N_\xi} E^{(N_\xi-1)} + D_{N_\xi} E^{(N_\xi)} = - U_{N_\xi} f^{(N_\xi+1)}_{\text{ex}}.
				\label{eq:Error_DKE_Legendre_Nxi}
			\end{align}
			Note that the system of equations constituted by (\ref{eq:Error_DKE_Legendre_k}) and (\ref{eq:Error_DKE_Legendre_Nxi}) for the error is identical to (\ref{eq:DKE_Legendre_expansion}) substituting $f^{(k)}$ by $E^{(k)}$ and $s^{(k)}$ by $- U_{N_\xi} f^{(N_\xi+1)}_{\text{ex}}$. Hence, by assumption, the solution to (\ref{eq:Error_DKE_Legendre_k}) and (\ref{eq:Error_DKE_Legendre_Nxi}) satisfying (\ref{eq:kernel_elimination_condition_Legendre}) is unique, implying that $E^{(k)}\ne 0$ unless $ {U}_{N_\xi} f^{(N_\xi+1)}_{\text{ex}} = 0$.

			To conclude this appendix we prove identities (\ref{eq:I2k_0_identity}), (\ref{eq:I2k_1_identity}) and  (\ref{eq:I2k_2_identity}). The differential equation (\ref{eq:Legendre_eigenvalues})
			and identities (\ref{eq:Legendre_Three_Term_Recurrence}) and (\ref{eq:Legendre_Differential_Recurrence}) are useful to compute the following indefinite integrals
			\begin{align}
				I_{2k}^{(0)}(x)
				& := 
				2\int_{0}^{x}
				P_{2k}(\xi) \dd{\xi},
				\label{eq:Legendre_I2k}
				\\
				I_{2k+1}^{(1)}(x)
				& := 
				2\int_{0}^{x}
				\xi
				P_{2k+1}(\xi) \dd{\xi},
				\\
				I_{2k}^{(2)}(x)
				& := 
				2\int_{0}^{x}
				\xi^2
				P_{2k}(\xi) \dd{\xi},
			\end{align}
			where $x\in[0,1]$. 
			
			\begin{enumerate}
				\item \textbf{Calculation of $I_{2k}$:} For this, we first integrate (\ref{eq:Legendre_eigenvalues}) to obtain $2\int_{0}^{x}2\Lorentz P_{2k} \dd{\xi} = - 2k (2k+1) I_{2k}(x)$. As $\Eval{ \dv*{P_{2k}}{\xi}}_{\xi=0}=0$, we have that $\int_{0}^{x}2\Lorentz P_{2k} \dd{\xi} = (1-x^2)\dv*{P_{2k}(x)}{x}$. Combining (\ref{eq:Legendre_Three_Term_Recurrence}) and (\ref{eq:Legendre_Differential_Recurrence}) gives $(1-x^2)\dv*{P_{2k}}{x} = 2k(2k+1)(P_{2k-1}-P_{2k+1})/(4k+1)$. Hence, 
				\begin{align*}
					I_{2k}^{(0)}(x)
					=
					\frac{2}{4k+1}
					\left(
					P_{2k+1}(x) - P_{2k-1}(x)
					\right).
				\end{align*}
				and as $P_k(1)=1$ for any positive integer $k$, $I_{2k}^{(0)}(1)=0$.
				
				\item \textbf{Calculation of $I_{2k+1}^{(1)}$:} Integrating (\ref{eq:Legendre_Three_Term_Recurrence}) we can easily write $I_{2k+1}^{(1)}$ in terms of $I_{2k}^{(0)}$ and $I_{2k+2}^{(0)}$. Namely,
				\begin{align*}
					(4k+3) I_{2k+1}^{(1)}(x)
					=
					(2k+2)I_{2k+2}^{(0)}(x)
					+
					(2k+1) I_{2k}^{(0)}(x).
				\end{align*}

				\item \textbf{Calculation of $I_{2k}^{(2)}$:} Using (\ref{eq:Legendre_Three_Term_Recurrence}) we can write $\xi^2 P_{2k} = ( (2k+1)\xi P_{2k+1} + 2k P_{2k-1})/(4k+1) $. Integrating this expression we obtain
				\begin{align}
					I_{2k}^{(2)}(x)
					=
					\frac{1}{4k+1}
					(
					(2k+1) I_{2k+1}^{(1)}(x)
					+
					2k I_{2k-1}^{(1)}(x)
					).
				\end{align}
			\end{enumerate}

			\chapter{Invertibility of the spatial differential operators}
			\label{sec:Appendix_Invertibility} 
			In this Appendix we will study
			the invertibility of the left-hand-side of (\ref{eq:DKE_Legendre_expansion}). We are only concerned in elucidating under which conditions the algorithm given in section \ref{chap:MONKES} can be applied to solve (\ref{eq:DKE_Legendre_expansion}). For instance, we will consider the possibility of the flux surface being rational despite of the fact that (among other things) it may be inconsistent with the assumption that thermodynamic forces are a flux-function. We will conclude that the solution to (\ref{eq:DKE_Legendre_expansion}) submitted to (\ref{eq:kernel_elimination_condition_Legendre}) is unique in ergodic flux surfaces and also on rational flux surfaces with $E_\psi\ne 0$ and can be obtained with the aforementioned algorithm. In order to do this, we view $L_k$, $D_k$ and $U_k$ as operators that act on $\mathcal{F}$, where $\mathcal{F}$ is the space of smooth functions on the flux surface equipped with the inner product
			\begin{align}
				\mean*{f,g}_{\mathcal{F}}=\frac{\Nfp}{4\pi^2}\oint\oint f\bar{g}\dd{\theta}\dd{\zeta},
				\label{eq:Fourier_inner_product}
			\end{align}
			where $\bar{z}$ denotes the complex conjugate of $z$ and the inner product induces a norm 
			\begin{align}
				\norm{f}_{\mathcal{F}}:=\sqrt{\mean*{f,f}_{\mathcal{F}}}.
				\label{eq:Fourier_norm}
			\end{align}
			In this setting $L_k$, $D_k$ and $U_k$ are operators from $\mathcal{F}$ to $\mathcal{F}$ as all of their coefficients are smooth on the flux surface. However, the operators $L_k$ and $U_k$ given by (\ref{eq:DKE_Legendre_expansion_Lower}) and (\ref{eq:DKE_Legendre_expansion_Upper}) do not have a uniquely defined inverse. This is a consequence of the fact that the parallel streaming operator $\xi \vb*{b}\cdot \nabla+\nabla \cdot \vb*{b} {(1-\xi^2)}/{2}  \pdv*{\xi}$ has a non trivial kernel comprised of functions $g((1-\xi^2)/B)$. On the other hand, the operator $D_k$ has a unique inverse for $k\ge 1$. For $k=0$, the operator $D_0$ is not invertible as it has a kernel comprised of functions $g(B_\theta\theta + B_\zeta\zeta)$.
			
			Whether $L_k$ and $U_k$ are or not invertible can be determined studying the uniqueness of continuous solutions (on the flux surface) to
			\begin{align}
				\vb*{B}\cdot \nabla f + \omega_k f = s B,
				\label{eq:Invertibility_Lk_Uk_MDE}
			\end{align}
			for some $s,\omega_k\in\mathcal{F}$. Note that equations $L_k f = ks/(2k-1)$ and $U_k f =(k+1)s/(2k+3)$ can be written in the form of equation (\ref{eq:Invertibility_Lk_Uk_MDE}) setting, respectively, $\omega_k=(k-1)\vb*{B}\cdot\nabla \ln B /2$ and $\omega_k=-(k+2)\vb*{B}\cdot\nabla \ln B /2$. We will determine a condition for $\omega_k$ which, if satisfied, equation (\ref{eq:Invertibility_Lk_Uk_MDE}) has a unique solution $f\in\mathcal{F}$.
			
			The solution to equation (\ref{eq:Invertibility_Lk_Uk_MDE}) can be written as
			\begin{align}
				f = (f_0 + K ) \Phi,
				\label{eq:Invertibility_Lk_Uk_Variation_of Constants}
			\end{align}
			where
			\begin{align}
				& \vb*{B}\cdot \nabla f_0 = 0,  \label{eq:Invertibility_Lk_Uk_constant}
				\\
				& \vb*{B}\cdot \nabla \Phi + \omega_k \Phi = 0,  \label{eq:Invertibility_Lk_Uk_Homogeneous}
				\\
				& \vb*{B}\cdot \nabla K = {sB}/{\Phi}.  \label{eq:Invertibility_Lk_Uk_Particular}
			\end{align}
			Equations (\ref{eq:Invertibility_Lk_Uk_Homogeneous}) and (\ref{eq:Invertibility_Lk_Uk_Particular}) are integrated (along a field line) imposing $\eval{\Phi}_{p}=1$ and $\eval{K}_{p}=0$ at a point $p$ of the field line. Note that $f_0=\eval{f}_{p}$ is an integration constant. Depending on the form of $\omega_k$, $f_0$ can or cannot be determined imposing continuity on the flux surface. The solution to equation (\ref{eq:Invertibility_Lk_Uk_Homogeneous}) can be written as
			\begin{align}
				\Phi = \exp(-W_k), 
				\label{eq:Invertibility_Lk_Uk_Homogeneous_solution}
			\end{align}
			where $\vb*{B}\cdot\nabla W_k = \omega_k$ and is integrated imposing $\eval{W_k}_{p}=0$. Note that this implies that $\Phi\ne 0$ and that 
			\begin{align}
				- \vb*{B}\cdot \nabla \left(\frac{1}{\Phi}\right) + \omega_k \frac{1}{\Phi} = 0.
				\label{eq:Invertibility_Lk_Uk_Adjoint_kernel}
			\end{align}
			When $\Phi\in\mathcal{F}$, the left-hand side of (\ref{eq:Invertibility_Lk_Uk_MDE}) has a non trivial kernel (as an operator from $\mathcal{F}$ to $\mathcal{F}$). In order to proceed further, we employ coordinates $(\alpha,l)$ where $\alpha:=\theta-\iota \zeta$ is a poloidal angle that labels field lines and $l$ is the length along magnetic field lines. Depending on the type of flux surface there are two possible situations
			\begin{enumerate}
				\item For ergodic flux surfaces, $\iota\in\mathbb{R}\backslash \mathbb{Q}$ and satisfying (\ref{eq:Invertibility_Lk_Uk_constant}) implies that $f_0$ is a flux-function. The solution $f$ to (\ref{eq:Invertibility_Lk_Uk_MDE}) is a differentiable function on the torus if $\mean*{\vb*{B}\cdot\nabla f} = 0$ (see appendix \ref{sec:FSA_MDE}). Applying $\mean*{\text{Eq. (\ref{eq:Invertibility_Lk_Uk_MDE})}}$ combined with splitting (\ref{eq:Invertibility_Lk_Uk_Variation_of Constants}) yields
				\begin{align}
					f_0\mean*{\omega_k \Phi}  & = \mean*{B s} - \mean*{K \omega_k \Phi} \nonumber
					\\
					& = \mean*{\vb*{B}\cdot\nabla(K \Phi)}. \label{eq:Invertibility_Lk_Uk_Ergodic_condition}
				\end{align}
				Hence, if $\mean*{\omega_k \Phi} \ne 0$, equation (\ref{eq:Invertibility_Lk_Uk_Ergodic_condition}) fixes the value of $f_0$ so that $f$ is continuous on the torus. Note that if $\mean*{\omega_k \Phi} \ne 0$, by virtue of (\ref{eq:Invertibility_Lk_Uk_Homogeneous}), $\Phi$ is not single valued and does not belong to $\mathcal{F}$. On the contrary, if $f_0$ is free, then $\Phi$ is a continuous function on the torus. Then, (\ref{eq:Invertibility_Lk_Uk_Ergodic_condition}) implies that $K\Phi$ is continuous on the torus when $\Phi$ is. The function $K$ is also continuous as long as $sB$ belongs to the image of $\vb*{B}\cdot \nabla + \omega_k$. Note that using (\ref{eq:Invertibility_Lk_Uk_Adjoint_kernel}) we can derive from $\mean*{\text{Eq. (\ref{eq:Invertibility_Lk_Uk_MDE})}/\Phi}$ the solvability condition $\mean*{sB/\Phi}=0$.
				
				\item For rational flux surfaces, $\iota\in \mathbb{Q}$ and satisfying (\ref{eq:Invertibility_Lk_Uk_constant}) implies that $f_0(\alpha)$ depends on the field line chosen. At these surfaces, the field line labelled by $\alpha$ closes on itself after a length $L_{\text{c}}(\alpha)$. If the solution $f$ is continuous on the flux surface, then $\int_{0}^{L_{\text{c}}} \vb*{B}\cdot\nabla f \dd{l}/B=0$ for each field line (see appendix \ref{sec:FSA_MDE}). Applying $\int_{0}^{L_{\text{c}}} \text{Eq. (\ref{eq:Invertibility_Lk_Uk_MDE})}\dd{l}/B$ combined with splitting (\ref{eq:Invertibility_Lk_Uk_Variation_of Constants}) yields
				\begin{align}
					f_0(\alpha)
					\int_{0}^{L_{\text{c}}} \omega_k \Phi \frac{\dd{l}}{B}  
					& =  
					\int_{0}^{L_{\text{c}}} s \dd{l} - \int_{0}^{L_{\text{c}}} \omega_k K \Phi \frac{\dd{l}}{B}
					\nonumber 
					\\
					& =  
					\int_{0}^{L_{\text{c}}} \vb*{B}\cdot\nabla(K \Phi) \frac{\dd{l}}{B}
					. \label{eq:Invertibility_Lk_Uk_Rational_condition}
				\end{align}
				If $\int_{0}^{L_{\text{c}}} \omega_k \Phi {\dd{l}}/{B} \ne 0$, condition (\ref{eq:Invertibility_Lk_Uk_Rational_condition}) fixes a unique value of $f_0(\alpha)$ (for each field line) for which $f$ is continuous on the torus. As for ergodic surfaces, if (\ref{eq:Invertibility_Lk_Uk_Rational_condition}) does not fix $f_0$, then $\Phi$ and $K\Phi$ are continuous along field lines. Again, $K$ is also continuous as long as $sB$ belongs to the image of $\vb*{B}\cdot \nabla + \omega_k$. Using (\ref{eq:Invertibility_Lk_Uk_Adjoint_kernel}) we can derive from $\int_{0}^{L_{\text{c}}}\text{Eq. (\ref{eq:Invertibility_Lk_Uk_MDE})}/\Phi{\dd{l}}/{B}$ the solvability condition $\int_{0}^{L_{\text{c}}}sB/\Phi{\dd{l}}/{B}=0$.
			\end{enumerate}
			
			Thus, we have seen that when $\mean*{\omega_k \Phi}=0$ or $\int_{0}^{L_{\text{c}}} \omega_k \Phi {\dd{l}}/{B}=0$, the operator $\vb*{B}\cdot\nabla + \omega_k$ from $\mathcal{F}$ to itself is not one-to-one (it has a non trivial kernel comprised of multiples of $\Phi$).  
			Moreover, we have the solvability conditions $\mean*{sB/\Phi}=0$ for ergodic surfaces and $\int_{0}^{L_\text{c}}sB/\Phi\dd{l}/B = 0$ for rational surfaces. The existence of a solvability condition implies that $\vb*{B}\cdot\nabla + \omega_k$ is not onto. We can derive a simpler and equivalent condition for $\omega_k$ from (\ref{eq:Invertibility_Lk_Uk_Homogeneous_solution}). Note that $\Phi$ is continuous on the torus only when $W_k$ is. As $\vb*{B}\cdot\nabla W_k = \omega_k$, continuity of $W_k$ along field lines imposes $\mean*{\omega_k}=0$ on ergodic flux surfaces and $\int_{0}^{L_{\text{c}}} \omega_k  {\dd{l}}/{B} =0$ on rational ones. Hence, the operator $\vb*{B}\cdot\nabla + \omega_k$ is invertible if $\mean*{\omega_k} \ne 0 $ or $\int_{0}^{L_{\text{c}}} \omega_k {\dd{l}}/{B} \ne 0$. 
			
			This result can be applied to determine that $L_k$ and $U_k$ are not invertible. For both $L_k$ and $U_k$, $\omega_k \propto \vb*{B}\cdot\nabla\ln B^\gamma$ for some rational exponent $\gamma$. As $B$ is continuous on the flux surface we have for $L_k$ and $U_k$ that $\int_{0}^{L_{\text{c}}} \omega_k  {\dd{l}}/{B} =  0$ or $\mean*{\omega_k }=0$, which means that neither $L_k$ nor $U_k$ are invertible. 
			
			Now we turn our attention to the invertibility of $D_k$ for $k\ge 1$. For $\widehat{E}_\psi =0$, $D_k$ is just a multiplicative operator and is clearly invertible when $\hat{\nu}, k\ne0$. For $\widehat{E}_\psi \ne 0$, the invertibility of $D_k$ can be proven by studying the uniqueness of solutions to
			\begin{align} 
				\vb*{B}\times\nabla\psi\cdot \nabla g - \hat{\nu}_k g  = - \frac{\mean*{B^2}}{\widehat{E}_\psi} s,
				\label{eq:Invertibility_Dk_MDE}
			\end{align}
			where $\hat{\nu}_k = \hat{\nu} k (k+1) {\mean*{B^2}}/{2\widehat{E}_\psi}$. The procedure is very similar to the one carried out for $L_k$ and $U_k$. First, we write the solution to equation (\ref{eq:Invertibility_Dk_MDE}) as
			\begin{align}
				g = ( g_0 + I ) \Psi,
				\label{eq:Invertibility_Dk_Variation_of_Constants}
			\end{align}
			where
			\begin{align}
				& \vb*{B}\times\nabla\psi\cdot \nabla g_0 = 0,  \label{eq:Invertibility_Dk_constant}
				\\
				& \vb*{B}\times\nabla\psi\cdot \nabla \Psi - \hat{\nu}_k \Psi = 0,  \label{eq:Invertibility_Dk_Homogeneous}
				\\
				& \vb*{B}\times\nabla\psi\cdot \nabla I = - \frac{\mean*{B^2}}{\widehat{E}_\psi}\frac{s}{\Psi}.  \label{eq:Invertibility_Dk_Particular}
			\end{align}
			Equations (\ref{eq:Invertibility_Dk_Homogeneous}) and (\ref{eq:Invertibility_Dk_Particular}) are integrated along a integral curve of $\vb*{B}\times\nabla\psi$ imposing $\eval{\Psi}_{p}=1$ and $\eval{I}_{p}=0$ at the initial point $p$ of integration. The integral curves of $\vb*{B}\times\nabla\psi$ are, in Boozer coordinates, straight lines $B_\theta \theta + B_\zeta \zeta = \text{constant}$. In order to proceed further, we change from Boozer angles $(\theta,\zeta)$ to a different set of magnetic coordinates $(\alpha,\phi)$ using the linear transformation
			\begin{align}
				\Matrix{c}{\theta \\ \zeta}
				=
				\Matrix{cc}
				{
					(1 + \iota \delta)^{-1}	 &  \iota\\ 
					-\delta(1 + \iota \delta)^{-1}	 &  1
				}
				\Matrix{c}{\alpha \\ \phi}
				\label{eq:Invertibility_Magnetic_Coordinates_Dk}
			\end{align}
			where $\delta = B_\theta/B_\zeta$. In these coordinates $\vb*{B} =\nabla \psi \times \nabla \alpha $, $B_\alpha = 0$ and
			\begin{align}
				\vb*{B}\times\nabla \psi \cdot \nabla = 
				B^2 \pdv{}{\alpha}.
				\label{eq:Invertibility_Dk_Localization_alpha}
			\end{align}
			Depending on the rationality or irrationality of $\delta$ we can distinguish two options
			\begin{enumerate}
				\item If $\delta \in \mathbb{R}\backslash\mathbb{Q}$, satisfying (\ref{eq:Invertibility_Dk_constant}) implies that $g_0$ is a flux-function (the integral curves trace out the whole flux surface). Note that if $g$ is a differentiable function on the torus $\mean{ \vb*{B} \times \nabla \psi \cdot \nabla g } = \mean{ \nabla\times (g\vb*{B}) \cdot\nabla \psi }=0$, where we have used $\nabla\times\vb*{B}\cdot\nabla\psi = 0$. Taking $\mean*{\text{Eq. (\ref{eq:Invertibility_Dk_MDE})}}$ assuming that $f$ is continuous on the flux surface, combined with (\ref{eq:Invertibility_Dk_Variation_of_Constants}) gives
				\begin{align} 
					\mean*{\Psi} g_0  
					& = 
					\frac{ \mean*{B^2} }{ \hat{\nu}_k \widehat{E}_\psi } 
					\mean*{s}
					- 
					\mean*{I \Psi}\nonumber
					\\
					& = 
					\frac{1}{\hat{\nu}_k}
					\mean*{\vb*{B}\times\nabla \psi \cdot \nabla(I \Psi)}. 
					\label{eq:Invertibility_Dk_Ergodic_condition}
				\end{align}
				Hence, if $\mean*{\Psi}\ne 0$, continuity of $g$ on the torus fixes the integration constant $g_0$. 
				
				\item If $\delta \in \mathbb{Q}$, satisfying (\ref{eq:Invertibility_Dk_constant}) implies that $g_0(\phi)$ is a function of $\phi$. Now the integral curves $\phi=\text{constant}$ close on itself after moving in $\alpha$ an arc-length $L_\alpha$. In this scenario, if $g$ is a differentiable function on the torus $\int_{0}^{L_\alpha}\vb*{B}\times \nabla\psi \cdot \nabla g \dd{\alpha}/ B^2  = 0$, where we have used (\ref{eq:Invertibility_Dk_Localization_alpha}). Thus, taking $\int_{0}^{L_\alpha}\text{Eq. (\ref{eq:Invertibility_Dk_MDE})} \dd{\alpha}/ B^2 $, combined with (\ref{eq:Invertibility_Dk_Variation_of_Constants}) gives
				\begin{align}
					g_0(\phi) \int_{0}^{L_\alpha} \Psi\frac{\dd{\alpha}}{ B^2 }
					& = 
					\frac{ \mean*{B^2} }{ \hat{\nu}_k \widehat{E}_\psi }
					\int_{0}^{L_\alpha} s    \frac{\dd{\alpha}}{ B^2 }
					\nonumber 
					-      
					\int_{0}^{L_\alpha}   I \Psi  \frac{\dd{\alpha}}{ B^2 }
					\\
					& =       
					\frac{1}{\hat{\nu}_k}
					\int_{0}^{L_\alpha}
					\vb*{B}\times\nabla \psi \cdot \nabla(I \Psi)
					\frac{\dd{\alpha}}{ B^2 }
					.
					\label{eq:Invertibility_Dk_Rational_condition}
				\end{align}
				Thus, if $\int_{0}^{L_\alpha} \Psi \dd{\alpha} / B^2 \ne 0 $ condition (\ref{eq:Invertibility_Dk_Rational_condition}) fixes the value of $g_0(\phi)$ so that $g$ is continuous on the flux surface. 
				
			\end{enumerate}
			Similarly to what happened to $\Phi$ when studying the invertibility of $L_k$ and $U_k$, continuity of the solution implies that $\Psi$ cannot be single valued. We can write $\Psi$ as
			\begin{align}
				\Psi = \exp(-A_k), 
				\label{eq:Invertibility_Dk_Exponential}
			\end{align}
			where $\vb*{B}\times \nabla\psi \cdot \nabla A_k = \hat{\nu}_k$ and is integrated along with condition $\eval{A_k}_{p}=0$. Using (\ref{eq:Invertibility_Dk_Localization_alpha}), we can write
			\begin{align}
				A_k(\alpha,\phi) = \hat{\nu}_k \int_{0}^{\alpha} \frac{\dd{\alpha'}}{ B^2(\alpha',\phi)}.
			\end{align}
			Note that $A_k$ is monotonically increasing with $\alpha$, which means that $\Psi $ cannot be single valued. Besides, (\ref{eq:Invertibility_Dk_Exponential}) implies $\Psi>0$, which means that $\mean*{\Psi}\ne 0 $ and $\int_{0}^{L_\alpha}\Psi   \dd{\alpha}/  B^2 \ne 0 $. Thus, there is a unique value of the constant $g_0$ which compensates the jumps in $\Psi$ and $I\Psi$ so that $g=g_0 \Psi + I \Psi$ is continuous on the flux surface. Hence, $D_k$ is an invertible operator from $\mathcal{F}$ to itself.

			The inverse of $D_k$ for $k \ge 1$ and $\widehat{E}_\psi \ne 0$ is defined by
			\begin{align}
				D_k^{-1} s :=  ( \mathcal{G}_0[s] + \mathcal{I}[s] ) \Psi,
			\end{align}
			where $\mathcal{G}_0[s]$ and $\mathcal{I}[s]$ denote the linear operators which define, respectively, the constant of integration and the solution to (\ref{eq:Invertibility_Dk_Particular}) with $\eval{I}_{p}=0$ for a given source term. Specifically,
			\begin{align}
				\mathcal{I}[s](\alpha,\phi)
				& :=
				- \frac{\mean*{B^2}}{\widehat{E}_\psi}
				\int_{0}^{\alpha}
				\frac{s(\alpha',\phi)}{\Psi(\alpha',\phi)}
				\frac{\dd{\alpha'}}{ B^2(\alpha',\phi)}  ,
			\end{align}
			and
			\begin{align}
				\mathcal{G}_0[s](\phi)
				& :=
				\begin{dcases}
					& \text{If }\delta\in\mathbb{R}\backslash\mathbb{Q}:\\
					& \frac{2}{ \hat{\nu} k (k+1) } \frac{\mean*{s}}{\mean*{\Psi} }
					- 
					\frac{\mean*{\mathcal{I}[s] \Psi}}{\mean*{\Psi} },  
					\vspace{0.5cm} \\ 
					& \text{If }\delta\in\mathbb{Q}:\\
					& \frac{2}{\hat{\nu} k(k+1)} \dfrac{\int_{0}^{L_\alpha} s \frac{\dd{\alpha}}{ B^2 }}{\int_{0}^{L_\alpha} \Psi\frac{\dd{\alpha}}{ B^2 }}
					-      
					\dfrac{\int_{0}^{L_\alpha}   \mathcal{I}[s] \Psi  \frac{\dd{\alpha}}{ B^2 }}{\int_{0}^{L_\alpha} \Psi\frac{\dd{\alpha}}{ B^2 }} .
				\end{dcases}  
			\end{align}

			Finally, we will study the invertibility of the operator $\Delta_k$ 
			\begin{align}
				\Delta_{k} = D_k - U_k \Delta_{k+1}^{-1} L_{k+1} 
				\label{eq:Invertibility_Delta}
			\end{align}
			assuming that $\Delta_{k+1}$ is an invertible operator from $\mathcal{F}$ to $\mathcal{F}$. For this, first, we note that in the space of functions of interest (smooth periodic functions on the torus), using a Fourier basis $\{e^{\ii (m \theta+ n\Nfp\zeta)}\}_{m,n\in\mathbb{Z}}$, we can approximate any function $f(\theta,\zeta)=\sum_{m,n\in \mathbb{Z}} \hat{f}_{mn} e^{\ii (m \theta+ n\Nfp\zeta)} \in \mathcal{F}$ using an approximant $\tilde{f}(\theta,\zeta)$
			\begin{align}
				\tilde{f}(\theta,\zeta)=\sum_{- N \le m,n\le N } \hat{f}_{mn} e^{\ii (m \theta+ n\Nfp\zeta)}
				\label{eq:Fourier_truncated}
			\end{align}
			truncating the modes with mode number greater than some positive integer $N $ where 
			\begin{align}
				\hat{f}_{mn} = \mean*{f,e^{\ii (m \theta+ n\Nfp\zeta)}}_{\mathcal{F}}  \norm{e^{\ii (m \theta+ n\Nfp\zeta)}}_{\mathcal{F}}^{-2}
			\end{align}
			are the Fourier modes of $f$. Thus, we approximate $\mathcal{F}$ using a finite dimensional subspace $\mathcal{F}^{N} \subset \mathcal{F}$ consisting on all the functions of the form given by equation (\ref{eq:Fourier_truncated}).

			Hence, we can approximate $D_k$, $U_k$, $\Delta_{k+1}$ and $L_{k+1} $ restricted to $\mathcal{F}^{N}$ (and therefore $\Delta_{k}$) in equation (\ref{eq:Invertibility_Delta}) by operators $D_k^N$, $U_k^N$, $\Delta_{k+1}^N$ and $L_{k+1}^N$ that map any $\tilde{f}\in\mathcal{F}^N$ to the projections of $D_k \tilde{f}$, $U_k \tilde{f}$, $\Delta_{k+1} \tilde{f}$ and $L_{k+1} \tilde{f}$ onto $\mathcal{F}^N$. The operators $D_k^N$, $U_k^N$, $\Delta_{k+1}^N$ and $L_{k+1}^N$ can be exactly represented (in a Fourier basis) by square matrices of size $\dim \mathcal{F}^N$. When the operators are invertible, these matrices are invertible aswell. Doing so, we can interpret the matrix representation of $\Delta_{k}$ as the Schur complement of the matrix
			\begin{align}
				M_k^N = 
				\Matrix{cc}
				{ D_k^N & U_k^N \\
					L_{k+1}^N & \Delta_{k+1}^N
				}.
				\label{eq:Invertibility_Delta_Schur}
			\end{align}
			It is well known from linear algebra that the determinant of $M_k^N$ satisfies
			\begin{align}
				\det(M_k^N)
				& =
				\det(\Delta_{k+1}^N)
				\det(\Delta_k^N)
				.
				\label{eq:Schur_complement_determinant}
			\end{align}
			When both $D_k$ and $\Delta_{k+1}$ are invertible, the matrix $M_k^N$ is invertible. Hence, note from (\ref{eq:Schur_complement_determinant}) that, for $k\ge1$, the matrix $\Delta_{k}^N$ can be inverted for any $N$, and therefore $\Delta_{k}$ (as an operator from $\mathcal{F}$ to $\mathcal{F}$) is invertible.

			
			The case $k=0$ requires special care. In this case $D_0$ is not invertible and the previous argument cannot be applied. In order to make the solution unique, we need to impose an additional constraint to $f^{(0)}$. On ergodic flux surfaces, condition (\ref{eq:kernel_elimination_condition_Legendre}) is sufficient to fix the value of $f^{(0)}$. However, this is not always the case when $\iota$ is rational. Condition (\ref{eq:kernel_elimination_condition_Legendre}) fixes the value of $f^{(0)}$ solely when the only functions that lie simultaneously at the kernels of $D_0 = -\widehat{E}_\psi \mean*{B^2}^{-1} \vb*{B}\times\nabla\psi\cdot\nabla$ and $L_1=\vb*{b}\cdot\nabla$ are constants (flux-functions). If $\widehat{E}_\psi \ne 0$, this occurs for any $\delta\ne -1/\iota $. However, the case $\delta= -1/\iota$ is unphysical as it would imply $\sqrt{g}=0$. Hence, in practice, when $\widehat{E}_\psi \ne 0$ condition (\ref{eq:kernel_elimination_condition_Legendre}) is sufficient to fix the value of $f^{(0)}$ even if the surface is not ergodic. For rational flux surfaces and $\widehat{E}_\psi = 0$, condition (\ref{eq:kernel_elimination_condition_Legendre}) is insufficient to fix $f^{(0)}$. In such case, we would need to fix the value of $f^{(0)}$ at a point of each field line as any function $g(\alpha)$ lies in the kernel of $\vb*{b}\cdot\nabla$. In order to clarify this assertion, let's try to obtain $f^{(0)}$ assuming that $f^{(1)}$ is known. Integrating the Legendre mode $k=1$ of equation (\ref{eq:DKE_Forward_elimination}) along a field line gives
			\begin{align}
				f^{(0)}(\alpha,l)
				& =
				f^{(0)}_0(\alpha)
				\label{eq:Invertibility_nullspace_condition}
				-
				\int_{0}^{l}
				\left(
				\sigma^{(1)}
				- \Delta_1 f^{(1)} 
				\right)\dd{l'}.
			\end{align} 
			If $\iota$ is irrational $f_0^{(0)}$ does not depend on $\alpha$. In this case, equation (\ref{eq:Invertibility_nullspace_condition}) and condition (\ref{eq:kernel_elimination_condition_Legendre}) fix $f^{(0)}$ for each $\sigma^{(1)}$, $f^{(1)}$. When $\iota$ is rational we need to distinguish between the case with and without radial electric field. 
			\begin{enumerate}
				\item For $E_\psi =0$, the constant $f^{(0)}_0$ is free as no other equation includes $f^{(0)}$. As $f_0^{(0)}$ depends on $\alpha$, condition (\ref{eq:kernel_elimination_condition_Legendre}) does not fix this integration constant. 
				%
				%
				\item For $E_\psi \ne 0$,	
				inserting (\ref{eq:Invertibility_nullspace_condition}) in the Legendre mode $k=0$ of equation (\ref{eq:DKE_Legendre_expansion}) gives
				\begin{align}  
					& -\frac{\widehat{E}_\psi}{\mean*{B^2}}
					B^2 \pdv{f_0^{(0)}}{\alpha} 
					=
					s^{(0)}
					- U_0 f^{(1)}
					\label{eq:Invertibility_Integration_constant_rationals_w_Er}
					-
					\frac{\widehat{E}_\psi}{\mean*{B^2}}
					B^2 \pdv{}{\alpha} 
					\int_{0}^{l}
					\left(
					\sigma^{(1)}
					- \Delta_1 f^{(1)} 
					\right)\dd{l'}.
				\end{align}
				Integrating $\int_{0}^{L_{\text{c}}} \text{ Eq. (\ref{eq:Invertibility_Integration_constant_rationals_w_Er})}\dd{l}$ gives a differential equation in $\alpha$ from which we can obtain $f_0^{(0)}$ up to a constant. Thus, (\ref{eq:Invertibility_nullspace_condition}), condition (\ref{eq:kernel_elimination_condition_Legendre}) and (\ref{eq:Invertibility_Integration_constant_rationals_w_Er}) fix $f^{(0)}$. 
			\end{enumerate} 
			
			Hence, in ergodic flux surfaces or rational flux surfaces with finite radial electric field, $M_0^N$ has a one-dimensional kernel. Thus, for $k=0$, it is necessary to substitute one of the rows of $[D_0^N \ \ U_0^N]$ by the condition (\ref{eq:kernel_elimination_condition_Legendre}) so that $M_0^N$ is invertible for any $N$ and as $\Delta_1^N$ can be inverted, also $\Delta_0^N$ constructed in this manner for any $N$, which implies that $\Delta_0$ (as the limit $\lim_{N\rightarrow\infty} \Delta_0^N$) is invertible.
			

			\chapter{Fourier collocation method}  \label{sec:Appendix_Fourier}
			
			In this appendix we describe the Fourier collocation (also called pseudospectral) method for discretizing the angles $\theta$ and $\zeta$. This discretization will be used to obtain the matrices $\vb*{L}_k$, $\vb*{D}_k$ and $\vb*{U}_k$. For convenience, we will use the complex version of the discretization method but for the discretization matrices we will just take their real part as the solutions to (\ref{eq:DKE}) are all real. We search for approximate solutions to equation (\ref{eq:DKE_Legendre_expansion}) of the form
			\begin{align}
				f^{(k)}(\theta,\zeta) 
				& = 
				\sum_{n=-N_{\zeta1}/2}^{N_{\zeta2}/2-1}
				\sum_{m=-N_{\theta1}/2}^{N_{\theta2}/2-1}
				\tilde{f}_{mn}^{(k)}
				e^{\ii(m\theta + nN_{p}\zeta)}
				\label{eq:Discrete_Fourier_Expansion}
			\end{align}
			where $N_{\theta1} = N_\theta - N_\theta\mod 2 $, $N_{\theta2} = N_\theta + N_\theta\mod 2 $, $N_{\zeta1} = N_\zeta - N_\zeta\mod 2 $, $N_{\zeta2} = N_\zeta + N_\zeta\mod 2 $ for some positive integers $N_\theta$, $N_\zeta$. The complex numbers 
			\begin{align}
				\tilde{f}_{mn}^{(k)}
				:=
				\mean*{f^{(k)}, e^{\ii(m\theta + nN_{p}\zeta)}}_{N_\theta N_\zeta}
				\norm{ e^{\ii(m\theta + nN_{p}\zeta)}}_{N_\theta N_\zeta}^{-2}
				\label{eq:Discrete_Fourier_Transform}
			\end{align}
			are the discrete Fourier modes (also called discrete Fourier transform), 
			\begin{align}
				\mean*{f,g}_{N_\theta N_\zeta}:= 
				\frac{1}{N_\theta N_\zeta}	
				\sum_{j'=0}^{N_{\zeta}-1}
				\sum_{i'=0}^{N_{\theta}-1}
				f(\theta_{i'},\zeta_{j'})
				\overline{g(\theta_{i'},\zeta_{j'})}
				\label{eq:Discrete_Fourier_Inner_product}
			\end{align} 
			is the discrete inner product associated to the equispaced grid points (\ref{eq:Theta_grid}), (\ref{eq:Zeta_grid}), $\norm{ f}_{N_\theta N_\zeta}:=\sqrt{\mean*{f,f}_{N_\theta N_\zeta}}$ its induced norm and $\bar{z}$ denotes the complex conjugate of $z$. We denote by $\mathcal{F}^{N_\theta N_\zeta}$ to the finite dimensional vector space (of dimension $N_\theta N_\zeta$) comprising all the functions that can be written in the form of expansion (\ref{eq:Discrete_Fourier_Expansion}).
			
			The set of functions $\{e^{\ii(m\theta + n\Nfp\zeta)}\}\subset \mathcal{F}^{N_\theta N_\zeta}$ forms an orthogonal basis for $\mathcal{F}^{N_\theta N_\zeta}$ equipped with the discrete inner product (\ref{eq:Discrete_Fourier_Inner_product}). Namely, 
			\begin{align}
				\mean*{e^{\ii(m\theta + nN_{p}\zeta)},e^{\ii(m'\theta + n'N_{p}\zeta)}}_{N_\theta N_\zeta} 
				\propto
				\delta_{mm'}\delta_{nn'}
			\end{align}
			for $-N_{\theta 1}/2\le m \le N_{\theta 2}/2$ and $-N_{\zeta 1}/2\le n \le N_{\zeta 2}/2$. Thus, for functions lying in $\mathcal{F}^{N_\theta N_\zeta}$, discrete expansions such as (\ref{eq:Discrete_Fourier_Expansion}) coincide with their (finite) Fourier series. The discrete Fourier modes (\ref{eq:Discrete_Fourier_Transform}) are chosen so that the expansion (\ref{eq:Discrete_Fourier_Expansion}) interpolates $f^{(k)}$ at grid points. Hence, there is a vector space isomorphism between the space of discrete Fourier modes and $f^{(k)}$ evaluated at the equispaced grid.

			Combining equations (\ref{eq:Discrete_Fourier_Expansion}), (\ref{eq:Discrete_Fourier_Transform}) and (\ref{eq:Discrete_Fourier_Inner_product}) we can write our Fourier interpolant as
			\begin{align}
				f^{(k)}(\theta,\zeta) 
				& = 
				\vb*{I}(\theta,\zeta) \cdot \vb*{f}^{(k)}
				\nonumber
				\\
				& =
				\sum_{j'=0}^{N_{\zeta}-1}
				\sum_{i'=0}^{N_{\theta}-1}
				I_{i'j'}(\theta,\zeta)
				f^{(k)}(\theta_{i'},\zeta_{j'})
				,
				\label{eq:Fourier_interpolant}
			\end{align}
			where $\vb*{f}^{(k)}\in\mathbb{R}^{N_{\text{fs}}}$ is the state vector containing $f^{(k)}(\theta_{i'},\zeta_{j'})$. The entries of the vector $\vb*{I}(\theta,\zeta)$ are the functions $I_{i'j'}(\theta,\zeta)$ given by, 
			\begin{align}
				& I_{i'j'}(\theta,\zeta)
				=
				I_{i'}^\theta(\theta)
				I_{j'}^\zeta(\zeta),
				\\
				I_{i'}^{\theta}(\theta) &= 
				\frac{1}{N_\theta}
				\sum_{m=-N_{\theta1}/2}^{N_{\theta2}/2-1}
				e^{{ \ii m (\theta-\theta_{i'})} },
				\\
				I_{j'}^{\zeta}(\zeta) &= 
				\frac{1}{N_\zeta}
				\sum_{n=-N_{\zeta1}/2}^{N_{\zeta2}/2-1}
				e^{{ \Nfp\ii n (\zeta-\zeta_{j'})} }
				.
			\end{align}
			Note that the interpolant is the only function in $\mathcal{F}^{N_\theta N_\zeta}$ which interpolates the data at the grid points, as $I_{i'}^\theta(\theta_i)=\delta_{ii'}$ and $I_{j'}^\zeta(\zeta_j)=\delta_{jj'}$. 
			
			Of course, our approximation (\ref{eq:Fourier_interpolant}) cannot (in general) be a solution to (\ref{eq:DKE_Legendre_expansion}) at all points $(\theta,\zeta)\in[0,2\pi)\times[0,2\pi/\Nfp)$. Instead, we will force that the interpolant (\ref{eq:Fourier_interpolant}) solves equation (\ref{eq:DKE_Legendre_expansion}) exactly at the equispaced grid points. Thanks to the vector space isomorphism (\ref{eq:Discrete_Fourier_Transform}) between $\vb*{f}^{(k)}$ and the discrete modes $\tilde{f}_{mn}^{(k)}$ this is equivalent to matching the discrete Fourier modes of the left and right-hand-sides of equation (\ref{eq:DKE_Legendre_expansion}).
			
			Inserting the interpolant (\ref{eq:Fourier_interpolant}) in the left-hand side of equation (\ref{eq:DKE_Legendre_expansion}) and evaluating the result at grid points gives
			\begin{align}
				& 
				\eval{\left(
					L_k f^{(k-1)} 
					+
					D_k f^{(k)}
					+
					U_k f^{(k+1)}
					\right)}_{(\theta_i,\zeta_j)}
				=
				\nonumber
				\\
				& 
				\eval{\left(
					L_k \vb*{I} \cdot \vb*{f}^{(k-1)} 
					+
					D_k \vb*{I} \cdot \vb*{f}^{(k)}
					+
					U_k \vb*{I} \cdot \vb*{f}^{(k+1)}
					\right)}_{(\theta_i,\zeta_j)}.
			\end{align}
			Here, $L_k \vb*{I}(\theta_i,\zeta_j)$, $D_k \vb*{I}(\theta_i,\zeta_j)$ and $U_k \vb*{I}(\theta_i,\zeta_j)$ are respectively the rows of $\vb*{L}_k$, $\vb*{D}_k$ and $\vb*{U}_k$ associated to the grid point $(\theta_i,\zeta_j)$. We can relate them to the actual positions they will occupy in the matrices choosing an ordenation of rows and columns. We use the ordenation that relates respectively the row $i_{\text{r}}$ and column $i_{\text{c}}$ to the grid points $(\theta_i,\zeta_j)$ and $(\theta_{i'},\zeta_{j'})$ as
			\begin{align}
				i_{\text{r}} & = 1 + i + j N_\theta,  \label{eq:Row_ordenation}\\ 
				i_{\text{c}} & = 1 + i' + j' N_\theta, \label{eq:Column_ordenation}
			\end{align}
			for $i,i'=0,1,\ldots,N_\theta-1$ and  $j,j'=0,1,\ldots,N_\zeta-1$. With this ordenation, we define the elements of the row $i_{\text{r}}$ and column $i_{\text{c}}$ given by (\ref{eq:Row_ordenation}) and (\ref{eq:Column_ordenation}) of the matrices $\vb*{L}_k$, $\vb*{D}_k$ and $\vb*{U}_k$ to be 
			\begin{align}
				\left(\vb*{L}_k\right)_{i_{\text{r}} i_{\text{c}}}
				& =
				{L_k I_{i'j'}}{(\theta_i,\zeta_j)},
				\\
				\left(\vb*{D}_k\right)_{i_{\text{r}} i_{\text{c}}}
				& =
				{D_k I_{i'j'}}{(\theta_i,\zeta_j)},
				\\
				\left(\vb*{U}_k\right)_{i_{\text{r}} i_{\text{c}}}
				& =
				{U_k I_{i'j'}}{(\theta_i,\zeta_j)}.
			\end{align}
			Explicitly,
			\begin{align}
				\eval{L_k I_{i'j'}}_{(\theta_i,\zeta_j)}
				& =
				\frac{k}{2k-1} 
				\left(
				\eval{\vb*{b} \cdot \nabla I_{i'j'}}_{(\theta_i,\zeta_j)}
				\nonumber
				\right.
				\\
				&
				+
				\frac{k-1}{2}
				\left.
				\eval{\vb*{b}\cdot\nabla \ln B}_{(\theta_i,\zeta_j)}	
				\delta_{ii'}\delta_{jj'}
				\right)
				,
				\\
				\eval{D_k I_{i'j'}}_{(\theta_i,\zeta_j)}
				& =
				-\frac{\widehat{E}_\psi}{\mean*{B^2}}
				\eval{\vb*{B}\times \nabla\psi  \cdot \nabla 
					I_{i'j'}}_{(\theta_i,\zeta_j)}
				\nonumber \\
				& +  
				\frac{k(k+1)}{2}
				\hat{\nu}\delta_{ii'}\delta_{jj'}
				,
				\\
				\eval{U_k I_{i'j'}}_{(\theta_i,\zeta_j)}
				& = 
				\frac{k+1}{2k+3} 
				\left(
				\eval{\vb*{b} \cdot \nabla  I_{i'j'}}_{(\theta_i,\zeta_j) } 
				\right. \nonumber
				\\
				& +
				\left.
				\frac{k+2}{2}
				\eval{\vb*{b}\cdot\nabla \ln B}_{(\theta_i,\zeta_j)}	
				\delta_{ii'}\delta_{jj'}
				\right)
				,
			\end{align}
			where we have used expressions (\ref{eq:Parallel_streaming_spatial_operator}) and (\ref{eq:ExB_spatial_operator}) to write
			\begin{align}
				& \eval{\vb*{b} \cdot \nabla  I_{i'j'}}_{(\theta_i,\zeta_j) }
				=
				\eval{\frac{B}{B_\zeta + \iota B_\theta}}_{(\theta_i,\zeta_j)}
				\nonumber\\
				& \qquad \times
				\left(
				\iota 
				\delta_{jj'}
				\eval{\dv{I_{i'}^{\theta}}{\theta}}_{\theta_i}
				\right.
				-
				\left.
				\delta_{ii'}
				\eval{\dv{I_{j'}^{\zeta}}{\zeta}}_{\zeta_j}
				\right),
				\\
				& \eval{\vb*{B}\times \nabla\psi  \cdot \nabla 
					I_{i'j'}}_{(\theta_i,\zeta_j)}
				=
				\eval{\frac{B^2}{B_\zeta + \iota B_\theta}}_{(\theta_i,\zeta_j)}
				\nonumber \\ 
				&
				\qquad
				\times\left(
				B_\zeta 
				\delta_{jj'}
				\eval{\dv{I_{i'}^{\theta}}{\theta}}_{\theta_i}
				\right. 
				-
				\left.
				B_\theta 
				\delta_{ii'}
				\eval{\dv{I_{j'}^{\zeta}}{\zeta}}_{\zeta_j}
				\right).
			\end{align}
			We remark that, for $k=0$, the rows of $\vb*{D}_0$ and $\vb*{U}_0$ associated to the grid point $(\theta_0,\zeta_0)=(0,0)$, are replaced by equation (\ref{eq:kernel_elimination_condition_Legendre}). Finally, each state vector $\vb*{f}^{(k)}$ for the Fourier interpolants contains the images $f^{(k)}(\theta_{i'},\zeta_{j'})$ at the grid points, ordered according to (\ref{eq:Column_ordenation}).
			
			\chapter{Convergence of monoenergetic coefficients calculated by {\DKES}}\label{sec:Appendix_DKES_Bounds}
			
			\begingroup
			
			\captionsetup[sub]{skip=-1.75pt, margin=110pt}
			
			The code {\DKES} gives an approximation to the monoenergetic coefficients as a semisum of two quantities $\widehat{D}_{ij}^- $ and $\widehat{D}_{ij}^+$ by solving a variational principle \cite{VanRij_1989}. For each coefficient, the output of {\DKES} consists on two quantities $\widehat{D}_{ij}^\mp K_{ij}$, where $K_{ij}$ are the normalization factors
			\begin{align}
				K_{ij} & :=\left(\dv{\psi}{r}\right)^{-2}, &\quad i,j \in\{1,2\},
				\\
				K_{i3} & :=  \left(\dv{\psi}{r} \right)^{-1}, &\quad i \in\{1,2\},
				\\
				K_{3j} & := \left(\dv{\psi}{r} \right)^{-1}, &\quad j \in\{1,2\},
				\\
				K_{33} & := 1, &
			\end{align} 
			to change from the radial coordinate $\psi$ to $r$. In table \ref{tab:DKES_normalization_factors}, the normalization factors for the configurations considered are listed. 
			\begin{table}[h]
				\centering
				\begin{tabular}{@{}lccc@{}}
					\toprule
					Configuration & $\dv*{\psi}{r}$ & $K_{11}$    & $K_{31}$    
					\\ \midrule
					W7X-EIM       & 0.5237  & 3.6462 & 1.9095 \\
					W7X-KJM       & 0.5132  & 3.7969 & 1.9486 \\
					CIEMAT-QI     & 0.4674  & 4.5774 & 2.1395 \\ \bottomrule
				\end{tabular}
				\caption{Normalization factors for {\DKES} results. $\dv*{\psi}{r}$ in $\text{T}\cdot\text{m}$, $K_{11}$ in $\text{T}^{-2}\cdot\text{m}^{-2}$ and $K_{31}$ in $\text{T}^{-1}\cdot\text{m}^{-1}$.}
				\label{tab:DKES_normalization_factors}
			\end{table}
			
			Apart from the normalization factors, there is still a nuance left for the parallel conductivity coefficient: the code {\DKES} computes this coefficient measured with respect to the one obtained by solving the Spitzer problem
			\begin{align}
				-\hat{\nu} \Lorentz f_{\text{Sp}} = s_3.
			\end{align}
			Using (\ref{eq:Legendre_eigenvalues}) is immediate to obtain the $1-$th Legendre mode of $f_{\text{Sp}}$
			\begin{align}
				f_{\text{Sp}}^{(1)} =  \frac{1}{\hat{\nu}} \frac{B}{B_0}
			\end{align}
			and using (\ref{eq:Gamma_33_Legendre}) we obtain its associated $\widehat{D}_{33}$ coefficient
			\begin{align}
				\widehat{D}_{33,\text{Sp}} & = \frac{2}{3\hat{\nu}} \mean*{\frac{B^2}{B_0^2}}.
			\end{align}
			Thus, the output of {\DKES} for the parallel conductivity coefficient has to be compared against the deviation $(\widehat{D}_{33} - \widehat{D}_{33,\text{Sp}})$.

			From the output of {\DKES}, the diagonal elements $\widehat{D}_{ii}^{\pm}$ satisfy $\widehat{D}_{ii}^{-} \ge \widehat{D}_{ii} \ge\widehat{D}_{ii}^{+}$ and allow to compute bounds for $\widehat{D}_{ij}$
			\begin{align}
				\frac{\widehat{D}_{ij}^{-} + \widehat{D}_{ij}^{+}}{2}
				-
				\Delta_{ij}
				\le
				\widehat{D}_{ij}
				\le
				\frac{\widehat{D}_{ij}^{-} + \widehat{D}_{ij}^{+}}{2}
				+
				\Delta_{ij}
			\end{align}
			and $\Delta_{ij} = \sqrt{(\widehat{D}_{ii}^{-} - \widehat{D}_{ii}^{+})(\widehat{D}_{jj}^{-} - \widehat{D}_{jj}^{+})} /2 $.  
			
			In figures \ref{fig:DKES_Convergence_W7X_EIM_Er_0}, \ref{fig:DKES_Convergence_W7X_EIM_Er_3e-4}, \ref{fig:DKES_Convergence_W7X_KJM_Er_0}, \ref{fig:DKES_Convergence_W7X_KJM_Er_3e-4}, \ref{fig:DKES_Convergence_CIEMAT_QI_Er_0} and \ref{fig:DKES_Convergence_CIEMAT_QI_Er_1e-3} the convergence study for selecting {\DKES} resolutions is shown. In the code {\DKES} the number of Legendre modes used are specified by $N_\xi$. In order to select the number of Fourier modes in the Boozer angles $(\theta,\zeta)$ that {\DKES} uses, an integer called ``coupling order'' must be specified. Using figures \ref{subfig:DKES_D31_convergence_Legendre_W7X_EIM_0200_Erho_0_Detail}, \ref{subfig:DKES_D31_convergence_Legendre_W7X_EIM_0200_Erho_3e-4_Detail}, \ref{subfig:DKES_D31_convergence_Legendre_W7X_KJM_0204_Erho_0_Detail}, \ref{subfig:DKES_D31_convergence_Legendre_W7X_KJM_0204_Erho_3e-4_Detail}, \ref{subfig:DKES_D31_convergence_Legendre_CIEMAT_QI_0250_Erho_0_Detail} and \ref{subfig:DKES_D31_convergence_Legendre_CIEMAT_QI_0250_Erho_1e-3_Detail}, the number of Legendre modes $N_\xi$ is selected so that it satisfies convergence condition (i) using the region $\mathcal{R}_\epsilon$ for each case. After that, using \ref{subfig:DKES_D31_convergence_Coupling_parameter_W7X_EIM_0200_Erho_0}, \ref{subfig:DKES_D31_convergence_Coupling_parameter_W7X_EIM_0200_Erho_3e-4}, \ref{subfig:DKES_D31_convergence_Coupling_parameter_W7X_KJM_0204_Erho_0}, \ref{subfig:DKES_D31_convergence_Coupling_parameter_W7X_KJM_0204_Erho_3e-4}, \ref{subfig:DKES_D31_convergence_Coupling_parameter_CIEMAT_QI_0250_Erho_0} and \ref{subfig:DKES_D31_convergence_Coupling_parameter_CIEMAT_QI_0250_Erho_1e-3}, we select the minimum value of the coupling order for which the calculation with the selected value of $N_\xi$ satisfies convergence condition (ii).

			\begin{figure}[h]
				\centering	
				\begin{subfigure}[t]{0.45\textwidth}
					\tikzsetnextfilename{DKES-Convergence-Legendre-W7X-EIM-s0200-Er-0-D31-Detail} 
					\includegraphics{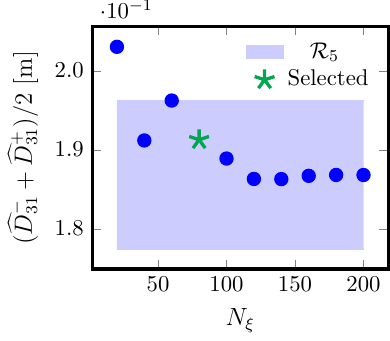} 
					\caption{}
					\label{subfig:DKES_D31_convergence_Legendre_W7X_EIM_0200_Erho_0_Detail}
				\end{subfigure}
				\begin{subfigure}[t]{0.45\textwidth}
					\tikzsetnextfilename{DKES-Convergence-theta-zeta-W7X-EIM-s0200-Er-0-D31}
					\includegraphics{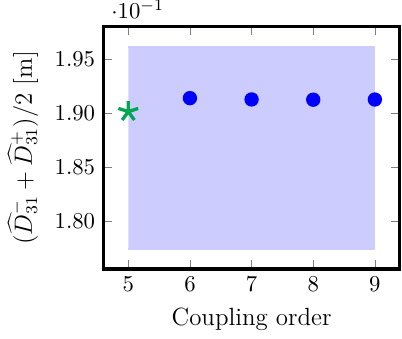} 
					\caption{}
					\label{subfig:DKES_D31_convergence_Coupling_parameter_W7X_EIM_0200_Erho_0}
				\end{subfigure}

				\caption{Convergence of $(\widehat{D}_{31}^- + \widehat{D}_{31}^+) /2$ computed with {\DKES} for W7X-EIM at the surface labelled by $\psi/\psi_{\text{lcfs}}=0.200$, for $\hat{\nu}(v)=10^{-5}$ $\text{m}^{-1}$ and $\widehat{E}_r(v)=0$ $\text{V}\cdot\text{s}/\text{m}^2$. (a) Convergence with $N_\xi$ for coupling order = 9. (b) Convergence with the coupling order for $N_\xi= 80$.}
				\label{fig:DKES_Convergence_W7X_EIM_Er_0}
			\end{figure}
			\begin{figure}[h]
				\centering
				\begin{subfigure}[t]{0.45\textwidth}
					\tikzsetnextfilename{DKES-Convergence-Legendre-W7X-EIM-s0200-Er-3e-4-D31-Detail}
					\includegraphics{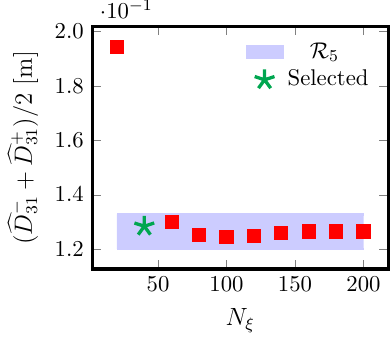}
					\caption{}
					\label{subfig:DKES_D31_convergence_Legendre_W7X_EIM_0200_Erho_3e-4_Detail}
				\end{subfigure}
				\begin{subfigure}[t]{0.45\textwidth}
					\tikzsetnextfilename{DKES-Convergence-theta-zeta-W7X-EIM-s0200-Er-3e-4-D31}
					\includegraphics{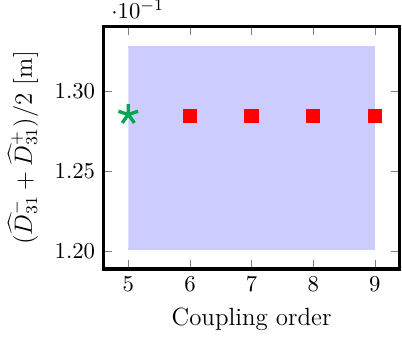}
					\caption{}
					\label{subfig:DKES_D31_convergence_Coupling_parameter_W7X_EIM_0200_Erho_3e-4}
				\end{subfigure}
				
				\caption{Convergence of $(\widehat{D}_{31}^- + \widehat{D}_{31}^+) /2$ computed with {\DKES} for W7X-EIM at the surface labelled by $\psi/\psi_{\text{lcfs}}=0.200$, for $\hat{\nu}(v)=10^{-5}$ $\text{m}^{-1}$ and $\widehat{E}_r(v)=3\cdot 10^{-4}$ $\text{V}\cdot\text{s}/\text{m}^2$. (a) Convergence with $N_\xi$ for coupling order = 9. (b) Convergence with the coupling order for $N_\xi= 40$.}
				\label{fig:DKES_Convergence_W7X_EIM_Er_3e-4}
			\end{figure}
			
			\begin{figure}[h]
				\centering	
				\begin{subfigure}[t]{0.45\textwidth}
					\tikzsetnextfilename{DKES-Convergence-Legendre-W7X-KJM-s0204-Er-0-D31-Detail}
					\includegraphics{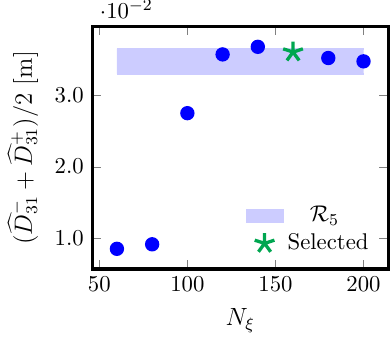}
					\caption{}
					\label{subfig:DKES_D31_convergence_Legendre_W7X_KJM_0204_Erho_0_Detail}
				\end{subfigure}
				\begin{subfigure}[t]{0.45\textwidth}
					\tikzsetnextfilename{DKES-Convergence-theta-zeta-W7X-KJM-s0204-Er-0-D31}
					\includegraphics{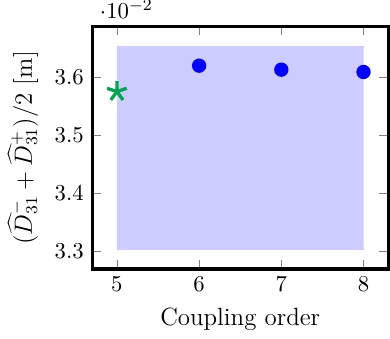}
					\caption{}
					\label{subfig:DKES_D31_convergence_Coupling_parameter_W7X_KJM_0204_Erho_0}
				\end{subfigure}

				\caption{Convergence of $(\widehat{D}_{31}^- + \widehat{D}_{31}^+) /2$ computed with {\DKES} for W7X-KJM at the surface labelled by $\psi/\psi_{\text{lcfs}}=0.204$, for $\hat{\nu}(v)=10^{-5}$ $\text{m}^{-1}$ and $\widehat{E}_r(v)=0$ $\text{V}\cdot\text{s}/\text{m}^2$. (a) Convergence with $N_\xi$ for coupling order = 8. (b) Convergence with the coupling order for $N_\xi=160 $.}
				\label{fig:DKES_Convergence_W7X_KJM_Er_0}
			\end{figure}
			\begin{figure}[h]
				\centering
				\begin{subfigure}[t]{0.45\textwidth}
					\tikzsetnextfilename{DKES-Convergence-Legendre-W7X-KJM-s0204-Er-3e-4-D31-Detail}
					\includegraphics{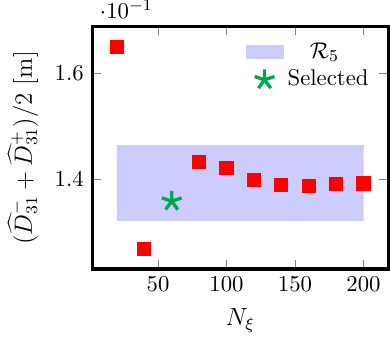}
					\caption{}
					\label{subfig:DKES_D31_convergence_Legendre_W7X_KJM_0204_Erho_3e-4_Detail}
				\end{subfigure}
				\begin{subfigure}[t]{0.45\textwidth}
					\tikzsetnextfilename{DKES-Convergence-theta-zeta-W7X-KJM-s0204-Er-3e-4-D31}
					\includegraphics{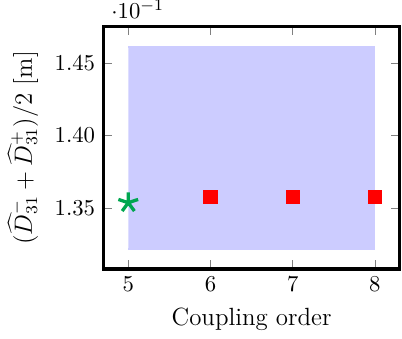}
					\caption{}
					\label{subfig:DKES_D31_convergence_Coupling_parameter_W7X_KJM_0204_Erho_3e-4}
				\end{subfigure}
				
				\caption{Convergence of $(\widehat{D}_{31}^- + \widehat{D}_{31}^+) /2$ computed with {\DKES} for W7X-KJM at the surface labelled by $\psi/\psi_{\text{lcfs}}=0.204$, for $\hat{\nu}(v)=10^{-5}$ $\text{m}^{-1}$ and $\widehat{E}_r(v)=3\cdot 10^{-4}$ $\text{V}\cdot\text{s}/\text{m}^2$. (a) Convergence with $N_\xi$ for coupling order = 7. (b) Convergence with the coupling order for $N_\xi= 60$.}
				\label{fig:DKES_Convergence_W7X_KJM_Er_3e-4}
			\end{figure}
			\begin{figure}[h]
				\centering	
				\begin{subfigure}[t]{0.45\textwidth}
					\tikzsetnextfilename{DKES-Convergence-Legendre-CIEMAT-QI-s0250-Er-0-D31-Detail}
					\includegraphics{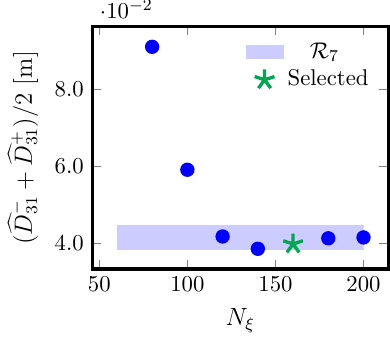}
					\caption{}
					\label{subfig:DKES_D31_convergence_Legendre_CIEMAT_QI_0250_Erho_0_Detail}
				\end{subfigure}
				\begin{subfigure}[t]{0.45\textwidth}
					\tikzsetnextfilename{DKES-Convergence-theta-zeta-CIEMAT-QI-s0250-Er-0-D31}
					\includegraphics{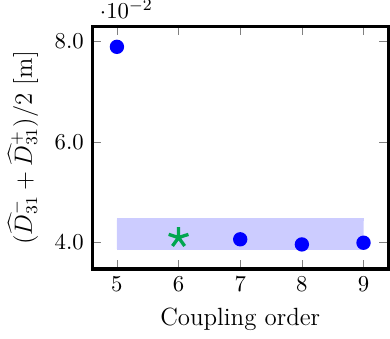}
					\caption{}
					\label{subfig:DKES_D31_convergence_Coupling_parameter_CIEMAT_QI_0250_Erho_0}
				\end{subfigure}

				\caption{Convergence of $(\widehat{D}_{31}^- + \widehat{D}_{31}^+) /2$ computed with {\DKES} for CIEMAT-QI at the surface labelled by $\psi/\psi_{\text{lcfs}}=0.250$, for $\hat{\nu}(v)=10^{-5}$ $\text{m}^{-1}$ and $\widehat{E}_r(v)=0$ $\text{V}\cdot\text{s}/\text{m}^2$. (a) Convergence with $N_\xi$ for coupling order = 9. (b) Convergence with the coupling order for $N_\xi= 160$.}
				\label{fig:DKES_Convergence_CIEMAT_QI_Er_0}
			\end{figure}
			\begin{figure}[h]
				\centering	
				\begin{subfigure}[t]{0.45\textwidth}
					\tikzsetnextfilename{DKES-Convergence-Legendre-CIEMAT-QI-s0250-Er-1e-3-D31-Detail}
					\includegraphics{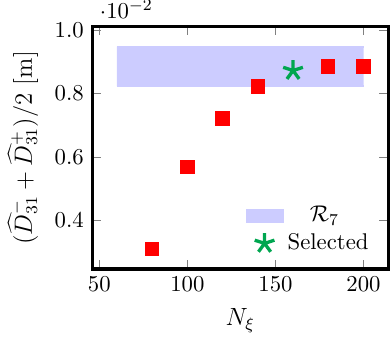}
					\caption{}
					\label{subfig:DKES_D31_convergence_Legendre_CIEMAT_QI_0250_Erho_1e-3_Detail}
				\end{subfigure}
				\begin{subfigure}[t]{0.45\textwidth}
					\tikzsetnextfilename{DKES-Convergence-theta-zeta-CIEMAT-QI-s0250-Er-1e-3-D31}
					\includegraphics{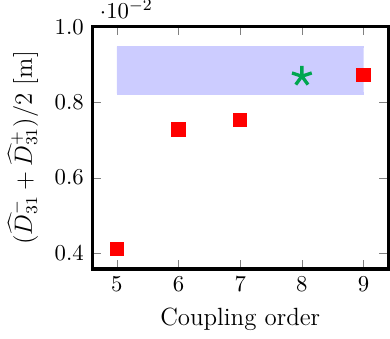}
					\caption{}
					\label{subfig:DKES_D31_convergence_Coupling_parameter_CIEMAT_QI_0250_Erho_1e-3}
				\end{subfigure}

				\caption{Convergence of $(\widehat{D}_{31}^- + \widehat{D}_{31}^+) /2$ computed with {\DKES} for CIEMAT-QI at the surface labelled by $\psi/\psi_{\text{lcfs}}=0.250$, for $\hat{\nu}(v)=10^{-5}$ $\text{m}^{-1}$ and $\widehat{E}_r(v)=10^{-3}$ $\text{V}\cdot\text{s}/\text{m}^2$. (a) Convergence with $N_\xi$ for coupling order = 9. (b) Convergence with the coupling order for $N_\xi= 160$.}
				\label{fig:DKES_Convergence_CIEMAT_QI_Er_1e-3}
			\end{figure}
			\endgroup

			\chapter{Derivatives of flux surface averaged quantities}\label{sec:Appendix_FSA_derivative}
			In this appendix, we will derive a useful expression for computing the derivative of a flux surface averaged quantity $\mean*{f}$ with respect to a parameter $\eta$ upon which $f$ and/or the flux surface average operation depends. First, we repeat the definition of the flux surface average operation in Boozer coordinates (\ref{eq:FSA_Boozer}). Namely,    
			\begin{align}
				\mean*{f}
				=
				\left(\dv{V}{\psi}\right)^{-1}
				\oint\oint
				f
				\sqrt{g}
				\dd{\theta}\dd{\zeta}
				.
				\tag{\ref{eq:FSA_Boozer}}
			\end{align}
			
			Deriving (\ref{eq:FSA_Boozer}) with respect to $\eta$ yields 
			\begin{align}
				\pdv{\eta}
				\mean*{f}
				& = 
				\mean*{\pdv{f}{\eta}}
				-
				2
				\mean*{
					\left(f-\mean*{f}\right)
					\pdv{\ln B}{\eta}
				}    
				\nonumber \\	
				& = 
				\mean*{\pdv{f}{\eta}}
				-
				2
				\mean*{
					\left(\pdv{\ln B}{\eta} -\mean*{\pdv{\ln B}{\eta} }\right)
					f
				},
				\label{eq:FSA_derivative_eta}
			\end{align}
			where we have used $\sqrt{g} =(B_\zeta + \iota B_\theta)/B^2$, 

			\begin{align*}
				\dv{V}{\psi}
				&
				=
				(B_\zeta + \iota B_\theta)
				\Mmean{ \frac{1}{B^2} } 
				,
			\end{align*}
			and
			\begin{align*}
				\pdv{\eta}
				\ln( 
				\sqrt{g}
				\left(\dv{V}{\psi}\right)^{-1}
				)
				& 
				=
				\pdv{\eta}
				\ln(\frac{1}{B^2})
				-
				\pdv{\eta}
				\ln(
				\Mmean{ \frac{1}{B^2} }    	
				)
				\\ 
				& 
				=
				-2
				\left(
				\pdv{\ln B}{\eta}
				-
				\mean*{
					\pdv{\ln B}{\eta}
				}
				\right)
				.
			\end{align*}
			
			Here, we have denoted
			\begin{align*}
				\mmean{f}:= \oint\oint f \dd{\theta}\dd{\zeta}.
			\end{align*}
			Note that, in spite of the fact that we have used Boozer coordinates for the intermediate steps, identity (\ref{eq:FSA_derivative_eta}) is valid for any set of coordinates which parametrize the flux surface. Besides, the dependence of the flux surface on the parameter $\eta$ is encapsulated on the dependence of $B$ on $\eta$.
			
			\FloatBarrier
			
			\chapter{Normalization of the monoenergetic coefficients}\label{sec:Appendix_monoenergetic_normalization}
			The monoenergetic coefficients $\Dij{11}$ and $\Dij{31}$ defined in section \ref{sec:DKE}, are related to their normalized versions $D_{11}^*$ and $D_{31}^*$ defined in \cite{Beidler_2011} as
			\begin{align}
				D_{11}^* & = \frac{ 8R  B_0^2 \iota}{\pi} K_{11} \Dij{11}, 
				\\
				D_{31}^* & =  \iota B_0 \sqrt{\frac{r_{\text{lcfs}}}{R}} K_{31} \Dij{31}. 
			\end{align}
			Here, $R$ and $r_{\text{lcfs}}$ are, respectively, the major and minor radius of the device. $B_0$ is a reference value for $B$ on the flux surface and $\iota$ is the rotational transform. The normalization factors $K_{31}=\dv*{r}{\psi}$, $K_{11}=K_{31}^2$ change from the flux surface label $\psi$ to $r=r_{\text{lcfs}}\sqrt{\psi/\psi_{\text{lcfs}}}$ where we recall that $2\pi\psi$ is the toroidal flux of $\vb*{B}$ enclosed by the flux surface and $\psi_{\text{lcfs}}$ is the label of the last closed flux surface. 
			
			\chapter{Evaluation of proxies at higher collisionalities}
			\label{sec:Appendix_Correlations_nu_scan}\begingroup
			
			\captionsetup[sub]{skip=-1.75pt, margin=90pt}
			In this appendix, we illustrate that the conclusions about the efficiency of the proxies discussed in section \ref{sec:Correlations} for $\hat{\nu}=\num{e-5}$ are applicable to higher collisionalities in the interval $\hat{\nu}\in[\num{e-5},\num{e-3}]$. We can check this applicability for the effective ripple $\epseff$ by comparing figures \ref{fig:Correlation_epseff_nu_scan_Er_0} and \ref{fig:Correlation_epseff_nu_scan_Er_1e-3} with figure \ref{fig:Correlation_epseff}. By comparing figures \ref{fig:Correlation_VBB_nu_scan_Er_0} and \ref{fig:Correlation_VBB_nu_scan_Er_1e-3} with figure \ref{fig:Correlation_VBB}, we can verify that the conclusions extracted for $\sigma^2\left(\Bmin^{\text{r}}\right)$ in section \ref{sec:Correlations} are applicable to higher collisionalities. We can check the similarity of the results for $\sigma^2\left(B(\theta,0)\right)$ at different collisionalities by comparing \ref{fig:Correlation_VB0_nu_scan_Er_0} and \ref{fig:Correlation_VB0_nu_scan_Er_1e-3} with figure \ref{fig:Correlation_VB0}. We conclude that the results for the fast ion proxy $\GammaC$ at $\hat{\nu}=\num{e-5}$ are representative of higher collisionalities by comparing figures \ref{fig:Correlation_GMC_nu_scan_Er_0} and \ref{fig:Correlation_GMC_nu_scan_Er_1e-3} with the left columns of, respectively, figures 
			\ref{fig:Correlation_GMC_GMA_Er_0} and \ref{fig:Correlation_GMC_GMA_Er_1e-3}. Complementarily, for $\GammaAlpha$ we can check the applicability of the conclusions for higher collisionalities by comparing figures \ref{fig:Correlation_GMC_nu_scan_Er_0} and \ref{fig:Correlation_GMC_nu_scan_Er_1e-3} with the right columns of, respectively, figures 
			\ref{fig:Correlation_GMC_GMA_Er_0} and \ref{fig:Correlation_GMC_GMA_Er_1e-3}.
			\begin{figure*}[h]
				\centering
				%
				
				%
				\tikzsetnextfilename{D31_vs_D11_epseff_0100_Er_0_nu_3}
				\begin{subfigure}[t]{0.45\textwidth}	 		
					\includegraphics{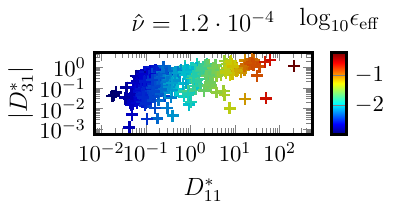}
					\caption{}
					\label{subfig:D31_vs_D11_epseff_0100_Er_0_nu_3}
				\end{subfigure}    
				\tikzsetnextfilename{D31_vs_epseff_0100_Er_0_nu_3}
				\begin{subfigure}[t]{0.45\textwidth}	 		
					\includegraphics{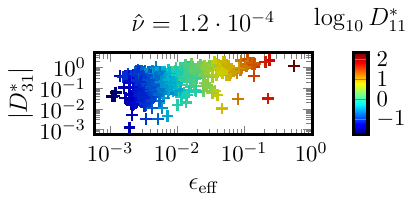}
					\caption{}
					\label{subfig:D31_vs_epseff_0100_Er_0_nu_3}
				\end{subfigure} 

				\tikzsetnextfilename{D31_vs_D11_epseff_0100_Er_0_nu_4}
				\begin{subfigure}[t]{0.45\textwidth}	 		
					\includegraphics{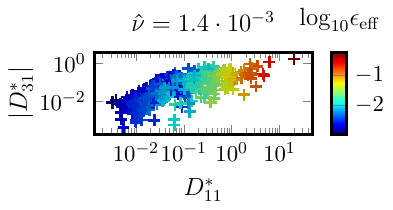}
					\caption{}
					\label{subfig:D31_vs_D11_epseff_0100_Er_0_nu_4}
				\end{subfigure}    
				\tikzsetnextfilename{D31_vs_epseff_0100_Er_0_nu_4}
				\begin{subfigure}[t]{0.45\textwidth}	 		
					\includegraphics{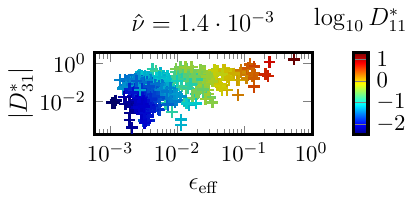}
					\caption{}
					\label{subfig:D31_vs_epseff_0100_Er_0_nu_4}
				\end{subfigure} 
				
				\caption{Relation of the radial transport $D_{11}^*$ and bootstrap current $D_{31}^*$ coefficients with $\epsilon_{\text{eff}}$ for several collisionalities $\hat{\nu}$ (in $\text{m}^{-1}$) and $\widehat{E}_r=0$.} 
				\label{fig:Correlation_epseff_nu_scan_Er_0}
			\end{figure*}%
			\begin{figure*}[h]
				\centering
				%
				
				%
				\tikzsetnextfilename{D31_vs_D11_epseff_0100_Er_1e-3_nu_3}
				\begin{subfigure}[t]{0.45\textwidth}	 		
					\includegraphics{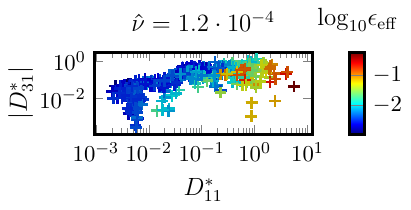}
					\caption{}
					\label{subfig:D31_vs_D11_epseff_0100_Er_1e-3_nu_3}
				\end{subfigure}    
				\tikzsetnextfilename{D31_vs_epseff_0100_Er_1e-3_nu_3}
				\begin{subfigure}[t]{0.45\textwidth}	 		
					\includegraphics{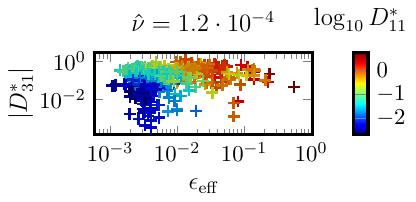}
					\caption{}
					\label{subfig:D31_vs_epseff_0100_Er_1e-3_nu_3}
				\end{subfigure} 

				\tikzsetnextfilename{D31_vs_D11_epseff_0100_Er_1e-3_nu_4}
				\begin{subfigure}[t]{0.45\textwidth}	 		
					\includegraphics{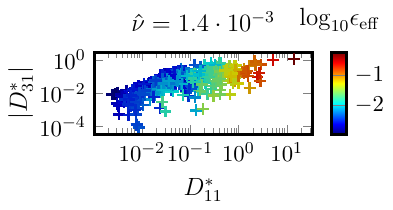}
					\caption{}
					\label{subfig:D31_vs_D11_epseff_0100_Er_1e-3_nu_4}
				\end{subfigure}    
				\tikzsetnextfilename{D31_vs_epseff_0100_Er_1e-3_nu_4}
				\begin{subfigure}[t]{0.45\textwidth}	 		
					\includegraphics{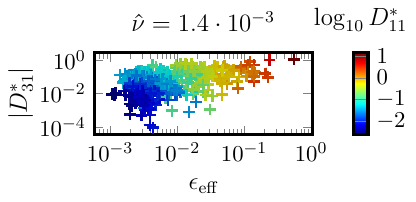}
					\caption{}
					\label{subfig:D31_vs_epseff_0100_Er_1e-3_nu_4}
				\end{subfigure} 
				\caption{Relation of the radial transport $D_{11}^*$ and bootstrap current $D_{31}^*$ coefficients with $\epsilon_{\text{eff}}$ for several collisionalities $\hat{\nu}$ (in $\text{m}^{-1}$) and $\widehat{E}_r\ne0$.} 
				\label{fig:Correlation_epseff_nu_scan_Er_1e-3}
			\end{figure*}
			%
			\begin{figure*}[h]
				\centering
				%
				
				%
				\tikzsetnextfilename{D31_vs_D11_KN_VBB_0100_Er_0_nu_3}
				\begin{subfigure}[t]{0.45\textwidth}	 		
					\includegraphics{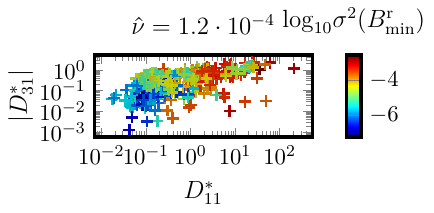}
					\caption{}
					\label{subfig:D31_vs_D11_KN_VBB_0100_Er_0_nu_3}
				\end{subfigure}    
				\tikzsetnextfilename{D31_vs_KN_VBB_0100_Er_0_nu_3}
				\begin{subfigure}[t]{0.45\textwidth}	 		
					\includegraphics{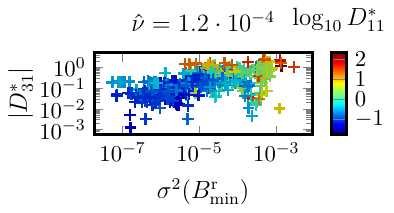}
					\caption{}
					\label{subfig:D31_vs_KN_VBB_0100_Er_0_nu_3}
				\end{subfigure} 

				\tikzsetnextfilename{D31_vs_D11_KN_VBB_0100_Er_0_nu_4}
				\begin{subfigure}[t]{0.45\textwidth}	 		
					\includegraphics{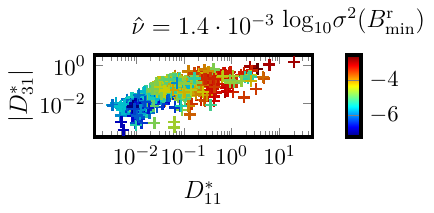}
					\caption{}
					\label{subfig:D31_vs_D11_KN_VBB_0100_Er_0_nu_4}
				\end{subfigure}    
				\tikzsetnextfilename{D31_vs_KN_VBB_0100_Er_0_nu_4}
				\begin{subfigure}[t]{0.45\textwidth}	 		
					\includegraphics{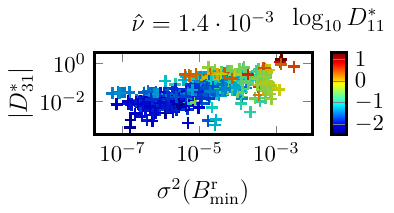}
					\caption{}
					\label{subfig:D31_vs_KN_VBB_0100_Er_0_nu_4}
				\end{subfigure} 
				
				\caption{Relation of the radial transport $D_{11}^*$ and bootstrap current $D_{31}^*$ coefficients with $\sigma^2(\Bmin^{\text{r}})$ for several collisionalities $\hat{\nu}$ (in $\text{m}^{-1}$) and $\widehat{E}_r=0$.} 
				\label{fig:Correlation_VBB_nu_scan_Er_0}
			\end{figure*} 
			\begin{figure*}[h]
				\centering
				%
				
				%
				\tikzsetnextfilename{D31_vs_D11_KN_VBB_0100_Er_1e-3_nu_3}
				\begin{subfigure}[t]{0.45\textwidth}	 		
					\includegraphics{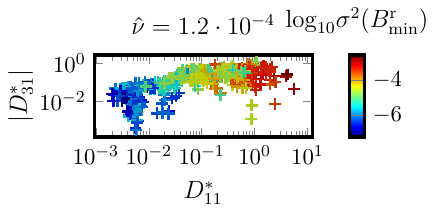}
					\caption{}
					\label{subfig:D31_vs_D11_KN_VBB_0100_Er_1e-3_nu_3}
				\end{subfigure}    
				\tikzsetnextfilename{D31_vs_KN_VBB_0100_Er_1e-3_nu_3}
				\begin{subfigure}[t]{0.45\textwidth}	 		
					\includegraphics{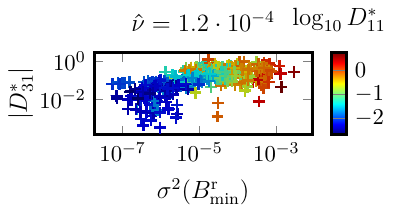}
					\caption{}
					\label{subfig:D31_vs_KN_VBB_0100_Er_1e-3_nu_3}
				\end{subfigure} 

				\tikzsetnextfilename{D31_vs_D11_KN_VBB_0100_Er_1e-3_nu_4}
				\begin{subfigure}[t]{0.45\textwidth}	 		
				    \includegraphics{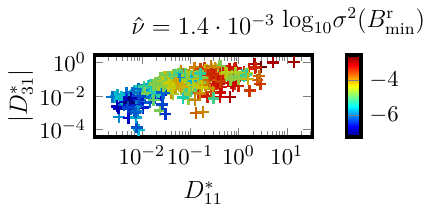}
					\caption{}
					\label{subfig:D31_vs_D11_KN_VBB_0100_Er_1e-3_nu_4}
				\end{subfigure}    
				\tikzsetnextfilename{D31_vs_KN_VBB_0100_Er_1e-3_nu_4}
				\begin{subfigure}[t]{0.45\textwidth}	 		
					\includegraphics{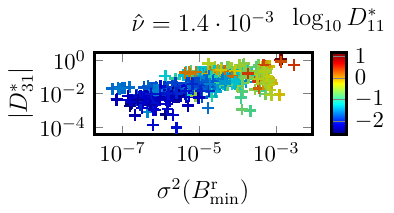}
					\caption{}
					\label{subfig:D31_vs_KN_VBB_0100_Er_1e-3_nu_4}
				\end{subfigure} 
				
				\caption{Relation of the radial transport $D_{11}^*$ and bootstrap current $D_{31}^*$ coefficients with $\sigma^2(\Bmin^{\text{r}})$ for several collisionalities $\hat{\nu}$ (in $\text{m}^{-1}$) and $\widehat{E}_r\ne0$.} 
				\label{fig:Correlation_VBB_nu_scan_Er_1e-3}
			\end{figure*}
			%
			\begin{figure*}[h]
				\centering
				%
				
				%
				\tikzsetnextfilename{D31_vs_D11_KN_VB0_0100_Er_0_nu_3}
				\begin{subfigure}[t]{0.45\textwidth}	 		
					\includegraphics{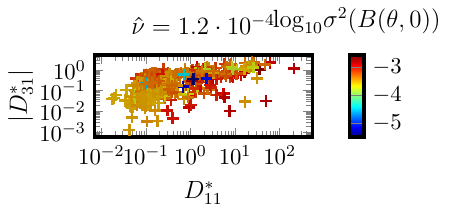}
					\caption{}
					\label{subfig:D31_vs_D11_KN_VB0_0100_Er_0_nu_3}
				\end{subfigure}    
				\tikzsetnextfilename{D31_vs_KN_VB0_0100_Er_0_nu_3}
				\begin{subfigure}[t]{0.45\textwidth}	 		
					\includegraphics{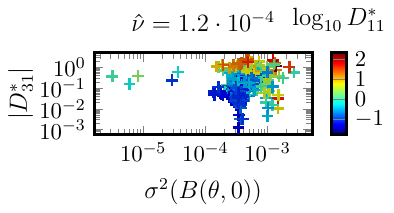}
					\caption{}
					\label{subfig:D31_vs_KN_VB0_0100_Er_0_nu_3}
				\end{subfigure} 

				\tikzsetnextfilename{D31_vs_D11_KN_VB0_0100_Er_0_nu_4}
				\begin{subfigure}[t]{0.45\textwidth}	 		
					\includegraphics{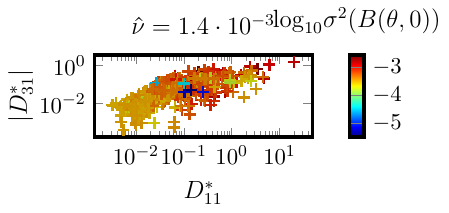}
					\caption{}
					\label{subfig:D31_vs_D11_KN_VB0_0100_Er_0_nu_4}
				\end{subfigure}    
				\tikzsetnextfilename{D31_vs_KN_VB0_0100_Er_0_nu_4}
				\begin{subfigure}[t]{0.45\textwidth}	 		
					\includegraphics{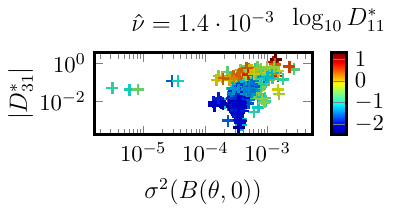}
					\caption{}
					\label{subfig:D31_vs_KN_VB0_0100_Er_0_nu_4}
				\end{subfigure} 
				
				\caption{Relation of the radial transport $D_{11}^*$ and bootstrap current $D_{31}^*$ coefficients with $\sigma^2(B(\theta,0))$ for several collisionalities $\hat{\nu}$ (in $\text{m}^{-1}$) and $\widehat{E}_r=0$.} 
				\label{fig:Correlation_VB0_nu_scan_Er_0}
			\end{figure*} 
			\begin{figure*}[h]
				\centering
				%
				
				%
				\tikzsetnextfilename{D31_vs_D11_KN_VB0_0100_Er_1e-3_nu_3}
				\begin{subfigure}[t]{0.45\textwidth}	 		
					\includegraphics{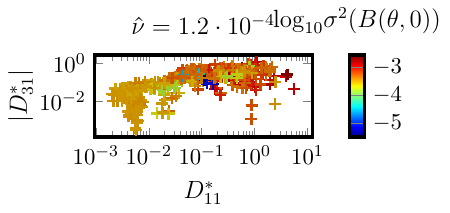}
					\caption{}
					\label{subfig:D31_vs_D11_KN_VB0_0100_Er_1e-3_nu_3}
				\end{subfigure}    
				\tikzsetnextfilename{D31_vs_KN_VB0_0100_Er_1e-3_nu_3}
				\begin{subfigure}[t]{0.45\textwidth}	 		
					\includegraphics{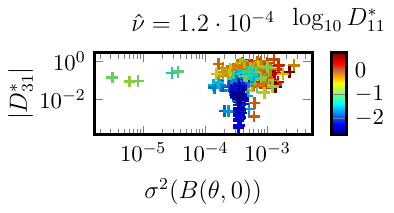}
					\caption{}
					\label{subfig:D31_vs_KN_VB0_0100_Er_1e-3_nu_3}
				\end{subfigure} 

				\tikzsetnextfilename{D31_vs_D11_KN_VB0_0100_Er_1e-3_nu_4}
				\begin{subfigure}[t]{0.45\textwidth}	 		
					\includegraphics{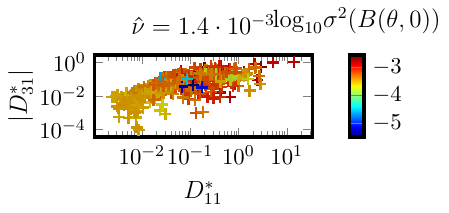}
					\caption{}
					\label{subfig:D31_vs_D11_KN_VB0_0100_Er_1e-3_nu_4}
				\end{subfigure}    
				\tikzsetnextfilename{D31_vs_KN_VB0_0100_Er_1e-3_nu_4}
				\begin{subfigure}[t]{0.45\textwidth}	 		
					\includegraphics{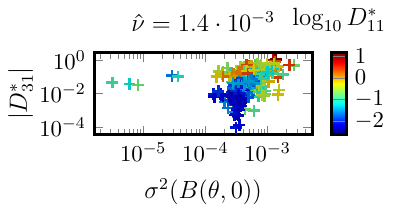}
					\caption{}
					\label{subfig:D31_vs_KN_VB0_0100_Er_1e-3_nu_4}
				\end{subfigure} 
				
				\caption{Relation of the radial transport $D_{11}^*$ and bootstrap current $D_{31}^*$ coefficients with $\sigma^2(B(\theta,0))$ for several collisionalities $\hat{\nu}$ (in $\text{m}^{-1}$) and $\widehat{E}_r\ne0$.} 
				\label{fig:Correlation_VB0_nu_scan_Er_1e-3}
			\end{figure*}
			%
			\begin{figure*}[h]
				\centering
				%
				
				%
				\tikzsetnextfilename{D31_vs_D11_GMC_0100_Er_0_nu_3}
				\begin{subfigure}[t]{0.45\textwidth}	 		
					\includegraphics{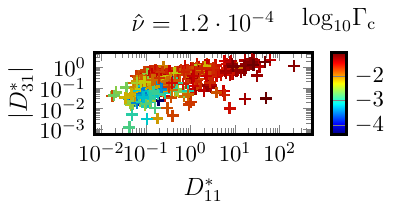}
					\caption{}
					\label{subfig:D31_vs_D11_GMC_0100_Er_0_nu_3}
				\end{subfigure}    
				\tikzsetnextfilename{D31_vs_GMC_0100_Er_0_nu_3}
				\begin{subfigure}[t]{0.45\textwidth}	 		
					\includegraphics{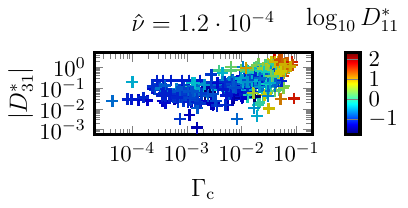}
					\caption{}
					\label{subfig:D31_vs_GMC_0100_Er_0_nu_3}
				\end{subfigure} 

				\tikzsetnextfilename{D31_vs_D11_GMC_0100_Er_0_nu_4}
				\begin{subfigure}[t]{0.45\textwidth}	 		
					\includegraphics{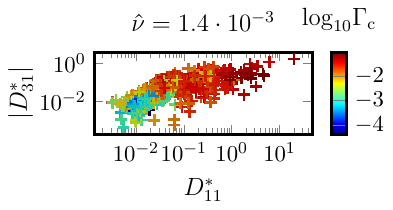}
					\caption{}
					\label{subfig:D31_vs_D11_GMC_0100_Er_0_nu_4}
				\end{subfigure}    
				\tikzsetnextfilename{D31_vs_GMC_0100_Er_0_nu_4}
				\begin{subfigure}[t]{0.45\textwidth}	 		
					\includegraphics{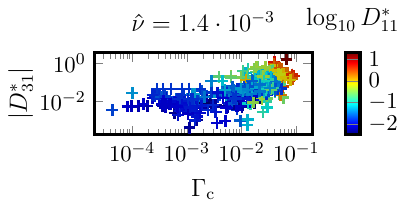}
					\caption{}
					\label{subfig:D31_vs_GMC_0100_Er_0_nu_4}
				\end{subfigure} 
				
				\caption{Relation of the radial transport $D_{11}^*$ and bootstrap current $D_{31}^*$ coefficients with $\GammaC$ for several collisionalities $\hat{\nu}$ (in $\text{m}^{-1}$) and $\widehat{E}_r=0$.} 
				\label{fig:Correlation_GMC_nu_scan_Er_0}
			\end{figure*} 
			\begin{figure*}[h]
				\centering
				%
				%
				%
				\tikzsetnextfilename{D31_vs_D11_GMC_0100_Er_1e-3_nu_3}
				\begin{subfigure}[t]{0.45\textwidth}	 		
					\includegraphics{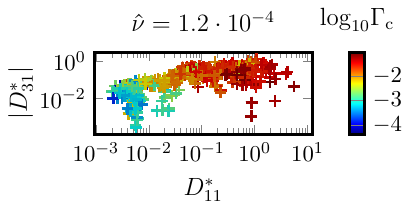}
					\caption{}
					\label{subfig:D31_vs_D11_GMC_0100_Er_1e-3_nu_3}
				\end{subfigure}    
				\tikzsetnextfilename{D31_vs_GMC_0100_Er_1e-3_nu_3}
				\begin{subfigure}[t]{0.45\textwidth}	 		
					\includegraphics{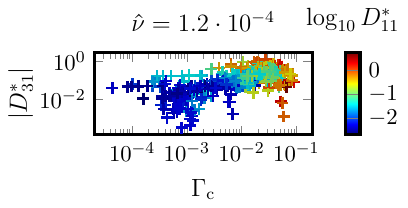}
					\caption{}
					\label{subfig:D31_vs_GMC_0100_Er_1e-3_nu_3}
				\end{subfigure} 

				\tikzsetnextfilename{D31_vs_D11_GMC_0100_Er_1e-3_nu_4}
				\begin{subfigure}[t]{0.45\textwidth}	 		
					\includegraphics{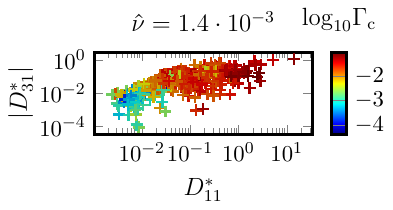}
					\caption{}
					\label{subfig:D31_vs_D11_GMC_0100_Er_1e-3_nu_4}
				\end{subfigure}    
				\tikzsetnextfilename{D31_vs_GMC_0100_Er_1e-3_nu_4}
				\begin{subfigure}[t]{0.45\textwidth}	 		
					\includegraphics{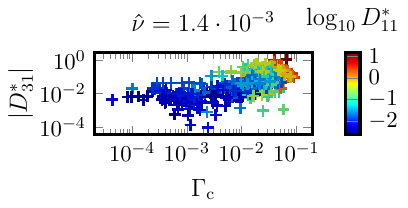}
					\caption{}
					\label{subfig:D31_vs_GMC_0100_Er_1e-3_nu_4}
				\end{subfigure} 
				
				\caption{Relation of the radial transport $D_{11}^*$ and bootstrap current $D_{31}^*$ coefficients with $\GammaC$ for several collisionalities $\hat{\nu}$ (in $\text{m}^{-1}$) and $\widehat{E}_r\ne0$.} 
				\label{fig:Correlation_GMC_nu_scan_Er_1e-3}
			\end{figure*}

			%
			\begin{figure*}[h]
				\centering
				%
				
				%
				\tikzsetnextfilename{D31_vs_D11_GMA_0100_Er_0_nu_3}
				\begin{subfigure}[t]{0.45\textwidth}	 		
					\includegraphics{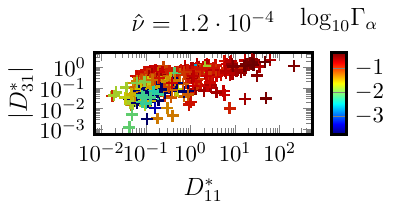}
					\caption{}
					\label{subfig:D31_vs_D11_GMA_0100_Er_0_nu_3}
				\end{subfigure}    
				\tikzsetnextfilename{D31_vs_GMA_0100_Er_0_nu_3}
				\begin{subfigure}[t]{0.45\textwidth}	 		
					\includegraphics{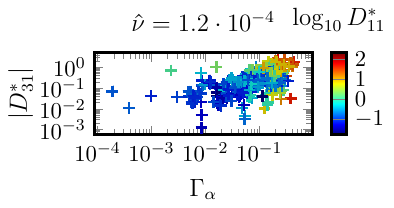}
					\caption{}
					\label{subfig:D31_vs_GMA_0100_Er_0_nu_3}
				\end{subfigure} 

				\tikzsetnextfilename{D31_vs_D11_GMA_0100_Er_0_nu_4}
				\begin{subfigure}[t]{0.45\textwidth}	 		
					\includegraphics{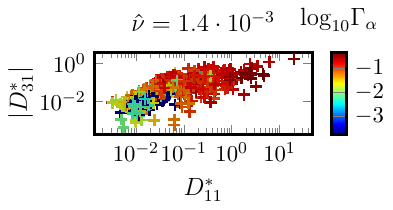}
					\caption{}
					\label{subfig:D31_vs_D11_GMA_0100_Er_0_nu_4}
				\end{subfigure}    
				\tikzsetnextfilename{D31_vs_GMA_0100_Er_0_nu_4}
				\begin{subfigure}[t]{0.45\textwidth}	 		
					\includegraphics{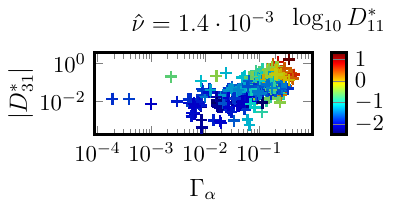} 
					\caption{}
					\label{subfig:D31_vs_GMA_0100_Er_0_nu_4}
				\end{subfigure} 
				
				\caption{Relation of the radial transport $D_{11}^*$ and bootstrap current $D_{31}^*$ coefficients with $\GammaAlpha$ for several collisionalities $\hat{\nu}$ (in $\text{m}^{-1}$) and $\widehat{E}_r=0$.} 
				\label{fig:Correlation_GMA_nu_scan_Er_0}
			\end{figure*} 
			\begin{figure*}[h]
				\centering
				%
				
				%
				\tikzsetnextfilename{D31_vs_D11_GMA_0100_Er_1e-3_nu_3}
				\begin{subfigure}[t]{0.45\textwidth}	 		
					\includegraphics{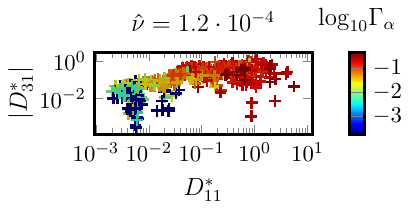} 
					\caption{}
					\label{subfig:D31_vs_D11_GMA_0100_Er_1e-3_nu_3}
				\end{subfigure}    
				\tikzsetnextfilename{D31_vs_GMA_0100_Er_1e-3_nu_3}
				\begin{subfigure}[t]{0.45\textwidth}	 		
					\includegraphics{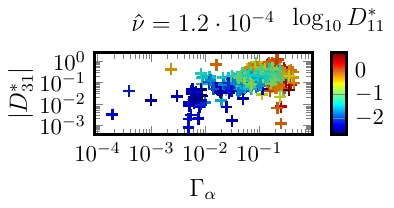} 
					\caption{}
					\label{subfig:D31_vs_GMA_0100_Er_1e-3_nu_3}
				\end{subfigure} 

				\tikzsetnextfilename{D31_vs_D11_GMA_0100_Er_1e-3_nu_4}
				\begin{subfigure}[t]{0.45\textwidth}	 		
					\includegraphics{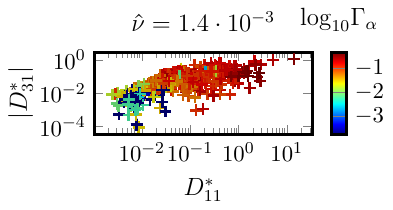} 
					\caption{}
					\label{subfig:D31_vs_D11_GMA_0100_Er_1e-3_nu_4}
				\end{subfigure}    
				\tikzsetnextfilename{D31_vs_GMA_0100_Er_1e-3_nu_4}
				\begin{subfigure}[t]{0.45\textwidth}	 		
					\includegraphics{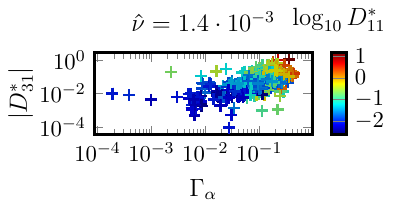} 
					\caption{}
					\label{subfig:D31_vs_GMA_0100_Er_1e-3_nu_4}
				\end{subfigure} 
				
				\caption{Relation of the radial transport $D_{11}^*$ and bootstrap current $D_{31}^*$ coefficients with $\GammaAlpha$ for several collisionalities $\hat{\nu}$ (in $\text{m}^{-1}$) and $\widehat{E}_r\ne0$.} 
				\label{fig:Correlation_GMA_nu_scan_Er_1e-3}
			\end{figure*}

			\endgroup
			
			\chapter{Equivalence between $\sigma^2(B(\theta,0))$ and $\sigma^2(\Bmax)$ for the optimization campaign}\label{sec:Appendix_VBM}
			\begingroup
			\captionsetup[sub]{skip=-1.75pt,aboveskip=-15pt, belowskip=5pt, margin=90pt}
			In this appendix we illustrate the equivalency of the proxies $\sigma^2(B(\theta,0))$ and $\sigma^2(\Bmax)$ for the optimization campaign considered in section \ref{sec:Correlations}. For this we plot the results of the evaluation against $\sigma^2(\Bmax)$. By comparing figures \ref{subfig:D31_vs_D11_KN_VBM_0100_Er_0}-\ref{subfig:D31_vs_KN_VBM_0100_Er_1e-3_restricted}, respectively, with \ref{subfig:D31_vs_D11_KN_VB0_0100_Er_0}-\ref{subfig:D31_vs_KN_VB0_0100_Er_1e-3_restricted} we can see that both the distribution of points and the colour pattern are quite similar. Thus, the conclusions that were extracted for $\sigma^2(B(\theta,0))$ in section \ref{sec:Correlations} are applicable to $\sigma^2(\Bmax)$.
			\begin{figure}[h]		 
				\tikzsetnextfilename{D31_vs_D11_KN_VBM_0100_Er_0}	
				\begin{subfigure}[t]{0.45\textwidth}	 		
					\includegraphics{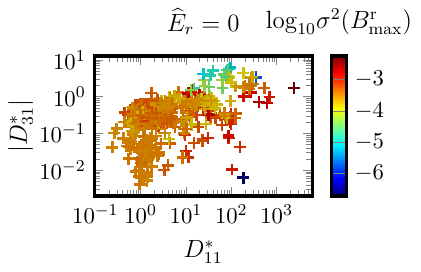}	
					{}	 		
					\caption{}
					\label{subfig:D31_vs_D11_KN_VBM_0100_Er_0}
				\end{subfigure} 
				\tikzsetnextfilename{D31_vs_D11_KN_VBM_0100_Er_1e-3}	
				\begin{subfigure}[t]{0.45\textwidth}	 		
					\includegraphics{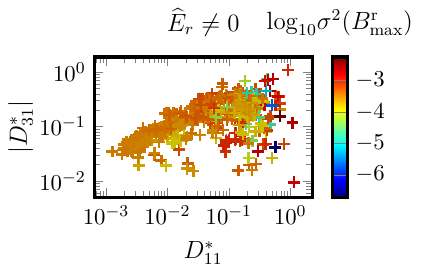}
					{}	 			
					\caption{}
					\label{subfig:D31_vs_D11_KN_VBM_0100_Er_1e-3}
				\end{subfigure} 

				\tikzsetnextfilename{D31_vs_KN_VBM_0100_Er_0}	
				\begin{subfigure}[t]{0.45\textwidth}	 		
					\includegraphics{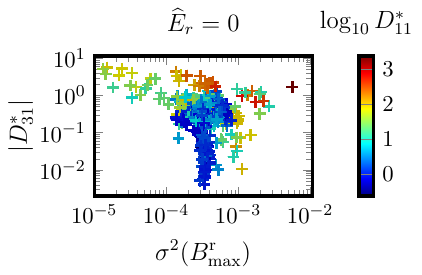}
					{}	 			
					\caption{}
					\label{subfig:D31_vs_KN_VBM_0100_Er_0}
				\end{subfigure} 
				\tikzsetnextfilename{D31_vs_KN_VBM_0100_Er_1e-3}	
				\begin{subfigure}[t]{0.45\textwidth}	 		
					\includegraphics{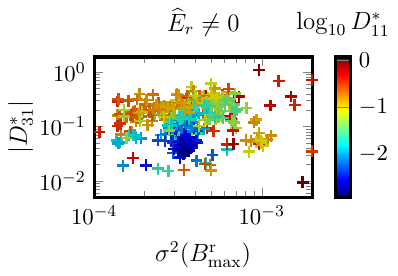}
					{}
					\caption{}
					\label{subfig:D31_vs_KN_VBM_0100_Er_1e-3}
				\end{subfigure} 

				\tikzsetnextfilename{D31_vs_KN_VBM_0100_Er_0_restricted}	
				\begin{subfigure}[t]{0.45\textwidth}
					\includegraphics{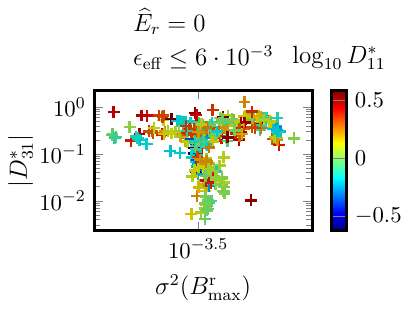}	
					{}	 		
					\caption{}
					\label{subfig:D31_vs_KN_VBM_0100_Er_0_restricted}
				\end{subfigure} 
				\tikzsetnextfilename{D31_vs_KN_VBM_0100_Er_1e-3_restricted}	
				\begin{subfigure}[t]{0.45\textwidth}	 		
					\includegraphics{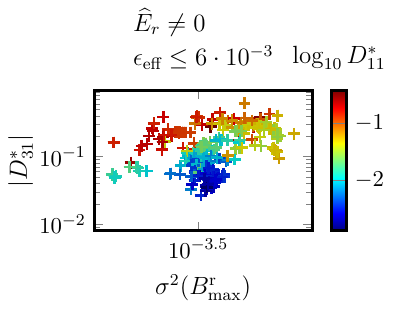}
					{}	
					\caption{}
					\label{subfig:D31_vs_KN_VBM_0100_Er_1e-3_restricted}
				\end{subfigure}

				\caption{Relation of the radial transport $D_{11}^*$ and bootstrap current $D_{31}^*$ coefficients with $\sigma^2(\Bmax^{\text{r}})$.}
				\label{fig:Correlation_VBM}
			\end{figure}

			\endgroup

			\chapter{Construction of the approximately pwO field using equations (\ref{eq:Exponential_pwO}) and (\ref{eq:iota_pwO})}\label{sec:Appendix_pwO_B}

			In this appendix, we will explain how the approximately pwO magnetic fields from section \ref{sec:pwO_QI} are constructed using (\ref{eq:Exponential_pwO}) and (\ref{eq:iota_pwO}). In section \ref{sec:pwO_QI} we mentioned that $\widetilde{B}$, as defined in (\ref{eq:Exponential_pwO}), can only define an exactly pwO field for $w_\theta\le \pi - |t_\zeta|w_\zeta$. Note that the inadequacy of $\widetilde{B}$ for $w_\theta> \pi -|t_\zeta|w_\zeta$ is inherited from the inadequacy of the function 
			\begin{align}
				\eta(\theta,\zeta)
				& := 
				\exp(
				- 
				\left(
				\frac{\theta-\theta_{\text{c}} - t_\zeta\left(\zeta-\zeta_{\text{c}}\right)}{w_\theta}
				\right)^{2p} 
				)
				\exp(
				-
				\left(
				\frac{\zeta-\zeta_{\text{c}}}{w_\zeta}
				\right)^{2p} 
				),
			\end{align}
			in this same region of the parameter space. 
			
			We can circumvent this problem by defining the following auxiliary functions
			\begin{align}
				\eta_k(\theta,\zeta) : = \eta(\theta + 2k\pi,\zeta)
			\end{align}
			for $k\in\mathbb{Z}$, 
			\begin{align}
				\eta^{\text{s}} : = 
				\sum_{k=-1}^{1}
				\eta_k
				, 
			\end{align}
			\begin{align}
				\eta^{\text{m}} : = 
				\max_{k\in\{-1,0,1\}}  	
				\{\eta_k\},
			\end{align}
			and 
			\begin{align}
				\eta^{H} : = 
				H
				\left(
				\eta^{\text{s}} -\Bmax
				\right)
				\eta^{\text{m}} 
				+
				H
				\left(
				\Bmax - \eta^{\text{s}} 
				\right)
				\eta^{\text{s}}
			\end{align}
			where $H(x)=1$ for $x\ge0$ and $H(x)=0$ otherwise. Note that $\eta^{\text{s}}$, $\eta^{\text{m}}$ and $\eta^H$ are equal in the limit $p\rightarrow\infty$ for all $w_\theta\le\pi$. In this limit we could define the magnetic field strength of an exactly pwO for $w_\theta > \pi - |t_\zeta| w_\zeta$ using e.g. $\eta^{\text{s}}$ as $B=\Bmin + (\Bmax-\Bmin)\eta^{\text{s}}$. 
			
			For finite $p$, $\eta^{\text{s}}$, $\eta^{\text{m}}$ and $\eta^H$ are different and we will need to choose which one use at each point of the $(\theta,\zeta)$ plane for different values of $w_\theta$. In addition, $w_\theta$ must be defined for values greater than $\pi$ (recall that for finite $p$ the isoline of $\Bmin$ does not close poloidally at $w_\theta=\pi$). For $w_\theta > \pi$, $\eta^{\text{m}}$ is not differentiable and $\eta^{\text{s}}$ can be larger than 1 at some points. On the other hand, the function $\eta^H$ is always smaller or equal than 1 at the expense of not being differentiable at a few points. Using these functions we can define 
			\begin{align}
				\eta_{\text{pwO}} = 
				\begin{dcases}
					\eta^{\text{m}}, & w_\theta < \pi,
					\\
					\eta^{\text{s}}, & w_\theta = \pi,
					\\
					\eta^{H}, & w_\theta > \pi,
				\end{dcases}  	
			\end{align}
			and 
			\begin{align}
				\BpwO = 
				\Bmin 
				+ 
				(\Bmax-\Bmin)
				\eta_{\text{pwO}}.
			\end{align}
			
			\begingroup 
			\captionsetup[sub]{skip=-1.75pt, margin=40pt}
			\begin{figure}[h]
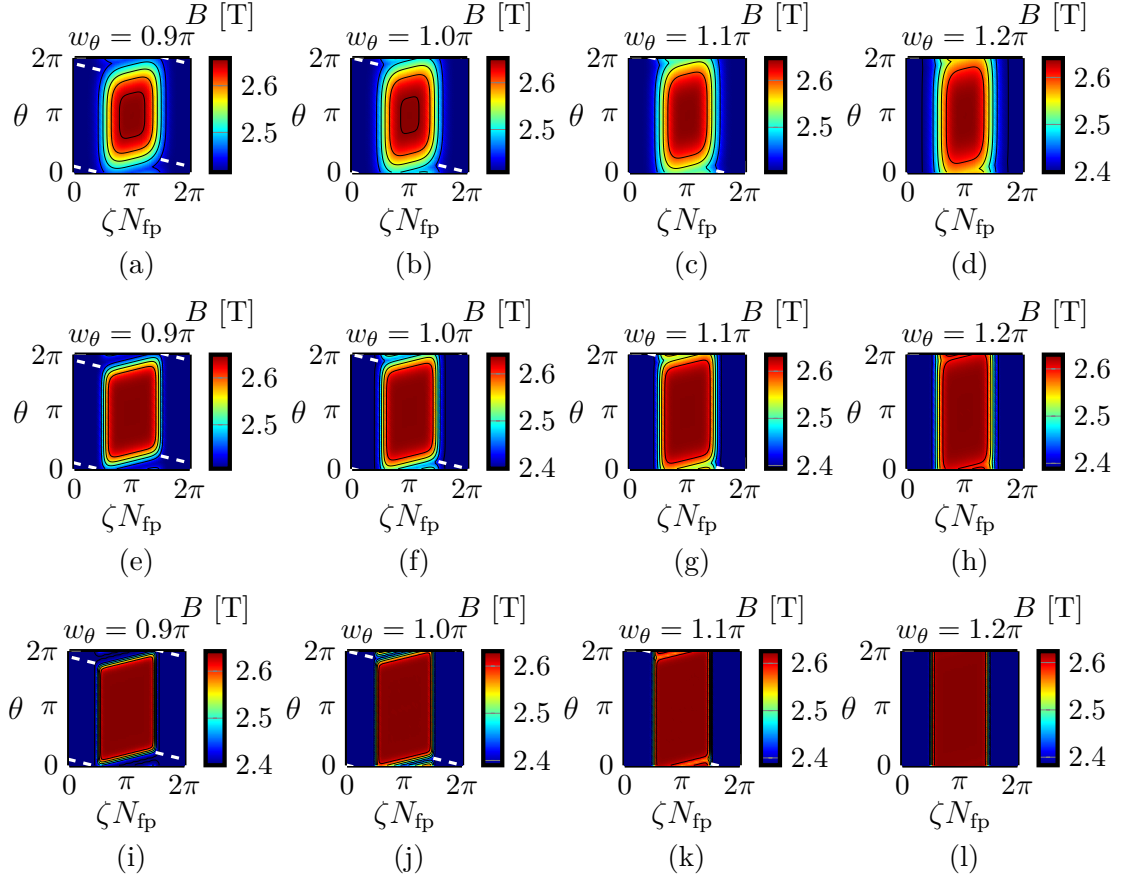
  
				\centering
				\foreach \p in {4, 8, 20}
				{ 
					\foreach \w in {0.9, 1.0, 1.1, 1.2}
					{%
						\tikzsetnextfilename{Appendix_pwQI_B_\w_\p} 
						\includepwOMagneticFieldAppendix{\p}{\w}%
					} 			
				} 
				\caption{Magnetic field strength $B$ for the parameter scan in pwO configuration space. $p=2$ (top row), $p=4$ (middle row) and $p=10$ (bottom row).}
				\label{fig:Appendix_Magnetic_field_strength_pwO_QI}
			\end{figure}
			\endgroup
			Finally, we approximate $\BpwO$ using a stellarator-symmetric Fourier series
			\begin{align}
				B = \sum_{m,n} B_{mn}\cos(m\theta + n\Nfp \zeta),
				\label{eq:Magnetic_field_strength_approximate_pwO}
			\end{align}
			where $ B_{mn}$ are the discrete Fourier modes of $\BpwO$. Hence, if we fix the parameters $\{\theta_{\text{c}},\zeta_{\text{c}},w_\theta,w_\zeta,t_\zeta\}$ that define $\eta$ and, additionally, $\Bmin$ and $\Bmax$, we can define the magnetic field strength $B$ using (\ref{eq:Magnetic_field_strength_approximate_pwO}) and $\iota$ using (\ref{eq:iota_pwO}) for each pair $(p,w_\theta)$. It is important to remark that, in the limit $p\rightarrow\infty$, representation (\ref{eq:Magnetic_field_strength_approximate_pwO}) will suffer from the Gibbs phenomenon due to the discontinuity at the perimeter of the parallelogram (where $\BpwO$ abruptly changes from $\BpwO=\Bmin$ to $\BpwO=\Bmax$). For finite $p$, Gibbs phenomenon does not appear around the parallelogram, but as $\BpwO$ still changes in a very short length scale, the modes $B_{mn}$ will have very large mode numbers $m$ and $n$ making the Fourier spectra extremely broad (in comparison with standard stellarator configurations). Finally we nuance that, for finite $p$, Gibbs phenomenon does appear as $\BpwO$ is not periodic in $\theta$ nor $\zeta$. However, this form of Gibbs phenomenon is benign. As the exponential $\eta_{\text{pwO}}$ is not periodic we can find values of $\zeta$ and $\theta$ where $\BpwO(0,\zeta)\ne\BpwO(2\pi,\zeta)$ and $\BpwO(\theta,0)\ne\BpwO(\theta,2\pi/\Nfp)$ respectively. However, due to the attenuation produced by the exponential, the differences $\BpwO(0,\zeta)-\BpwO(2\pi,\zeta)$ and $\BpwO(\theta,0)-\BpwO(\theta,2\pi/\Nfp)$ are of the order of the round-off error and have no significant impact on the modes $B_{mn}$.
			
			The parameters required for defining $\eta$ have been selected so that the magnetic configuration resembles that of Wendelstein 7-X KJM at $s=0.2$. For each pair $(p,w_\theta)$, the values of $\Bmax$ and $\Bmin$ are selected so that the discrete Fourier mode $B_{00}$ of $B$ matches that of the KJM configuration (however $\Bmax$ and $\Bmin$ vary very little between fields). This Fourier mode is also the reference value of $B$ on the flux surface, which we denote by $B_0$. In table \ref{tab:pwO_parameters} the values of the remaining parameters that define $\eta$ for each pair $(w_\theta,p)$ are shown. Note that $\theta_{\text{c}}$ and $\zeta_{\text{c}}$ are selected so that $\eta$ (and therefore $B$) satisfies stellarator-symmetry. Also note that fixing $t_\zeta$ and $w_\zeta \Nfp$ also determines $\iota=-t_\zeta$ via constraint (\ref{eq:iota_pwO}). 
			
			In order to compute the monoenergetic coefficients $\Dij{ij}$ we also need to specify $\{\Nfp,B_\theta,B_\zeta\}$ where we recall that $B_\theta$ and $B_\zeta$ are the covariant components of $\vb*{B}$ in Boozer coordinates. In addition, for computing their normalized versions $D_{ij}^*$ we need to specify the minor radius $r_{\text{lcfs}}$ and major radius $R$ along with the radial derivative of the toroidal flux (divided by $2\pi$) $\dv*{\psi}{r}$. The quantities $\dv*{\psi}{r}$, $B_\theta$ and $B_\zeta$ are those of Wendelstein 7-X KJM at $s=0.2$. In table \ref{tab:pwO_parameters_KJM}, the remaining parameters required to compute the normalized monoenergetic coefficients $D_{ij}^*$ are listed. The minor and major radius are approximated employing, respectively, estimates $r_{\text{lcfs,lar}}$ and $R_{\text{lar}}$, which are valid for a large aspect ratio stellarator 
			\begin{align}
				r_{\text{lcfs,lar}} & :=  \frac{ \left| \dv*{\psi}{r} \right|}{B_0} , 
				\\
				R_{\text{lar}} & :=  \frac{ \left|B_\zeta \right|}{B_0 }.
			\end{align}
			
			\begin{table}[]
				\centering
				\begin{tabular}{@{}ccccc@{}}
					\toprule
					$\Nfp$ &    $w_\zeta \Nfp$ & $t_\zeta$    & $\theta_{\text{c}}$ & $\zeta_{\text{c}}$ \\ \midrule
					5&  $\pi/2$ & $1.242$ & $\pi$      & $\pi/N_{\text{fp}}$     \\ \bottomrule
				\end{tabular}
				\caption{Parameters selected for defining $\widetilde{B}$ . }
				\label{tab:pwO_parameters}
			\end{table}

			\begin{table}[]
				\centering
				\begin{tabular}{@{}cccccc@{}}
					\toprule
					$B_0$ & $\dv*{\psi}{r}$   & $B_\theta$ & $B_\zeta$ & $r_{\text{lcfs,lar}}$ & $R_{\text{lar}}$\\ \midrule
					2.5003 &   0.5132   &  0    &   $-14.4$    & 0.205 & 5.76 \\ \bottomrule
				\end{tabular}
				\caption{Parameters that define the rescaling of the pwO magnetic field. $B_0$ is given in T, $\dv*{\psi}{r}$, $B_\theta$ and $B_\zeta$ are given in $\text{T}\cdot\text{m}$. The minor $r_{\text{lcfs,lar}}$ and major radius $R_{\text{lar}}$ are given in m.}
				\label{tab:pwO_parameters_KJM}
			\end{table}

			In figure \ref{fig:Appendix_Magnetic_field_strength_pwO_QI}, the magnetic field strength $B$ is represented for $p\in\{2,5,10\}$ for the values $w_\theta/\pi\in\{0.9,1.0,1.1,1.2\}$, which are those of the transition from pwO to QI. The effect of increasing $p$ can be observed by looking the columns of figure \ref{fig:Appendix_Magnetic_field_strength_pwO_QI} from the top row ($p=2$) to the bottom row ($p=10$). As was mentioned in section \ref{sec:pwO_QI}, we can verify that increasing $p$ compresses the isolines between $\Bmin$ and $\Bmax$ and thus, the gradient of $B$ on the flux surface is maximum in the surroundings of the perimeter of the parallelogram. The effect of increasing $w_\theta$ can be observed in figure \ref{fig:Appendix_Magnetic_field_strength_pwO_QI}, inspecting each row from the leftmost column ($w_\theta=0.9\pi$) to the rightmost one ($w_\theta=1.2\pi$). When $w_\theta<\pi$ the parallelogram fits in a single poloidal period. When $w_\theta$ is increased beyond $\pi$, the isolines of $B$ begin to close poloidally as expected. For $w_\theta \sim \pi$, we can see on figures \ref{subfig:Appendix_pwOMagneticField_pow_4_wa1.1pi}, \ref{subfig:Appendix_pwOMagneticField_pow_4_wa1.2pi}, \ref{subfig:Appendix_pwOMagneticField_pow_8_wa1.1pi}, \ref{subfig:Appendix_pwOMagneticField_pow_8_wa1.2pi}, \ref{subfig:Appendix_pwOMagneticField_pow_20_wa1.1pi} and \ref{subfig:Appendix_pwOMagneticField_pow_20_wa1.2pi} that the growth of the parallelogram with $w_\theta$ is periodic in the interval $\theta\in[0,2\pi]$. Thus, in the limit $w_\theta\rightarrow\infty$ (even for finite $p$), all isolines close poloidally and the magnetic field becomes quasi-poloidally symmetric. A particular case of quasi-poloidal symmetry with discontinuous $B$ can be attained if, for sufficiently large $w_\theta$, we take the limit $p\rightarrow\infty$. An approximation to this type of quasi-poloidal symmetry is shown in figure \ref{subfig:Appendix_pwOMagneticField_pow_20_wa1.2pi}, consisting of a central poloidally closed \qmarks{stripe} of width $w_\zeta$ where $B \approx\Bmax$ and on the rest of the flux surface $B\approx \Bmin$. Thus, as required, this scan permits to approach quasi-isodynamicity from pwO in a controlled manner by increasing $w_\theta$ and/or $p$.

			\pagestyle{empty}
			\bibliographystyle{unsrturl}
			\bibliography{refs}{}
\end{document}